\documentclass[twoside,11pt]{report}
\usepackage{amsfonts,amssymb,amsmath}
\usepackage{amsthm,multibox,array}
\usepackage{pstricks}
\usepackage{latexsym}
\usepackage{graphicx}
\usepackage{epic}
\usepackage{hyperref}
\usepackage{epsfig}
\usepackage{color}

\def\a{\alpha}
\def\b{\beta}
\def\g{\gamma}
\def\d{\delta}
\def\e{\eta}
\def\h{\eta}
\def\l{\lambda}
\def\G{\Gamma}
\def\D{\Delta}
\def\m{\mu}
\def\n{\nu}
\def\r{\rho}
\def\o{\omega}

\def\s{\sigma}
\def\S{\Sigma}

\def\th{\theta}
\def\p{\pi}
\def\e{\varepsilon}

\def\mf{\mathfrak}
\def\beq{\begin{eqnarray}}
\def\eeq{\end{eqnarray}}
\def\nn{\nonumber}
\def\ft{\footnotesize}
\newcommand{\be}{\begin{equation}}
\newcommand{\ee}{\end{equation}}
\newcommand{\bea}{\begin{eqnarray}}
\newcommand{\eea}{\end{eqnarray}}

\def\cb{{\cal B}}

\def\ce{{\cal E}}
\def\cF{{\cal F}}
\def\cf{{\cal F}}
\def\cG{{\cal G}}
\def\ch{{\cal H}}

\def\ck{{\cal K}}
\def\cl{{\cal L}}
\def\cm{{\cal M}}
\def\cn{{\cal N}}

\def\cP{{\cal P}}

\def\cQ{{\cal Q}}

\def\cv{{\cal V}}
\def\cV{{\cal V}}

\def\cH{{\cal H}}

\def\cA{{\cal A}}
\def\cN{{\cal N}}
\def\cK{{\cal K}}
\def\cL{{\cal L}}

\def\td{\tilde}
\newcommand{\bpsi}{\overline{\psi}}
\newcommand{\hpsi}{\hat{\psi}}
\newcommand{\bhpsi}{\overline{\hat{\psi}}}

\newcommand{\he}{\hat{e}}
\newcommand{\hgamma}{\hat{\gamma}}

\def\hg{\hat{\gamma}}
\def\hm{\hat{\mu}}
\def\hn{\hat{\nu}}

\def\hs{\hat{s}}
\def\hb{\hat{b}}
\newcommand{\te}{\tilde{e}}
\newcommand{\tf}{\tilde{f}}
\newcommand{\tk}{\tilde{k}}

\def\oneone{\rlap 1\mkern4mu{\rm l}}
\def\fft#1#2{{#1 \over #2}}
\def\sst#1{{\scriptscriptstyle #1}}
\def\Dm{{{D_{\sst{max}}}}}
\newcommand{\aaa}{{\sst{(a)} }}
\newcommand{\bb}{{\sst{(b)}}}
\newcommand{\ccc}{{\sst{(c)}}}
\newcommand{\dd}{{\sst{(d)}}}
\newcommand{\mm}{\sst{(\m)}}
\newcommand{\nnn}{\sst{(\n)}}
\newcommand{\rrr}{\sst{(\rho)}}
\newcommand{\ddd}{\sst{(\d)}}
\newcommand{\sss}{\sst{(\s)}}
\newcommand{\ii}{{\sst{(i)}}}
\newcommand{\jj}{{\sst{(j)}}}
\newcommand{\kk}{{\sst{(k)}}}
\newcommand{\0}{{\sst{(0)}}}
\newcommand{\1}{{\sst{(1)}}}
\newcommand{\2}{{\sst{(2)}}}
\newcommand{\3}{{\sst{(3)}}}
\newcommand{\4}{{\sst{(4)}}}
\newcommand{\5}{{\sst{(5)}}}

\newcommand{\w}{\wedge}
\newcommand{\Sh}[1]{#1\hskip-9.5pt \diagup}
\def\ZZ{{\mathbb{Z}}}
\def\CC{{\mathbb{C}}}
\def\RR{{\mathbb{R}}}

\newsavebox{\uuunit}
\sbox{\uuunit}
    {\setlength{\unitlength}{0.825em}
     \begin{picture}(0.6,0.7)
        \thinlines
        \put(0,0){\line(1,0){0.5}}
        \put(0.15,0){\line(0,1){0.7}}
        \put(0.35,0){\line(0,1){0.8}}
       \multiput(0.3,0.8)(-0.04,-0.02){12}{\rule{0.5pt}{0.5pt}}
     \end {picture}}
\newcommand {\unity}{\mathord{\!\usebox{\uuunit}}}
\newcommand{\ud}{\mathrm{d}}

\newcommand{\ie}{\emph{i.e. }}

\begin{document}
\pagestyle{empty}
\bibliographystyle{plain}
\begin{titlepage}
\begin{center}
{\small Universit\'e Libre de Bruxelles \\
Facult\'e des Sciences \\
Service de Physique Th\'eorique et Math\'ematique\\}
\vspace{5cm}

{\Huge \bf
Kac-Moody Algebras in M-theory}

\vspace{4.5cm}

{\it Sophie de Buyl} \\
{\small Aspirant F.N.R.S.} \\

\vspace{7.5cm}
{\small Ann\'ee acad\'emique 2005--2006}

\end{center}
\end{titlepage}
%%%%%%%%%%% Dos de la page de garde
\pagestyle{empty}
\mbox{}
\newpage
\mbox{}
\vfill
\newpage
\mbox{}
\vfill
\newpage
%%%%%%%%%%%% 
\mbox{}

\vfill

\setcounter{page}{1}

\emph{Mes   remerciements s'adressent tout d'abord \`a Marc Henneaux, mon directeur de th\`ese. Quand j'\'etais petite, ses cours sur les repr\'esentations des groupes finis et sur la Relativit\'e G\'en\'erale m'ont fortement s\'eduite. Je garde un excellent souvenir de ses expos\'es d'une clart\'e remarquable. J'ai donc os\'e frapper \`a sa porte et c'est avec joie que j'ai entrepris  un m\'emoire, puis la pr\'esente th\`ese dans le cadre de la gravitation d'Einstein,  non plus sur les groupes finis, mais sur des structures beaucoup plus grandes: les alg\`ebres de Kac--Moody infini dimensionnelles. 
 Il m'a propos\'e des probl\`emes stimulants et bien pos\'es me permettant d'entrer rapidement dans des sujets pointus.  J'ai beaucoup appr\'eci\'e sa mani\`ere positive de voir les choses, ainsi que son intuition qui  ---\`a ma connaissance--- n'a jamais \'et\'e mise en d\'efaut. J'ai \'et\'e  impressionn\'ee par son efficacit\'e: 
 d\`es qu'il se d\'ecide \`a terminer un projet, c'est comme si c'\'etait fait!
 Et malgr\'e sa pr\'esence euh... disons fort al\'eatoire, il a \'et\'e pour moi un promoteur 
exemplaire. Maintenant que  je suis grande, j'esp\`ere \^etre \`a la hauteur de  ses attentes  et  pouvoir voler de mes propres ailes.}

\emph{ Je remercie de tout coeur 
Christiane Schomblond. Elle a guid\'e mes premiers pas, parfois trop hatifs, dans le monde de la recherche scientifique et a \'et\'e une source de r\'eponses  indispensable tout au long de l'\'elaboration de cette th\`ese. Je lui en suis fort reconnaissante et suis tr\`es sensible \`a sa grande disponibilit\'e. 
C'est avec grand plaisir que j'ai travaill\'e avec Laurent Houart. La porte de son bureau  toujours ouverte, et  \`a deux pas de la mienne, a contribu\'e au partage de son  enthousiasme (d\'ebordant certains jours
et ayant le don d'acc\'el\'erer le d\'ebit de ses propos) pour les very interesting Kac--Moody algebras. Je le remercie pour son attention face \`a mes angoisses diverses, dont celles de parler en anglais devant des monsieurs s\'erieux.}

\emph{Louis Paulot a \'et\'e un collaborateur hors pair pour affronter le monde sans piti\'e des fermions et les diagrammes de Satake, je lui suis reconnaissante pour les nombreuses choses qu'il m'a apprises.  Je remercie \'egalement Nassiba Tabti et Ga\"ia Pinardi pour des collaborations pass\'ees ainsi que Geoffrey Comp\`ere pour collaboration sur le feu...  Je remercie St\'ephane Detournay pour les mille et un projets qu'il m'a propos\'es et que je compte bien partager avec lui malgr\'e les emb\^uches sur notre chemin de post--doctorants.}

\emph{Sandrine Cnockaert occupe certainement une place particuli\`ere dans ces remerciements. Nous avons effectivement suivi  un parcours tr\`es proche (mais dans des \'etats de stress plus ou moins oppos\'es). Je la remercie pour plein de petites choses qui ont contribu\'e au bon d\'eroulement de cette th\`ese. Aujourd'hui nos voies semblent s'\'eloigner mais je suis s\^ure qu'elles garderont des points communs.}

\emph{J'ai eu la chance de rencontrer diverses personnes qui m'ont accompagn\'ee le long de mon parcours scientifique.  Je pense ici \`a Philippe L\'eonard qui a \'eveill\'e mon int\'eret pour la physique. Puis \`a Kim Claes, Laura Lopez--Honorez, Sandrine Cnockaert,  Claire No\"el, Georges Champagne et Yannick Kerckx [par ordre de taille] qui ont contribu\'e \`a cr\'eer une atmosph\`ere stimulante pendant nos quatre ann\'ees d'\'etudes universitaires.
 J'ai eu la chance de faire partie d'un service fort dynamique qui semble grandir de plus en plus. J'en remercie tous les membres, les  secr\'etaires si efficaces, les habitu\'es de la salle caf\'e et les doctorants (et ex--doctorant) du couloir N du 6i\`eme \'etage et puis tous ceux que j'oublie. J'ai profit\'e de l'expr\'erience de mes ain\'es, Xavier Bekaert et Nicolas Boulanger, que je tiens beaucoup \`a remercier.
L'enthousiasme contagieux de Jarah Evslin, Daniel Persson, Carlo Maccafferi, Nazim Bouatta, Geoffrey Comp\`ere, Mauricio Leston, St\'ephane Detournay et Stanislav Kuperstein [l'interpretation de l'ordre de pr\'esentation est laiss\'ee au lecteur] lors des ``Weinberg lectures'' dont j'ai malheureusement manqu\'e nombre d'\'episodes  ainsi que le bruit qu'ils font en discutant de physique est extr\^ement motivant! Je leur souhaite \`a tous les meilleures d\'ecouvertes ... Pour diverses discussions, je remercie [dans un ordre al\'eatoire]  Philippe Spindel, Glenn Barnich, Fran\c{c}ois Englert, Thomas Fischbacher, Pierre Bieliavsky, Axel Kleinschmidt, Arjan Keurentjes, Riccardo Argurio et Nicolas Boulanger. Il me tient \`a coeur d'\'evoquer la  premi\`ere 
``Modave Summer School in Mathematical Physics'' et tous ses participants si  avides de savoir.}

\emph{Etant de nature plut\^ot anxieuse et ne ma\^itrisant pas vraiment  la langue de Shakespeare, j'ai beaucoup de remerciements \`a adresser concernant ma r\'edaction. Ma th\`ese a subi  d'innom-brables am\'eliorations suite aux lectures et re--lectures de 
Christiane Schomblond. Je lui suis fort reconnaissante pour sa patience, ses conseils avis\'es et son 
exigence face \`a mon style parfois invontairement peu clair... 
Je remercie vivement Marc Henneaux pour sa lecture tr\`es attentive et les pr\'ecisions apport\'ees. Daniel Persson m'a encourag\'ee, aid\'ee et a ponctu\'e diff\'erentes \'etapes de ma r\'edaction  de ``I think this is a good job!".\footnote{
Daniel Persson har st\"ott och hj\"alpt mig och godk\"ande alla avsnitt av min avhandling med ``I think this is a good job!".}
Geoffrey Comp\`ere a grandement contribu\'e \`a la clart\'e de cet expos\'e, je lui en suis fort reconnaissante. Je remercie \'egalement St\'ephane Detournay, mon fabricant de phrases pr\'ef\'er\'e, ainsi 
que Sandrine Cnockaert pour leur aide indispensable dans la phase finale de r\'edaction.  Je n'oublie pas non plus ici Mauricio Leston, Louis Paulot, Laurent Houart et Paola Aliani. Finalement, je remercie le cnrs pour son dictionnaire en ligne   $http://dico.isc.cnrs.fr/fr/index \_ tr.html$ dont j'ai us\'e et abus\'e, je l'esp\`ere \`a bon escient.}

\emph{Je remercie Thomas Hertog,  Laurent Houart,  Bernard Julia,  Bernhard M\"uhlherr, Christiane Schomblond, Philippe Spindel et Michel Tytgat pour avoir accept\'e de faire partie de mon 
jury.}

\emph{Pour les longues heures pass\'ees pendue au fil du t\'el\'ephone \`a g\'emir en leur assurant que non je n'y arriverais jamais, que  je n'\'etais un pauvre petit lapin, je remercie tout particuli\`erement ma maman, mon fr\`ere Pierre et Sophie Van den Broeck. Ils ont eu
l'oreille plus qu'attentive \`a mes nombreux doutes quant au bon fonctionnement de mes neurones...  Pour, lorsque j'habitais chez mes parents, avoir \'et\'e d'une  compr\'ehension totale et plus tard pour avoir continu\'e \`a satisfaire mes petits caprices, je remercie mon papa. Pour me rappeler que la physique peut avoir des applications pratiques et ne pas c\'eder \`a mes quatre volont\'es, je remercie mon fr\`ere Martin. Je pense ici aussi \`a Aur\'elie Feron, Fran\c{c}oise de Halleux et encore \`a Sophie Van den Broeck dont l'amiti\'e m'est particuli\`erement ch\`ere. A ma famille et mes amis: j'oublie pas  que j'vous aime! }

\emph{Je remercie beaucoup ma grand--m\`ere, Grand--Mamy,  pour son support  informatique! Pour sa disponibilit\'e et ses connaissances en linux, \LaTeX\ , mathematica, maple et bugs divers et vari\'es des nombreux ordinateurs que j'ai u(tili)s\'es, je remercie Pierre de Buyl. Glenn Barnich a aussi  \'et\'e d'une patience exemplaire lors des r\'einstallations de diff\'erents syst\`emes d'exploitation ainsi que pour divers probl\`emes quotidiens.
}

\emph{
La physique m\`ene \`a des choses  diverses et vari\'ees ainsi qu' \`a de nombreuses rencontres, je vous laisse deviner celle qui  m'a combl\'ee au-del\`a de toute esp\'erance.}
%%%%%%%%%%%%%
\newpage
\mbox{}
%%%%%%%%%%%%%%%%%%%%%%%%%%%%%%
\newpage
\cleardoublepage
\thispagestyle{empty}
\tableofcontents
\cleardoublepage
\addcontentsline{toc}{part}{Introduction}
\part*{Introduction}
\markboth{INTRODUCTION}{}
\pagestyle{myheadings} 
\cleardoublepage
%%%%%%%%%%intro%%%%%%%%%%%%%%%%%
%%%%%%%%%%%%%%%%%%%%%%%%%%%%%%
%%\include{intro}

\vspace*{8cm}

One of the main challenges of contemporary physics is the unification of the four fundamental interactions. On the one hand, the electromagnetic, weak and strong interactions have been unified in the general framework of quantum field theory, within the 
\emph{Standard Model}.  This theory provides a \emph{quantum }description of these interactions in agreement with special relativity but restricted to a regime where gravity can be neglected.  
Numerous predictions of the Standard Model have been verified with an impressive accuracy through experiments with particles colliding in accelerators. On the other hand, the  gravitational interaction has a very different status since, according to  Einstein's General Relativity which remains its current best description, gravity is encoded in the curvature of spacetime induced by the presence of matter. In this context, spacetime becomes dynamical, which is different to what happens in quantum field theory where particles propagate in a \emph{fixed} background.  General Relativity  successfully computed  the perihelion precession of Mercury, the deflection of light by the sun, etc... and  it is  even used in our everyday life through the {global positioning system}.
But it is a \emph{classical }theory, and cannot be applied to energy scales larger than the Planck energy. 
In some extreme situations such as those encountered in the center of black holes or at the origin of the universe, all interactions become simultaneously relevant. 
Unfortunately, direct attempts to express General Relativity
in quantum mechanical terms have led to a web of contradictions,
basically because the non--linear mathematics necessary
to describe the curvature of spacetime clashes with the delicate 
requirements of quantum mechanics \cite{Goroff:1985th}. 
This situation is clearly not satisfactory and a theory encompassing all fundamental interactions must be found. This hypothetic theory should reproduce the Standard Model and General Relativity in their 
respective domains of validity. 

So far, the most promising candidate for such a unification is string theory. 
 The basic idea of string theory is to replace the point particles by one--dimensional objects.  These 
one--dimensional objects can vibrate and the different vibration modes correspond to various particles. A very exciting feature of string theories is that their particle spectra contain a graviton (the \emph{hypothetic} particle mediating the quantum gravitational interaction) and may contain particles mediating the other three forces.  String theory 
predicts additional spatial dimensions, and  supersymmetry --- mixing bosons and fermions --- is a key ingredient.  
A drawback of string theories is that a
satisfactory formulation based on first principles is still missing. Another drawback at first sight is that 
there exist five coherent supersymmetric string theories: type I, type IIA, type IIB, heterotic $E_8 \times E_8$  and heterotic $SO(32)$. One may wonder how to choose the one which is the most fundamental?  Fortunately,  relations between these theories, named \emph{dualities}, have been discovered during the nineties. There are essentially two types of dualities: the 
\emph{T--dualities }and the \emph{S--dualities}.

In order to understand how T--duality acts, some knowledge of dimensional reduction is necessary.  If one of the space--like dimensions is assumed to take its values on a circle of radius $R$,  a closed string can wrap  $\o$ times around this circle. Winding requires energy because the string must be stretched against its tension and the amount of energy is $\o R / \ell_{st}^2$, where $\ell_{st}$ is the string length. This energy is quantised since $\o$ is an integer. 
 A string travelling around this circle has a quantised momentum around the circle; indeed,  its momentum is proportional to the inverse of  its wavelength $\l= 2\pi R / n$, where $n$ is an integer. The momentum around the circle --- and the contribution to its energy --- goes like $n/R$. 
 At large $R$ there are many more momentum states than winding states (for a given maximum energy), and conversely at small $R$. 
A theory with large $R$ and a theory with small $R$ can be equivalent, if the role of momentum modes in the first is played by the winding modes in the second, and vice versa. Such a T--duality relates type IIA superstring theory to type IIB superstring theory. This means that if we take type IIA and type IIB theories and compactify them on a circles, then switching the momentum and winding modes as well as inverting the radii, changes one theory into the other. The same is also true for the two heterotic theories.

In contrast with standard quantum field theories, the coupling constant of the string  depends on a spacetime field called the dilaton. Replacing the dilaton field by minus itself exchanges a very large coupling constant for a very small one. An \emph{S--duality }is a relation exchanging the strong coupling of one theory with the weak coupling of another one.
 If two string theories are related by \emph{S--duality}, then the first theory with a strong coupling constant is the same as the other theory with a weak coupling constant. 
Superstring theories related by S--duality are: type I superstring theory with heterotic $SO(32)$ superstring theory, and type IIB theory with itself.
 
  It turns out that T--duality of compactified string theories and S--duality do not commute in general but generate the discrete \emph{U--duality group} \cite{Hull:1994ys}.  In particular, when compactifying type II string theories on a $d$--torus, the conjectured U--duality group is the discrete group  $E_{d+1(d+1)}(\ZZ)$.

The discovery of the various dualities between the five $D=10$ superstring theories  suggested that each of them could be a part of a more fundamental  theory that has been named \emph{M--theory} \cite{Witten:1995ex}.   Much of the knowledge about M--theory comes from the  study of the $D=10$ maximal supergravity theories which encode the complete low energy behaviour of all the known superstring theories. Moreover,
the $D=11$ supergravity \cite{Cremmer:1978km} has been argued to be the low energy limit of M--theory \cite{Witten:1995ex}.   
This  theory has a special status: it is the highest dimensional supergravity theory and it is  unique in $D=11$ \cite{Nahm:1977tg}.

\begin{figure}[h]
\caption{\label{e8+++def} {\small The Dynkin diagram of $\mf{e}_8^{+++}$ }}
\begin{center}
\scalebox{.5}{
\begin{picture}(180,60)
%nom des racines
\put(-95,-5){1}
\put(-55,-5){$2$} \put(-15,-5){$3$}
%\put(85,-5){$\alpha_1$}
% \put(125,-5){$\alpha_2$}
%  \put(165,-5){$\alpha_3$} \put(205,-5){$\alpha_4$}
%  \put(245,-5){$\alpha_5$}   \put(285,-5){$\alpha_6$}
%  \put(325,-5){$\alpha_7$}
%  \put(260,45){$\alpha_8$}
%10 vertex + lignes simples
\thicklines  \multiput(-90,10)(40,0){10}{\circle{10}}
 \multiput(-85,10)(40,0){9}{\line(1,0){30}}
%1 vertex du dessus
\put(190,50){\circle{10}} \put(190,15){\line(0,1){30}}
\end{picture}
} 
\end{center}
\end{figure}
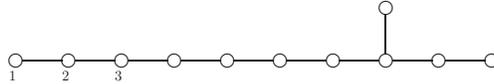

Although very little is known about M--theory, it is believed that M--theory should be a background independent theory that would give rise to spacetime as a secondary concept. It is expect that M--theory should also possess a rich symmetry structure. Since symmetries have always been a powerful guide in the formulation of fundamental interactions, the knowledge of the symmetry group of 
M--theory would certainly constitute an important step towards an understanding of its underlying structure. The following statement is therefore very intriguing: 
the infinite--dimensional Kac--Moody group
$E_{11(11)}$ has been conjectured to be the  symmetry group of M--theory \cite{West:2001as,Lambert:2001gk}.\footnote{Here, Kac--Moody groups are understood to be the formal exponentiation of their algebras. \label{kmgroup}} The corresponding Kac--Moody algebra $\mf{e}_{11(11)}$  is the very extension of the finite 
dimensional Lie algebra $\mf{e}_{8(8)}$. [ More precisely $\mf{e}_{8(8)}^{+++}$ is the split real form of the complex Kac--Moody algebra $\mf{e}_8^{+++}$ defined by  the Dynkin diagram depicted
in Figure \ref{e8+++def},  obtained from those of $\mf{e}_8$ by adding three
nodes to its Dynkin diagram \cite{Gaberdiel:2002db}. 
 One first adds the affine node, labelled 3 in the
figure, then a second node labelled 2,  connected to it by a single line and
defining the overextended ${\mf{g}}^{++}$ algebra,   then  a third
one labelled 1, connected  by a single line to the overextended node.]

\noindent 
This conjecture is supported by the fact that evidence has been given according to which the bosonic sector of $D=11$ supergravity could be reformulated as a non--linear
realisation based on the Kac--Moody group $E_{11(11)}$ and the conformal group \cite{West:2000ga,West:2001as}.  Theories of 
 type IIA and IIB can also be reformulated as non--linear realisations based on $E_{11(11)}$ \cite{Schnakenburg:2001ya, West:2000ga}. 
The $E_{11(11)}$ group contains the Lorentz group which is evidently a symmetry 
of the perturbative theory but it also contains symmetries which are non--perturbative. 
The simplest example is provided by the case of IIB theory: the $E_{11(11)}$ symmetry contains its 
$SL(2,\RR)$ symmetry \cite{Schwarz:1983wa} which acts on the dilaton and therefore  mixes 
perturbative with ``non--perturbative'' phenomena.

The relevance of infinite--dimensional Kac--Moody algebras as symmetries of gravitational theories has already been pointed in \cite{Julia:1981wc,Julia:1982gx,Julia:1980gr}: 
it  has been conjectured  
 that the dimensional reduction of $D=11$ supergravity down to one dimension would lead to the
appearance of the Kac--Moody group 
$E_{8(8)}^{++}=E_{10(10)}$. In this context, $E_{10(10)}$ is called a \emph{hidden symmetry }of $D=11$ supergravity since it was thought to appear only \emph{after }dimensional reduction (and/or 
dualisations of certain fields). 
The emergence of \emph{hidden  symmetries} is not particular to $D=11$ supergravity: they also appear in other supergravities and in pure gravity. Their full implications  are
to a large extent still mysterious. 
Before concentrating on dimensional reduction of $D=11$ supergravity, let us  recall the first example of \emph{hidden symmetries} and comment about the $N=8$, $D=4$ supergravity. 

\noindent This first example comes from the study of solutions of the Einstein equations of pure gravity in $D=4$ admitting Killing vectors. 
The Ehlers $SL(2, \RR)$ group is a symmetry group acting on certain solutions possessing one Killing vector  \cite{Ehlers}. When combined with the Matzner--Misner group, it leads to 
an infinite--dimensional symmetry --- the Geroch group --- acting on solutions of Einstein's equations with two commuting Killing vectors (axisymmetric stationary solutions) \cite{Geroch:1970nt}. The Geroch group has been identified with the affine extension of $SL(2,\RR)$, namely the affine Kac--Moody group $SL(2,\RR)^+$
\cite{Julia:1981wc}. 
These 
results also provided a direct link with the integrability of these theories in 
the reduction to two dimensions, \ie the existence of Lax pairs for the 
corresponding equations of motion \cite{Dietz,Kramer,Breitenlohner:1986um,Maison:1978es,Belinsky:1971nt}.

\noindent Hidden symmetries have further been  discovered in various supergravities  \cite{Cremmer:1978ds,Cremmer:1979up}.  The maximal $N = 8$ supergravity in $D=4$  was shown to possess an  $E_{7(7)}$ symmetry; the $D>4$ theories which reproduce this theory upon dimensional reduction(s) \emph{\`a la Kaluza--Klein }also possess
hidden symmetries which are displayed in Table \ref{magic}. Non--maximal ($N<8$) supergravities in $D=4$ --- obtained by consistent truncations of the fields decreasing the number of supersymmetries of the   $N=8$ supergravity in $D=4$ --- and their higher dimensional parents again exhibit hidden symmetries. [The $D=3$ theories obtained from the $D=4$ ones also give rise to hidden symmetries.] All these symmetries enter very nicely in the so--called \emph{magic triangle} of Table \ref{magic}. For each $N$, the highest dimensional theory is said to possess hidden symmetries since it possesses symmetries revealed only \emph{after} dimensional reduction and suitable dualisations of fields. 
\begin{table}[h]
\caption{\label{magic} {\small Real magic triangle Cosets: $N$ is the number of supersymmetries in $D=4$, $D$ the spacetime dimension and a $+$ means that a theory exists but possesses no scalars. The remarkable property of the magic triangle is its symmetry --- up to the particular real form --- with respect to the diagonal (see \cite{Henry-Labordere:2002dk} for more details).}}
\begin{flushleft}
\scalebox{0.74}{
\begin{tabular}{|c||c|c|c|c|c|c|c|c|}
\hline
 &$N =8$&$N =6$&$N =5$&$N =4$&$N =3$&$N =2$&$N =1$ & $N =0$\\
\hline
{\small $D=11$} & $+$ \\
\cline{1-2}
{\small $D=10$} &$\RR$ \\
\cline{1-2} {\small$D=9$} & ${SL(2) \over SO(2)} \times \RR $  \\
\cline{1-2}{\small $D=8$ }& ${SL(3) \times SL(2) \over SO(3) \times SO(2)}$
 \\
\cline{1-2}{\small $D=7$ }& ${SL(5) \over SO(5)}$\\
\cline{1-3} {\small$D=6$} & ${SO(5,5) \over SO(5) \times SO(5)}$ & ${SO(5,1) \times SO(3) \over SO(5) \times SO(3)}$  \\
         \cline{1-3}
{\small$D=5$} & ${E_{6(6)} \over Usp(8)}$ & ${SU^\star(6) \over
Usp(6) }$   \\
       \hline
{\small$D=4$} & ${E_{7(7)} \over SU(8)}$ & ${SO^\star(12) \over
U(6) }$ & ${SU(5,1) \over U(5) }$ &
${SU(4) \times SU(1,1) \over SU(4) \times SO(2) }$ &
${U(3) \over U(3)}$ & ${U(2) \over U(2)}$ & ${U(1) \over U(1)}$
& +   \\
 \hline
{\small$D=3$} & ${E_{8(8)} \over Spin(16)/\ZZ_2}$ & ${E_{7(-5)} \over SO(12)
\times SO(3) }$ & ${E_{6(-14)} \over SO(10) \times SO(2)
}$ & ${SO(8,2)  \over SO(8)/\ZZ_2
 \times SO(2)}$ & ${SU(4,1)  \over SO(6) \times SO(2)}$
        & ${SU(2,1) \times SU(2) \over SO(4)\times SO(2)}$
        & ${SL(2) \times SO(2) \over SO(2) \times SO(2)}$
& ${SL(2) \over SO(2)}$ \\
\hline
\end{tabular}
}
\end{flushleft}
\end{table}

\noindent Let us now focus on the $D=11$ supergravity, which is of particular interest in the context of M--theory, 
and its successive dimensional reductions (for excellent lecture notes on the subject see \cite{reddim}). More precisely, we will consider the scalar sector of the reduced theories
since it can be shown that the symmetry of the scalar sector extends to the entire Lagrangian \cite{reddim}. Performing successive dimensional reductions on an circle, a torus, ... an $n$--torus increases the number of scalar fields. Indeed, from the $(D-1)$--dimensional point of view, the $D$--dimensional metric is interpreted as a metric, a 1--form called  \emph{graviphoton  }and a scalar field called \emph{dilaton} while a $D$--dimensional $p$--form potential yields a $p$--form potential and a $(p-1)$--form potential. 
The scalar fields appearing in successive dimensional reduction of the $D=11$ supergravity down to $D<11$, including those that arise from dualisation of forms, combine into  a non--linear realisation of a group $\cG$. More precisely, they parametrise a coset space $\cG / \cK$ where $\cK$ is the maximally compact subgroup of $\cG$, see Table \ref{hiddene11}. These results deserve some comments: 
\begin{quote}
-- When  a gravity theory coupled to forms is reduced on 
an $n$--torus, one expects that the reduced Lagrangian possesses a $GL(n,\RR)$ symmetry since this is the symmetry group of the $n$--torus and one assumes that the reduced fields do \emph{not }depend on the reduced dimensions. In the case at hand, this is what happens in $D=10$ and $D=9$.    $D=8$ is the first dimension in which  one gets  a scalar from the reduction of the $D=11$ 3--form. The symmetry in $D=8$ is an $SL(3,\RR) \times SL(2,\RR)$ symmetry instead of the expected $GL(3,\RR) = SL(3,\RR) \times \RR$ symmetry: this is because there is a congruence between the scalar field coming from the $D=11$ 3--form potential and the scalar fields arising from the reduction of the metric. The same kind of phenomenon happens in lower dimensions.\newline 

-- Three--dimensional spacetimes are special because there all
physical (bosonic) degrees of freedom can be converted into
scalars\footnote{Remember that a $p$--form potential in $D=d+1$ dimensions is dual to a $(d-p-1)$--form potential.}: the scalar fields arising from the dualisation of the 1--form potential contribute to the $E_{8(8)}$ symmetry in $D=3$. 
\end{quote}
Going down bellow three dimensions, one obtains Kac--Moody extensions of the exceptional $E$--series. The two--dimensional
symmetry $E_{9(9)}$ is, like the Geroch group, an affine symmetry \cite{Julia:1982gx,Breitenlohner:1986um,Nicolai:1987kz}.  In one--dimensional spacetimes,
the expected symmetry is the hyperbolic extension
$E_{10(10)}$ \cite{Julia:1981wc,Julia:1982gx,Julia:1980gr} and in  zero--dimensional spacetimes, one formally finds the Lorentzian
group $E_{11(11)}$. 
\begin{table}[h]
\caption{\label{hiddene11}  {\small The symmetry groups of the $D=11$ supergravity reduced on an $n$--torus are given in this table.  The scalar sector in $D= 11-n$ fits into a 
non--linear realisation based on the coset spaces $\cG/\cK$ where $\cK$ is the maximally compact subgroup of $\cG$.}}
\begin{center}
\scalebox{0.8}{
\begin{tabular}{|c|c|c|}
\hline  $D$ & $\cG$ & $\cK$ \\
\hline $10$ & $\RR$ & 1 \\ 
\hline  $9$ & $SL(2,\RR) \times \RR$ & $SO(2)$  \\
\hline $8$ &  $SL(3,\RR) \times SL(2,\RR)$  & $SO(3) \times SO(2) $  \\
\hline $7$ &$SL(5,\RR)$  & $ SO(5)$  \\
\hline $6$& $Spin(5,5,\RR)$  & $Ê(Sp(2)  \times Sp(2))/\ZZ_2$  \\
\hline $5$ & $E_{6(6)}$ &  $ÊSp(4)$  \\
 \hline $4$&
$E_{7(7)}$& $SU(8)/ \ZZ_2$\\
\hline $3$&
$E_{8(8)}$& $Spin(16) / \ZZ_2$ \\
\hline  \hline $2$&
$E_{9(9)}$ & $K(E_{9(9)})$ \\
\hline $1$ &$E_{10(10)}$ & $K(E_{10(10)})$ \\
\hline $0$ &$E_{11(11)}$ & $K(E_{11(11)})$ \\
\hline
\end{tabular}
}
\end{center}
\end{table}

The hidden symmetries encompass the U--duality \cite{Obers:1998fb}: in the case of type II string theories compactified to $n$ dimensions on an $(11-n)$--torus the conjectured U--duality group is the discrete group $E_{11-n(11-n)}(\ZZ)$, which is a subgroup of the hidden group $E_{11-n(11-n)}$ in $D=11- n$ dimensions.

Another  reason to think that  infinite--dimensional Kac--Moody algebras might be symmetries of gravitational theories comes from \emph{Cosmological Billiards}, which shed a new light on the  work of Belinskii, Khalatnikov and Lifshitz (BKL) \cite{Belinsky:1970ew,Belinsky:1982pk,BKL2}. BKL gave a description of the  asymptotic behaviour, near 
a space--like singularity, of the general solution of  Einstein's empty spacetime equations  in $D=4$. They argued that, 
in  the vicinity of a space--like singularity, 
 the Einstein's equations --- which are a system of partial differential equations in 4 variables $(t, x_i )$ --- can be approximated
by a 3--dimensional family, parametrised by $(x_i ) \in \RR^3$, of ordinary differential equations with respect to the time variable $t$. This means that the spatial points effectively decouple. The coefficients 
entering the non--linear terms of these  ordinary differential equations depend on the spatial point 
$x_i$ but are the same, at each given $x_i$, as those that arise in the  Bianchi type IX or VIII spatially 
homogeneous models.
References \cite{KLL,Bar,Belinsky:1970ew,Belinsky:1982pk,BKL2} provide a description of this general asymptotic solution in terms of chaotic successions of generalised Kasner  solutions  [a Kasner solution is given by a metric $ds^2 = -dt^2 +  t^{2p^1} (dx^1)^2 +  t^{2p^2} (dx^2)^2+ t^{2p^3} (dx^3)^2$ where $p^1,p^2, p^3$ are constants subject to 
$p^1+p^2+p^3 = (p^1)^2 +(p^2)^2 + (p^3)^2 = 1$; a generalised Kasner solution can be obtained from the Kasner solution by performing a linear transformation on the metric in order to ``un--diagonalise" it]. Using Hamiltonian methods, these chaotic successions have been shown to possess an interpretation in terms of a billiard motion on the Poincar\'e disk  \cite{Chitre,Misnerb}. 

\noindent 
The emergence of Kac--Moody structure \emph{\`a la limite BKL }has given a new impulse to this field \cite{Damour:2000hv,Damour:2000th,Damour:2000wm,Damour:2002et}. This approach allows to tackle  theories of gravity coupled to $p$--forms and dilatons in any spacetime dimensions and generalises the previous works.  In reference \cite{Damour:2002et} a self--consistent description of the asymptotic behaviour of all fields in the vicinity of a space--like singularity as a billiard motion is given. It is also shown how to systematically derive the billiard shape, in particular the position and orientation of the walls,
 from the field content and the explicit form of the Lagrangian. 
The striking result is that,  when one studies the asymptotic dynamics of the gravitational field and the dilatons, the shape of the billiard happens to be the fundamental Weyl chamber of 
some Lorentzian Kac--Moody algebra for pure gravity in $D\leq10$ as well as for all the (bosonic sector of the) low energy limits 
of the superstring theories and $D=11$ supergravity \cite{Damour:2000th,Damour:2000hv,Damour:2000wm}.

\noindent Depending on the theory under consideration, there are essentially two types of behaviours.  Either the billiard volume is finite and the asymptotic dynamics is mimicked  by a monotonic Kasner--like regime or the billiard volume 
is infinite and the asymptotic dynamics is a \emph{chaotic }succession of Kasner epochs. When the billiard is associated with a Kac--Moody algebra, the criterium for the behaviour  to be  chaotic is that the Kac--Moody algebra be \emph{hyperbolic}  \cite{Damour:2001sa}. 
The spacetime dimension 10 is critical from this point of view. Indeed, it has been shown that pure gravity in $D$ 
dimensions is related, in this context, to the over--extended Kac--Moody algebra $A_{D-3}^{++}$. This algebra is hyperbolic for $D\leq 10$ and therefore the asymptotic dynamics are chaotic \cite{Demaret:1986ys,Demaret:1988sg}. The Kac--Moody algebras relevant for the (bosonic sector of the) low energy limits of various superstring theories and for the $D=11$ supergravity
are also hyperbolic. Note that the $D=11$ supergravity asymptotic dynamics is chaotic in spite of the fact that $D>10$, thanks to the crucial presence of the 3--form potential. 

\noindent The question arose whether the appearance of these Kac--Moody algebras  in the asymptotic regime  was a manifestation of the actual symmetry of the \emph{full }theory. 
A way to tackle this question is to write an action explicitly invariant under the relevant Kac--Moody algebra and compare the equations of motion of this action with the ones 
of the gravitational theory. Let us specify this discussion to the bosonic sector of the $D=11$ supergravity for which the relevant Kac--Moody group is $E_{10(10)} = E_{8(8)}^{++}$. An action $S_{E_{8(8)}^{++}}$ invariant under this group can be constructed by considering a geodesic motion, depending on time, on the 
coset space  $E_{10(10)}/ K(E_{10(10)})$, where  $K(E_{10(10)})$ is the maximally compact subgroup of $E_{10(10)}$ \cite{Damour:2002cu}. This coset space being infinite--dimensional, the motion is parametrised by an infinite number of fields and a 
``level" is introduced to allow a recursive approach. On the $D=11$ supergravity side, one considers a ``gradient" expansion of the bosonic equations of motion. It has been shown that, up to a certain level, the level expansion and   the gradient expansion match perfectly provided one adopts a dictionary mapping spacetime bosonic  fields at a given spatial point to time dependent geometrical quantities entering the coset construction \cite{Damour:2002cu}. A key ingredient to establish this dictionary is that the level relies on the choice of a $\mf{sl}(10,\RR)$ subalgebra of the Kac--Moody algebra $\mf{e}_{10(10)}$ the representations of which can be identified with spacetime fields. Unfortunately, the Kac--Moody algebras are poorly understood and this causes problems to continue the comparison.

Hidden symmetries and Cosmological Billiards have been invoked above. One should note that they are intimately connected. For instance, pure gravity in $D$ spacetime dimensions possesses the 
hidden symmetry $ SL(D-2,\RR)$ [\ie when reduced to $D=3$, this theory is a non--linear sigma model based on the coset space $SL(D-2,\RR)/SO(D-2)$ coupled to gravity] and its asymptotic dynamics in the vicinity of a space--like singularity is controlled by the Weyl chamber of the Kac--Moody algebra $\mf{sl}(D-2,\RR)^{++}$. This suggests that when the hidden symmetry of a theory is given by a simple Lie group $\cG$ with Lie algebra $\mf{g}$, the asymptotic dynamics of this theory is controlled by the over--extended algebra $\mf{g}^{++}$.   
This is indeed what happens  for the (bosonic sector of the) low--energy limits of various string theories and the bosonic sector of $D=11$ supergravity. 
There exists a constructive approach, based on the \emph{oxidation }procedure, to find theories possessing a hidden symmetry given by a simple Lie group $\cG$ and whose asympotic dynamics is controlled by the over--extended Lie algebra $\mf{g}^{++}$.   
The \emph{oxidation }of a theory refers to the procedure inverse to dimensional reduction \cite{Keurentjes:2002xc,Keurentjes:2002rc,Keurentjes:2002vx}. The \emph{oxidation endpoint } of a theory is the highest dimensional theory the reduction of which reproduces the theory we started with. In some cases, there can be more than one oxidation endpoint. Reference \cite{Cremmer:1999du} considers the
three--dimensional non--linear $\s$--models based on a coset spaces $\cG/\cK$ coupled to gravity for each simple Lie group $\cG$, where $\cK$ is the maximally compact subgroup of 
$\cG$.\footnote{See also \cite{Breitenlohner:1987dg} for an earlier work.} And it provides the oxidation endpoints of these theories, which we will refer to in the sequel as the \emph{maximally oxidised theories }$\cG$.  These theories include in particular pure gravity in $D$ dimensions, the bosonic sector of the low energy limits of the various $D=10$ superstring theories and
of  M--theory, which all admit a supersymmetric extension. But it also contains many more theories which do \emph{not }admit a supersymmetric version such as  the low energy effective action of the bosonic string in $D=26$.  Therefore, hidden symmetries  appear to 
have a wider scope than supersymmetry.

Generalising the work \cite{West:2001as}, it has been conjectured that the maximally oxidised theories, or some extensions of them,
possess the very--extended Kac--Moody symmetry
$\cG^{+++}$ \cite{Englert:2003zs}. 
The possible existence of this Kac--Moody symmetry $\cG^{+++}$ motivated
the construction of an action $S_{\cG^{+++}}$ explicitly invariant
under $\cG^{+++}$  \cite{Englert:2003py}. The action $S_{\cG^{+++}}$ is defined in a
reparametrisation invariant way on a world--line, a priori
unrelated to spacetime, in terms of fields depending on an evolution parameter $\xi$ and parameterising the
coset $\cG^{+++}/\td \cK^{+++}$. 
A
 level
decomposition of $\cG^{+++}$ with respect to a preferred $\mf{sl}(D,\RR)$ subalgebra is performed, where $D$ can be
identified to the spacetime dimension\footnote{ Level expansions
of very--extended algebras in terms of the subalgebra $\mf{sl}(D,\RR)$
have been considered in \cite{West:2002jj, Nicolai:2003fw, Kleinschmidt:2003mf}.}. 
The
subalgebra  $\td \cK^{+++}$ is defined to be the subalgebra of $\cG^{+++}$ invariant under a ``temporal'' involution, which is chosen such that the 
 action is $SO(1,D-1)$ invariant at each
level. In this formulation, spacetime is expected to be generated dynamically since it is not included as a  basic ingredient.
Note that this construction is similar to the one of $S_{E_{8(8)}^{++}}$ described above. In fact, this action can be found again by performing a consistent truncation of some fields
in $S_{E_{8(8)}^{+++}}$. More generally,   
each action $S_{\cG^{+++}}$ encompasses two distinct actions $S_{\cG^{++}_C}$ and $S_{\cG^{++}_B}$ invariant under the overextended Kac-Moody subalgebra $\cG^{++}$
 \cite{Englert:2004ph}. $S_{\cG^{++}_C}$ carries an Euclidean signature and is the generalisation to all $\cG^{++}$ of $S_{E_{8(8)}^{++}}$. The second action $S_{\cG^{++}_B}$ carries various Lorentzian signatures revealed through various equivalent formulations related by Weyl transformations of fields. The signatures found in the analysis of references \cite{Keurentjes:2004bv,Keurentjes:2004xx}
and in the context of $S_{E^{++}_{8 \, B}}$ in \cite{Englert:2004ph} match perfectly with the signature changing dualities and the 
exotic phases of M--theories discussed in \cite{Hull:1998vg,Hull:1998fh,Hull:1998ym}. 
The action $S_{\cG^{++}_B}$  admits exact solutions that can be identified to solutions of the maximally oxidised theories, which describe intersecting extremal branes smeared in all directions but one. For a very pedagogical introduction on intersecting branes, we refer to \cite{Argurio:1998cp}. Moreover, 
the intersection rules for extremal branes \cite{Argurio:1997gt} translate elegantly, in the Kac--Moody formulation,  into orthogonality conditions between roots \cite{Englert:2004it}.
The dualities of M--theory are interpreted in the present context as Weyl reflections. 

Let us mention that there exists an approach of $M$--theory based on Borcherds algebras and del Pezzo surfaces. Borcherds algebras encode nice algebraic structures and generalise the Kac--Moody algebras \cite{Henry-Labordere:2002dk,Henry-Labordere:2002xh,Henry-Labordere:2003rd}.

\clearpage

\subsubsection{Plan of the thesis}

This thesis is divided into two parts. The first one is devoted to the Cosmological Billiards and tries to provide a 
better understanding of their connection with Kac--Moody algebras. The appearance of  these algebras rises numerous questions like: 
What property of a theory is responsible for  a chaotic 
behaviour in the asymptotic regime? 
What does the asymptotic regime tell us about the 
full theory?
Is there a signal of a deeper importance of Kac--Moody algebras?
There are  lots of billiard walls, some of them are not relevant to determine the billiard shape. What is the significance of these non--dominant
walls? 
In the second part, we will focus on the attempts to reformulate  gravitational theories as  non--linear realisations that are proposed in \cite{Damour:2002cu} and \cite{Englert:2003py}. 

In the first chapter, an intuitive introduction is given. The purpose is
to shed light on the beautiful connection between the dynamics of the gravitational field in the vicinity 
of a space--like singularity [described as the free motion of a ball within a particular billiard plane] and Kac--Moody algebras. 
Chapter 2 presents the general setting of Cosmological Billiards within the Hamiltonian formalism; the Iwasawa decomposition plays an important role here. We also refer to \cite{Damour:2002et,Damour:2005ef,Damour:2005pe,Damour:2004gn,Damour:2005mr,Nicolai:2005su} for introductions on Cosmological Billiards.

Spatially homogeneous cosmological models play an important role in our understanding of the universe \cite{Ryan:1975jw}. It is therefore natural to investigate how the BKL behaviour is 
modified (or not) in this simplified context. Conversely, one might hope to get insight about conjectures about the BKL limit by testing them in this simpler context. This question has already been addressed in \cite{Demaret:1985js,Demaret:1988sg} and references therein. In chapter 3 we reconsider this problem with the enlightening approach of \cite{Damour:2002et}. We analyse the Einstein and Einstein--Maxwell billiards for all spatially homogeneous cosmological models corresponding to 3 and 4--dimensional real unimodular Lie algebras and we provide the list of the models that are chaotic in the BKL limit. Through the billiard picture, we confirm that, in $D=5$ spacetime dimensions, chaos is present if off--diagonal metric elements are kept: the finite volume billiards can be identified with the fundamental Weyl chambers of hyperbolic Kac--Moody algebras. The most generic cases bring in the same algebras as in the inhomogeneous case, but other algebras appear through special initial conditions. These ``new'' algebras are subalgebras of the Kac--Moody algebras characterising the generic, \ie non homogeneous, case. Therefore the study of homogeneous cosmologies naturally leads to tackle Lorentzian subalgebras of the Kac--Moody algebras appearing in the generic case.

In chapter 4, we quit the simplified context of homogeneous cosmologies and focus on questions concerning oxidation. We used the billiard analysis to set constraints on the field content and maximal dimension a theory must possess to exhibit a given duality group in $D=3$.  We concentrate on billiards  controlled by the restricted root system of a given (non--split) real form of any complex simple Lie algebra. 
The three--dimensional non--linear $\s$--model based on the coset spaces $\cG/\cK$ coupled to gravity, where $\cG$ is a Lie group, the Lie algebra of which is a given real form of 
$\mf{g}$, have been shown to exhibit these 
billiards \cite{Henneaux:2003kk}, \ie the restricted root system controls the billiard. We show how the properties of the Cosmological Billiards provide useful information (spacetime dimension and $p$--form spectrum) on the oxidation endpoint of these $\s$--models. We compare this approach to other methods dealing with $GL(n,\RR)$ subgroups and the superalgebras of dualities.

Hyperbolic Kac--Moody algebras seem to play a distinguished role since there is a deep connection between the hyperbolicity of the algebra and the chaotic behaviour of spacetime
in the vicinity of a space--like singularity \cite{Damour:2001sa}. A complete classification of hyperbolic algebras is known \cite{S,WZX}; it is natural to wonder if these algebras 
are related to some Lagrangian. 
We answer completely this question in chapter 5: we identify the hyperbolic Kac--Moody algebras for which there exists a Lagrangian of gravity, dilatons and $p$--forms which produces a billiard  identifiable with the fundamental Weyl chamber of the hyperbolic Kac--Moody algebra in question. Because of the invariance of the billiard upon toroidal dimensional reduction  \cite{Damour:2002fz}, the list of admissible algebras is determined by the existence of a Lagrangian in three spacetime dimensions, where a systematic analysis can be carried out since only zero-forms are involved. We provide all highest dimensional parent Lagrangians with their full spectrum of $p$--forms and dilaton couplings. We confirm, in particular, that for the rank 10 hyperbolic algebra, $CE_{10} = A_{15}^{(2)\wedge}$, also known as the dual of $B_8^{\wedge\wedge}$, the maximally oxidised Lagrangian is 9 dimensional and involves besides gravity, 2 dilatons, a 2--form, a 1--form and a 0--form. We insist on the fact that we are interested, in this chapter, in the \emph{minimal }field content that is necessary to exhibit a given hyperbolic Weyl group  in the BKL limit. In particular we are not interested in the Chern--Simons terms nor the $p$--forms that would not lead to dominant walls.

The second part of the thesis deals with non--linear realisation based on coset spaces $\cG^{+++}Ê/ \td \cK^{+++}$ and $\cG^{++}/ \cK^{++}$. Chapter 6 presents an introduction to non--linear realisations based
on the coset spaces $\cG^{+++}/ \td \cK^{+++}$. The level decomposition, which permits a recursive approach,  is  explained in detail as well as the choice of the subgroup 
$\td \cK^{+++}$. The link with gravitational theories coupled to $p$--forms is, as far as it is understood, explained. In this perspective, the two non equivalent truncations of 
$S_{\cG^{+++}/\td \cK^{+++}}$ are recalled.  

Chapters 7 and 8 deal with the inclusion of fermions in this context. 

 Dirac fermions are considered in chapter 7. Before confronting the difficulties coming from infinite--dimensional algebras, we analyse the compatibility of Dirac fermions with the hidden duality symmetries which appear in the toroidal compactification of gravitational theories down to three spacetime dimensions. We show that the Pauli couplings to the $p$--forms can be adjusted, for all simple (split) groups, so that the fermions transform in a spinorial representation of the maximal compact subgroup of the symmetry group $\cG$ in three dimensions. 
Then we investigate how the Dirac fermions fit in the conjectured hidden overextended symmetry $\cG^{++}$. We show compatibility with this symmetry up to the same level as in the pure bosonic case. We also investigate the BKL behaviour of the Einstein--Dirac--$p$--form systems and provide a group theoretical interpretation of the Belinskii--Khalatnikov result that the Dirac field removes chaos.

The gravitino field is next envisaged in chapter 8. Recall that the hyperbolic Kac--Moody algebra $E_{10(10)}$ has repeatedly been suggested to play a crucial role in the symmetry structure of M--theory. This attempt, in line with the established result that the scalar fields which appear in the toroidal compactification down to three spacetime dimensions form the coset $E_{8(8)}/(Spin(16)/\ZZ_2)$, was verified for the first bosonic levels in a level expansion of the theory \cite{Damour:2002cu}. We show that the same features remain valid when one includes the gravitino field.

In chapter 9,  we turned to the $S_{\cG^{++}_B}$ --theories which are obtained by a consistent truncation of $S_{\cG^{+++}}$ --theories \emph{after }a Weyl reflection with respect to the very--extended root. The actions $S_{\cG^{++}_B}$ are invariant under $\cG^{++}$ and in particular under Weyl reflections of $\mf{g}^{++}$ which happen to change the signature. The $\cG^{++}$ content of the formulation of gravity and M--theories as very--extended Kac--Moody invariant theories is further analysed. The different exotic phases of all the $\cG_B^{++}$ theories, which admit exact solutions describing intersecting branes smeared in all directions but one, are derived. This is achieved by analysing for all $\cG^{++}$ the signatures which are related to the conventional one $(1,D-1)$ by ``dualities'' generated by the Weyl reflections.

Brief conclusions and perspectives are given in the last chapter. \newline

\noindent The results presented in this thesis have been published in references \cite{deBuyl:2004md,deBuyl:2003za,deBuyl:2003ub,deBuyl:2005it,deBuyl:2005zy,deBuyl:2005mt}.

%%%%%%%%%%%%%%%%%%%%%%%%%%%%%%%
%%%%%%%%%%%%%%%%%%%%%%%%%%%%%%%
\cleardoublepage
\part{Cosmological Billiards}
\cleardoublepage
\pagestyle{myheadings}
%%%%%%%%%%%%%%%%%%%%%%%%%%%%%%%%
%%%%%%%%%%%%%%%%%%%%%%%%%%%%%%%%
%%\include{intro_billards}

\chapter{A First Approach}
 \markboth{A {F}IRST {A}PPROACH}{}

\begin{flushright}
\emph{
The intention of this first chapter is to introduce \\
the Cosmological Billiard picture in an intuitive way.\\
 Precise statements are given in the next chapter.}
\end{flushright}

The Einstein equations are second order partial differential equations. Their
complexity is such that no general solution of them is known. Under 
certain conditions, the appearance of singularities is a generic property of the solutions
of the Einstein equations \cite{Hawking:1969sw}.  Reference \cite{Hawking:1969sw} does not 
provide a detailed description of how spacetime becomes singular. 
The work of BKL is therefore 
remarkable. Indeed, they described the general behaviour of the $D=4$ 
empty spacetime in the vicinity of a space--like singularity ($t$ = cst) \cite{Belinsky:1970ew,BKL2,Belinsky:1982pk}.\footnote{ This simplification has not been rigourously justified although it is strongly supported.} The essential 
simplification of the Einstein equations \emph{in this regime} follows from the 
fact that 
 the partial differential equations for the metric 
components become ordinary differential equations with respect to the time $t=x^0$. Accordingly, there is an effective decoupling of the spatial points in the asymptotic regime. BKL further argued that the 
general behaviour of spacetime is a never ending chaotic succession of Kasner 
epochs at each spatial point. They also showed that chaos disappears in the 
presence of a mass--less scalar field \cite{Andersson:2000cv,BK}. 
Before entering into the details of the asymptotic dynamics of the gravitational field 
 in the vicinity of a space--like singularity, we will 
focus on the empty $D=4$ spacetime and 
(i) review the Kasner metric, 
(ii) describe the BKL solution \emph{with words}, 
(iii) reinterpret this solution as a billiard motion (thanks to the 
enlightening approach of \cite{Damour:2002et}) and 
(iv) show how Kac--Moody algebras come in. 
Finally, we make general comments which extend beyond the  example 
of pure $D=4$ gravity. 

\subsubsection{Kasner Solution} 

\begin{figure}[h]
  \centering
\begin{picture}(0,0)%
\epsfig{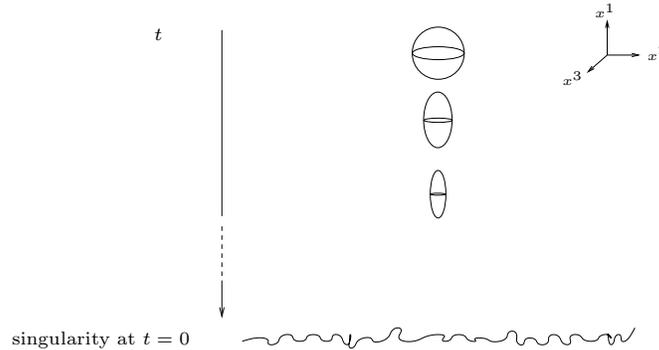}%
\end{picture}%
\setlength{\unitlength}{1865sp}%
\begingroup\makeatletter\ifx\SetFigFont\undefined%
\gdef\SetFigFont#1#2#3#4#5{%
  \reset@font\fontsize{#1}{#2pt}%
  \fontfamily{#3}\fontseries{#4}\fontshape{#5}%
  \selectfont}%
\fi\endgroup%
\begin{picture}(8921,4552)(-1259,-7018)
\put(631,-2896){\makebox(0,0)[lb]{\smash{{\SetFigFont{7}{8.4}{\familydefault}{\mddefault}{\updefault}{\color[rgb]{0,0,0}$t$}%
}}}}
\put(6487,-2594){\makebox(0,0)[lb]{\smash{{\SetFigFont{5}{6.0}{\familydefault}{\mddefault}{\updefault}{\color[rgb]{0,0,0}$x^1$}%
}}}}
\put(7185,-3159){\makebox(0,0)[lb]{\smash{{\SetFigFont{5}{6.0}{\familydefault}{\mddefault}{\updefault}{\color[rgb]{0,0,0}$x^2$}%
}}}}
\put(6054,-3492){\makebox(0,0)[lb]{\smash{{\SetFigFont{5}{6.0}{\familydefault}{\mddefault}{\updefault}{\color[rgb]{0,0,0}$x^3$}%
}}}}
\put(-1259,-6946){\makebox(0,0)[lb]{\smash{{\SetFigFont{7}{8.4}{\familydefault}{\mddefault}{\updefault}{\color[rgb]{0,0,0}singularity at $t=0$}%
}}}}
\end{picture}%
  \caption{\small{A ``sphere'' located at one \emph{fixed} spatial point is deformed as one goes to the singularity in a 
``Kasner spacetime''.}}
  \label{spheres}
\end{figure}

\noindent The Kasner metric is the solution of the $D=4$ 
Einstein equations $R_{\m\n}(g_{\r\s})=0$ depending only on time (see
\cite{LL} paragraph 117) which reads
\beq ds^2 &=& -e^{2\tau} d\tau^2 + e^{\b^1} (dx^1)^2 + e^{\b^2}(dx^2)^2 + e^{\b^3}
(dx^3)^2  \label{Kasner4} \\
\b^i &=& 2p_i \tau  \nn \, , 
\eeq
where $i=1,2,3$ and $ p_1, \ p_2, \ p_3$ are constants such that
\beq p_1 + p_2 + p_3 &=& 1  \label{lapsch} \ , \\
\overrightarrow{p}^2 = p_1^2+p_2^2+p_3^2 -( p_1+p_2+p_3 )^2&=& 0 \ . 
\label{metricKasner4} 
 \eeq
Eq.(\ref{lapsch}) is a gauge choice, namely the proper time $t = e^\tau$ obeys $g = t^2 $ ($g$ is the 
determinant of the spatial metric), Eq.(\ref{metricKasner4}) comes from the invariance 
under time reparametrization. 
Important properties of the Kasner solution are the  spatial homogeneity  and 
the anisotropy.
This metric is singular at $t=0$ and this singularity cannot 
be removed by a coordinate change. The  scalar $R^{\a\b\g\d}R_{\a\b\g\d}$ becomes infinite when 
$t$ goes to zero and the spatial volume vanishes as $t \rightarrow
0$. The only exception is when the $p$'s take their values in the set
$\{1,0,0\}$ because it corresponds to flat spacetime\footnote{This is clear after performing the coordinate 
change $(\tau, x^1,x^2,x^3) \rightarrow (\xi, \rho, x^2 , x^3)$ given by the equations  $e^\tau \sinh x^1 = \xi$ and  $e^\tau \cosh x^1 = \rho$. }. 
One can infer from Eqs.(\ref{lapsch} \& \ref{metricKasner4}) that one of 
the $p$'s is negative 
while the other two are positive. For definiteness, take $p_1$ negative. 
To gain a better understanding of the Kasner metric, let us imagine how  a sphere ---located at a \emph{fixed }spatial point-- gets deformed as one goes to  
 the singularity. More 
precisely, the ``sphere'' will be elongated in the direction corresponding 
to the coordinate $x^1$ and contracted in the directions corresponding to 
the coordinates $x^2$ and $x^3$ as depicted in 
Figure \ref{spheres}.

\subsubsection{BKL Solution}

\begin{figure}[h]
  \centering
\begin{picture}(0,0)%
\epsfig{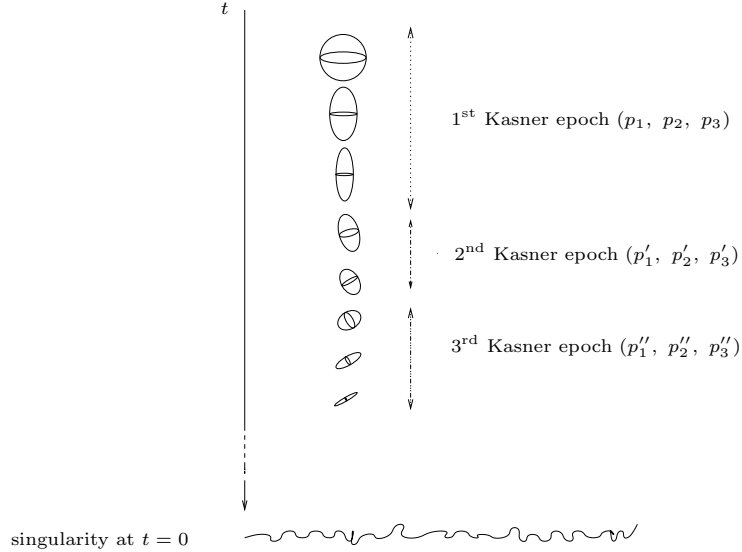}%
\end{picture}%
\setlength{\unitlength}{1865sp}%
\begingroup\makeatletter\ifx\SetFigFont\undefined%
\gdef\SetFigFont#1#2#3#4#5{%
  \reset@font\fontsize{#1}{#2pt}%
  \fontfamily{#3}\fontseries{#4}\fontshape{#5}%
  \selectfont}%
\fi\endgroup%
\begin{picture}(8337,7293)(-269,-6523)
\put(-269,-6451){\makebox(0,0)[lb]{\smash{{\SetFigFont{7}{8.4}{\familydefault}{\mddefault}{\updefault}{\color[rgb]{0,0,0}singularity at $t=0$}%
}}}}
\put(5626,-2671){\makebox(0,0)[lb]{\smash{{\SetFigFont{7}{8.4}{\familydefault}{\mddefault}{\updefault}{\color[rgb]{0,0,0}$2^{\mathrm{nd}}$ Kasner epoch $(p_1', \ p_2', \ p_3')$}%
}}}}
\put(5581,-3886){\makebox(0,0)[lb]{\smash{{\SetFigFont{7}{8.4}{\familydefault}{\mddefault}{\updefault}{\color[rgb]{0,0,0}$3^{\mathrm{rd}}$ Kasner epoch $(p_1'', \ p_2'', \  p_3'')$}%
}}}}
\put(5581,-871){\makebox(0,0)[lb]{\smash{{\SetFigFont{7}{8.4}{\familydefault}{\mddefault}{\updefault}{\color[rgb]{0,0,0}$1^{\mathrm{st}}$ Kasner epoch $(p_1, \ p_2, \ p_3)$}%
}}}}
\put(2521,614){\makebox(0,0)[lb]{\smash{{\SetFigFont{7}{8.4}{\familydefault}{\mddefault}{\updefault}{\color[rgb]{0,0,0}$t$}%
}}}}
\end{picture}%
  \caption{\small{A ``sphere'' located at one \emph{fixed }spatial point is deformed as one goes a space--like singularity in a \emph{generic }empty $D=4$ spacetime.}}
  \label{spheres2}
\end{figure}

\noindent The general behaviour of empty $D=4$ spacetime in the vicinity of a 
space--like singularity is, as explained by BKL and depicted in Figure \ref{spheres2},  an infinite succession of Kasner epochs at each spatial
point. The decoupling of the spatial points follows from the dominance of 
the temporal gradients compared to the spatial ones and the changes from one Kasner metric to another one due to the spatial curvature (and reflects the non--linearity of the Einstein's equations).  Moreover, as shown by BKL,  these never ending successions are 
\emph{chaotic}. 
\begin{quote}
{\small \emph{Remark }: The never ending succession of Kasner epochs is the 
\emph{generic} behaviour of the empty $D=4$ spacetime in the vicinity of a space--like singularity. If 
symmetry conditions are imposed, as for the Schwarzschild solution, the behaviour can be different. The Schwarzschild metric is, 
\beq
ds^2_{\footnotesize{ Schwarschild}} =- (1 - {2m \over r}) dt^2 + (1- {2m \over r})^{-1}dr^2 + r^2(d\theta^2 + \sin^2(\th) d\phi^2) \, ,  
\nn 
\eeq
and possesses a space--like singulartiy located in $r=0$ (inside the horizon, the coordinate $r$
is \emph{time--like}). In the vicinity of the singularity the metric takes the following form, 
\beq ds^2_{\footnotesize{ Schwarschild}} \underset{ r \rightarrow 0}{\rightarrow} ds^2 = {2m \over r} dt^2 - {r \over 2m} dr^2 + r^2(d\theta^2 + \sin^2(\th) d\phi^2) \, .
\nn \eeq
If one sets $\tau = {2 r^{3/2} \over 3 \sqrt{2m}}, \, \sigma = (4 m / 3)^{1/3} t, \, \bar{\th} =  (9 m/2)^{1/3}\th, $ and  
$\bar{\phi} = (9m /2)^{1/3} \phi$ this metric reads 
\beq ds^2 = - d\tau^2 + \tau^{-2/3}Êd\sigma^2 + \tau^{4/3} (\sin (\th) d\bar{\phi})^2 \, ,  \nn 
\eeq 
and is a Kasner metric. The asymptotic dynamics of the gravitational field is therefore \emph{not}
a chaotic succession of Kasner epochs but a single Kasner epoch. }
\end{quote}
This analysis has been done in \cite{Belinsky:1970ew,BKL2,Belinsky:1982pk,
Bar,KLL}. 
 
\subsubsection{Billiard Picture}

\begin{figure}[h]
  \centering
\begin{picture}(0,0)%
\epsfig{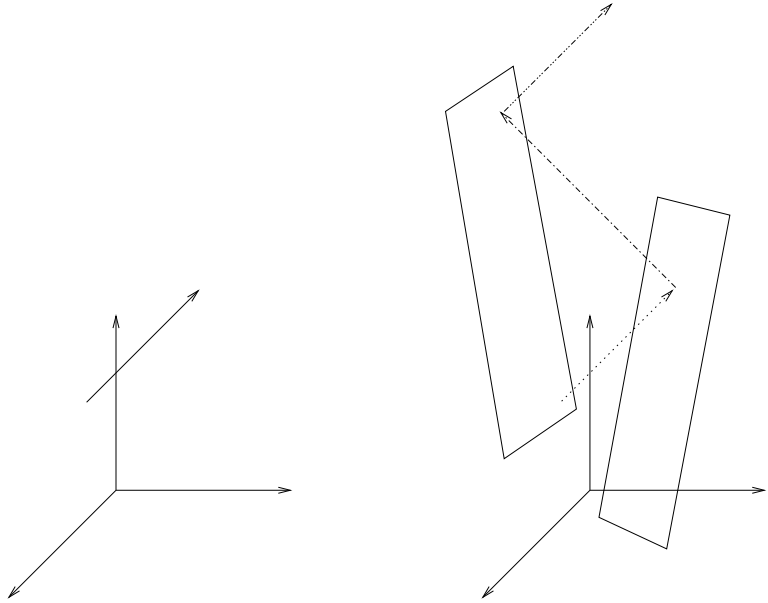}%
\end{picture}%
\setlength{\unitlength}{2486sp}%
\begingroup\makeatletter\ifx\SetFigFont\undefined%
\gdef\SetFigFont#1#2#3#4#5{%
  \reset@font\fontsize{#1}{#2pt}%
  \fontfamily{#3}\fontseries{#4}\fontshape{#5}%
  \selectfont}%
\fi\endgroup%
\begin{picture}(7584,5964)(2284,-5563)
\end{picture}%
    \caption{\small{The Kasner solution (\ref{Kasner4}) can be represented as a free ball moving on the 
  $\b$--space with a null velocity (\ref{metricKasner4})  as depicted in the left
  panel [the origin of the line depends on initial conditions]. The right panel 
  illustrates a BKL motion in the auxiliary Lorentz space. }}
 \label{kasner}
\end{figure}

\begin{figure}[h]
  \centering
\begin{picture}(0,0)%
\epsfig{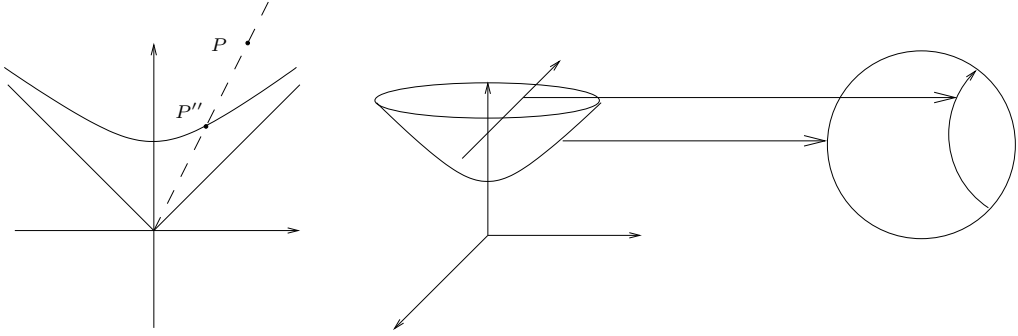}%
\end{picture}%
\setlength{\unitlength}{2486sp}%
\begingroup\makeatletter\ifx\SetFigFont\undefined%
\gdef\SetFigFont#1#2#3#4#5{%
  \reset@font\fontsize{#1}{#2pt}%
  \fontfamily{#3}\fontseries{#4}\fontshape{#5}%
  \selectfont}%
\fi\endgroup%
\begin{picture}(10100,3371)(1078,-5023)
\put(2791,-2896){\makebox(0,0)[lb]{\smash{{\SetFigFont{7}{8.4}{\familydefault}{\mddefault}{\updefault}{\color[rgb]{0,0,0}$P''$}%
}}}}
\put(3151,-2221){\makebox(0,0)[lb]{\smash{{\SetFigFont{7}{8.4}{\familydefault}{\mddefault}{\updefault}{\color[rgb]{0,0,0}$P$}%
}}}}
\end{picture}%
  \caption{ \label{hypproj}   \small{The projection of the auxiliary Lorentz space onto the upper sheet of the 
  hyperbole on a two--dimensional example is illustrated in the left panel. In this picture the point $P$ in the Lorentz space, as all the points of the dashed line,  are projected on the 
  point $P'$ of the upper sheet of the hyperbole. \newline
A Kasner regime depicted in the Lorentzian auxiliary space and its representation on the Poincar\'e disk are illustrated in the right panel. The light--like line in the Lorentz space, \ie the Kasner regime, is first projected onto an hyperbole of 
the upper sheet of the hyperboloid and then represented by a geodesic in the Poincar\'e disk.}}
\end{figure}

\noindent The question addressed here is:
\emph{What is the link between the BKL solution and a billiard motion?} 
To apprehend the connection it is useful to represent a Kasner solution
(\ref{Kasner4}) by a null line in a 3--dimensional auxiliary Lorentz space
with coordinates $\{ \b^1, \b^2, \b^3 \}$, which are the ``scale factors'' introduced in (\ref{Kasner4}).\footnote{The $\b$'s will be called the scale factors although they more precisely refer to their logarithms.} The Lorentzian metric, given by the 
quadratic form $\overrightarrow{\b}^2 = (\b^1)^2+(\b^2)^2+(\b^3)^2 -( \b^1+\b^2+\b^3 )^2$ (see 
(\ref{metricKasner4})), comes from the De Witt supermetric ---restricted to the diagonal components of the metric--- entering the Hamiltonian of General Relativity [this Hamiltonian being quatratic in the momenta indeed encodes  a metric]. The null  line is depicted in the left panel of Figure \ref{kasner}.
The BKL solution being a succession of Kasner epochs will be 
represented on the auxiliary Lorentz space as a succession of null lines. 
The BKL solution depicted in Figure \ref{spheres2} is illustrated by 
Figure \ref{kasner} in the auxiliary Lorentz space. Notice that the Kasner
epoch $(p_1,p_2,p_3)$ ends when the corresponding null line meets a
hyperplane. $2^{nd}$ Kasner epoch $(p_1',p_2',p_3')$ follows and lasts until 
the collision with the next hyperplane;  $3^{rd}$ Kasner 
epoch $(p_1'',p_2'',p_3'')$ begins and... \newline
The hyperplanes are called 
the \emph{walls}. As will later be explained, these walls originate from 
the spatial curvature and the non-diagonal terms of the metric (also from the matter if one
considers non empty spacetime). 
The succession of null lines  \emph{also} describes
a massless ball moving freely between the walls on which it bounces elastically: this is how 
the \emph{billiard picture} emerges. In the case at hand (empty $D=4$ spacetime), the ball 
will always meet some wall and there will be an endless succession of Kasner
epochs. \newline
This description is redundant since the $\b$'s (and their conjugate momenta) are subject to the Hamiltonian constraint, which expresses the invariance by time reparametrisation. 
To avoid this, the auxiliary Lorentz space can be 
radially projected on the unit hyperboloid.   
Let $\b^i = \rho \g^i$ such that $\overrightarrow{\g}^2 = -1$. Then the motion can 
be projected on the upper sheet of the unit hyperboloid $\overrightarrow{\g}^2 = 
-1$, $\cH_2$. This hyperboloid $\cH_2$ can be represented  on the 
Poincar\'e disk \cite{Ratcliffe}, see Figure \ref{hypproj} for a representation of a Kasner epoch on the 
Poincar\'e disk (after its projection onto the upper sheet of the hyperboloid). 

\begin{figure}[h]
  \centering
\begin{picture}(0,0)%
\epsfig{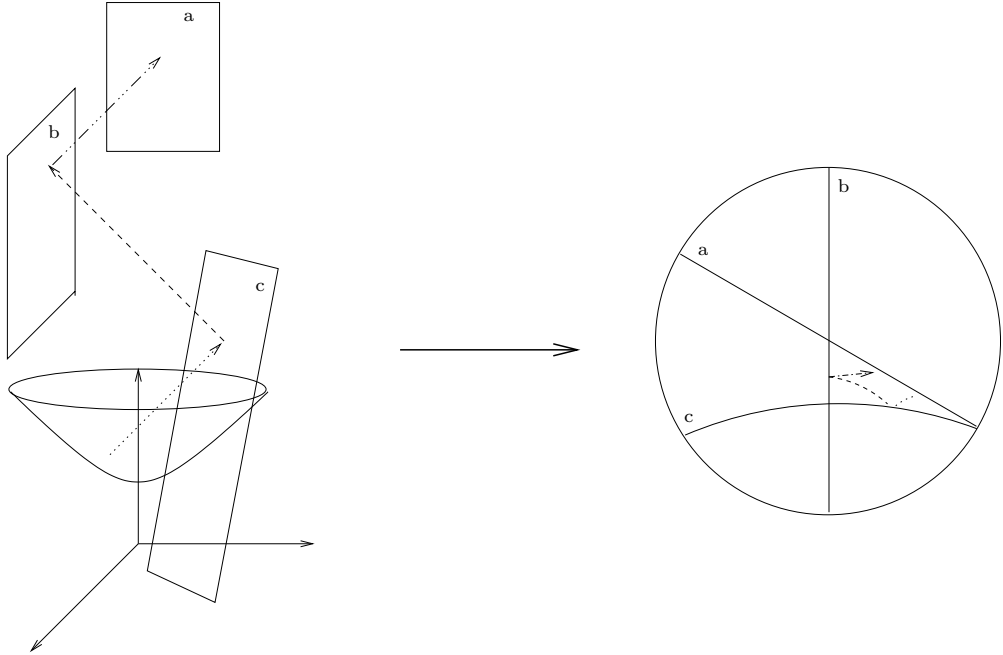}%
\end{picture}%
\setlength{\unitlength}{2486sp}%
\begingroup\makeatletter\ifx\SetFigFont\undefined%
\gdef\SetFigFont#1#2#3#4#5{%
  \reset@font\fontsize{#1}{#2pt}%
  \fontfamily{#3}\fontseries{#4}\fontshape{#5}%
  \selectfont}%
\fi\endgroup%
\begin{picture}(9921,6504)(664,-6373)
\put(3151,-2761){\makebox(0,0)[lb]{\smash{{\SetFigFont{7}{8.4}{\familydefault}{\mddefault}{\updefault}{\color[rgb]{0,0,0}c}%
}}}}
\put(7426,-4066){\makebox(0,0)[lb]{\smash{{\SetFigFont{7}{8.4}{\familydefault}{\mddefault}{\updefault}{\color[rgb]{0,0,0}c}%
}}}}
\put(7561,-2401){\makebox(0,0)[lb]{\smash{{\SetFigFont{7}{8.4}{\familydefault}{\mddefault}{\updefault}{\color[rgb]{0,0,0}a}%
}}}}
\put(8956,-1771){\makebox(0,0)[lb]{\smash{{\SetFigFont{7}{8.4}{\familydefault}{\mddefault}{\updefault}{\color[rgb]{0,0,0}b}%
}}}}
\put(2431,-61){\makebox(0,0)[lb]{\smash{{\SetFigFont{7}{8.4}{\familydefault}{\mddefault}{\updefault}{\color[rgb]{0,0,0}a}%
}}}}
\put(1081,-1231){\makebox(0,0)[lb]{\smash{{\SetFigFont{7}{8.4}{\familydefault}{\mddefault}{\updefault}{\color[rgb]{0,0,0}b}%
}}}}
\end{picture}%
  \caption{  \label{pcdisk2}\small{Lorentz space and projection on Poincar\'e disk: a succession 
of Kasner epochs. The time--like hyperplanes in the Lorentz space intersect the upper sheet of the hyperboloid along hyperboles; these hyperboles are represented by geodesics on the Poincar\'e disk, \ie 
the three plain lines in the right panel. The Kasner epochs depicted in the Lorentz space are projected onto sections of 
hyperboles of the upper sheet of the hyperboloid; these sections of hyperboles are themselves represented by sections
 of geodesics of the Poincar\'e disk, \ie the dashed line in right panel.}}
\end{figure}

\noindent The Figure \ref{pcdisk2} illustrates the nomenclature used on the Poincar\'e disk and 
given here under, \newline

\begin{tabular}{ccp{8cm}}
The walls & = & The walls in the auxiliary Lorentz space are projected on 
hyperplanes in the upper sheet of the hyperbolid (or geodesics of the Poincar\'e disk) which are also called \emph{walls}.  \\
The billiard  & = &  The section of the Poincar\'e disk delimited by the 
 walls is called in this context the \emph{billiard table} or simply 
the \emph{billiard}. \\
The ball motion &= & The projection of a null line in the auxiliary Lorentz 
space is a geodesic onto the hyperbolic plane. The 
succession of geodesics is the interrupted free motion of a \emph{ball} on 
this billiard.
\end{tabular}
  
 The billiard picture was first obtained for the homogeneous (Bianchi 
IX) four--dimensional case \cite {Chitre,Misnerb} and latter extended to higher 
spacetime dimensions with $p$--forms and dilatons \cite{Kirillov1993, 
Kirillov:1994fc,Ivashchuk:1994tu,Ivashchuk:1994fg,Andersson:2000cv,
Damour:2000hv,Damour:2002cu,Damour:2002tc,Damour:2002et}. 

\subsubsection{The $A_1^{++}$ Kac--Moody Algebra}

\noindent 
There are  lots of walls but only few of them can really be hit by the ball, the other 
walls standing behind the former. The notion of \emph{dominant} wall refers to 
the walls seen by the ball. The metric on the auxiliary space, \ie the Poincar\'e disk, is known and  from this
one can compute the angles between the dominant walls and get information about the billiard volume. This reveals the 
shape of the billiard. \emph{This shape appears to be that of the 
fundamental Weyl chamber of $A_1^{++}$}. 
Recall that the roots of the Kac--Moody
algebra $A_1^{++}$ live in an hyperbolic space (the Poincar\'e disk is an
hyperbolic space) and that the region of this space delimited by the hyperplanes
orthogonal to the simple roots define the fundamental Weyl chamber.  The statement is that
the 
simple roots of $A_1^{++}$ can be identified with the vectors normal to the dominant walls. 

The connection between the asymptotic behaviour of the gravitational field
dynamics and Kac--Moody algebra $A_1^{++}$ is very surprising. Indeed, the 
billiard can be identified with the  Weyl chamber of some 
corresponding Lorentzian Kac--Moody algebra only when 
many conditions
are simultaneously met. In particular, (i) the billiard table must be a Coxeter polyhedron
(the dihedral angles between adjacent walls must be integer submultiples of
$\pi$) and (ii) the billiard must be a simplex \cite{Ratcliffe,Damour:2002fz}. There
is no reason, {\it a priori},  why these conditions are satisfied. But in the case at hand, namely pure gravity 
in $D=4$, they are. 

\subsubsection{More Kac--Moody Algebras}

\noindent
In this brief introduction to billiards, we focussed on the empty $D=4$ 
spacetime. The more general action we are interested in is the one that describes gravity
coupled to an assortment of $p$--forms and dilatons in any spacetime 
dimension $D$. A natural question 
to address is which changes are brought by the matter fields. 
The \emph{dilatons play the same role as the scale factors}, therefore for $n$ dilatons the dimension of the auxiliary Lorentz space becomes
$n+d$. On the contrary, in the BKL limit the \emph{$p$--forms get frozen}  and 
their effect is simply to add new walls to the billiards.  The billiard picture is still valid for these more general theories but the shape of the billiard will generically \emph{not }be associated with a Kac--Moody algebra. The appearance these algebras requires indeed a very particular set of  $p$--forms and dilatons  with extremely fine tuning of the coupling constants. There are essentially two types of behaviour of the gravitational field in the vicinity of a space--like singularity: either the billiard volume is infinite and the asymptotic dynamics is mimicked by a monotonic 
Kasner--like regime or the billiard volume is finite and the asymptotic dynamics is a chaotic 
succession of Kasner epochs. The criterion to get a chaotic behaviour, \ie an infinite succession of Kasner 
epochs,  is that the volume of the billiard be finite. For the billiards identifiable with a Weyl chamber
of a Kac--Moody algebra, reference \cite{Damour:2001sa} showed that the criterion becomes the following one:  to get a chaotic behaviour the Kac--Moody algebra must be an  \emph{hyperbolic} Kac--Moody algebra  --- see appendix \ref{km} for the definition of hyperbolic.

The astonishing fact is that 
\emph{for all ``physically relevant'' theories, the billiards are 
characterised by  hyperbolic Kac--Moody algebras and therefore the asymptotic 
dynamics is chaotic} \cite{Damour:2000wm,Damour:2000th,Damour:2000hv,
Damour:2001sa,Damour:2002fz,Damour:2002cu,Damour:2002et}. 
``Physically relevant theories'' refered to  here are pure gravity and 
the low energy limits of string theories and M-theory.

A general class of theories happens to offer a chaotic asymptotic  
behaviour of the gravitational field. This class corresponds to theories 
which, upon dimensional reduction to $D=3$, describe gravity coupled to a non 
linear sigma model $\cG/\cK$, where $\cG$ is a finite dimensional simple 
Lie group (the Lie algebra of which is the maximally non compact real form of one of the finite dimensional simple 
Lie algebras $ \mf{g} = \{ \mf{a}_n, \,Ê\mf{b}_n, \, \mf{c}_n, \,Ê\mf{d}_n, \,Ê\mf{g}_2, \,Ê\mf{f}_4, \, \mf{e}_6 , \,Ê\mf{e}_7, \, \mf{e}_8 \}$) and $\cK$ its maximal 
compact subgroup (see \cite{deBuyl:2004md} and references therein). In these cases, the billiard's shape is the fundamental 
Weyl chamber of the Kac--Moody algebra $\mf{g}^{++}$. Other real forms are considered in reference \cite{Henneaux:2003kk}.  
Notice that all the 
``physically relevant theories'' mentioned above fall into this class, see Table \ref{kmth}. 

Remember that the oxidation procedure of a given $D$ dimensional theory consists in finding a theory of dimension $D' > D$ such that the reduction of this theory gives back the $D$ dimensional theory. A oxidation endpoint or \emph{maximally oxidised theory,} which may be not unique, refers to a theory which 
possess no higher dimensional parent. The general term \emph{maximally oxidised theory $\cG$} refers here to the 
oxidation endpoint of $D=3$ gravity coupled to a non--linear sigma model based on the coset space 
$\cG / \cK$ where $\cG$ is a finite dimensional simple Lie group and $\cK$ its maximal compact subgroup.

The appearance of these Kac--Moody algebras rises numerous questions like: 
 What properties do share the theories which lead to a chaotic 
behaviour in the asymptotic regime? 
What can we infer from  the asymptotic regime about the 
full theory?
 Is there a signal of a deeper importance of Kac--Moody algebras?
 There are  lots of walls: what is the signifiance of the non-dominant
ones? This connection deserves to be studied in depth.

\noindent One should notice that unfortunately rigorous mathematical proofs 
\cite{Andersson:2000cv,Damour:2002tc,Rendall:2001nx,Isenberg:2002jg}
concerning the connection between the Einstein equations in the vicinity of a space--like singularity and the equations of motion of a ball moving freely within a billiard plane are only available
for ``non chaotic'' billiards.
\newpage

\begin{table}
\caption{Theories exhibiting Kac--Moody Billiards}
\label{kmth}
\end{table}
{\small \noindent This table lists theories of gravity coupled to dilatons and $p$--forms which exhibit Kac--Moody 
algebras  when the dynamics of the gravitational field in the vicinity of a space--like singularity is 
studied. }
\noindent {\footnotesize 
\begin{itemize} 
\item[$\star$]
As mentioned in the text, a wide class of theories shearing this property is the set of all  \emph{maximally oxidised theories}.  The over--extended  Kac--Moody algebras are listed here under together with the \emph{maximally 
oxidised theory} exhibiting them. 

\item[$\mf{a}_r^{++}$] The oxidation endpoint of gravity in $D=3$ coupled to a non--linear $\sigma$--model based on the coset space $SL(r+1,\RR)/O(r+1)$ --- the Lie algebra of $SL(r+1,\RR)$ is $\mf{sl}(r+1,\RR) = \mf{a}_r$ --- is the pure gravity in $D=r+3$, 
\beq \cL = R \star 1 \nn
\eeq 
which exhibits the Weyl chamber of the over--extended Kac-Moody algebra $A_r^{++}$. 
\item[$\mf{b}_r^{++}$]  The oxidation endpoint of $D=3$ gravity coupled to the non--linear $\s$-model 
based on the coset space $SO(r,r+1)/(SO(r)\times SO(r+1))$ --- the Lie algebra of $SO(r,r+1)$ is $\mf{so}(r,r+1) = \mf{b}_r$ --- is  a $D = r+2$ dimensional  theory which comprises
the metric, a
dilaton, a $2$--form $B$, and a $1$--form $A$.  The
Lagrangian reads
\beq 
\cL_D = R\star 1 - \star d\phi\wedge d\phi - \frac{1}{2}
e^{a\sqrt{2}\phi}\star G\wedge G -\frac{1}{2} e^{
a\frac{\sqrt{2}}{2}\phi} \star F\wedge F,
\nn 
\eeq 
where $a^2 = 8/r$, $G = dB + \frac{1}{2}A\wedge
dA$ and $F = dA$. The Kac--Moody algebras $B_r^{++}$ are hyperbolic for $r \leq 9$. The 
$r=8$ case corresponds to the 
bosonic sector of the low energy limit of the \emph{heterotic String Theory} and 
\emph{type I String Theory}. 
\item[$\mf{c}_r^{++}$] The oxidation endpoint of $D=3$ gravity coupled to the non--linear $\s$-model 
based on the coset space $Sp(r,\RR)/U(r)$ --- the Lie algebra of $Sp(r)$ is $\mf{sp}(r) = \mf{c}_r$ ---
 is the $D=4$ theory whose Lagrangian is
given by  
\beq {\cal L}_4 &=& R\star 1 - \star d\vec\phi\wedge
d\vec\phi -\frac{1}{2} \sum_{\alpha} e^{2 \vec
\sigma_\alpha.\vec\phi} \star (d\chi^\alpha + \cdots)\wedge
(d\chi^\alpha+ \cdots)
\nn \\
&&- \frac{1}{2}\sum_{a=1}^{n-1} e^{\vec e_a.\vec\phi\sqrt{2}}\star
dA^a_{(1)}\wedge dA^a_{(1)} \, , 
\nn 
\eeq
 where the dots in  the brackets 
complete the ``curvatures'' of the $\chi$'s \cite{Cremmer:1999du}.  The
$(r-1)$ dilatons $\vec \phi = (\phi^1,...,\phi^{r-1})$ are
associated with the Cartan subalgebra of  $C_{2r-2}$ and the
$\frac{1}{2}r(r-1)$ axions $\chi^\a$ are associated with the
positive roots of $C_{2r-2}$. The fields $A^a_{(1)}$ are
one--forms. The $\vec \sigma_\alpha$ are the positive roots of
$Sp(2n-2,\RR)$; these can be written in terms of an orthonormalised
basis of $(n-1)$ vectors in Euclidean space ($\vec e_a  \cdot \vec e_b =
\d_{a b}$) $\vec e_a$ ($a = 1, \dots, n-1$) as \be
\vec\sigma_\alpha = \{ \sqrt{2}\vec e_a, \frac{1}{\sqrt{2}} (\vec
e_a \pm \vec e_b),\; a>b\}. \nn
\ee 
\item[$\mf{d}_r^{++}$] The oxidation endpoint of $D=3$ gravity coupled to the non--linear $\s$-model 
based on the coset space $SO(r,r)/(SO(r)\times SO(r))$ --- the Lie algebra of $SO(r,r)$ is $\mf{so}(r,r) = \mf{d}_r$ --- is a  $D=r+2$ dimensional theory whose Lagrangian is  
\beq {\cal L} = R\star 1 - \star d\phi\w d\phi -
\frac{1}{2} \,e^{a\sqrt{2}\phi}\star dB\wedge dB \, , 
\nn 
\eeq 
where $B$ is a
$2$--form and $a^2 = 8/n$. For $r=8$, one gets the last hyperbolic algebra in this family,
namely $DE_{10} \equiv D_8^{++}$, \cite{Damour:2000hv}. For
$r=24$, which is the case relevant to the \emph{Bosonic String}, one gets
$D_{24}^{++}$. 
\item[$\mf{e}_6^{++}$] The oxidation endpoint of  $D=3$ gravity coupled to the non--linear $\s$-model 
based on the coset space $E_6/Sp(4)$ is a theory in  dimension $D=8$. This theory is the smallest obtainable as a truncation of
maximal supergravity in which the $3$--form potential is retained.
It comprises the metric, a dilaton and an axion, $\chi$, together
with the $3$--form, $C$ \cite{Cremmer:1999du}. The $8$--dimensional Lagrangian
is given by 
\beq {\cal L} = R\star 1 - \star d\phi\wedge d\phi
-\frac{1}{2} e^{2\sqrt{2}\phi}\star d\chi\wedge d\chi -
\frac{1}{2}e^{-\sqrt{2}\phi}\star G\wedge G + \chi\,G\wedge G \, , 
\nn
\eeq
where $G=dC$. 
\item[$\mf{e}_7^{++}$] The oxidation endpoint  of $D=3$ gravity coupled to the non--linear $\s$-model 
based on the coset space $E_7/SU(8)$ is  a consistent (albeit non
supersymmetric) truncation of $D=9$ maximal supergravity to the
theory whose bosonic sector comprises the metric, a dilaton, a
$1$--form, $A$, and a $3$--form potential $C$ \cite{Cremmer:1999du}. The
Lagrangian reads as 
\beq {\cal L}_9 = && R\star 1 - \star
d\phi\wedge d\phi -\frac{1}{2}
e^{\frac{2\sqrt{2}}{\sqrt{7}}\phi}\star dC\wedge dC \nn 
\\ && -\frac{1}{2} e^{-\frac{4\sqrt{2}}{\sqrt{7}}\phi}\star
dA\wedge dA -\frac{1}{2} dC\wedge dC\wedge
A.\nn
\eeq
\item[$\mf{e}_8^{++}$] The oxidation endpoint  of $D=3$ gravity coupled to the non--linear $\s$-model 
based on the coset space $E_8/SO(16)$  is $D=11$--dimensional supergravity whose
bosonic sector is given by 
\beq 
{\cal L} = R\star 1
-\frac{1}{2}\star dC\wedge dC -\frac{1}{6} dC\wedge dC\wedge C \, , 
\nn 
\eeq
$C$ is a $3$--form. This theory  has been postulated to be the low 
energy limit of the \emph{M-theory}. As pointed out in \cite{Damour:2000hv}, $E_8^{++}$ is
also relevant in the asymptotic dynamics of the gravitational field in the vicinity of a 
space--like singularity for the
\emph{type IIA supergravity} in ten
dimensions as well as \emph{type IIB} \cite{Romans:1985tz}  \cite{Bergshoeff:1996ui}. 
The relevance of $E_{10}$ in the supergravity context was
first conjectured in \cite{Julia:1980gr}.
\item[$\mf{f}_4^{++}$] The oxidation endpoint  of $D=3$ gravity coupled to the non--linear $\s$-model 
based on the coset space $F_4/(Sp(3)\times SU(2))$   is a $D=6$
dimensional theory containing the metric, a dilaton $\phi$, an
axion $\chi$, two one--forms $A^{\pm}$, a two--form $B$ and a
self-dual $3$--form field strength $G$ \cite{Cremmer:1999du}. The
Lagrangian is given by 
\beq {\cal L}_6 = && R\star 1 - \star
d\phi\wedge d\phi -\frac{1}{2} e^{2\phi}\star d\chi\wedge d\chi -
\frac{1}{2} e^{-2\phi} \star H\wedge H
\nn \\ 
&& - \frac{1}{2} \star G \wedge G -\frac{1}{2}   e^{\phi}
\star F^{+}\wedge F^{+} -\frac{1}{2}    e^{-\phi} \star
F^{-}\wedge F^{-} \nn \\ 
&& -\frac{1}{\sqrt{2}}\chi \,H\wedge G
-\frac{1}{2} A^+\wedge F^+\wedge H -\frac{1}{2} A^+\wedge
F^-\wedge G \, . \nn 
\eeq 
The field strengths are given in terms
of potentials as follows: 
\beq 
&& F^+ = d A^+ + \frac{1}{\sqrt{2}}
\chi\,dA^- \nn \\ 
&& F^- = dA^- \nn \\ 
&& H = dB + \frac{1}{2} A^-\wedge dA^- \nn \\ 
&& G = d C - \frac{1}{\sqrt{2}}\,\chi\,H - \frac{1}{2}
A^+\wedge d A^-.
\nn
\eeq
\item[$\mf{g}_2^{++}$] 
The oxidation endpoint  of $D=3$ gravity coupled to the non--linear $\s$-model 
based on the coset space $G_2/SU(8)$   is the Einstein-Maxwell system in $D=5$
with an extra $FFA$ Chern--Simons term \cite{Cremmer:1999du}: 
\beq 
{\cal L}_5 = R\star1 -
\frac{1}{2}\star F\wedge F + \frac{1}{3\sqrt{3}} F\wedge F\wedge
A,\quad\quad F=dA \, . \nn 
\eeq
\item[$\star$]  also twisted
overextensions \cite{Henneaux:2003kk} - are associated to gravitational models
that reduce to $\cG/ \cK$ coset models upon toroidal dimensional
reduction to $D=3$. 
\item[$\star$] Several other hyperbolic algebras also appear
in the billiard analysis of $D=4$ and $D=5$ spatially homogeneous
cosmological models \cite{deBuyl:2003za}, see chapter \ref{homogeneous}.  
\item[$\star$] The \emph{hyperbolic} algebras play a special role since they are 
associated with \emph{chaotic} behaviour. The 
question of whether or not there exists a Lagrangian such that  the billiard is 
the fundamental Weyl chamber of a given hyperbolic algebra is addressed in chapter \ref{hyperbolic}. 
\end{itemize}
}

\cleardoublepage
\chapter{General Framework}
\markboth{GENERAL  {F}RAMEWORK}{}
\label{billiard}

The general analysis of \cite{Damour:2002et}, which 
is sketched in the previous pages, is reviewed in detail
in this chapter.  The general models considered are
the Einstein--dilaton--$p$--form systems. Reference \cite{Damour:2002et} (and references
therein) 
establishes that  
 the dynamics of these systems in the vicinity of a space--like singularity can be 
 asymptotically described, at a generic spatial point, as a billiard motion in a 
region of the Poincar\'e disk. 
The following points will be reviewed in detail,
\begin{itemize}
\item Near the space--like singularity, $t \to 0$, due to the decoupling of space 
points,  Einstein's PDE equations become ODE's with respect to time. 
\item The study of these ODE's near $t\to 0$, shows that the $d \equiv D-1$ 
diagonal spatial metric components ``$g_{ii}$'' and the dilaton $\phi$ move on 
a billiard in an auxiliary $d+1\equiv D$ dimensional Lorentz space.
\item All the other field variables ($g_{ij}, i\neq j, A_{i_1...i_p})$  freeze as $t \rightarrow 0$. 
\item In many interesting cases, the billiard tables
can be identified with the fundamental Weyl chamber of  hyperbolic 
Kac--Moody algebra.
\end{itemize}
In this perspective, the following two ingredients are very powerful,

 \emph{Iwasawa decomposition}: In the previous chapter, the Kasner exponents $p$'s are 
the variables relevant for the description of the BKL solution. For a deeper 
analysis of the asymptotic dynamics the most convenient variables are the 
\emph{Iwasawa} ones \cite{Damour:2002et}. More precisely, one uses a 
decomposition of the spatial
metric g given by  g$= \, ^T\cN \cA^2 \cN$ where $\cN$ is an upper triangular matrix
with 1's on the diagonal and $\cA =$ diag$(e^{-\b^1},\cdots,e^{-\b^d})$. The 
scale factors $\b$ are the interesting dynamical variables, with non trivial dynamics in the asymptotic regime;
the $\cN$'s get ``frozen'' in the vicinity of the singularity. 

\emph{Hamiltonian formalism} : The presence of a space--like singularity 
naturally leads to a splitting of spacetime into space and time so the 
Hamiltonian formalism is well adapted to the BKL limit.

\section{General Models}

The general systems considered here are of the following form
\beq &&S[ g_{\m\n}, \phi, A^{\sst{(p)}}] = \int d^D x \, \sqrt{- 
^{(D)}g} \;
\Bigg[R (g) - \partial_\m \phi \partial^\m \phi \nn \\
&& \hspace{2.5cm} - \frac{1}{2} \sum_p \frac{1}{(p+1)!} e^{\l_p
\phi} F^{\sst{(p)}}_{\m_1 \cdots \m_{p+1}} F^{\sst{(p)}  \, \m_1 \cdots \m_{p+1}}
\Bigg] + \dots .~~~~~~~ \label{keyaction} \eeq Units are
chosen such that $16 \pi G_N = 1$,  $G_N$ is Newton's
constant and the spacetime dimension $D \equiv d+1$ is left
unspecified. Besides the standard Einstein--Hilbert term the above
Lagrangian contains a dilaton 
field $\phi$ and a number of $p$--form fields $A^{\sst{(p)}}_{\m_1 \cdots \m_p}$ (for 
$p\geq 0$).  The generalisation to any number of dilatons is
straightforward. The $p$--form field strengths $F^{\sst{(p)}} = dA^{\sst{(p)}}$ are
normalised as \be F^{\sst{(p)}}_{\m_1 \cdots \m_{p+1}} = (p+1)
\partial_{[\m_1} A^{\sst{(p)}}_{\m_2 \cdots \m_{p+1}]} \equiv
\partial_{\m_1} A^{\sst{(p)}}_{\m_2 \cdots \m_{p+1}} \pm p \hbox{
permutations }. 
\nn
\ee As a
convenient common formulation we have adopted the Einstein conformal
frame and normalised the kinetic term of the dilaton $\phi$ with
weight one with respect to the Ricci scalar. The Einstein metric
$ g_{MN}$ has Lorentz signature $(- + \cdots +)$ and is used
to lower or raise the indices; its determinant is denoted by $
g$. The dots in the action (\ref{keyaction}) above
indicate possible modifications of the field strength by
additional Yang--Mills or Chapline--Manton-type couplings
\cite{Bergshoeff:1981um,Chapline:1982ww}. 
The real parameter $\l_p$ measures the strength
of the coupling of $A^{\sst{(p)}}$ to the dilaton, sometimes we will use $\td \l_p =  \l_2 /2$. When $p=0$, we assume
that $\l_0\neq 0$ so that there is only one dilaton  in order to simplify the notations.

At the singularity, which is chosen to lie in the past for definiteness, the proper time $t$  is assumed to remain finite and to decrease towards $0^+$.  
Irrespectively of the choice of coordinates, the spatial volume density is assumed to collapse to zero at each spatial point as one goes to the singularity. 

\section{Arnowitt--Deser--Misner Hamiltonian Formalism}

To focus on the features relevant to the billiard picture, we assume here that
there are no Chern--Simons and no Chapline--Manton terms and that the
curvatures $F^{\sst{(p)}}$ are abelian, $F^{\sst{(p)}}= d A^{\sst{(p)}}$. That such additional 
terms do not alter the asymptotic dynamical analysis has been proven in \cite{Damour:2002et}. In 
any \emph{pseudo-Gaussian gauge }($N_i = g_{0i}=0$ ) and in the \emph{temporal gauge }(
$A^{\sst{(p)}}_{0 i_2...i_p}=0$, $\forall p$), the Arnowitt-Deser-Misner Hamiltonian
action \cite{Arnowitt:1962hi}, reads ( see appendix \ref{hamilton} for details) 
\beq && S\left[ g_{ij}, \pi^{ij}, \phi,
\pi_\phi, A^{\sst{(p)}}_{j_1 \cdots j_p},
\pi_{\sst{(p)}}^{j_1 \cdots j_p}\right] = \nn \\
&& \hspace{1cm} \int dx^0 \int d^d x \left( \pi^{ij} \dot{g_{ij}}
+ \pi_\phi \dot{\phi} + \frac{1}{p!}\sum_p \pi_{\sst{(p)}}^{j_1 \cdots
j_p} \dot{A}^{\sst{(p)}}_{j_1 \cdots j_p} - H \right)\,,
\label{GaussAction} \eeq  where the Hamiltonian density $H$ is
\beq
\label{Ham}
H &\equiv&  \tilde{N} \ch \, ,\\[2mm]
\label{Ham1}
\ch &=& \ck + \cm \, ,\\[2mm]
\ck &=& \pi^{ij}\pi_{ij} - \frac{1}{d-1} \pi^i_{\;i} \pi^j_{\;j}
+ \frac{1}{4} \pi_\phi^2 
+ \sum_p \frac{e^{- \lambda_p \phi}} {2 \, p!} \, \pi_{\sst{(p)}}^{j_1
\cdots j_p}
\pi_{\sst{(p)} \, j_1 \cdots j_p} \, , 
\label{kinetic} \\[2mm]
\cm &=& - g R + g g^{ij} \partial_i \phi \partial_j \phi + \sum_p
\frac{e^{ \lambda_p \phi}}{2 \; (p+1)!} \, g \, F^{\sst{(p)}}_{j_1
\cdots j_{p+1}} F^{\sst{(p)} \, j_1 \cdots j_{p+1}}\,, \label{hamm}
\eeq

\noindent where $R$ is the spatial curvature scalar, $\tilde{N} = 
N/\sqrt{g}$ is the rescaled lapse and $g$ is the determinant of the 
spatial metric. The dynamical
equations of motion are obtained by varying the above action with
respect to the spatial metric components, the dilaton, the spatial
$p$--form components and their conjugate momenta. In addition,
there are constraints on the dynamical variables,
\beq
\ch &\approx& 0  \; \; \; \; \; \; \hbox{(``Hamiltonian constraint")}, \nn \\[2mm]
\ch_i &\approx& 0  \; \; \; \; \; \; \hbox{(``momentum constraint")}, \nn \\[2mm]
\varphi_{\sst{(p)}}^{j_1 \cdots j_{p-1}} &\approx& 0 \; \; \; \;\; \;
\hbox{(``Gauss law" for each $p$--form), } \nn \eeq with
\beq \ch_i &=& -2 {\pi^j}_{i|j} + \pi_\phi
\partial_i \phi + \sum_p \frac1{p!} \
\pi_{\sst{(p)}}^{j_1 \cdots j_p} F^{\sst{(p)}}_{i j_1 \cdots j_{p}} \,, \nn \\[2mm]
\varphi_{\sst{(p)}}^{j_1 \cdots j_{p-1}} &=& {\pi_{\sst{(p)}}^{j_1 \cdots
j_{p-1} j_p}}_{\vert j_p}\,, \nn  \eeq where the subscript $|j$ stands
for spatially covariant derivative. 

\section{Iwasawa Decomposition of the Spatial Metric}

The metric $g_{ab}$ can be build out of the co--vielbeins  matrix $e^{\sst{(c)}}{}_a$ as
$g_{ab}Ê= e^{\sst{(c)}}{}_a e^{\sst{(d)}}{}_b \eta_{\sst{(c)(d)}}$, 
see appendix \ref{cartan}.  This matrix $e^{\aaa}{}_{b}$ defines the co--vielbeins $\th^\aaa$ trough  
$\th^\aaa= e^{\aaa}{}_{b} dx^b$ and 
 can be seen as an element of 
$GL(D,\RR)$. Every element of $GL(D,\RR)$ can be written as the product of a 
matrix proportionnal to the identity and an element of $SL(D,\RR)$. The Iwasawa decomposition of $SL(D,\RR)$ allows to decompose each
element of this group as the product of three elements $\cK$, $\cA$ and 
$\cN$ such that $\cK$ belongs to $SO(D,\RR)$, $\cA$ is a traceless diagonal 
matrix and $\cN$ is an upper triangular matrix with 1's on the
diagonal. One chooses the gauge $\cK = 1$. 
In terms of the Iwasawa decomposition the veilbein and the metric are,
\beq 
e^{-1} &=& \cA \, \cN \hspace{4cm} e^{\sst{(c)}}{}_a  = e^{-\b^{\sst{(c)}}} \cN^{\sst{(c)}} {}_a \nn \\
g &=& \ ^t\cN \cA^2 \cN \hspace{3.5cm} g_{ab}Ê= \cN_{a}{}^{\ccc} \cA_{\ccc \dd} \, \cN^\dd{}_b
\label{Iwasawaex} \eeq with
\beq \cA = \left( \begin{array}{ccccc} 
e^{-\b^1} & 0 & 0 & \cdots & 0 \\
0 & e^{-\b^2} & 0 & \cdots &  0\\
\vdots &  & \ddots &  &  \\
0 & &  &  &e^{-\b^d} 
\end{array} \right)  
%\label{beta} 
\nn
\eeq 
and 
\beq
 \cN = \left( \begin{array}{cccccc} 
1 & \cN^1{}_2 &     &   &  \cdots & \cN^1{}_d \\
0 & 1         & \cN^2{}_3 &   & \cdots & \cN^2{}_d \\
0 & 0         &  1        & \cN^3{}_d & \cdots & \cN^3{}_d \\
\vdots &  &  &  &   \ddots &  \\
0 &  & \cdots &  &  & 1
\end{array} \right) \ . 
\label{ns}   
\nn
\eeq
One will also need the Iwasawa coframe $\{ \theta^\aaa_{iw} \}$,
\be
\label{Iwasawa1} 
\theta^\aaa_{iw} = {\cN^\aaa}_i \, dx^i\,, 
\ee 
as well as
the  frame $\{ e_{iw \, \aaa} \}$ ---also called Iwasawa frame---  dual to the coframe $\{ \theta_{iw}^\aaa
\}$, 
\be
%\label{Iwasawa2} 
e_{iw \, \aaa} = \cN^{i}{}_\aaa \frac{\partial}{\partial
x^i} \, .
\nn 
\ee 
The matrix ${\cN^{i}}_\aaa$ is
again an upper triangular matrix with 1's on the diagonal since it is the inverse of $\cN^\aaa{}_i$. 

\section{Hamiltonian in the Iwasawa Variables \label{moredetails}}

The Hamiltonian action gets transformed when one
performs, at each spatial point, the Iwasawa decomposition
of the spatial metric (for more details see appendix \ref{hamilton}). The change of 
variables $(g_{ij}\to \beta^\aaa, 
{\cn^\aaa}_i )$ corresponds
to a point transformation and can be extended to the momenta as a canonical 
transformation in the standard way via 
\be
\label{cantra}
\p^{ij}\dot{g}_{ij} \equiv \sum_a \pi_\aaa \dot{\b}^\aaa + \sum_{a}
{P^i}_\aaa \dot{{\mathcal N}^\aaa}_{i} \,\,. 
\ee 
To avoid the excess of parentheses, we will denote $\b^\aaa$ 
simply by $\b^a$ in the exponentials. The momenta ${P^i}_\aaa$ and $\pi_\aaa$ are
(see Eqs.(\ref{NNmomenta})),
\beq   
%\label{Nmomenta} 
\nn
{P^i}_\aaa 
&=& \sum_{b<a} e^{2(\beta^b - \beta^a)}
{\dot\cN}^\aaa{}_j {\cN^{-1j}}_\bb {\cN^{-1i}}_\bb \\
%\label{Bmomenta} 
\nn 
\pi_\aaa &=&  2 \ \tilde{N}^{-1}  G_{\aaa \bb}\dot{\b}^\bb \ .
\eeq 
where $G_{\aaa \bb}$ is the quatratic form given by
\beq 
G_{\aaa \bb}d\b^\aaa d\b^\bb = \sum_c (d\b^\ccc)^2 - (\sum_c d\b^\ccc)^2 \ . 
\label{metricb} 
\eeq 
Note that the momenta conjugate to the nonconstant off-diagonal Iwasawa 
components ${\cN^\aaa}_i$ are only defined for $a<i$; hence the second sum in 
(\ref{cantra}) receives only contributions from $a<i$.
In the following, 
the $\b^\aaa$ and $\phi$ are collectively denoted by $\b^\m = (\b^\aaa,\phi)$.
The metric (\ref{metricb}) generalises to 
\beq 
G_{\m\n} d\b^\m d\b^\n = \sum_c (d\b^\ccc)^2 - 
(\sum_c d\b^\ccc)^2 + d\phi^2 \hspace{1cm} \b^\m = (\b^\aaa,\phi) \ . 
\label{metric}
\eeq
$ \pi_\mu \equiv (\pi_\aaa, \pi_\phi)$ are the momenta
conjugate to $\beta^\aaa$ and $\phi$, respectively, i.e. 
\be 
\pi_\m =
2 \tilde{N}^{-1} G_{\m \n} \dot{\beta}^\n = 2 G_{\m \n} \frac { d
{\beta}^\n}{d\tau}\, . 
\nn
\ee
Reexpressing the Hamiltonian $H$ (\ref{Ham}) in terms of the Iwasawa
variables modifies ``non trivially'' only the first two terms of 
$\cK$ (\ref{kinetic}) and the first term of $\cm$ (\ref{hamm}),  
{\it i.e.} pure gravity terms. The Hamilton $H$ (\ref{Ham}) reads
(see Eq.(\ref{hamiwa})
of appendix \ref{hamilton}), 
\beq
H &=& 
\frac{\tilde{N} }{4} G^{\aaa \bb}\pi_\aaa \pi_\bb  
+{1\over 4 \tilde{N}} \sum_{b<d} e^{2(\b^d-\b^b)} (\cN^\dd{}_iP^i{}_\bb)^2 
+ \frac{\td{N}}{4} \pi_\phi^2 \nn \\
&+&   \frac{\td{N}}{2 \, p!} \sum_{a_1, a_2, \cdots, a_p} e^{-2 e_{a_1
\cdots a_p}(\b)} (\ce^{\sst{(a_1)}  \cdots \sst{(a_p)}})^2 \nn \\
&- &  \td{N} g^{(d)} R  + \td{N} \sum_a e^{-\mu_a(\beta)} (\mathcal{N}^{-1 \, i}{}_\aaa \partial_i
\phi)^2\nn \\ 
 &+ &  \frac{\td{N}}{2 \, (p+1)!} \sum_{a_1, a_2, \cdots, a_{p+1}} e^{-2
m_{a_{1}  \cdots a_{p+1}}(\b)} (\cf_{\sst{(a_1)}  \cdots \sst{\sst{(a_{p+1})}}})^2 
\, ,
\label{kiwa}
\eeq 
where the various variables are expressed in the Iwasawa basis, 
\beq 
\ce^{ \sst{(a_1)} \cdots \sst{(a_p)}} &\equiv& {\cn^{\sst{(a_1)}}}_{j_1} {\cn^{(a_2)}}_{j_2}
\cdots {\cn^{\sst{(a_p)}}}_{j_p} \pi^{j_1 \cdots j_p}\,, \nn \\
\cf_{\sst{(a_1)} \cdots \sst{(a_{p+1})}} & \equiv& \cN^{ j_1}{}_{\sst{(a_1)}} \cdots
{\cN^{ j_{p+1}}}_{\sst{(a_{p+1})}} F_{j_1 \cdots j_{p+1}}\,  , \nn 
\eeq 
and 
\beq 
& &e_{a_1 \cdots
a_p}(\b) = \b^{a_1} + \cdots + \b^{a_p} + \frac{\l_p}{2} \phi \, , \nn  \\
& &m_{a_{1} \cdots a_{p+1}}(\b) = \sum_{b \notin \{a_1,a_2,\cdots
a_{p+1}\}} \!\b^b - \frac{\l_p}{2}\, \phi
\nn \, , \\
& & \mu_a(\b) = \sum_e \b^e - \b^a \nn
\, .
\eeq 
The indices in the Iwasawa basis are raised and lowered with the diagonal metric $\cA^2$. The expression of the spatial scalar curvature $^{(d)}R$ in terms of 
the Iwasawa variables is given in appendix \ref{cartan}.

\section{Splitting of the Hamiltonian}

The Hamiltonian $\ch$  (\ref{Ham}) density is next split in two
parts: ${\mathcal H}_0$, which is the kinetic term
for the local scale factors and the dilaton $\beta^\mu= (\beta^\aaa, \phi)$, and
$\cv$, a ``potential density'' (of
weight 2) , which contains everything else. The
analysis below will show why it makes sense to put the kinetic
terms of both the off-diagonal metric components and the $p$--forms
together with the usual potential terms, i.e. the term $\mathcal M$ in
(\ref{Ham1}).  Thus,  one writes
\be 
\ch =  {\mathcal H}_0 + \cv
\label{HplusV} 
\ee 
with the kinetic term of the $\b$ variables
\be
\label{eq3.23} 
{\mathcal H}_0 = \frac{1}{4}\, G^{\mu\nu}
\pi_\mu \pi_\nu\,, 
\ee 
where $G^{\mu\nu}$ denotes the inverse of
the metric $G_{\mu\nu}$ of Eq.~(\ref{metric}) ($\cH_0$ includes the 
first term of (\ref{kiwa}) and the third of (\ref{kinetic})). In other words,
the right hand side of Eq.~(\ref{eq3.23}) is defined by
\be
\label{Gmunuup}
G^{\mu \nu} \pi_\mu \pi_\nu \equiv \sum_{a=1}^d \pi_\aaa^2 -
\frac{1}{d-1} \left(\sum_{a=1}^d \pi_\aaa\right)^2 + \pi_\phi^2\,,
\ee
The total (weight 2) potential density,
\be 
\cv = \cv_S + \cv_G + \sum_p \cv_{p}  + \cv_\phi\, , 
\nn
\ee 
is
naturally split into a ``centrifugal'' part $\cv_S$ linked to the kinetic
energy of the off-diagonal components (the index $S$ referring to
``symmetry,''), a ``gravitational'' part $\cv_G$, a term
from the $p$--forms, $\sum_p \cv_{p}$, which is a sum of an ``electric'' and a
``magnetic'' contribution and also a  contribution to
the potential coming from the spatial gradients of the dilaton
$\cv_\phi$.
\subsubsection{``Centrifugal'' Potential}
\beq 
\nn 
\cv_S = \frac{1}{2} \sum_{a<b}
e^{-2(\beta^b - \beta^a)} \left( {P^j}_\bb {{\mathcal
N}^\aaa}_j\right)^2, 
\eeq

\subsubsection{ ``Gravitational'' (or ``Curvature'') Potential}
\beq 
\nn 
\cv_G =  - g R\, = \frac{1}{4}
{\sum_{a\neq b \neq c}} e^{-2\a_{abc}(\beta)} (C^a_{\; \; bc})^2 -
\sum_a e^{-2 \m_a(\beta)} F_a\,, 
\eeq 
where  
\beq 
d\theta^\aaa =-\frac{1}{2}C^a_{\; \; bc}\theta^\bb \wedge\theta^\ccc
\nn
\eeq 
and the only property of the $F_a$ that will be of importance here is that they are polynomials of
degree two in the first derivatives $\partial \beta$ and of degree one
in the second derivatives $\partial^2 \beta$. 
The 
$\a_{abc}(\b)$ are given in the next section.  
\subsubsection{$p$--form Potential}
\beq
 \cv_{(p)} &=& \cv_{(p)}^{el} + \cv_{(p)}^{magn}\,,
\nn \\
\cv_{(p)}^{el} &=& \frac{1}{2 \, p!} \sum_{a_1, a_2, \cdots, a_p} e^{-2 e_{a_1
\cdots a_p}(\b)} (\ce^{\sst{(a_1)}  \cdots \sst{(a_p)}})^2 \,, \nn \\
\cv_{(p)}^{magn} &=&
\frac{1}{2 \, (p+1)!} \sum_{a_1, a_2, \cdots, a_{p+1}} e^{-2
m_{a_{1}  \cdots a_{p+1}}(\b)} (\cf_{\sst{(a_1)}  \cdots \sst{(a_{p+1})}})^2 \,.
\nn
\eeq 
\subsubsection{Dilaton Potential}
\beq 
\cv_\phi  &=& g g^{ij} \partial_i \phi \partial_j \phi\ \nn \\
&=& \sum_a e^{-\mu_a(\beta)} (\mathcal{N}^{-1 \, i}{}_\aaa \partial_i
\phi)^2 \, . \nn 
\eeq 

\section{Appearance of Sharp Walls in the BKL Limit}

In the decomposition of the hamiltonian as $\mathcal{H} = \mathcal{H}_0 +
\cv$, $\mathcal{H}_0$ is the kinetic term for the $\beta^\mu$'s while all 
other variables now only appear through the potential $\cv$ which is
schematically of the form 
\beq 
\label{V1} 
\cv(
 \b^\m, \partial_x \b^{\m}, P,Q)
=\sum_A c_A( \partial_x \b^{\m}, P,Q) \exp\big(-
2 w_A (\beta) \big)\,, 
\eeq  
where $(P,Q) = ({\cn^\aaa}_i, {P^i}_\aaa,
\ce^{\sst{(a_1)} \cdots \sst{(a_p)}},\cf_{\sst{(a_1)} \cdots \sst{(a_{p+1})}})$. Here $w_A (\beta)
= w_{A \m} \b^\m$ are the linear wall forms already introduced above: 
\beq 
\mbox{symmetry walls}&:& w^S_{ab}\equiv \beta^b - \beta^a; \quad a<b
\label{centrifugal}\\ 
\mbox{gravitational walls}&:& \a_{abc}(\b) \equiv \sum_e \beta^e + \beta^a-
\beta^b-\beta^c,\, a\neq b, b\neq c, c\neq a\label{gravitational}\\ 
&\,& \m_a(\beta)\equiv\sum_e \beta^e -\beta^a,\label{gravitational2}\\ 
\mbox{electric walls}&:& e_{a_1 \cdots a_p}(\b)\equiv\beta^{a_1}+...+
\beta^{a_p} + \frac{1}{2}\lambda_p\phi,\label{electric}\\ 
\mbox{magnetic walls}&:& 
m_{a_1 \cdots a_{p+1}}(\b)\equiv\sum_e \beta^e -\beta^{a_1}-...-
\beta^{a_{p+1}}-   \frac{1}{2}\lambda_p\phi. 
\label{magnetic}
\eeq  
In order to take the 
limit $t\rightarrow 0$ which corresponds to the squared norm $G_{\m\n}\b^\m \b^\n$ of the vector
$\b^{\m} $ tending to infinity and this vector pointing toward the future, one decomposes $\b^{\m}$
into hyperbolic polar coordinates $(\rho,\g^{\m})$, i.e. 
\beq
\b^{\m} = \rho \g^{\m} 
\nn
\eeq 
where $\g^{\m}$ are coordinates on the
future sheet of the unit hyperboloid, constrained by 
\beq
G_{\m\n} \g^{\m} \g^{\n} \equiv \g^{\m} \g_{\m} = -1
\nn
\eeq 
and
$\rho$ is the time--like variable defined by 
\beq 
\rho^2 \equiv - G_{\m \n}
\b^{\m} \b^{\n} \equiv - \b_{\m} \b^{\m} >0, 
\nn
\eeq 
which behaves like $\rho
\sim -\ln t \to +\infty$ at the BKL limit. In terms of these
variables, the potential term looks like
\beq 
\sum_A c_A(  \partial_x \g^{\m}, P,Q) \rho^2 \exp\big(- 2
\rho w_A (\g) \big)\,. 
\nn
\eeq
The essential point now is that, since $\r \to +
\infty$, each term $\rho^2 \exp\big(- 2 \rho w_A (\gamma) \big)$
becomes a {\it sharp wall potential}, i.e. a function of  $w_A
(\gamma)$ which is zero when $w_A (\gamma) >0$, and  $+\infty$
when $w_A (\gamma) < 0$. To formalise this behaviour one defines the
sharp wall $\Theta$-function\,\footnote{One should more properly
write $\Theta_\infty(x)$, but since this is the only step function
encountered here, one uses the simpler notation
$\Theta(x)$ which usually satisfies $\Theta(x) = 1$ for $x >0$. } as \be \Theta (x) := \left\{ \begin{array}{ll}
                      0  & \mbox{if $x<0$} \,,\\[1mm]
                      +\infty & \mbox{if $x>0$}\,.
                      \end{array}
                      \right.
\nn
\ee 
A basic formal property of this $\Theta$-function is its
invariance under multiplication by a positive quantity. Because
all the relevant prefactors $c_A( \partial_x \b^{\m}, P,Q)$ are
generically {\it positive} near each leading wall [the $c_A$'s for the gravitational walls $\m_a$ are potentially dangerous since one has no  clear control on their signs. But since they are \emph{behind } others they can be asymptotically neglected], one can formally
write
\beq
\lim_{\rho\rightarrow\infty} && \Big[ c_A( \partial_x \b^{\m},Q,P) \rho^2
\exp\big(-\rho w_A (\gamma) ) \Big] = c_A(Q,P)\Theta\big(-2
w_A (\gamma) \big) \nn\\[2mm]
&\equiv &\Theta\big(- 2 w_A (\gamma) \big) \, 
\nn
\eeq
valid in spite of the increasing of the spatial gradients \cite{Damour:2002et}.
Therefore, the limiting dynamics is equivalent to a free motion in
the $\b$-space interrupted by reflections against hyperplanes in
this $\b$-space given by $w_A (\b) = 0$ which correspond to a
potential described by the sum of infinitely high step functions
\beq 
\cv(\b, P,Q) = \sum_A \Theta\big(-2 w_A (\g) \big) 
\nn
\eeq 
The
other dynamical variables (all variables but the $\b^\mu$'s) disappear  from this
limiting Hamiltonian and therefore get frozen as $t\rightarrow
0$. Indeed, their equations of motion are schematically $\dot{f} = [f,H] = 0$. The volume of the region where 
$\cv = 0$, \ie the area given by the inequalities $w_A(\b) > 0 \quad \forall A $ and delimited by the hyperplanes $w_A=0$, is 
called the \emph{billiard}.
This derivation is explained in more details in \cite{Damour:2002et}. 

\section{Cosmological Singularities and Kac--Moody Algebras}

\begin{figure}[h]
  \centering
\begin{picture}(0,0)%
\epsfig{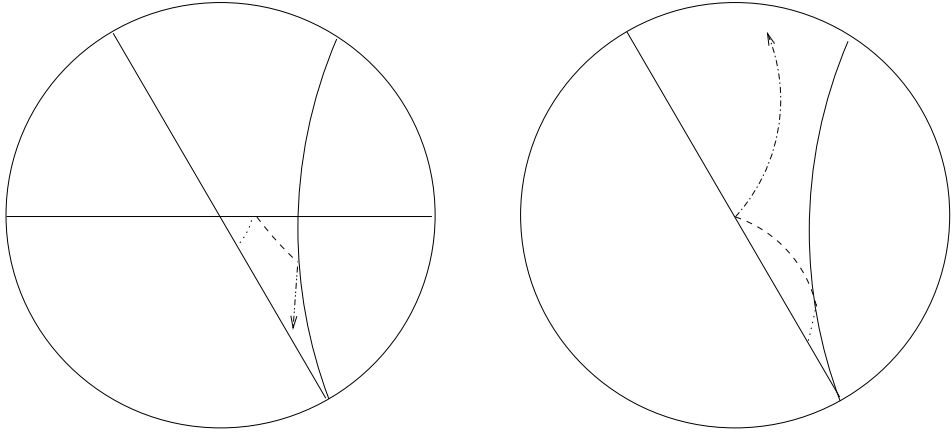}%
\end{picture}%
\setlength{\unitlength}{2486sp}%
\begingroup\makeatletter\ifx\SetFigFont\undefined%
\gdef\SetFigFont#1#2#3#4#5{%
  \reset@font\fontsize{#1}{#2pt}%
  \fontfamily{#3}\fontseries{#4}\fontshape{#5}%
  \selectfont}%
\fi\endgroup%
\begin{picture}(9424,4256)(732,-5322)
\end{picture}%
  \caption{\small{The panels represent the billiard tables 
  (and billiard motions) after projection onto
hyperbolic space  $H_{2}$ (as in the $D=4$ pure gravity case). The hyperbolic 
space $H_{2}$ is
represented by its image in the Poincar\'e disk. The billiard volume of the left pannel is
finite and therefore the billiard motion is chaotic. On the contrary, the billiard volume 
of the right pannel is infinite and after a finite number of reflections the ball freely move .  
 }}
  \label{projet2}
\end{figure}

Recall that the billiard is determined by the conditions,
\beq 
w_A(\b) > 0 \quad \forall A . 
\nn
\eeq
Two kinds of motion are possible according to whether the volume of the
\emph{projected billiard }is finite or infinite. This depends on the fields present in the Lagrangian, on their dilaton-couplings 
and on the spacetime dimension. The
finite volume case corresponds to never--ending, chaotic oscillations for the 
$\beta$'s while in the infinite volume case, after a finite number of 
reflections off the walls, they tend to an asymptotically monotonic 
Kasner--like behavior, see Figure \ref{projet2}. The \emph{billiard} refers to the area of 
the $\b$--space bounded by the hyperplanes  $w_A(\b) = 0 $ on which the
billiard ball undergoes spe\-cular reflections; the  \emph{projected 
billiard} corresponds to this area projected on the future light cone of the unit 
hyperboloid $\g^\m \g_\m = -1$.  

In fact, not all the walls are relevant for determining
the billiard table. Some of the walls stay behind the
others and are not met by the billiard ball. Only a subset of the walls 
$w_A(\b)$, called \emph{dominant walls} and here denoted $\{w_i(\b)\}$ are needed to 
delimit the hyperbolic domain. Once the dominant walls $\{w_i(\b)\}$ are
found, one can compute the following matrix
\beq 
A_{ij} \equiv 2 { (w_i, w_j) \over (w_{i},  w_{i}) } 
\label{cartanmatrix}
\eeq 
where
$(w_i,  w_j )= G^{\m\n} w_{i\m} w_{j\n }$. By definition,  the diagonal elements  
are all equal to 2. Moreover, in many interesting cases, the off-diagonal 
elements happen to be non positive integers. These are precisely the 
characteristics of a generalised Cartan matrix, namely that  of an infinite 
Kac--Moody algebra. The hyperbolic Kac--Moody algebras are those relevant for chaotic billiards 
since their fundamental Weyl chamber has a finite volume. 
As previously emphasised, for the \emph{physically relevant
theories}, the matrix (\ref{cartanmatrix}) \emph{is} the Cartan matrix of a Kac--Moody algebra, see
the list at the end of the first chapter. Here we recall in Table \ref{cordes} the hyperbolic 
Kac--Moody algebras relevant in the context of pure gravity, string theories and M-theory. 

\begin{table}[h]
\label{cordes}
\begin{center}
\begin{tabular}{|c|p{6.5cm}|}
\hline Theory & Hyperbolic Kac--Moody algebra \\  \hline Pure gravity in
$D \leq 10$ & \scalebox{.5} {
\begin{picture}(180,60)
%nom des racines
\put(5,-5){$\alpha_{1}$} \put(45,-5){$\alpha_2$}
 \put(125,-5){$\alpha_3$}  \put(50,45){$\alpha_{D-1}$}\put(85,-5){$\alpha_4$}
  \put(140,45){$\alpha_5$}
%quatre vertex + lignes simples
\thicklines \multiput(10,10)(40,0){4}{\circle{10}}
\multiput(15,10)(40,0){3}{\line(1,0){30}}
%deux vertex du dessus
\multiput(90,50)(40,0){2}{\circle{10}}
\put(130,15){\line(0,1){30}} \put(50,15){\line(1,1){35}}
\dashline[0]{2}(95,50)(105,50)(115,50)(125,50)
\end{picture}
}
 \\ \hline
 M-theory, IIA and  IIB Strings & \scalebox{.5}{
\begin{picture}(180,60)
%nom des racines
\put(5,-5){$\alpha_{1}$} \put(45,-5){$\alpha_2$}
\put(85,-5){$\alpha_3$}
 \put(125,-5){$\alpha_4$}
  \put(165,-5){$\alpha_5$} \put(205,-5){$\alpha_6$}
  \put(245,-5){$\alpha_7$}   \put(285,-5){$\alpha_8$}
  \put(325,-5){$\alpha_9$}
  \put(260,45){$\alpha_{10}$}
%9 vertex + lignes simples
\thicklines \multiput(10,10)(40,0){9}{\circle{10}}
\multiput(15,10)(40,0){8}{\line(1,0){30}}
%1 vertex du dessus
\put(250,50){\circle{10}} \put(250,15){\line(0,1){30}}
\end{picture}
 }
 \\ \hline
 type I and heterotic Strings & \scalebox{.5}{
\begin{picture}(180,60)
%nom des racines
\put(5,-5){$\alpha_{1}$}
\put(45,-5){$\alpha_2$}\put(85,-5){$\alpha_3$}
 \put(125,-5){$\alpha_{4}$}
  \put(165,-5){$\alpha_{5}$}
\put(205,-5){$\alpha_{6}$} \put(245,-5){$\alpha_{7}$}
\put(285,-5){$\alpha_{8}$} \put(325,-5){$\alpha_{9}$}
  \put(70,45){$\alpha_{10}$}
%5 vertex + lignes simples
\thicklines \multiput(10,10)(40,0){9}{\circle{10}}
\multiput(15,10)(40,0){7}{\line(1,0){30}}
\dashline[0]{2}(95,10)(105,10)(115,10)(125,10)
%double derni\`{A}re ligne
\put(295,7.5){\line(1,0){30}}\put(295,12.5){\line(1,0){30}}
%fl\`{A}che vers la droite
\put(305,0){\line(1,1){10}} \put(305,20){\line(1,-1){10}}
%un vertex du dessus
\put(90,50){\circle{10}} \put(90,15){\line(0,1){30}}
\end{picture}
} \\ \hline
closed bosonic string in $D=10$ & \scalebox{.5}{
\begin{picture}(180,60)
%nom des racines
\put(5,-5){$\alpha_{1}$} \put(45,-5){$\alpha_2$}
\put(85,-5){$\alpha_3$}
 \put(125,-5){$\alpha_4$}
  \put(165,-5){$\alpha_5$} \put(205,-5){$\alpha_6$}
  \put(245,-5){$\alpha_7$}   \put(285,-5){$\alpha_8$}
  \put(100,45){$\alpha_9$}
  \put(260,45){$\alpha_{10}$}
%8 vertex + lignes simples
\thicklines \multiput(10,10)(40,0){8}{\circle{10}}
\multiput(15,10)(40,0){7}{\line(1,0){30}}
%2 vertex du dessus
\put(250,50){\circle{10}} \put(250,15){\line(0,1){30}}
\put(90,50){\circle{10}} \put(90,15){\line(0,1){30}}
\end{picture} }\\ \hline
\end{tabular}
\end{center}
\caption{\footnotesize{This table displays the Coxeter--Dynkin diagrams which 
encode the geometry of the billiard tables describing the asymptotic 
cosmological behavior of General Relativity and of
three blocks of string theories: ${\mathcal B}_2 = \{$$M$-theory,
type IIA and type IIB superstring theories$\}$, ${\mathcal B}_1 =
\{$type I and the two heterotic superstring theories$\}$, and
${\mathcal B}_0 = \{$closed bosonic string theory in $D=10\}$.
Each node of the diagrams represents a dominant wall of the
cosmological billiard. Each Coxeter diagram of a billiard table corresponds to the Dynkin diagram of a (hyperbolic) Kac--Moody algebra: 
$E_{10}$, $BE_{10}$ and $DE_{10}$.}} 
\end{table}

\noindent The precise links between a chaotic billiard and its corresponding 
Kac--Moody algebra can be summarized as follows
\begin{itemize}
\item the scale factors $\b^{\mu}$ parametrize a Cartan element $h = 
\sum_{\m=1}^{r} \b^{\m}h_{\mu}$,
\item  the dominant walls $w_i(\b), (i=1,...,r)$ correspond to the simple roots
$\a_i$ of the Kac--Moody algebra,
 \item the group of reflections in the cosmological billiard is the Weyl group 
of the Kac--Moody algebra, and
\item the billiard table can be identified with the Weyl chamber
of the Kac--Moody algebra.
\end{itemize}
%%%%%%%%%%%%%%%%%%%%%%%%%%%%%%%%
%%%%%%%%%%%%%%%%%%%%%%%%%%%%%%%%
\cleardoublepage
%%%%%%%%%%%%%%%%%%%%%%%%%%%%%%%%%%
%%%%%%%%%%%%%%%%%%%%%%%%%%%%%%%%%%
%%\include{chap_homogene}

\chapter{Homogeneous Cosmologies}
\markboth{HOMOGENEOUS COSMOLOGIES}{}
\label{homogeneous}

The BKL equations describing the asymptotic dynamics of the 
gravitational field in the vicinity of a spacelike singularity
 coincide with the dynamical equations of some spatially
homogeneous cosmological models which exhibit therefore the main qualitative
properties of more generic solutions. For
$D=4$, the spatially homogeneous vacuum models that share the chaotic
behaviour of the more general inhomogeneous solutions are labelled as
Bianchi type IX and VIII; their homogeneity groups are respectively
$SU(2)$ and
$SL(2,\mathbb{R})$. In higher spacetime dimensions, i.e. for $5\leq D\leq
10$, one also knows chaotic spatially homogeneous cosmological models but
none of
them is diagonal \cite{Barrow:1985hy,Barrow:1998rm,Dabrowski:2001zk}. 
In fact, diagonal models are too restrictive to be
able to reproduce the general oscillatory behaviour but, as shown e.g. in
\cite{Demaret:1988sg}, chaos is restored when non-diagonal metric elements are taken into
account.

The purpose of this chapter is to analyse the billiard evolution of
spatially homogeneous non-diagonal cosmological models in $D=4$ or
$5$ spacetime dimensions, in the Hamiltonian formalism. 
Since one knows that the full field content of the
theory is important in the characterisation of the billiard, we
compare the pure Einstein gravity construction to that of the
coupled Einstein--Maxwell system. The gravitational models we are
interested in are in a one--to--one correspondence with the real Lie
algebras -- a complete classification based on their structure
constants exists for $d=3$ and $d=4$, $(D=d+1)$\cite{LL,
MacC2,Patera:1977hg}\footnote{We will use the notations of MacCallum, we refer to
\cite{MacC2} for translation to other notations} -- and we restrict
our analysis to the unimodular ones, because only for such models
can the symmetries of the metric be prescribed at the level of the
action. For the non--unimodular algebras, the addition of boundary terms to the Einstein--Hilbert action is  
necessary to impose the symmetry at the level of the action \cite{Sneddon,Mc}. We proceed along the same lines  as in the chapter \ref{billiard} but with
a special concern about
\begin{itemize}
\item the presence or absence of the each curvature walls, which depends on the structure constant of the particular homogeneity algebra; 
\item  the r\^ole played by the constraints.
Indeed, while in the general inhomogeneous case, the constraints
essentially assign limitations on the spatial gradients of the
fields without having an influence on the generic form of the BKL
Hamiltonian, in the present situation, they precisely relate the
coefficients that control the walls in the potential.
Consequently, the question arises whether they can enforce the
disappearing of some (symmetry, gravitational, electric or magnetic) walls. Because of this, they could prevent the generic oscillatory behaviour of the
scale factors. The answer evidently depends on the Lie algebra
considered and on its dimension: for example, while going from the
Bianchi IX model in $d=3$ to the corresponding $U3S3$ model in
$d=4$, the structure constants remain the same but the momentum
constraints get less restrictive. Hence generic behaviour is
easier to reach when more variables enter the relations. 
\end{itemize}
We find
that, except for the Bianchi IX and VIII cases in $D=4$, symmetry
walls (hence off-diagonal elements) are needed to close the
billiard table: thereby confirming, in the billiard picture,
previous results about chaos restoration. Moreover, we find that
when the billiard has a finite volume in hyperbolic space, it can
again be identified with the fundamental Weyl chamber of one of
the hyperbolic Kac--Moody algebras. In the most generic situation,
these algebras coincide with those already relevant in the general
inhomogeneous case. However, in special cases, new rank 3 or 4
simply laced algebras are exhibited.

The chapter is organised as follows. We first adapt to the spatially
homogeneous case that part of the general Hamiltonian formalism reviewed in the 
previous chapter 
 necessary to understand how, at the BKL limit, the billiard
walls arise in the potential. We explicitly write down the form of the
momentum constraints in the generalised Iwasawa variables and
analyse their meaning for each of the 3 and 4 dimensional real unimodular Lie
algebras as well as their impact on the billiard shape.
For the finite volume billiards, we compute the scalar products
of the gradients of the dominant walls using the metric defined by the kinetic
energy and show that the matrix
 \be A_{AB} =
2 \frac{(w_A\vert w_B)}{(w_A\vert w_A)}\quad\mbox{where}(w_A\vert w_B) =
G^{ab}\,w_{Aa} w_{Bb}\ee
 is the generalised Cartan matrix of an hyperbolic
Kac--Moody algebra.

\section{General Setting }

\subsection{Spatially Homogeneous Models, Hamiltonian}

In this chapter, we are specially interested in $d=3$ and
$d=4$ dimensional spatially homogeneous models equipped with a
homogeneity group simply transitively acting; these models are known to be in a
one-to-one correspondence with the $3$ and
$4$ dimensional real Lie algebras and have been completely
classified \cite{MacC2,Ryan:1975jw}. We restrict our analysis to the
unimodular algebras, i.e. those whose adjoint representation is
traceless\footnote{The group Adjoint representation is unimodular.}, that is
$C^i_{\,\,ik}=0$ , since only for these homogeneous models do the
equations of motion follow from a reduced Hamiltonian action in which the
symmetry of the metric is enforced before taking
variational derivatives \cite{Sneddon,Mc}.

We work in a pseudo-Gaussian gauge
defined by vanishing shift $N^i = 0$ and assume the $D=d+1$ dimensional
spacetime metric
of the form
\be ds^2 = - (N
dx^0)^2 + s_{\ii\jj}(x^0)\,
\omega^\ii\,\omega^\jj \nn \ee where $x^0$ is the time coordinate, $t$ is
the proper time, $dt = -N \,dx^0$, and $N$ is the
lapse. For definiteness,
we will assume that the spatial singularity occurs in the past, for $t=0$.

The gravitational dynamical variables
$s_{\ii\jj}$ are the components of the $d$ dimensional spatial metric in the
time-independent co--frame
$\{\omega^\ii =
\omega^\ii_{\,\, j}\,dx^j\}$ invariant
under the group transformations
\be d\omega^\ii = 
\frac{1}{2}\, \td C^\ii_{\,\,\,\jj\kk}\,\omega^\jj\wedge \omega^\kk \, ; \nn \ee
the $\td C^\ii_{\,\,\, \jj \kk}$'s are the group structure constants. The metric
$s_{\ii \jj}(x^0)$ depends only on time and may contain
off-diagonal elements. With use of $g \equiv det\,s_{ij}$, one defines
the rescaled lapse as $\tilde N = N/\sqrt{g}$.

In the spatially homogeneous Einstein--Maxwell system, there is besides the
metric, an electromagnetic
$1$--form potential $A$ and its $2$--form field strength $F=dA$. In
the temporal gauge $A_0=0$, the potential reduces to
\be A =  A_\jj\,\omega^\jj \, . \nn 
\ee 
In the Hamiltonian framework, we assume the
potential itself to be spatially homogeneous\footnote{This is more restrictive
than requiring spatial homogeneity of the field strength; in the
present analysis the difference only arises with regard to the magnetic walls
which are always subdominant.} so that its space components in the
$\omega^\jj$ frame are functions of
$x^0$ only: $A_\jj = A_\jj (x^0)$. Accordingly, its field strength takes the
special form
\be F= dA = \partial_0\, A_\jj\,dx^0\wedge \omega^\jj -
\frac{1}{2}\,A_\ii\, \td C^\ii_{\,\,\, \jj\kk}\,\omega^\jj\wedge \omega^\kk 
\nn 
\ee 
which shows the
links between the components of the magnetic field and the structure
constants. Hence, from the Jacobi identity, one infers that 
\be
F_{\ii [ \jj}\, \td C^\ii{}_{ \kk\sst{(l)}]} = 0.\label{cmag}\ee

The first order action for the homogeneous Einstein--Maxwell
system can be obtained from the $D$ dimensional Hilbert-Einstein action in
ADM form after space integration has been carried out; this operation
brings in a constant space volume factor that will be ignored hereafter.
The action is given by
\be
S[s_{\ii \jj},\pi^{\ii\jj},A_\jj,\pi^\jj] = \int dx^0 \big( \pi^{\ii\jj}{\dot s}_{\ii\jj} +
\pi_F^\jj{\dot A}_\jj - {\tilde N} H
\big).\label{action}\ee The Hamiltonian ${\tilde N}H$ reads as
\beq
{ H} &=& {K} + { M} \nn \\ 
{K} &=&
\pi^{\ii\jj}\pi_{\ii\jj}-\frac{1}{d-1}\pi^\ii_{\,\,\ii}\pi^\jj_{\,\,\jj}+\frac{1}{2}\pi_F^\jj
\pi_{F\, \jj} \nn \\ 
{ M} &=&-gR +
\frac{1}{4} F_{\ii\jj}F^{\ii\jj}
\nn
\eeq
where $R$ is the spatial curvature scalar defined, in the unimodular
cases, by the following combination of structure constants and metric
coefficients
\be R = -\frac{1}{2}\,( \td C^{\ii\jj\kk}\, \td C_{\jj\ii\kk} +
\frac{1}{2}\, \td C^{\ii\jj\kk}\, \td C_{\ii\jj\kk}\,) \, , \nn \ee 
where
\be 
\td C_{\ii\jj\kk} = s_{\ii \sst{(\ell)}} \, \td C^{\sst{(\ell)}}_{\,\,\,\jj\kk}\quad\mbox{and}
\quad \td C^{\ii\jj\kk} =
s^{\jj \sst{(\ell)}} \,s^{\kk \sst{(m)}} \, \td C^\ii{}_{\sst{(\ell)(m)}} \, .\label{struc2}\ee

The equations of motion
are obtained by varying the action (\ref{action}) with respect to the
spatial metric components $s_{\ii\jj}$, the spatial $1$--form components $A_\jj$
and their respective conjugate momenta $\pi^{\ii\jj}$ and $\pi^\jj$. The
dynamical variables still obey the following constraints:
\beq {H}&\approx & 0\quad \mbox{(Hamiltonian constraint)}\\ 
{H}_i&=&
 - \td C^\jj_{\,\,\, \ii \kk}\,\pi^{\kk}_{\,\, \jj}+
\pi_F^\jj F_{\ii\jj}\approx 0\quad \mbox{(momentum constraints)};\label{momentum}
\eeq
notice that the Gauss law for the electric
field is identically satisfied on account of the unimodularity condition.

\subsection{Generalised Iwasawa Variables}

In order to develop the billiard analysis, it was seen in the previous chapter to be  necessary to
change the variables and uses the Iwasawa decompostion of the spatial metric. This Iwasawa decomposition now is done in the 
invariant frames. More precisely,  the metric components $s_{\ii\jj}$ are replaced by
the new variables
$\beta^a$ and ${\cal  \td N}^\aaa{}_{\ii}$, defined through the Iwasawa
matrix decomposition 
\begin{equation}  
s = {\cal \td N}^T {\cal A}^2{\cal
\td N}\label{Iwas}
\end{equation}
where ${\cal \td N}$ is an upper triangular matrix with
$1$'s on the diagonal and ${\cal A}$ is a diagonal matrix with positive
entries parametrized as
\begin{equation} 
{\cal A} = exp (-\beta),
\qquad
\beta = diag (\beta^1, \beta^2,..., \beta^d) \, . \nn \end{equation} 
The explicit
form of (\ref{Iwas}) reads 
\be 
s_{\ii\jj} = \sum_{a=1}^d\,e^{-2\beta^a}{ \cal \td N}^{\aaa}{}_{\ii}{ \cal \td N}^{\aaa}{}_{\jj} \, .
\label{Iwasa}
\ee 
 The ${ \cal  \td N}^\aaa{}_{\ii}$'s measure the
strenght of the off-diagonal metric components and define how to pass
from the invariant $\{\omega^\ii \}$ co--frame to the Iwasawa co--frame
$\{\th_{iw}^\aaa\}$ -- see Eq. (\ref{Iwasawa1}) -- in which the metric is purely diagonal 
\be
\theta_{iw}^\aaa = {\cal \td
N}^\aaa{}_{\jj}
\omega^\jj \, . \nn
\ee 
In this basis, one has for the components of the $1$--form ${\cal A}_{\aaa}$,
\be 
A_\jj\equiv {\cal A}_{\aaa}{\cal
\td N}^{\aaa}{}_{\jj} \, .
\label{AjAa}\ee 
As in (\ref{cantra}), the
changes of variables (\ref{Iwasa}) and (\ref{AjAa}) are extended to the
momenta as canonical point transformations in the standard way via 
\be
\pi^{\ii\jj}{\dot s}_{\ii\jj} +
\pi^{\jj}{\dot A}_{\jj} = \pi_\aaa
\,{\dot\beta}^\aaa + \sum_{a<j}{\cal \td  P}^\jj{}_{\aaa}\,{\dot{\cal \td
N}}^\aaa{}_{\jj} + {\cal
E}^{a}\,{\dot {\cal A}}_{a} \, . \nn
\ee 
In this expression,
${\cal \td P}^\jj{}_{\aaa}$ denotes the momentum conjugated to ${\cal \td N}^\aaa{}_{\jj}$
and is defined for $a<j$, ${\cal E}^{a}$ denotes the momentum
conjugated to ${\cal A}_{a}$.  The Iwasawa components $ {\cal E}^{\aaa}$ and ${\cal F}_{\aaa \bb}$ of
the electric and magnetic fields are given by
\be {\cal E}^{\aaa}\equiv {\cal \td
N}^{ \aaa}{}_{\jj}\,\pi^{\jj}\quad,\quad {\cal F}_{\aaa \bb}
\equiv F_{\ii\jj}\,{\cal \td  N}^{-1 \, \ii}{}_{\aaa}{\cal \td 
N}^{-1 \, \jj}{}_{\bb} \, .  \nn
\ee 
The vectorial frame
$\{ e_{iw \, \aaa} \}$ dual to the co--frame $\{ \theta_{iw}^\aaa \}$ is given by
\be e_{iw \, \aaa} = {X}_\jj\, {\cal \td  N}^{-1 \, \jj}{}_{\aaa} \, , 
\label{cbas}
\ee 
where $X_\jj$ are the vectorial frame dual to $\o^\ii$. 
While shifting to
the Iwasawa basis and co-basis, the structure constants of the group,
which also define the Lie brackets of the vectorial frame
$\{X_\ii\}$ dual to the invariant co--frame $\{\omega^\ii\}$
\be [ X_\ii, X_\jj] = - X_\kk \, {\td C}^\kk{}_{\ii\jj} \, , 
\nn 
\ee 
transform as the components of a
$(^1_2)$-tensor so that
\be 
[e_{iw \, \bb}, e_{iw \, \ccc} ] = -e_{iw \, \aaa} \,C^{' \aaa}{}_{\bb\ccc},
\quad\mbox{with}\quad C^{'\aaa}{}_{\bb\ccc} = {\cal \td N}^{\aaa}{}_{\ii}\, {\cal
\td N}^{-1 \jj}{}_{\bb}\,{\cal \td N}^{-1 \kk}{}_{\ccc}\, \td C^\ii{}_{\jj \kk} \, .
\label{struc}
\ee

\subsection{BKL Analysis}

The general analysis of the previous chapter can be repeated in the homogeneous 
context. The Hamiltonian controlling the dynamics is, 
\be H = H_0 + V_S + V_G + V^{el} + V^{magn} \ee
where $H_0$ is given by (\ref{eq3.23}); we rewrite $V_S$ as
\be V_S = \frac{1}{2}\,\sum_{a<b}\,e^{-2(\beta^b -
\beta^a)}\, c_{ab}
\label{pre}
\ee 
where the $c_{ab}$ are explicitly written in the next sections for $D=4$ and $D=5$; the 
``gravitational potential'' now simply reads
\be V_G = - gR =
\frac{1}{2}\,e^{-2\sum_d
\beta^d}\sum_{a,b,c}( e^{2\beta^c}\,C^{\prime \aaa}{}_{\bb\ccc}\,C^{\prime
\bb}{}_{\aaa\ccc} +
\frac{1}{2}\,e^{-2\beta^a +2\beta^b + 2\beta^c}\,(C^{\prime
\aaa}{}_{\bb\ccc})^2\,) \nn 
\ee 
and one considers also the presence of a $1$--form, 
\beq 
V^{el}&=& 
\frac{1}{2}\,e^{-2e_{a}}\,({\cal
E}^{\aaa})^2 \nn 
\\ V^{magn} &=& \frac{1}{4}\,e^{-2 m_{a b}}\,({\cal
F}_{\aaa \bb})^2 \nn 
\eeq where the notations of  Eqs. (\ref{electric}, \ref{magnetic}) are used. 

 \noindent For the $D=4$ and $D=5$ homogeneous cosmologies, we will systematically
investigate,\newline
-- the
prefactors controling the presence of the curvature walls which crucially
depend on the homogeneity group under consideration: the more the group
"looks" abelian, the less curvature walls are present. \newline
-- the  momentum constraints, which in this spatial homogeneity
context, 
can influence the billiard shape. Indeed, they establish linear relations between the wall
coefficients
$c_A$'s in which all together the structure constants, the $\{\td \cN, \, \td \cP, \, \cal E, \, \cal F \}$'s and
even the variables
$\{\beta, \pi\}$ are mixed up; so the
question arises whether, asymptotically, they are equivalent to the
condition that some of the
$c_A$'s vanish forcing the
corresponding walls to disappear.

\section{$d=3$ Homogeneous
Models}

\subsection{Iwasawa Variables}
In spatial
dimension
$d=3$, using the simplified notations
\beq
{\cal \td N}^\1{}_{\2} &=& n_1,\quad {\cal \td N}^\1{}_{\3} = n_2,\quad {\cal \td
N}^\2{}_{\3} = n_3 \nn \\ 
{\cal \td P}^\2{}_{\1} &=& p_1,\quad {\cal \td P}^\3{}_{\1} =
p_2, \quad{\cal \td P}^\3{}_{\2} = p_3 \, ,
\nn 
\eeq 
the prefactors 
$({\cal \td P}^\jj{}_{\aaa}{\cal \td  N}^\bb{}_{\jj})^2$ of the possible symmetry walls
$e^{-2(\beta^b - \beta^a)}, b>a$, in (\ref{pre}), read
\beq
\mbox{for}\quad a=1, b=2 &:&\quad c_{12} =  (p_1 + n_3 p_2)^2
\label{coef1}\\
\mbox{for}\quad a=1, b=3 &:&\quad c_{13} =
(p_2)^2\label{coef2}\\ \mbox{for}\quad a=2, b=3 &:&\quad
c_{23} =  (p_3)^2.\label{coef3}\eeq The Iwasawa decomposition
(\ref{Iwas}) provides explicitly
 \beq s_{\1\1} &=&
e^{-2\beta^1},\quad s_{\1\2} = n_1 e^{-2\beta^1}, \quad s_{\1\3} = n_2
e^{-2\beta^1}\label{g11}\\ s_{\2\2} &=&n_1^2 e^{-2\beta^1} +
e^{-2\beta^2},\quad s_{\2\3} = n_1 n_2 e^{-2\beta^1} + n_3
e^{-2\beta^2}\label{g22}\\ s_{\3\3} &=& n_2^2
e^{-2\beta^1} + n_3^2 e^{-2\beta^2} +
e^{-2\beta^3} \, . \label{g33}\eeq The momenta conjugate to the $s_{ab}$  read
\beq
2\pi^{\1\1} &=& -(\pi_\1 + 2 n_1 p_1 + 2 n_2 p_2) e^{2\beta^1} - (n_1^2
\pi_\2 + 2 n_1^2 n_3 p_3 - 2 n_1 n_2 p_3)e^{2\beta^2} \nonumber \\ 
&-&
(n_2^2+ n_1^2 n_3^2 - 2 n_1 n_2 n_3) \pi_\3 e^{2\beta^3} \nn \\ 
2\pi^{\1\2}
&=& p_1 e^{2\beta^1} + (n_1 \pi_\2 + 2 n_1 n_3 p_3 - n_2 p_3)
e^{2\beta^2} \nonumber\\ 
&+& (n_1 n_3^2 - n_2 n_3)\pi_\3 e^{2\beta^3}
\nn \\
2\pi^{\1\3} &=& p_2 e^{2\beta^1} - n_1 p_3 e^{2\beta^2} + (n_2 - n_1
n_3)\pi_\3 e^{2\beta^3} \nn \\ 2\pi^{\2\2} &=& -(\pi_\2 + 2n_3 p_3)
e^{2\beta^2} - n_3^2 \pi_\3 e^{2\beta^3} \nn \\ 
2\pi^{\2\3} &=& p_3
e^{2\beta^2} + n_3 \pi_\3 e^{2\beta^3} \nn \\ 2\pi^{\3\3} &=& -\pi_\3
e^{2\beta^3}\, . \nn
\eeq 
In order to easily translate the constraints in
terms of the Iwasawa variables, we also mention the following
usefull formulae 
\beq  2\pi^\2{}_{\1} &=& p_1 \nn \\ 
2\pi^\3{}_{\1} &=& p_2 \nn  \\
2\pi^\3{}_{\2} &=& n_1 p_2 + p_3 \nn \\ 
 2\pi^\1{}_{\2} &=& n_1 (\pi_\2 -\pi_\1) +
(e^{-2(\beta^2-\beta^1)}-n_1^2) p_1 + (n_3
e^{-2(\beta^2-\beta^1)} -
 n_1 n_2) p_2 \nonumber\\
  &+& (n_1 n_3 - n_2) p_3 \nn \\ 
  2\pi^\1{}_{\3} &=&
n_2(\pi_\3-\pi_\1) + n_1 n_3(\pi_\2-\pi_\3) + [n_3
e^{-2(\beta^2-\beta^1)} - n_1 n_2] p_1
\nonumber \\
&+& [e^{-2(\beta^3-\beta^1)} + n_3^2 e^{-2(\beta^2-\beta^1)}-
n_2^2] p_2 \nonumber \\
&+& [n_1 n_3^2 -n_2 n_3 - n_1
e^{-2(\beta^3-\beta^2)}] p_3 \nn 
\\ 2\pi^\2{}_{\3} &=& -n_3 (\pi_\2-\pi_\3) + n_2 p_1 +
(e^{-2(\beta^3-\beta^2)}-n_3^2)p_3 \nn \, . \eeq

\subsection{$d=3$ Pure Gravity Billiards}
These spatially homogeneous models are known in the literature as the
class-A Bianchi models; they are classified according to their real,
unimodular, isometry Lie algebra \cite{LL,Ryan:1975jw}.
\newline

\noindent 1. {\bf Bianchi--type I}: $\td C^k_{\,ij} = 0, \forall i,j,k$.

\begin{figure}[h]
\centerline{\includegraphics[scale=0.6]{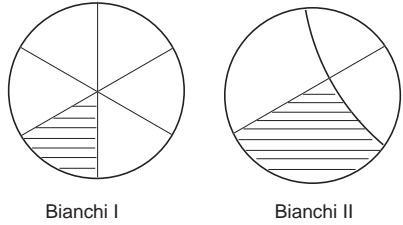}}
\caption{\label{bianchideuxx} \small{ Bianchi I and II billiards are the shaded areas limited by curved
lines
or curvature walls and right lines or symmetry walls.}}
\end{figure}

\noindent This is the abelian algebra. There is no spatial curvature and the
constraints are identically verified. Accordingly, the billiard
walls are only made of symmetry walls $\beta^i - \beta^j, i>j$,
among which the two dominant ones are $w_{32}=\beta^3-\beta^2$ and
$w_{21}=\beta^2-\beta^1$. This is the infinite volume non-diagonal
Kasner billiard. Its projection on the Poincar\'e disc is
represented by the shaded area in Figure \ref{bianchideuxx}.
\newline

\noindent 2. {\bf Bianchi--type II}: $\td C^1_{\,23} = 1$.

\noindent This case is particularly simple because the constraints are easy to analyse.
Indeed, the momentum constraints read \be \pi^\2{}_{\1} = 0\quad\mbox{and}\quad
\pi^\3{}_{\1}=0\quad
\Longleftrightarrow\quad p_1 = 0\quad\mbox{and}\quad p_2=0 \, .  \nn \ee That
means, referring to (\ref{coef1}), (\ref{coef2}) and (\ref{coef3}), that
$c_{12} =
0$ and $c_{13}=0$ and that they clearly eliminate the symmetry walls
$w_{21}=\beta^2 -
\beta^1$ and $w_{31}=\beta^3 -
\beta^1$; the last $p_3$ remains free so the
symmetry wall $w_{32} = \beta^3-\beta^2$ is present. Moreover, the only
non zero structure constants being
$ C^{\prime 1}{}_{23} = 1$, one single curvature wall
survives which is
$2\beta^1$. Since two walls can never close the billiard, its volume is
also infinite. Its projection on the Poincar\'e disc is given by the shaded
area in Figure \ref{bianchideuxx}.

\noindent 3. {\bf Bianchi--type VI$_0$}: $ \td C^1_{\,23} = 1 = \td C^2_{\,13}$.

\noindent The momentum constraints read 
\be \pi^\3{}_{\2} = 0\quad,\quad
\pi^\3{}_{\1}=0\quad\mbox{and}\quad \pi^\2{}_{\1} + \pi^\1{}_{\2} =
0.\label{cos}\ee They are equivalent to \beq 0&=& p_2 \label{p1} \\ 0 &=&
p_3
\label{p2}\\ 0 &=& n_1 (\pi_\2 -\pi_\1) + (e^{-2(\beta^2-\beta^1)}-n_1^2 +2)
p_1 .\label{p3}\eeq 
The first two, according to (\ref{coef2}) and
(\ref{coef3}), clearly tell that the symmetry walls
$w_{32}=\beta^3-\beta^2$ and
$w_{31}=\beta^3-\beta^1$ are absent from the potential. The billiard volume is therefore  infinite (whether or not the last symmetry wall is present) and the asymptotic dynamics is described by a monotonic
Kasner--like  regime. With
(\ref{p2}) put in (\ref{coef1}), one sees that the coefficient of the
third symmetry wall
$w_{21}$ becomes $c_{12}=(p_1)^2$. Since the 
constraint (\ref{p3}) does not imply that this coefficient vanishes, the 
corresponding wall will be generically present, \ie for generic initial conditions.

\begin{figure}[h]
\centerline{\includegraphics[scale=0.6]{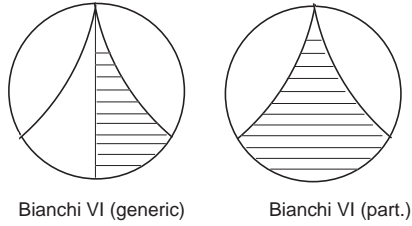}}
\caption{\label{bianchisix}\small{Bianchi VI Billiards}}
\end{figure}

Summarising, the 
 symmetry wall $w_{21}=\beta^2-\beta^1$ is present,
beside the two curvature walls $2\beta^1$ and $2\beta^2$; the
dominant walls being $w_{21}=\beta^2-\beta^1$ and $2\beta^1$, they
do not close the billiard, see Figure \ref{bianchisix}.

\noindent One can take advantage of the gauge freedom \cite{Coussaert:1993ti} to assign the initial value of 
$n^0_1$ to be zero, the constraint then implies that $p^0_1 =0$. According to the equations of motion, $n_1$ and $p_1$ remain null. In particular, the last symmetry wall is absent and the billiard is  delimited by  two curvature walls depicted in Figure \ref{bianchisix}. 
The billiard shape will not depend on gauge choices for $d>3$.\footnote{The number of off--diagonal components of a $d$--dimensional spatial  metric is given by 
$d(d-1)/2$ while the number of momentum constraints is $d$. For $d>3$, a gauge transformation cannot diagonalise this metric. } 

These examples show that the presence/absence of symmetry walls may depend
on initial conditions or on gauge conditions. This is not very satisfactory and suggest to analyse deeply the spatial diffeomorphisms  in the generic case, \ie non homogeneous.  Nevertheless, in the
cases mentionned hereabove, the finite/infinite nature of the billiard volume is not modified  by a gauge choice. 
\newline

\noindent 4. {\bf Bianchi--type VII}: $\td C^1_{\,23} = 1,\quad \td C^2_{\,13}=-1$.

\noindent This case is very similar to the preceding one. The only changes are
i) that the third constraint in (\ref{cos}) has to be replaced by
($\pi^\2{}_{\1} -
\pi^\1{}_{\2} = 0$) and ii) that its translation into the Iwasawa variables
in (\ref{p1}) now reads
\be 0 = n_1 (\pi_\2
-\pi_1) + (e^{-2(\beta^2-\beta^1)}-n_1^2 -2) p_1\, . \nn
\ee 
Accordingly,
we can apply the same reasoning as before.
\newline

\noindent 5. {\bf Bianchi-type IX}: $\td C^1_{\,23} = 1,\quad \td C^2_{\,31} =
1,\quad \td C^3_{\,12} = 1$.

\noindent This case and the next one deserve a particular treatment because the structure
constants are such that all curvature walls, namely $2\beta^1$, $2\beta^2$,
$2\beta^3$, appear  and these three gravitational walls already form a
finite\footnote{This situation is very specific to the homogeneous models in
$D=4$; in higher spacetime dimensions, the number of curvature walls allowed by
the structure constants is not sufficient to produce a finite volume
billiard.} volume billiard.

Let us first discuss the generic case. The momentum
constraints take the form
\be
\pi^\3{}_{\2} -\pi^\2{}_{\3}= 0,\quad
\pi^\3{}_{\1}-\pi^1{}_{\3} =0,\quad\pi^\2{}_{\1} - \pi^\1{}_{\2}
= 0 \, ; \nn 
\ee 
and in terms of the Iwasawa variables, they become
\beq
 n_3 (\pi_2 &-&\pi_\3)  - n_2 p_1 + n_1 p_2 \nonumber\\ &+&
[-e^{-2(\beta^3-\beta^2)}+n_3^2 +1]p_3 = 0 \label{b91}\eeq 
\beq
n_2(\pi_\3-\pi_\1) &+& n_1 n_3(\pi_\2 -\pi_\3)  + [n_3
e^{-2(\beta^2-\beta^1)} - n_1 n_2] p_1
\nonumber \\&+&[e^{-2(\beta^3-\beta^1)} + n_3^2 e^{-2(\beta^2-\beta^1)}-
n_2^2 - 1] p_2 \nonumber \\&+& [n_1 n_3^2 -n_2 n_3 - n_1
e^{-2(\beta^3-\beta^2)}] p_3 =0 \label{b92}\eeq  \beq  n_1 (\pi_\2
&-&\pi_\1) + [e^{-2(\beta^2-\beta^1)}-n_1^2- 2] p_1\nonumber \\&+& [n_3
e^{-2(\beta^2-\beta^1)}-
 n_1 n_2] p_2 \nonumber\\ &+& [n_1 n_3 - n_2] p_3=0.\label{b93}
\eeq Remember that we already know that the billiard has a finite volume, hence
the question is no longer to state between chaos or non chaos but rather to
define
more precisely the shape of the billiard. 

In order to study the implications of the constraints (\ref{b91}) -
(\ref{b93}),
we shall rely on the heuristic estimates made in \cite{Damour:2002et}, where the
asymptotic
behaviour of the variables is analysed in the BKL limit. From that analysis, it
follows that, when $\rho\to \infty$: i) the $\pi_a$'s go to zero as powers of
$1/\rho$, ii) the $n_i$'s and the $p_i$'s tend to constants $n_i^\infty$ and
$p_i^\infty$ up to additive terms which also go to zero as powers of
$1/\rho$ and
iii) that the exponentials either vanish (if the walls are present) or
oscillate
between zero and
$\infty$ (if they are absent).

If the symmetry walls are absent, the constraints cannot be 
satisfied at any time. These walls are therefore generically present, \ie they are present 
except if the coefficients in front of the exponential appearing in the constraints (\ref{b91}) -
(\ref{b93}) are identically zero.

\begin{figure}[h]
\centerline{\includegraphics[scale=0.6]{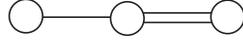}}
\caption{ \label{dda2} \small{Dynkin diagrams of the $A_1^{\wedge\wedge}$ Kac--Moody algebra}}
\end{figure}

\begin{figure}[h]
\centerline{\includegraphics[scale=0.6]{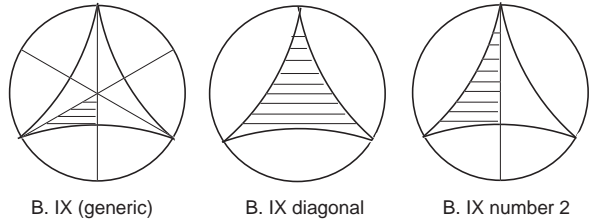}}
\caption{\label{bianchineuf} \small{Bianchi IX Billiards}}
\end{figure}

\begin{figure}[h]
\centerline{\includegraphics[scale=0.6]{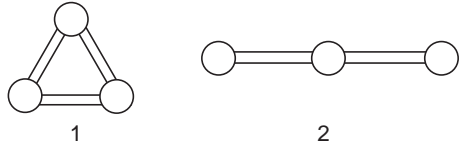}}
\caption{\label{partddbianchineuf} \small{Dynkin diagrams of special algebras met in Bianchi
IX models}}
\end{figure}

Accordingly, the billiard's edge is formed by the leading symmetry walls
$w_{32}$ and $w_{21}$ and by the curvature wall $2\beta^1$. On the
Poincar\'e disc, it is one of the six small triangles included in the larger
one bordered by the curvature walls. Its Cartan matrix is that of the
hyperbolic
Kac--Moody algebra
$AE_3 = A_1^{\wedge\wedge}$ \be \left
(\begin{array}{ccc}2&-1&0\\-1&2&-2\\0&-2&2\end{array}\right ) \nn \ee already
relevant in the general inhomogeneous case. Its Dynkin diagram is given in
Figure \ref{dda2}.
Other interesting cases exist with less symmetry walls, which require specific
initial conditions, see Figure \ref{bianchineuf}.
\begin{enumerate}
\item{No symmetry wall at all.} This situation is the one mentionned above;
it happens when the metric is assumed diagonal always, hence for
$n_1=0, n_2=0, n_3 = 0$; the solution of the momentum constraints is then
$p_1=0, p_2=0, p_3=0$. This assumption is consistent with the equations of
motion. The Cartan matrix of this billiard is given by
\be \left (\begin{array}{ccc}2&-2&-2\\-2&2&-2\\-2&-2&2\end{array}\right
) \nn \ee 
and its corresponding Dynkin diagram is number 1 in Figure \ref{partddbianchineuf} ; the
associated Kac--Moody algebra is hyperbolic, it has number 7 in the
enumeration provided in reference \cite{S}. The billiard shape is depicted in Figure 
\ref{bianchineuf} (diagonal case).

\item{One symmetry wall.} This happens when one chooses the initial data
such that
$n_1=0, n_2=0$ and $p_1=0$, $p_2=0$; then, according to the equations of
motion, these variables remains zero all the time. The billiard is closed
by two curvature walls
$2\beta^1$ and
$2\beta^2$ and the symmetry wall $w_{32}$. On the Poincar\'e disc, its
volume is half of that of the triangle made of curvature walls. The
Cartan matrix is given by
\be \left (\begin{array}{ccc}2&0&-2\\0&2&-2\\-2&-2&2\end{array}\right
) \nn \ee and its Dynkin diagram is number 2 in Figure \ref{partddbianchineuf} . 
It characterises the
third rank 3 Lorentzian Kac--Moody algebra in the classification given in
\cite{S}.
\end{enumerate}
These last two Kac--Moody algebras are subalgebras of $A_1^{\wedge\wedge}$
\cite{Feingold:2003es}. The billiard shape is depicted in Figure 
\ref{bianchineuf} (case number 2).

\noindent \textbf{Remark}: let us recall that the reflections on the faces of the billiard
generate a Coxeter group which is identified with the Weyl group of the
associated Kac--Moody algebra. The larger the set of walls, the
larger the number of generators. Since in the non
generic cases the set of walls is a subset of the one in the generic case,
the associate Coxeter group is a subgroup of the generic one.
\newline

\noindent 6. {\bf Bianchi--type VIII}: $ \td C^1_{\,23} = 1,\quad \td C^2_{\,31} =
1,\quad \td C^3_{\,12} = -1$.

\noindent The analysis of this case follows closely the
previous one. The sign change in the structure constants
modifies the constraints as follows
\be
\pi^\3{}_{\2} +\pi^\2{}_{\3}= 0,\quad
\pi^\3{}_{\1}+\pi^\1{}_{\3} =0,\quad\pi^\2{}_{\1} - \pi^\1{}_{\2}
= 0 \nn \ee 
and induces some sign changes in their Iwasawa counterparts.
The conclusions are those of the Bianchi IX model: the generic billiard is the
one of the algebra $A_1^{\wedge\wedge}$.

\subsection{$d=3$ Einstein--Maxwell Billiards}

The momentum constraints (\ref{momentum}) of the Einstein--Maxwell
homogeneous models generally (except for the abelian Bianchi I) mix
gravitational and one--form variables; in comparison with the pure
gravity case, i) no constraint remains which clearly forces the
prefactor of a symmetry wall to be zero and ii) additional terms of the type
$\pi_{F \, \ii } F^{\ii \jj }$ enter in the constraints system.

\begin{figure}[h]
\centerline{\includegraphics[scale=0.6]{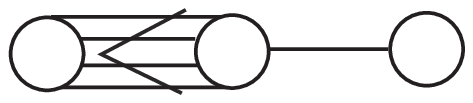}}
\caption{\label{einsteinmaxdtrois} \small{Dynkin diagram of the $A_2^{(2)\wedge}$ algebra}}
\end{figure}

Accordingly, generically, besides the curvature walls allowed by
the structure constants, one expects all symmetry, electric and
magnetic walls to be present. The dominant ones are the symmetry
walls $w_{21}= \beta^2-\beta^1, w_{32}= \beta^3-\beta^2$ and the
electric wall $e_1=\beta^1$ which replaces the curvature wall of
the pure gravity case. They close the billiard whose Cartan matrix
is \be \left
(\begin{array}{ccc}2&-4&0\\-1&2&-1\\0&-1&2\end{array}\right ) \, . \nn\ee
The Dynkin diagram is dispayed in Figure \ref{einsteinmaxdtrois}; the corresponding
Kac--Moody algebra is the hyperbolic $A_2^{(2)\wedge}$ algebra
which is the Lorentzian extension of the twisted affine algebra
$A_2^{(2)}$ also encountered in \cite{Henneaux:2003kk}.

\section{$d=4$ Homogeneous Models}

Our billiard analysis of the four dimensional spatially homogeneous
cosmological models has been carried out along the same lines as for
$d=3$: results relative to pure gravity models and those relative to the
Einstein--Maxwell homogeneous system will be given separately.

\subsection{Extension of the Notations and Iwasawa Variables}

In spatial dimension $d=4$, we introduce the matrix variables \be
{\mathcal N} =
\left (\begin{array}{cccc}
1&n_{12}&n_{13}&n_{14}\\0&1&n_{23}&n_{24}\\
0&0&1&n_{34}\\0&0&0&1\end{array}\right)\, . \nn \ee The Iwasawa decomposition
of the metric extends beyond the 3-dimensional formulae listed in
(\ref{g11}), (\ref{g22}), (\ref{g33}), through the additional
components  
\beq s_{\1\4} &=& n_{14}\,e^{-2\beta^1} \nn \\ 
s_{\2\4} &=&
n_{12}\,n_{14}\,e^{-2\beta^1} + n_{24}\,e^{-2\beta^2} \nn \\ 
s_{\3\4} &=&
n_{13}\,n_{14}\,e^{-2\beta^1} + n_{23}\,n_{24}\,e^{-2\beta^2}+
n_{34}\,e^{-2\beta^3}\nn \\ 
s_{\4\4}&=& n_{14}^2\,e^{-2\beta^1} +
n_{24}^2\,e^{-2\beta^2} + n_{34}^2\,e^{-2\beta^3} + e^{-2\beta^4}\, . \nn
\eeq
The momentum conjugate to $\beta^a$ is written as $\pi_\aaa$, as before; the
momentum conjugate to
${\cal \td N}^\aaa{}_{\jj} = n_{aj}$ is denoted
${\cal \td P}^\jj{}_{\aaa} = p^{ja}$ and is only defined for $j>a$.
The prefactors $c_{ab}=({\cal
\td P}^\jj{}_{\aaa}{\cal \td  N}^\bb{}_{\jj})^2$ of the symmetry walls $e^{-2(\beta^b
-\beta^a)}, b>a,$
in the Hamiltonian
 are explicitly given by \beq c_{12} &=& (p^{21} +
p^{31}\,n_{23} + p^{41}\,n_{24})^2 \nn
\\c_{13}&=& (p^{31} + p^{41}\,n_{34})^2 \nn \\c_{14}&=&
(p^{41})^2\nn \\c_{23}&=& (p^{32} + p^{42}\,n_{34})^2\nn \\c_{24}&=& (p^{42})^2 \nn \\
c_{34}&=& (p^{43})^2\, . \nn \eeq
Once the change of dynamical variables has been continued in a
canonical point transformation, the momentum contraints
still express linear relations among the $\pi^\ii{}_{\jj}$'s which
translate into linear relations on the momenta
$\pi_\aaa, p^{ia}$; their
coefficients are polynomials in the $n_{ai}$'s times
exponentials of the type
$e^{-2(\beta_b -\beta_a)}$, with $b>a$. The explicit form of the
constraints depends on the model considered.

\subsection{$d=4$ Pure Gravity}

We label the various 4 dimensional
spatially homogeneous models according to the classification of the 4
dimensional real unimodular Lie algebras given by M. MacCallum \cite{MacC1,MacC2}:
they are

\begin{enumerate}
\item{\bf Class $U1[1,1,1]$}: $\td C^1_{\,\,14}=\lambda,\,
\td C^2_{\,\,24}=\mu,\, \td C^3_{\,\,34}=\nu, \mbox{with}\,\,\lambda+\mu+\nu=0$.
One can still set $\lambda=1$ except in the abelian case where
$\lambda=\mu=\nu=0$.

\item{\bf Class $U1[Z,\bar Z,1]$}: $\td C^1_{\,\,14}=-\frac{\mu}{2},\,
\td C^2_{\,\,14}=1, \td C^1_{\,\,24}=-1,\, \td C^2_{\,\,24}= -\frac{\mu}{2},\
\td C^3_{\,\,34}=\mu.$ $\mu=0$ is a special case.

\item{\bf Class $U1[2,1]$}: $\td C^1_{\,\,14}=-\frac{\mu}{2},\,
\td C^2_{\,\,14}=1, \, \td C^2_{\,\,24}= -\frac{\mu}{2},\, \td C^3_{\,\,34}=
\mu $ and $\mu$ is $0$ or
$1$.
\item{\bf Class $U1[3]$}:
$\td C^2_{\,\,14}=1, \, \td C^3_{\,\,24}= 1$.
\item{\bf Class $U3I0$}: $\td C^4_{\,\,23}=1,\,
\td C^2_{\,\,31}=1, \, \td C^3_{\,\,12}= -1.$
\item{\bf Class $U3I2$}: $\td C^4_{\,\,23}=-1,\,
\td C^2_{\,\,31}=1, \, \td C^3_{\,\,12}= 1.$
\item{\bf Class $U3S1$} or $\mf{sl}(2)\oplus \mf{u}(1)$: $\td C^1_{\,\,23}=1,\,
\td C^2_{\,\,31}=1, \, \td C^3_{\,\,12}= -1.$
\item{\bf Class $U3S3$} or $\mf{su}(2)\oplus \mf{u}(1)$: $\td C^1_{\,\,23}=1,\,
\td C^2_{\,\,31}=1, \, \td C^3_{\,\,12}= 1.$

\end{enumerate}
Let us mention that for all of the four dimensional algebras, except of
course the abelian one, the structure constant $\ C^{\prime
1}_{\,\,\,34}\ne 0$ and consequently that the curvature wall
$2\beta^1+\beta^2$ is always present.

Our analysis leads to the conclusion that, from the billiard point of
view, the previous models can be collected into two sets: the first one
has an open billiard, the second one has a finite volume billiard
whose Cartan matrix is that of the hyperbolic Kac--Moody algebra
$A_2^{\wedge\wedge}$ exactly as in the general inhomogeneous situation.

The first set contains the abelian algebra and $U1[1,1,1]_{\mu\ne
0,-1}$, $U1[2,1]$ and $U1[Z,\bar Z,1]_{\mu\ne 0}$; all the others belong to
the second set. Because explicit developments soon become lenghty and
since the reasonings always rest on similar arguments, we have chosen
not to review systematically all cases as for $d=3$ but rather to
illustrate the
results on examples taken in both sets:
\newline

\noindent{1. As a representative of the first set, we take
$U1[Z,\bar Z,1]_{\mu\ne 0}$}.

\noindent The momentum constraints read \be \pi^\4{}_{\1}=\pi^\4{}_{\2} =
\mu\,\pi^\4{}_{\3} = -\frac{\mu}{2}(\pi^\1{}_{\1} + \pi^\2{}_{\2}
-2\,\pi^\3{}_{\3}\,) + \pi^\1{}_{\2} - \pi^\2{}_{\1} = 0 \, ;\label{co1}
\ee in terms of the Iwasawa variables, they become
\be p^{41} = p^{42} = \mu\,p^{43} = 0\label{co2}\ee and \beq
&-&\frac{\mu}{2} (2\pi_\3 - \pi_\1
-\pi_\2) + n_{12} (\pi_\2 -\pi_\1) +
\nonumber\\ &+&(e^{-2(\beta^2-\beta^1)}-n_{12}^2-1) \, p^{21}
+(n_{23}e^{-2(\beta^2-\beta^1)}- n_{12}n_{13}+\frac{3 n_{13}\mu}{2}  )
p^{31}
\nonumber \\ &+& (n_{12}n_{23}-
n_{13}+\frac{3 n_{23}\mu}{2}
)\,p^{32}=0 \, .\label{co3}\eeq
The first three constraints (\ref{co2}) clearly indicate that the
symmetry walls $\beta^4-\beta^3$,  $\beta^4-\beta^2$,
$\beta^4-\beta^1$ are absent from the potential. The fourth constraint (\ref{co3}) does not generically imply that $c_{23} = (p^{32})^2$ vanishes. 
Consequently,  in the
generic case, the dominant walls of the billiard are the symmetry walls
$w_{32}=\beta^3-\beta^2$,
$w_{21}=\beta^2-\beta^1$ and  the curvature wall $2\beta^1+\beta^2$; it is
indeed an open billiard.

Notice that in all cases of the first set, the non vanishing  structure
constants are all of the form $C^{\prime a}_{\hspace{.15cm}4b}$  with
$a,b=1,2,3$. It is easy to check that this forbids all
curvature wall containing $\beta^4$ and therefore that the remaining curvature
walls cannot be expected to close the billiard.
\newline

\noindent{2. As a first representative of the second set, we take
$U1[Z,\bar Z,1]_{\mu = 0}$}

\begin{figure}[h]
\centerline{\includegraphics[scale=0.5]{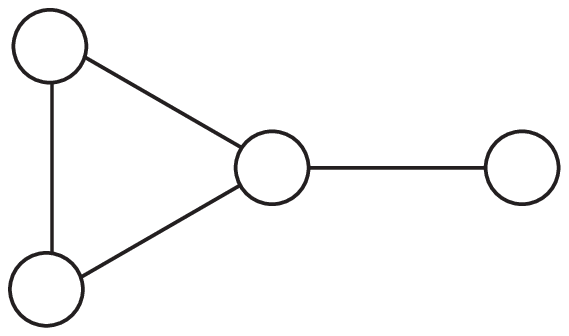}}
\caption{\label{ddadeux} \small{Dynkin diagram of the $A^{\wedge\wedge}_2$ algebra}}
\end{figure}

The constraints are given by (\ref{co1}), (\ref{co2}) and
(\ref{co3}) for $\mu=0$ and one immediately sees that, compared to
the preceding case, one constraint drops out leaving the symmetry
wall $w_{43}=\beta^4-\beta^3$ in place. The last constraint does not generically imply the vanishing of other prefactors $c_{ab}$. Again, we end up
here with the following list of dominant walls:
$w_{43}=\beta^4-\beta^3$, $w_{32}=\beta^3-\beta^2$,
$w_{21}=\beta^2-\beta^1$ and $2\beta^1+\beta^2$. The Cartan matrix
is that of the algebra $A_2^{\wedge\wedge}$ as previously
announced. Its Dynkin diagram is depicted in Figure \ref{ddadeux}. 

\noindent{3. Another interesting example of the second set is provided by
$U3S3$}

Its homogeneity group is the
direct product
$SU(2)\times U(1)$ so that this model appears as the four dimensional
trivial extension of the Bianchi IX model: since the structure constants
are the same as in the 3-dimensional case, the momentum constraints do not
change, when expressed in terms of the metric components and their momenta
\be
\pi^\1{}_\3-
\pi^\3{}_\1=0,\quad
\pi^\1{}_\2-\pi^\2{}_\1=0,\quad
\pi^\2{}_\3- \pi^\3{}_\1=0 \, ; \nn \ee their number does not change either but, in
terms of Iwasawa variables, they involve much more terms than their $d=3$
counterpart given in (\ref{b91}), (\ref{b92}) and (\ref{b93})
\beq &-&n_{13}(\pi_\1-\pi_\3) + n_{12} n_{23}(\pi_\2  -\pi_\3)
-(n_{12}n_{13}-n_{23}e^{-2(\beta^2-\beta^1)})p^{21}\nonumber \\ &-&
(n_{13}^2-n_{23}^2e^{-2(\beta^2-\beta^1)}-e^{-2(\beta^3-\beta^1)}-1)
p^{31}\nonumber\\
&+&(n_{12}n_{23}^2-n_{13}n_{23}-n_{12}e^{-2(\beta^3-\beta^2)})p^{32}\nonumber
\\ &-&(n_{13}n_{14}-n_{23}e^{-2(\beta^2-\beta^1)}n_{24}-n_{34}
e^{-2(\beta^3-\beta^1)})p^{41}\nonumber \\
&+&(n_{12}n_{23}n_{24}-n_{14}n_{23}-n_{12}n_{34}e^{-2(\beta^3-\beta^2)})
p^{42}\nonumber\\
&-&(n_{12}n_{23}n_{34}-n_{12}n_{24}-n_{13}n_{34}+n_{14})p^{43}=0 \nn \\ 
& & \nn \\
&-& n_{12} (\pi_\1- \pi_\2)  - (n_{12}^2-e^{-2(\beta^2-\beta^1)}-1 )p^{21}
\nonumber
\\&-& (  n_{12}  n_{13}- n_{23}e^{-2(\beta^2-\beta^1)} )  p^{31}
+ (n_{12}  n_{23}- n_{13} ) p^{32} \nonumber\\ &-& ( n_{12} n_{14}-
n_{24} e^{-2(\beta^2-\beta^1)})p^{41}+ (  n_{12} n_{24}- n_{14}  )
p^{42}=0 \nn \\ & & \nn \\ &-& n_{23} (\pi_\2- \pi_\3)+ n_{13}
p^{21}- n_{12}  p^{31}-( n_{23}^ 2 - e^{-2(\beta^3-\beta^2)}+1
)p^{32}\nonumber \\ &-& (  n_{23}  n_{24} - n_{34}e^{-2(\beta^3-\beta^2)})
p^{42}+ ( n_{23} n_{34}- n_{24} )  p^{43}=0 \,  . \nn \eeq
These constraints are of course i) linear in the momenta, ii) the
only exponentials which enter these expressions are build of
$w_{32}=\beta^3-\beta^2$,
$w_{31}=\beta^3-\beta^1$ and
$w_{21}=\beta^2-\beta^1$ and iii) as before, their coefficients are
exactly given by the square root of the corresponding $c_A$'s in the
potential, namely
$\sqrt{c_{12}},
\sqrt{c_{13}}$ and
$\sqrt{c_{23}}$. Accordingly, the question arises whether these equations are
equivalent to $c_{12}=0$, $c_{23}=0$ and $c_{13}=0$. If, asymptotically, the
exponentials go to zero and  can be dropped out of the constraints, the
remaining equations can be solved for $p^{21}, p^{31}, p^{32}$ in
terms of the other variables among which figure now, not only the
$\pi_\ii-\pi_\jj$ already present in the 3-dimensional case which
asymptotically go to
zero, but also the asymptotic values of the
$p^{4i}$'s which remain unconstrained. It follows that none of the
solutions of the above system is generically forced to vanish so that all the
symmetry walls are expected to be present. The absence of a wall can only
happen
in non generic situations with very peculiar initial conditions.

We  expect this result to become the rule
in higher dimensions for trivial extensions like $SU(2)\times
SU(2)\times...\times U(1)
\times...\times U(1)$; the
billiard will then become that of the Kac--Moody algebra
$A_n^{\wedge\wedge}$ relevant in the general Einstein theory.

\begin{figure}[h]
\centerline{\includegraphics[scale=0.5]{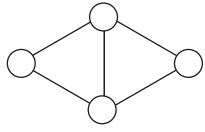}}
\caption{\label{ddquatre} \small{Dynkin diagram of algebra number 2 in the subset of rank 4}}
\end{figure}

We can nevertheless provide a particular solution with initial data obeying
$n_{12}=n_{13}=n_{14}=0$ and $p^{21}=p^{31} = p^{41}=0$. These values are
conserved in the time evolution. In this case, the leading walls are the
symmetry walls $w_{43}, w_{32}$ and the curvature ones $2\beta^2+\beta^3,
2\beta^1+\beta^2$. The Cartan matrix is
\be \left( \begin{array}{cccc}
2 & -1&-1&0\\
-1 & 2 & -1&-1 \\
-1&-1&2&-1\\
0&-1&-1&2\\
\end{array}  \right) \, . \nn \ee The billiard is characterised by the rank
4 Lorentzian Kac--Moody algebra which bears number 2 in the classification
given in
\cite{S} and which is a subalgebra of $A_2^{++}$; its Dynkin diagram is drawn in Figure \ref{ddquatre}.
The $D=5$ results are summarized in the following table:
\newline

\begin{center}

\begin{tabular}{|p{3cm}|c|}
\hline  $U1 [ 1,1,1 ] _{\mu = \nu = \lambda =0}$ $U1[1,1,1]_{\mu
\neq 0,  -1}$
$U1[Z,\bar{Z},1]_{\mu \neq 0}$& non chaotic \\
\hline  $U1[1,1,1]_{\mu = -1}$ $U1[Z,\bar{Z},1]_{\mu = 0}$
$U1[2,1]$\newline $U1[3]$
\newline
$U3I0$  \newline $U3I2$ \newline $U3S3$ \newline $U3S1$ & chaotic  \\
\hline
\end{tabular}
\end{center}
\centerline{Chaos or non chaos for $D=5$ models}

\subsection{$d=4$ Einstein--Maxwell Models}

\begin{figure}[h]
\centerline{\includegraphics[scale=0.6]{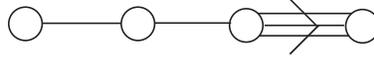}}
\caption{\label{ddg2} \small{Dynkin diagram of the $G^{\wedge\wedge}_2$ algebra}}
\end{figure}

\begin{figure}[h]
\centerline{\includegraphics[scale=0.6]{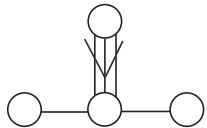}}
\caption{\label{ddemquatrew} \small{Dynkin diagram of the  algebra relevant for the Einstein
+ $2$--form system}}
\end{figure}

As in the three dimensional case, because of the presence of the
electromagnetic field in the expression of the momentum  constraint
(\ref{momentum}), no symmetry wall can be eliminated. Moreover, the electric
walls always prevail over the curvature ones if any.
The billiard is in all cases
characterised by the following set of dominant walls
$w_{43}=\beta^4-\beta^3$, $w_{32}=\beta^3-\beta^2$,
$w_{21}=\beta^2-\beta^1$ and
$e_1=\beta^1$. Its Cartan matrix is that of the hyperbolic
Kac--Moody algebra
$G^{\wedge\wedge}_2$ with Dynkin diagram depicted in Figure \ref{ddg2}.
In the general analysis of the billiards attached to coupled
gravity + p--forms systems, one could assume $2p<d$ without loss of
generality, because the complete set of walls is invariant under
electric--magnetic duality. This invariance may however not remain
in some spatially homogeneous cases due to the vanishing of some
structure constants which lead to incomplete sets of walls so that
other, a priori unexpected Kac--Moody algebras, might appear. An
illustrative and interesting example is given in $D=5$ by the
Einstein + $2$--form system governed by the abelian algebra. Here,
the dominant walls are  i) the symmetry walls $\beta^4-\beta^3$,
$\beta^3-\beta^2$ and $\beta^2-\beta^1$ and ii) the magnetic wall
$\beta^1+\beta^2$. This billiard brings in the new hyperbolic
Kac--Moody algebra carrying number 20 in \cite{S}, whose Dynkin
diagram is given in Figure \ref{ddemquatrew} hereafter.

\section{Conclusions}

In this chapter, we  have analysed the Einstein and Einstein--Maxwell
billiards for  all the  spatially homogeneous cosmological models
in 3 and 4 dimensions. In the billiard picture, we confirm that in
spacetime dimensions $5\leq D\leq10$, diagonal models are not rich enough
to produce the never ending oscillatory behaviour of the generic solution of
Einstein's equations and that chaos is restored when off-diagonal metric
elements
are kept. Chaotic models are characterised by a finite volume billiard
which can
be identified with the fundamental Weyl chamber of an hyperbolic Kac--Moody
algebra: in the most generic chaotic situation, the algebra coincides with
the one
already relevant in the inhomogeneous case; this remains true after the
addition of an generic homogeneous electromagnetic field. Other algebras
can also
appear for  special initial data or gauge choices: in fact, these are all the
simply-laced known hyperbolic Kac--Moody algebras of ranks 3 and 4, except
the one
which has number 3 among those of rank 4 listed in \cite{S}. This analysis furnishes a nice way to tackle infinite--dimensional hyperbolic  subalgebras of those appearing in the non--homogeneous context. 
%%%%%%%%%%%%%%%%%%%%%%%%%%%%%%%%%%
%%%%%%%%%%%%%%%%%%%%%%%%%%%%%%%%%%
\cleardoublepage
%%%%%%%%%%%%%%%%%%%%%%%%%%%%%%%%%%%
%%%%%%%%%%%%%%%%%%%%%%%%%%%%%%%%%%%%
%%\include{chap_oxidation}

\chapter{Oxidation \label{oxidation}}
\markboth{OXIDATION}{}

An interesting feature of the billiard region is that it is
invariant under toroidal dimensional reduction to any dimension
$D\geq 3$ \cite{Damour:2002fz}. The knowledge the billiard region in $D=3$
dimensions can then be used as a tool to determine the possible
higher dimensional parents of the theory.  This problem of going
up in dimension is known as the oxidation or disintegration
problem and goes back to the early days of supergravity and has
been studied repeatedly \cite{J,Julia:1980gr, J''}. It has been thoroughly
investigated recently by means of group theory techniques
\cite{Keurentjes:2002xc,Keurentjes:2002rc,Keurentjes:2002vx}. 
We show here that the billiard approach
gives direct information on possible obstructions to oxidation
and, when oxidation is possible, restricts efficiently  the
maximal dimension(s) of oxidation as well as determines the full
$p$--form spectrum of the maximally oxidised theory.  We consider
both split and non--split ${\cal U}_3$--groups.  For non maximally
non--compact groups (\ie non--split groups), this approach is complementary
to the general method of general linear subgroups of \cite{J''}
known as the ``A--chain'' method.  The same
information can be extracted from the superalgebra approach to the
problem \cite{deBuyl:2003ub, Henry-Labordere:2003rd, Henry-Labordere:2002xh, Henry-Labordere:2002dk}.

\section{Walls associated with simple roots and oxidation}

The walls bounding the billiards have different origins: they can be symmetry 
(\ref{centrifugal}) or curvature walls (\ref{gravitational} \& 
\ref{gravitational2}), 
related
to the Einstein--Hilbert action; or they can be $p$--form walls
(electric (\ref{electric}) or magnetic (\ref{magnetic})), related to the 
$p$--form part of the
action.  The key to the derivation of the oxidation constraints is
to investigate how the billiard walls behave upon dimensional
reduction. Although the billiard region is invariant, the formal
origin of the walls may  change. E.g., a symmetry wall in $D$
dimensions may appear as a $1$--form wall associated with the
Kaluza--Klein graviphoton(s) in lower dimensions \cite{Damour:2000hv}.

The translation rules relating walls in higher dimensions to walls
in lower dimensions have been worked out in 
\cite{Damour:2002fz,Henneaux:2003kk}.  To
recall them, we denote by  $\bar{B}$ the restricted root system of
the real form ${\cal U}_3$.  This restricted root system may be
reduced or non reduced, in the latter case it is of $BC_r$--type
\cite{Helgason}. The billiard region is the fundamental Weyl
chamber of the overextension $\bar{B}^{\wedge \wedge}$ \cite{Henneaux:2003kk}
obtained by adding to the simple roots of the finite root system
$\bar{B}$, the affine root $\a_0$ and the ``overextended root''
$\a_{-1}$, which is attached to the affine root with a single
link. There are at most two different root lengths, except when
the underlying finite root system is non reduced, in which case
one has three different root lengths.  The highest root is always
a long root (``very long'' root in the $BC_r$--case) \cite{Henneaux:2003kk}.

{}From the point of view of the three--dimensional action
(Einstein--Hilbert action + ${\cal U}_3$--non linear $\sigma$-model
action), these roots have the following interpretation: 
(i) the overextended root is the symmetry
wall $\b^2 - \b^1$ and involves only gravitational variables; 
(ii)
the affine root corresponds to the dominant magnetic wall $\b^1 -
\theta(\phi)$ where $\theta(\phi)$ is the highest root of
${\bar{B}}$; it involves both the dilatons and the gravitational
variable $\b^1$; and 
(iii) the simple roots of $\bar{B}$
correspond to the dominant electric walls defined by the $0$--forms
(axions) and involve the dilatons only.

In $D= d+1  >3 $ dimensions, these walls appear as follows. 
(i)
The overextended root remains a symmetry wall\footnote{Scale
factors get redefined as one changes the spacetime dimension but
we shall keep the same notation since the context is in each case
clear.}, namely $\b^{d} - \b^{d-1}$; 
(ii) The affine root becomes
the symmetry wall $\b^{d-1} - \b^{d-2}$. 
(iii) The (long) roots of
${\bar B}$ connected to the affine root by a chain of single links
form a chain of symmetry walls $\b^{d-2} - \b^{d-3}$, $\cdots$,
$\b^2 - \b^1$.  The other roots correspond to $p$--form walls of
the form $\b^1 + \cdots +  \b^p + \frac{1}{2} \sum_\a \l^{(p)}_\a
\phi^\a $ (electric walls) or $\b^1 + \cdots +  \b^{d-p-1} -
\frac{1}{2} \sum_\a \l^{(p)}_\a \phi^\a$ (magnetic
walls)\footnote{For pure gravity, one curvature wall, namely $2
\b^1 + \b^2 + \cdots + \b^{d-2}$ defines also a simple root.  This
case is covered in \cite{Damour:2001sa}.}.  In these equations, the
$\l^{(p)}_\a$ are the couplings to the dilatons $\phi^\a$ present
in $D$ dimensions (if any).

\section{Obstructions to oxidation}
A crucial feature of the gravitational walls (symmetry and
curvature walls) is that they are non degenerate.  By contrast,
the $p$--form walls may come with a non trivial multiplicity since
one can include many $p$--forms with the same degree and same
dilaton couplings.

It follows from this observation that a ``necessary condition''
for oxidability of a $3$--dimensional theory to $D>3$ dimensions is
that the (very) long roots of $\bar{B}$ must be non degenerate,
since the affine root, which is the magnetic wall $\b^1 -
\theta(\phi^\Delta)$ in three dimensions, becomes a symmetry wall
in $D>3$ dimensions (and the degeneracy of $\b^1 -
\theta(\phi^\Delta)$ is the degeneracy of the (very) long highest
root $\theta(\phi^\Delta)$).  This necessary condition on the
highest root was spelled out in \cite{Henneaux:2003kk}. It provides an
obstruction to oxidation to $D>3$ dimensions for the following
cases (see \cite{Helgason}) for conventions):

\vspace{.2cm} \noindent  $A\; II$ ($SU^*(2n)/Sp(n)$) for which $
m_\theta = 4$; $B\; II$ and $D\; II$ ($SO(p,1)$) for which
$m_\theta = p -1$; $C\; II$ ($Sp(p,q)/Sp(p)\times Sp(q))$ for
which $m_\theta = 3$; $E\; IV$ ($E_{6 (-28)}/F_4$) for which
$m_\theta = 8$; and $F\; II$ ($F_{4(-20)}/SO(9)$) for which
$m_\theta = 7$. Here, $m_\theta$ is the multiplicity of the
highest root $\theta$.  That these non--split theories cannot be
oxidised agrees with the findings of \cite{Brown:2004jb,Keurentjes:2002rc}.

Note that this list is precisely the list of
real forms whose Satake diagrams have a ``compact'' Cartan
generator at the simple root(s) that connects (or connect) to the
affinizing root or equivalently that is not orthogonal to the most
negative root of the simple Lie algebra under consideration.
Affinization occurs in 2 dimensions; in the build-up of the
$SL(D_{max}-3,R) \times R $ chain one would expect the dilation of
the fourth coordinate to be a noncompact Cartan generator at that
vertex as soon as $D_{max} \ge 4$  \cite{J''}.
Conversely the fact that the root before the affinizing one is
noncompact, to use approximate language, seems to imply that the
affine root is noncompact too and can participate in a
$GL(D_{max}-2,R)$ group of duality symmetries \cite{J''}.

Finally, we note that the $C\; II$ (with $q=2$) and $F\; II$ cases
have a supersymmetric extension with an odd number of
$3$--dimensional supersymmetries ($N= 5$ and $N=9$, respectively)
\cite{deWit:1992up}. These supersymmetries cannot come from four dimensions
since a four--dimensional theory yields an even number of
supersymmetries in $D=3$. Our obstruction to oxidation given above
does not rely on supersymmetry, however,  and prevents even
non-supersymmetric ``parents''.

\section{Maximal dimensions}
When the theory can be oxidised, one can infer the maximal
dimension(s) in which it can be formulated from the rules recalled
above: this is $D_{max} = k + 2$ where $k$ is the length of the
symmetry wall chain, i.e. the length of the single-linked chain of
nodes in the Dynkin diagram of the restricted root system
$\bar{B}^{\wedge \wedge}$ that can be constructed starting from
the overextended root, without loop (hereafter called ``A--chain'').
If there is a fork, each branch yields an independent maximal
dimension. By mere inspection of the tables and Satake diagrams in
\cite{Helgason} and of the Dynkin diagrams of the overextensions
given e.g. in \cite{Damour:2002fz,Henneaux:2003kk}, one easily gets the results
collected in the Table \ref{realforms}. 
\vspace{3cm}
\begin{center}
\begin{table}[h]
\caption{\label{realforms}}
\begin{tabular}{|l|l|c|}
\hline  Class & Noncompact Symmetric Space D=3 & $D_{max}$ \\
\hline  $A \; I$ &$SL(r+1)/SO(r+1)$ & $r+3$  \\
\hline  $A\; III$ and
$A\; IV$ &$SU(p,q)/(S(U_p \times U_q))$  & $4$  \\
\hline  $B\; I$ and $D\; I$ &$SO(p,q) / (SO(p) \times SO(q))$,
$p \geq q >1$  & $q+2$ and $6$ ($q \geq 3$) \\
\hline $C\; I$ &$Sp(n,R)/U(n)$  & $4$ \\
\hline $D\; III$& $SO^*(2n)/U(n)$  & $4$  \\
\hline $E\; I$ & $E_{6(6)}/Sp(4)$ &  $8$ \\
 \hline $E\; II$&
$E_{6(2)}/(SU(6) \times SU(2))$& $6$ \\
\hline $E\; III$&
$E_{6(-14)}/(SO(10) \times U(1))$& 4\\
\hline $E\; V$&
$E_{7(7)}/SU(8)$ & $8$ and $10$ \\
\hline $E\;
VI$ &$E_{7 (-5)}/(SO(12) \times SU(2))$ & 6 \\
\hline $E\; VII$ & $E_{7 (-25)}/ (E_{6 (-78)} \times U(1))$ & 4 \\
\hline $E\; VIII$ & $E_{8(8)}/SO(16) $ & 10 and 11 \\
\hline $E\; IX$ & $E_{8(-24)}/(E_{7(-133)} \times SU(2))$ & 6 \\
\hline  $F\; I$ &$F_{4(4)}/(Sp(3) \times SU(2))$ &6\\
\hline   $G$ & $G_{2(2)}/SO(4)$& 5 \\
\hline
\end{tabular}
\end{table}
\end{center}

The ``necessary condition'' that we used is \emph{\`a priori }not sufficient but it is supported by the inspection of the literature. Indeed, in all maximal oxidation dimensions, a theory which correctly
reduces to $D=3$ actually exists.  These theories are listed in
\cite{Keurentjes:2002rc} in terms of previously constructed models
(\cite{Brown:2004jb,Cremmer:1999du,Sagnotti:1992qw,deWit:1991nm,D'Auria:1997cz,deWit:1992wf}).

\section{$p$--form spectrum and oxidation}
The billiard provides also information on the $p$--forms that must
be present in the maximal oxidation dimension because one knows
how the $p$--form walls must connect to the A--chain.  A simple
electric $p$--form wall connects to the $p$--th root in the A--chain
(counting now from the root at the end of the A--chain opposite to
the overextended root). A simple magnetic $p$--form wall connects
to the $(d-p-1)$--th root. One also knows the multiplicities.
Together with the Weyl group, this can be used to construct the
$p$--form spectrum and determine the dilaton couplings.

Rather than deriving explicitly all cases, we shall focus on $E\,
\; IX$, i.e., $E_{8(-24)}/(E_{7(-133)} \times SU(2))$, which has a
restricted root system of $F_4$--type

\begin{center}
\scalebox{1}{

\begin{picture}(220,20)

\thicklines \multiput(10,10)(40,0){6}{\circle{10}}
\multiput(15,10)(40,0){3}{\line(1,0){30}}
\put(175,10){\line(1,0){30}} \put(135,12.5){\line(1,0){30}}
\put(135,7.5){\line(1,0){30}} \put(145,0){\line(1,1){10}}

\put(145,20){\line(1,-1){10}} \put(-5,-5){$\alpha_{-1}$}
\put(45,-5){$\alpha_{0}$} \put(85,-5){$\alpha_{1}$}
\put(125,-5){$\alpha_{2}$}
\put(165,-5){$\alpha_{3}$}\put(205,-5){$\alpha_{4}$}
\end{picture}
}
\end{center}
The other cases are treated similarly and reproduce known results.
The root $\a_{-1}$ in the above Dynkin diagram is the overextended
root, the root $\a_0$ is the affine root, while $\a_1$ and $\a_2$
are the long roots of $F_4$. The roots $\a_{-1}, \a_0, \a_1, \a_2$
have multiplicity $1$. The roots $\a_3$ and $\a_4$ are the short
roots and have multiplicity $8$ \cite{Helgason}. The A--chain is
given by the roots $\a_{-1}, \a_0, \a_1, \a_2$, which read, in
$D_{max}=6$ dimensions, $\a_2 = \b^2 - \b^1$, $\a_1= \b^3 - \b^2$,
$\a_0 = \b^4 - \b^3$ and $\a_{-1} = \b^5 - \b^4$. Since $F^{\wedge
\wedge}_4$ has rank $6$ and since there is only five logarithmic
scale factors, one needs one dilaton.

Because the short root $\a_3$ is attached to the first root $\a_2$
of the A--chain, it corresponds to the electric wall of a $1$--form.
Requiring that the root has length squared equal to one (one has
$(\a_2 \vert \a_2) = 2$ and $(\a_2 \vert \a_3) = -1$) fixes the
dilaton coupling of the one--form to $\lambda^{(1+)} = - 1$, i.e.,
$\a_3 = \b^1 - (\phi/2)$. The last simple root $\a_4$ is not
attached to the A--chain: it corresponds therefore to an axion,
with dilaton coupling equal to $\lambda^{(0)} =  2$, i.e.,
electric $0$--form wall $\a_4 = \phi$. The degeneracy of the short
roots is $8$; hence, at this stage, we need eight $1$--forms and
eight $0$--forms.

This is not the entire spectrum of $p$--forms because we have only
accounted so far for the simple roots.  There are other positive
roots in the theory which correspond also to walls of the $p$--form
type, as can be seen by acting with the finite Weyl group of $F_4$
(= restricted Weyl group of ${\cal U}_3$) on the roots $\a_1$,
$\a_2$, $\a_3$ and $\a_4$. These roots must be included by
incorporating the corresponding $p$--forms.

By Weyl--reflecting the short root $\a_3 = \b^1 - (\phi/2)$ in
$\a_4$, we get the root $\b^1 + (\phi/2)$. This is also an
electric $1$--form wall, with degeneracy $8$. We thus need eight
further $1$--forms, with dilaton couplings $\lambda^{(1-)} =  1$.
We then observe that the symmetry wall reflection in $\b^2 - \b^1$
replaces $\b^1 - (\phi/2)$ by $\b^2 - (\phi/2)$. Reflecting this
root in $\b^1 + (\phi/2)$ yields $\b^1 + \b^2$, which is the
electric (= magnetic) wall of a chiral $2$--form with zero dilaton
coupling. This root is short, so again degenerate $8$ times: we
need eight such chiral $2$--forms (or four non--chiral $2$--forms
with zero dilaton couplings). Finally, by reflecting in $\a_3$ the
long root $\b^2 - \b^1$, we generate the long root $\b^1 + \b^2 -
\phi$. This is an electric wall for a $2$--form with dilaton
coupling $\lambda^{(2)} = - 2$, which we must include.  A
reflection in $\a_4$ yields $\b^1 + \b^2 + \phi$ but this is just
a magnetic root for the same $2$--form, so we do not need any
further additional field to get these walls in the Lagrangian.
Similarly, the short roots $\b^1 + \b^2 + \b^3 \pm (\phi/2) $
obtained by reflecting $\b^3 \pm (\phi/2)$ in $\b^1 + \b^2$ are
the magnetic walls of the $1$--forms and do not need new fields
either. [Note in passing the obvious misprints in formulas (6.21)
and (6.22) for the magnetic walls of the $2$--forms in
\cite{Damour:2002fz}. (Also the magnetic walls of $G_2$ are miswritten
there).]

To summarise: by acting with the finite Weyl group of $F_4$ on the
simple roots $\a_1$, $\a_2$, $\a_3$ and $\a_4$, we get all the
roots of $F_4$.  The positive roots, which must have a term in the
Lagrangian, are $\b^i - \b^j$ ($i>j)$ (long), $\phi$ (short),
$\b^i \pm (\phi/2)$ (short), $\b^i + \b^j$ ($i <j$) (short), $\b^i
+ \b^j \pm \phi$ ($i <j$) (long), $\b^i + \b^j + \b^k \pm
(\phi/2)$ ($i<j<k$) (short) and $2 \b^i + \b^j + \b^k$ ($i \not=
j$, $i \not=k$, $j<k$) (long). Here, $i,j,k \in \{1,2,3\}$.  These
are all the $24$ positive roots of $F_4$.  They are all accounted
for by the $p$--form walls, except the last ones, which are
curvature walls following from the Einstein--Hilbert action (and
the symmetry walls).  By acting with the $F_4^{\wedge
\wedge}$--Weyl reflections associated with the other symmetry
walls, one covariantizes the above expressions (i.e., one
generates the same walls but with $i,j,k \in \{1,2,3,4,5\}$).
Continuing, \ie acting with the other elements of the infinite
Weyl group of $F_4^{\wedge \wedge}$, one generates new walls but
these are not of the $p$--form type because they necessarily
contain a $\b^i$ with a coefficient at least 2 (except the
magnetic walls $\b^i + \b^j + \b^k + \b^m - \phi$ ($i<j<k<m$) of
the $0$--form and the null magnetic walls $\b^i + \b^j + \b^k +
\b^m$ of the dilaton, but these are already accounted for). All
the $p$--form roots have been exhausted.

We mention that the superalgebra approach gives an alternative procedure to
generate the $p$--spectrum from the Borcherds--Chevalley--Serre
relations and also provides detailed information on the
Chern--Simons terms in the Lagrangian (as the group theory approach
\cite{Keurentjes:2002xc,Keurentjes:2002rc,Keurentjes:2002vx} and the principles given in \cite{Englert:2003pd}
do). See \cite{Henry-Labordere:2003rd,Henry-Labordere:2002dk} for the split case and
\cite{Henry-Labordere:2002xh,Henry-Labordere:2002dk} for the non split case. 

\section{Results and comments}
We have shown in this chapter how the billiard analysis of
gravitational theories, related to Satake diagrams and restricted
root systems \cite{Henneaux:2003kk},  provides useful information on their
oxidation endpoint both in the split and in the non--split cases.

It is noteworthy that the split real forms allow oxidation by at
least one dimension, to $D \geq 4$. One is stuck to $D=3$ only for
certain non--split real forms, because of the non trivial
multiplicities of the (very) long roots (which are always
non-degenerate in the split case). The constraint on
multiplicities provides a rather powerful insight. {}Finally, we
mention the paper \cite{Fre:2003ep} which
also deals with ${\cal U}_3$, billiards and oxidation.
%%%%%%%%%%%%%%%%%%%%%%%%%%%%%%%%%%%%%
%%%%%%%%%%%%%%%%%%%%%%%%%%%%%%%%%
\cleardoublepage
%%%%%%%%%%%%%%%%%%%%%%%%%%%%%%%%%%%%
%%%%%%%%%%%%%%%%%%%%%%%%%%%%%%%%%%%%
%%\include{chap_hyperbolic}

\chapter{Hyperbolic Algebras}
\markboth{HYPERBOLIC ALGEBRAS }{}
\label{hyperbolic}

A criterion for the gravitational dynamics
to be chaotic is that the billiard has a finite volume. This in
turn stems from the remarkable property that the billiard can be
identified with the fundamental Weyl chamber of an hyperbolic Kac--Moody
algebra.
The purpose of this chapter is twofold: first we determine all
hyperbolic Kac--Moody algebras for which a billiard description
exists and then we explicitly construct all Lagrangians describing
gravity coupled to dilatons and $p$--forms producing these
billiards.

We are able to give exhaustive results because (i) the hyperbolic
algebras are all known and classified\footnote{Note however six
missing cases in \cite{S}, two with rank 3, two with rank 4 and
two with rank 5; their Dynkin diagrams are displayed at the end of
the chapter.} \cite{S}, and (ii) only the finite number of algebras
with rank $r$ between $3$ and $10$ are relevant in this context.
Note that there are infinitely many hyperbolic algebras of rank
two and that there exists no hyperbolic algebra of rank $r>10$.
The analysis is considerably simplified because of the invariance
of the billiard under toroidal dimensional reduction to dimensions $D\geq 3$,
see \cite{Damour:2002fz}.
Indeed, the billiard region remains
the same, but a symmetry wall in $D$ dimensions may become an
electric or magnetic $p$--form wall in a lower dimension (see previous chapter). The
invariance under dimensional reduction implies in particular that
the determination of algebras with a billiard description can be
performed by analysing Lagrangians in $D=3$ dimensions.

Simplifications
in $D=3$ occur because only $0$--forms are present: indeed, via
appropriate dualisations, all
$p$--forms can be reduced to $0$--forms. To be concrete, for the
hyperbolic algebras of real rank $r$ between
$3$ and $6$, we first try to reproduce their Dynkin diagram with a
set of $r$ dominant walls comprising one symmetry wall (\ref{centrifugal})  
($\beta^2-\beta^1)$ and
$(r-1)$ scalar walls (Eqs. (\ref{electric}) $\&$ (\ref{magnetic}) for $p=0$). 
If this can be
done, we still have to check that the remaining walls are
subdominant, \ie that they can be written as linear combinations of
the dominant ones with positive coefficients. In particular, this
analysis requires that any dominant set necessarily involves one
magnetic wall and
$(r-2)$ electric walls, as will become clear by focussing on an example. 
%Note that our search for gravitational
%Lagrangians in $D=3$ is systematic although no symmetry is required.
To deal with the hyperbolic algebras of ranks
$7$ to $10$, it is actually not necessary to first reduce to 3
dimensions: the overextensions of finite simple Lie
algebras have already been associated to billiards of some
Lagrangians and for the remaining four algebras, the rules we have
found in the previous cases allow to straightforwardly construct the
Lagrangian in the maximal oxidation dimension.

We then analyse which three--dimensional system admits parents in
higher dimensions and construct the Lagrangian in the maximal
oxidation dimension. In order to do so, we take an algebra in the
previous list and we determine successively the maximal spacetime
dimension, the dilaton number, the $p$--form content and the
dilaton couplings:

\begin{enumerate}
\item One considers the Dynkin diagram of the selected algebra and
looks at the length of its "A--chain"\footnote{An "A--chain" of
length $k$ is a chain of $k$ vertices with norm squared equal to
$2$ and simply laced.}, starting with the only symmetry root in $D=3$. Our
analysis produces the following oxidation rule:  if the A--chain has
length $k$, the theory can be oxidised up to

\begin{enumerate}
\item{$D_{max}= k+2$}, if the root(s) attached to the end of the A--chain  have a norm
squared smaller than
$2$,
\item{$D_{max}=k+1$}, if the root(s) attached to the end of the A--chain have  a norm squared greater
than
$2$.
\end{enumerate}

This generalises the oxidation rule by \cite{deBuyl:2003ub} and
\cite{Keurentjes:2002xc,Keurentjes:2002rc,Keurentjes:2002vx}, obtained by 
group theoretical arguments applied to coset models.

\item For given space dimension $d=D-1$ and rank $r$ of the
algebra, the number of dilatons is given by $n=r-d$ because the
dominant walls are required to be $r$ independent linear forms in
the $d$ scale factors $\{\beta^1,...,\beta^d\}$ and the $n$
dilatons.

\item Because it is known how the $p$--form walls connect
to the A--chain \cite{Damour:2002fz}, one can read on the Dynkin diagram
which $p$--forms\footnote{or their dual $(d-p-1)$--forms} appear in the
maximal oxidation dimension.

\item The dilaton couplings of the $p$--forms are computed
from the norms and scalar products of the walls which have to
generate the Cartan matrix of the hyperbolic algebra. This means
in particular that, even if the nature of the walls changes during
the oxidation procedure, their norms and scalar products remain
unchanged. Note also that in all dimensions $D>3$ the subdominant
conditions are always satisfied.

\end{enumerate}

As a byproduct of our analysis, we note that, for each billiard
identifiable as the fundamental Weyl chamber of an hyperbolic
algebra, the positive linear combinations of the dominant walls
representing the subdominant ones only contain integer
coefficients. Hence, the dominant walls of the Lagrangian correspond to
the simple roots of the hyperbolic algebra, while the subdominant
ones correspond to non simple positive roots. \newline
Note however that  even the 3--dimensional scalar Lagrangians do
not describe coset spaces in general. Nevertheless, the reflections
relative to the simple roots generate the Weyl group of the
hyperbolic algebra; this group in turn gives an access to other
positive roots and suggests that a Lagrangian capable to produce
these roots via billiard walls needs more exotic fields than just
$p$--forms.

This chapter is organized as follows. In the four sections, 
we deal with hyperbolic
algebras of rank 3 to 6. First, in $D=3$ spacetime dimensions, we
compute the 3--dimensional dilaton couplings needed to reproduce
the Dynkin diagram and check the status of the subdominant walls.
This is how we select the admissible algebras. Next, for each of
them, we determine which Lagrangian can be oxidised and we produce
it in the maximal oxidation dimension. The 18 hyperbolic algebras
of ranks 7 to 10 are reviewed in the last section; as explained
before, they are singled out for special treatment because 14 of
them are overextensions of finite dimensional simple Lie algebras
and the remaining 4 are dual to the overtextension $B_n^{\wedge\wedge}$
(with $n= 5,6,7,8$). We explicitly write down the $D_{max}=9$ Lagrangian
system obtained previously in \cite{MHBJ} the billiard of which is
the fundamental Weyl chamber of the algebra $CE_{10}$. Among the
four hyperbolic algebras of rank 10, $CE_{10}$ is special because,
unlike $E_{10}, BE_{10}$ and $DE_{10}$, it does not stem from
supergravities.

\section{Rank 3, 4, 5  and 6 Hyperbolic Algebras}

\subsection*{Rank 3 Hyperbolic algebras: $D=3$}

In space dimension $d=2$, one has a single symmetry wall (\ref{centrifugal}), 
namely
\beq 
\alpha_1 =\beta^2- \beta^1
\nn
\eeq
and $n=r-d=3-2=1$ dilaton denoted as $\phi$. Since we do not consider 
pure gravity (as it is already known), the dominant walls are
$p$--form walls, {\it i.e.} in the present $d=2$ case $0$--form walls. It
is obvious that only a $0$--form magnetic wall (\ref{magnetic}) can be 
connected to the symmetry wall (see the form of the metric (\ref{metric}) and
of the 0--form electric walls (\ref{electric})), say
\beq 
\alpha_2 = \beta^1-\td \lambda \phi \ .
\nn
\eeq
Let us show that the last dominant wall has to be an electric one
denoted by
\beq
\alpha_3=\td \lambda^{\prime} \phi  \ .
\label{ewa}
\eeq
Indeed, had we taken for dominant the magnetic wall
$\tilde\alpha_3=\beta^1-\td \lambda^{\prime}\phi$ instead of (\ref{ewa}), then
its corresponding electric wall (which is precisely
$\alpha_3=\td \lambda^{\prime}\phi$) would be dominant too because of the
impossibility to write it as a linear combination with positive coefficients
of the three others $\alpha_1, \alpha_2$ and
$\tilde\alpha_3$. Since we are interested in $r=3$ algebras, we only want
three dominant walls.

\noindent Using the metric (\ref{metric}) adapted to
$d=2$, we build the matrix (\ref{cartanmatrix})
\beq A_{ij} = 2 { (\alpha_i \vert \alpha_j) \over (\alpha_i \vert \alpha_i)} 
\nn
\eeq
and obtain
\beq
 A = \left(
\begin{array}{ccc}
2 & -1 & 0 \\
-{2 \over \td \lambda^2} & 2 & -2 {\td \lambda^\prime \over \td \lambda} \\
0 & -2{ \td \lambda \over \td \lambda^\prime} & 2\\
 \end{array}
\right) 
\label{Amatrix}
\eeq
which has to be identified with the generalised Cartan matrix of an
hyperbolic Kac--Moody algebra of rank 3. Because $\phi$ can be
changed into $-\phi$,
$\td \lambda$ and $\td \lambda'$ can be chosen positive.

Since in such a matrix i) the non zero off-diagonal entries are
negative integers and ii) not any finite or affine Lie algebra of
rank 2 has an off-diagonal negative integer $<-4$, one immediately
infers from the expression of $A_{21}$ in (\ref{Amatrix}) that the
allowed values for $\td \lambda$ are 
\beq \td \lambda \in
\{\sqrt{2},\, 1,\, \sqrt{2/3},\, 1/2\} \ .
\nn
\eeq 
Being a symmetry wall,
$\alpha_1$ has norm squared equal to $2$; $\alpha_2$ has norm
squared $\td \lambda^2\leq 2$, so that, if the Dynkin diagram has an
arrow between $\alpha_1$ and $\alpha_2$, this arrow must be
directed towards $\alpha_2$. Once the value of $\td \lambda$ has been
fixed, one needs to find $\td \lambda^\prime$ such that both
$2\td \lambda'/\td \lambda$ and $2\td \lambda/\td \lambda'$ are positive integers:
this leaves $\td \lambda = \td \lambda'/2, \td \lambda', 2\td \lambda'$. These
values are further constrained by the condition that the
subdominant walls $\tilde\alpha_2=\td \lambda \phi$ and
$\tilde\alpha_3=\beta^1 - \td \lambda^\prime \phi$, stay really behind
the others \ie, that there exist $k>0$ and $\ell\geq 0$ such that
\begin{eqnarray}\tilde\alpha_2 &=& k\alpha_3
\Longrightarrow \td \lambda/\td \lambda' = k\\ \tilde\alpha_3 &=&
\alpha_2+\ell\alpha_3
\Longrightarrow \td \lambda/\td \lambda' = \ell+1\end{eqnarray} which implies
$k=\ell+1\geq 1$. Hence, the subdominant conditions require
\beq
\td \lambda^\prime = \td \lambda\quad\mbox{or}\quad \td \lambda^\prime =
\td \lambda /2 \ .
\nn
\eeq  
Let us summarize the 8 different
possibilities for the pairs
$(\td \lambda, \td \lambda')$ that lead to Cartan matrices and draw the
corresponding Dynkin diagrams:

(i) for $\td \lambda = \sqrt{2}$ and $\td \lambda^\prime =
\sqrt{2}$, the Dynkin diagram describes the overextension
$A_1^{\wedge\wedge}$

\begin{center}
\scalebox{.5}{
\begin{picture}(180,60)
%nom
\put(-45,10){3-1}
%trois vertex
\thicklines \multiput(10,10)(40,0){3}{\circle{10}}
%premiere ligne
\put(15,10){\line(1,0){30}}
%deux lignes
\put(55,7.5){\line(1,0){30}} \put(55,12.5){\line(1,0){30}}
\end{picture}
}
\end{center}
and for $\td \lambda=\sqrt{2}$ and $\td \lambda^\prime= 1/\sqrt{2}$, the
Dynkin diagram corresponds to the twisted overextension \cite{Henneaux:2003kk}
$A_2^{(2)\wedge}$
\begin{center}
\scalebox{.5}{
\begin{picture}(180,60)
%nom
\put(-45,10){3-2}
%trois vertex
\thicklines \multiput(10,10)(40,0){3}{\circle{10}}
%premiere ligne
\put(15,10){\line(1,0){30}}
%quatre lignes
\put(55,8.75){\line(1,0){30}} \put(55,11.25){\line(1,0){30}}
\put(55,6.25){\line(1,0){30}} \put(55,13.75){\line(1,0){30}}
%fleche vers la droite (deuxieme ligne)
\put(65,0){\line(1,1){10}} \put(65,20){\line(1,-1){10}}
\end{picture}
}
\end{center}

(ii) $\td \lambda = 1$; the two possibilities are $\td \lambda^\prime = 1$ and
$\td \lambda^\prime= 1/2$. The Dynkin diagrams are respectively
\begin{center}
\scalebox{.5}{
\begin{picture}(180,60)
%nom
\put(-45,10){3-3}
%trois vertex
\thicklines \multiput(10,10)(40,0){3}{\circle{10}}
%deux premieres lignes
\put(15,7.5){\line(1,0){30}} \put(15,12.5){\line(1,0){30}}
%fleche vers la droite
\put(25,0){\line(1,1){10}} \put(25,20){\line(1,-1){10}}
%deux lignes
\put(55,7.5){\line(1,0){30}} \put(55,12.5){\line(1,0){30}}
\end{picture}
}
\end{center}
\begin{center}
\scalebox{.5}{
\begin{picture}(180,60)
%nom
\put(-45,10){3-4}
%trois vertex
\thicklines \multiput(10,10)(40,0){3}{\circle{10}}
%deux premieres lignes
\put(15,12.5){\line(1,0){30}} \put(15,7.5){\line(1,0){30}}
%fleche vers la droite
\put(25,0){\line(1,1){10}} \put(25,20){\line(1,-1){10}}
%quatre lignes
\put(55,8.75){\line(1,0){30}} \put(55,11.25){\line(1,0){30}}
\put(55,6.25){\line(1,0){30}} \put(55,13.75){\line(1,0){30}}
%fleche vers la droite (deuxieme ligne)
\put(65,0){\line(1,1){10}} \put(65,20){\line(1,-1){10}}
\end{picture}
}
\end{center}

(iii) $\td \lambda = \sqrt{2/3}$; the two possibilities are
$\td \lambda^\prime = \sqrt{2/3}$ and $\td \lambda^\prime= 1/\sqrt{6}$
with Dynkin diagrams given by,
\begin{center}
\scalebox{.5}{
\begin{picture}(180,60)
%nom
\put(-45,10){3-5}
%trois vertex
\thicklines \multiput(10,10)(40,0){3}{\circle{10}}
%trois premieres lignes
\put(15,7.5){\line(1,0){30}} \put(15,12.5){\line(1,0){30}}
\put(15,10){\line(1,0){30}}
%fleche vers la droite
\put(25,0){\line(1,1){10}} \put(25,20){\line(1,-1){10}}
%deux lignes
\put(55,7.5){\line(1,0){30}} \put(55,12.5){\line(1,0){30}}
\end{picture}
}
\end{center}
\begin{center}
\scalebox{.5}{
\begin{picture}(180,60)
%nom
\put(-45,10){3-6}
%trois vertex
\thicklines \multiput(10,10)(40,0){3}{\circle{10}}
%trois premieres lignes
\put(15,12.5){\line(1,0){30}} \put(15,7.5){\line(1,0){30}}
\put(15,10){\line(1,0){30}}
%fleche vers la droite
\put(25,0){\line(1,1){10}} \put(25,20){\line(1,-1){10}}
%quatre lignes
\put(55,8.75){\line(1,0){30}} \put(55,11.25){\line(1,0){30}}
\put(55,6.25){\line(1,0){30}} \put(55,13.75){\line(1,0){30}}
%fleche vers la droite (deuxieme ligne)
\put(65,0){\line(1,1){10}} \put(65,20){\line(1,-1){10}}
\end{picture}
}
\end{center}

(iv) $\td \lambda = 1/2$; the two possibilities are $\td \lambda^\prime = 1/2$
and
$\td \lambda^\prime= 1/4$. The Dynkin diagrams are respectively,
\begin{center}
\scalebox{.5}{
\begin{picture}(180,60)
%nom
\put(-45,10){3-7}
%trois vertex
\thicklines \multiput(10,10)(40,0){3}{\circle{10}}
%quatre premieres lignes
\put(15,8.75){\line(1,0){30}} \put(15,11.25){\line(1,0){30}}
\put(15,6.25){\line(1,0){30}} \put(15,13.75){\line(1,0){30}}
%fleche vers la droite
\put(25,0){\line(1,1){10}} \put(25,20){\line(1,-1){10}}
%deux lignes
\put(55,7.5){\line(1,0){30}} \put(55,12.5){\line(1,0){30}}
\end{picture}
}
\end{center}
\begin{center}
\scalebox{.5}{
\begin{picture}(180,60)
%nom
\put(-45,10){3-8}
%trois vertex
\thicklines \multiput(10,10)(40,0){3}{\circle{10}}
%quatre premieres lignes
\put(15,8.75){\line(1,0){30}} \put(15,11.25){\line(1,0){30}}
\put(15,6.25){\line(1,0){30}} \put(15,13.75){\line(1,0){30}}
%fleche vers la droite
\put(25,0){\line(1,1){10}} \put(25,20){\line(1,-1){10}}
%quatre lignes
\put(55,8.75){\line(1,0){30}} \put(55,11.25){\line(1,0){30}}
\put(55,6.25){\line(1,0){30}} \put(55,13.75){\line(1,0){30}}
%fleche vers la droite (deuxieme ligne)
\put(65,0){\line(1,1){10}} \put(65,20){\line(1,-1){10}}
\end{picture}
}
\end{center}

\textbf{Comments}

1) When $\td \lambda^\prime=\td \lambda$, $\alpha_2$ and $\alpha_3$ have
to be assigned to a single scalar field; when
$\td \lambda^\prime\ne\td \lambda$, two scalars are needed in the 3--dimensional
Lagrangian.

2) The algebra $(3-8)$ is missing in table 2 of reference
\cite{S}. The subalgebra obtained when removing the first or the
last root is the affine $A_2^{(2)}$; the removal of the middle
root gives $A_1\times A_1$ so that this algebra satisfies indeed
the criterion of hyperbolicity.

3) Remark that none of the 8 algebras above is strictly hyperbolic
\footnote{A strictly hyperbolic algebra is such that upon removal of a
simple root of its Dynkin diagram, only finite Lie algebras Dynkin diagram remain. A Kac--Moody \emph{projected }billiard associated with a strictly hyperbolic aglebra is compact. }. The latter are listed in table 1 of \cite{S}.

\subsection*{Rank 3 Hyperbolic Algebras: Oxidation}

The 4--dimensional Lagrangian will have no dilaton in it since $n=r-d=0$;
hence, if such a Lagrangian exists, it cannot stem from an higher dimensional
parent and $D_{max}=4$. When looking at the algebras of rank 3 selected
above, one sees that only
$(3-1)$ and
$(3-2)$ have an A--chain of length $k=2$ and allow, a priori, a
second symmetry wall. We start with 
\beq
\alpha_1=\beta^3-\beta^2\quad\mbox{and}\quad
\alpha_2=\beta^2-\beta^1.
\nn
\eeq
The third root may only contain $\beta^1$ and can be associated to
\begin{enumerate}
\item{the curvature wall
$\alpha_3 = 2\beta^1$ in the case of 4--dimensional pure gravity.}
The Dynkin diagram bears number $(3-1)$ above and is the
overextension $A_1^{\wedge\wedge}$.
\item{the electric/magnetic wall of a $1$--form: $\alpha_3 = \beta^1$.}
This case leads to diagram $(3-2)$ which belongs to the twisted overextension
$A_2^{(2)\wedge}$.
\end{enumerate}
One sees immediately that the regions of hyperbolic space
delimited by both sets of walls coincide; the difference is
entirely due to the normalization of the third wall which is thus
responsible for the emergence of two distinct Cartan matrices.

\subsection*{Rank 4 Hyperbolic Algebras: $D=3$}

In order to reproduce through walls the four roots of such an
algebra, besides the scale factors $\beta^1$ and $\beta^2$, one
needs $n=2$ dilatons; they will be denoted as
$\phi^1=\phi,\,\,\phi^2 = \varphi$. There is one symmetry wall,
\ie $\alpha_1 = \beta^2-\beta^1$ and, a priori, two choices can
be made for the next three dominant walls: either (i) one takes
one magnetic wall and two electric ones or (ii) one takes one
electric wall and two magnetic ones. We will start with case (i)
and show later how case (ii) is eliminated on account of the
subdominant conditions.

\subsubsection*{One magnetic wall and two electric ones}

The dominant walls are thus the symmetry wall 
\beq 
\alpha_1 =
\beta^2-\beta^1,
\nn 
\eeq 
the magnetic wall, written as\footnote{This
ansatz represents no loss of generality because starting from the
more general expression $\alpha_2 = \beta^1 - \td \lambda \phi +\mu
\varphi$, one can redefine the dilatons via an
orthogonal transformation - leaving the dilaton
Lagrangian invariant - to get the simpler expression used above.} 
\beq
\alpha_2 = \beta^1 - \td \lambda \phi 
\nn
\eeq 
and the two electric ones
\beq 
\alpha_3=\td \lambda'\phi - \td \mu' \varphi 
\nn
\eeq 
respectively 
\beq
\alpha_4=\td \lambda'' \phi + \td \mu'' \varphi \ .
\nn
\eeq 
As before, the signs
have already been distributed to account for the negative signs of
the off-diagonal Cartan matrix elements when allowing the
parameters to be either all $\geq 0$ or all $\leq 0$; that they
can further be chosen positive is due the possibility to change
$\phi^\alpha$ into $-\phi^\alpha$.
The general structure of the Dynkin diagram is therefore the
following
\begin{center}
\scalebox{.5}{
\begin{picture}(180,60)
\thicklines \multiput(10,10)(40,0){2}{\circle{10}}
\put(90,30){\circle{10}} \put(90,-10){\circle{10}}
\put(90,-5){\line(0,1){30}} \put(15,10){\line(1,0){30}}
\put(50,15){\line(2,1){35}} \put(50,5){\line(2,-1){35}}
\put(30,15){$q$} \put(100,10){$r$} \put(65,30){$m$}
\put(65,-15){$p$} \put(5,20){$\alpha_1$} \put(43,20){$\alpha_2$}
 \put(85,-25){$\alpha_4$} \put(85,40){$\alpha_3$}
\end{picture}
}
\end{center}\vspace{1cm}
where we have not drawn the arrows and $m, \ q, \ p, \ s$ are
integers which count the number of lines joining two vertices.

What are the possible values that can be assigned to $q$, $m$, $s$
and $p$? Since this diagram has to become the Dynkin diagram of an
hyperbolic algebra, the maximal value of each of these integers is
$3$, because there is no finite or affine algebra of rank 3 with
off-diagonal Cartan matrix elements smaller than $-3$. Another
point is that if there were an arrow between $\alpha_1$ and
$\alpha_2$ it necessarily points towards $\alpha_2$: one has
indeed $(\alpha_1,\alpha_2)=-1$, $(\alpha_1,\alpha_1)=2$ (it is a
symmetry wall), $(\alpha_2,\alpha_2) = \td \lambda^2$ and $A_{21}
=-2/\td \lambda^2$ can only be $-1,-2$ or $-3$. We may also state that
if $A_{ij}$ is neither $0$ nor $-1$ then $A_{ji} = -1$, because
this is a common property of all finite or affine algebras of rank
3. Taking all these restrictions into account, one has to consider
three different situations characterized respectively by (i) $m$,
$s$, $p$ are all different from zero, (ii) $s=0$ and $m$, $p$ are
not zero (iii) $p=0$ and $s$, $m$ are not zero.

(i) If $m$, $s$ and $p$ are all non zero, then they must all be equal to 1
because, upon removal of the root $\alpha_1$, one obtains a triangular
diagram; now, in the set of the finite or
affine algebras, there is only one such triangular Dynkin diagram and it is
simply laced. That leaves, a priori, three cases labelled by the values
$q=1,2,3$. The corresponding dilaton couplings are
\beq
\td \lambda =\sqrt{ {2 \over q}};\quad
\td \lambda' = {1 \over \sqrt{2q}} ;\quad \td \mu' = \sqrt{ {3 \over 2 q}};\quad
\td \lambda''= {1 \over \sqrt{2q}} ;\quad  \td \mu'' = \sqrt{ {3 \over 2
q}}.
\label{i} 
\eeq
The Dynkin diagrams corresponding to $q=1$, $2$ and $3$ are
respectively,
% diagramme correspondant \~{o} q=1
\begin{center}
\scalebox{.5}{
\begin{picture}(180,60)
%nom
\put(-45,10){4-1}
%trois vertex + une ligne entre chaque
\thicklines \multiput(10,10)(40,0){3}{\circle{10}}
\multiput(15,10)(40,0){2}{\line(1,0){30}}
%vertex au dessus
\put(50,50){\circle{10}} \put(50,15){\line(0,1){30}}
%une ligne fermant le triangle
\put(55,50){\line(1,-1){35}}
\end{picture}
}
\end{center}
which is the overextension $A_2^{\wedge\wedge}$ and
% diagramme correspondant \~{o} q=2
\begin{center}
\scalebox{.5}{
\begin{picture}(180,60)
%nom
\put(-45,10){4-2}
%trois vertex + deux ligne et fleche entre les deux premiers + ligne
\thicklines \put(15,8){\line(1,0){30}} \put(15,12){\line(1,0){30}}
\multiput(10,10)(40,0){3}{\circle{10}} \put(25,0){\line(1,1){10}}
\put(25,20){\line(1,-1){10}} \put(55,10){\line(1,0){30}}
%vertex au dessus
\put(50,50){\circle{10}} \put(50,15){\line(0,1){30}}
%une ligne fermant le triangle
\put(55,50){\line(1,-1){35}}
\end{picture}
}
\end{center}
% diagramme correspondant \~{o} q=3
\begin{center}
\scalebox{.5}{
\begin{picture}(180,60)
%nom
\put(-45,10){4-3}
%trois vertex + trois ligne et fleche entre les deux premiers + ligne
 \thicklines \put(15,7.5){\line(1,0){30}}
\put(15,12.5){\line(1,0){30}}
\multiput(10,10)(40,0){3}{\circle{10}} \put(25,0){\line(1,1){10}}
\put(25,20){\line(1,-1){10}}
\multiput(15,10)(40,0){2}{\line(1,0){30}}
%vertex au dessus
\put(50,50){\circle{10}} \put(50,15){\line(0,1){30}}
%une ligne fermant le triangle
\put(55,50){\line(1,-1){35}}
\end{picture}
}
\end{center}
The subdominant conditions are satisfied in all cases; let us show
this explicitly. With the couplings in (\ref{i}), the dominant
walls other than the symmetry wall read 
\beq 
\alpha_2 = \beta^1 -
2\,\frac{\phi}{\sqrt{2q}}\quad,\quad \alpha_3 =
\frac{\phi}{\sqrt{2q}} -
\varphi\sqrt{\frac{3}{2q}}\quad,\quad\alpha_4 =
\frac{\phi}{\sqrt{2q}} + \varphi\sqrt{\frac{3}{2q}} \ .
\nn
\eeq 
The
corresponding subdominant ones are 
\beq 
\tilde\alpha_2 =
 2\,\frac{\phi}{\sqrt{2q}}\quad,\quad
\tilde\alpha_3 = \beta^1-\frac{\phi}{\sqrt{2q}} +
\varphi\sqrt{\frac{3}{2q}}
\quad,\quad \tilde\alpha_4=\beta^1-\frac{\phi}{\sqrt{2q}} -
\varphi\sqrt{\frac{3}{2q}}
\nn
\eeq 
and they obey 
\beq 
\tilde\alpha_2 =
\alpha_3+\alpha_4\quad,\quad \tilde\alpha_3 =
\alpha_2+\alpha_4\quad,\quad\tilde\alpha_4 = \alpha_2+\alpha_3 \ .
\nn
\eeq
(ii) if $s=0$ and $m$, $p$ are not zero, the structure of the
Dynkin diagram is the following
\begin{center}
\scalebox{.5}{
\begin{picture}(180,60)
%trois premiers vertex + 1 ligne entre deux premiers
\thicklines \multiput(10,10)(40,0){3}{\circle{10}}
\put(15,10){\line(1,0){30}}
%une lignes
\put(55,10){\line(1,0){30}}
% dernier vertex
\put(50,50){\circle{10}}
%1 ligne vers le haut
\put(50,15){\line(0,1){30}}
%noms des racines
\put(5,-5){$\alpha_1$} \put(45,-5){$\alpha_2$}
 \put(60,47.5){$\alpha_4$} \put(85,-5){$\alpha_3$}
\end{picture}
}
\end{center}
Comparison with the similar graphs of \cite{S} impose (i) $m=p=2$,
(ii) $q=1$ or $q=2$ and (iii) an arrow pointing from
$\alpha_2$ to $\alpha_3$ and another arrow from $\alpha_2$ to $\alpha_4$.
Accordingly, the dilaton couplings producing them are given by
\beq
\td \lambda = \sqrt{ {2 \over q}};\quad
\td \lambda' =  {1 \over \sqrt{2q}} ;\quad \td \mu' = {1 \over \sqrt{2q}} ;\quad
\td \lambda''= {1 \over \sqrt{2q}} ;\quad \td \mu'' = {1 \over \sqrt{2q}}.
\label{ii}
\eeq
The Dynkin diagrams corresponding to $q=1$ and $2$ are respectively,
%q = 1
\begin{center}
\scalebox{.5}{
\begin{picture}(180,60)
%nom
\put(-45,10){4-4}
%trois premiers vertex + 1 ligne entre deux premiers
\thicklines \multiput(10,10)(40,0){3}{\circle{10}}
\put(15,10){\line(1,0){30}}
%deux lignes
\put(55,8){\line(1,0){30}} \put(55,12){\line(1,0){30}}
% dernier vertex
\put(50,50){\circle{10}}
%2 lignes vers le haut
\put(48,15){\line(0,1){30}} \put(52,15){\line(0,1){30}}
%fleche vers la droite (deuxieme ligne)
\put(65,0){\line(1,1){10}} \put(65,20){\line(1,-1){10}}
%fleche vers le haut(dernier vertex)
\put(40,25){\line(1,1){10}} \put(50,35){\line(1,-1){10}}
\end{picture}
}
\end{center}
%q=2
\begin{center}
\scalebox{.5}{
\begin{picture}(180,60)
%nom
\put(-45,10){4-5}
%trois vertex + deux ligne et fleche entre les deux premiers +  2 lignes
\thicklines \put(15,8){\line(1,0){30}} \put(15,12){\line(1,0){30}}
\multiput(10,10)(40,0){3}{\circle{10}} \put(25,0){\line(1,1){10}}
\put(25,20){\line(1,-1){10}} \put(55,12){\line(1,0){30}}
\put(55,8){\line(1,0){30}}
% dernier vertex
\put(50,50){\circle{10}}
%2 lignes vers le haut
\put(48,15){\line(0,1){30}} \put(52,15){\line(0,1){30}}
%fleche vers la droite (deuxieme ligne)
\put(65,0){\line(1,1){10}} \put(65,20){\line(1,-1){10}}
%fleche vers le haut(dernier vertex)
\put(40,25){\line(1,1){10}} \put(50,35){\line(1,-1){10}}
\end{picture}
}
\end{center}
Again, the subdominant conditions are fulfilled: indeed, one gets
$\tilde\alpha_2=\alpha_3+\alpha_4$, $\tilde\alpha_3=\alpha_2+\alpha_4$,
$\tilde\alpha_4=\alpha_2+\alpha_3$.
(iii) if $p=0$ and $s$, $m$ are not zero, the structure of the
Dynkin diagram is the following
\begin{center}
\scalebox{.5}{
\begin{picture}(180,60)
%trois premiers vertex + 1 ligne entre chaque
\thicklines \multiput(10,10)(40,0){4}{\circle{10}}
\multiput(15,10)(40,0){3}{\line(1,0){30}}
%nom des racines
\put(5,-5){$\alpha_1$} \put(45,-5){$\alpha_2$}
 \put(125,-5){$\alpha_4$} \put(85,-5){$\alpha_3$}
\end{picture}
}
\end{center}
The dominant walls now simplify as
\beq
\alpha_1 =
\beta^2-\beta^1 \quad,\quad \alpha_2 = \beta^1 - \td \lambda
\phi
\quad,\quad \alpha_3 =\td \lambda'\phi - \td \mu'
\varphi\quad,\quad \alpha_4 = \td \mu'' \varphi \ .
\nn
\eeq 
We want the
corresponding magnetic and electric walls to be effectively
subdominant: this is indeed satisfied when 1) $\td \lambda$ and
therefore $\td \lambda'$ are positive; 2) $\td \lambda' / \td \lambda \leq 1$
(that is $\td \lambda' / \td \lambda = 1$ or $\td \lambda' / \td \lambda =1/2$)
and 3) $\td \mu' / \td \mu'' \geq \td \lambda' / \td \lambda$. Accordingly, the
remaining possibilities for $\td \lambda$, $\td \lambda'$ and $\td \mu'$ are,
a priori, those given in the following table:
\begin{center}
\begin{tabular}{|c|c|c|c|}
\hline & $\td \lambda$ & $\td \lambda'$  & $ \td \mu'$ \\
\hline 1 &  $\sqrt{2}$ & $\sqrt{2}$ & $\sqrt{2}$  \\
\hline 2.a &  $\sqrt{2}$ & ${1 / \sqrt{2}}$ & $\sqrt{3 / 2}$  \\
\hline 2.b &  $\sqrt{2}$ & ${1 / \sqrt{2}}$ & ${1 / \sqrt{2}}$  \\
\hline 2.c &  $\sqrt{2}$ & ${1 / \sqrt{2}}$ & ${1 / \sqrt{6}}$  \\
\hline 3 &1 & 1& 1  \\
\hline 4.a  &1 & ${1/ 2}$ & ${  \sqrt{3}/ 2}$  \\
\hline  4.b &1 & ${1/ 2}$ & ${ 1 / 2}$  \\
\hline 4.c  &1 & ${1/2}$ & ${ 1 /\sqrt{12}}$  \\
\hline 5 & $ \sqrt{ {2 / 3}}$ &  $ \sqrt{ {2 / 3}}$ & $ \sqrt{ {2 /
3}}$  \\
\hline 6.a &  $ \sqrt{ {2 / 3}}$ &  $  {1 / \sqrt{6}}$ &
${1 / \sqrt{2}}$ \\
\hline 6.b&  $ \sqrt{ 2 / 3}$ &  $  {1 / \sqrt{6}}$ & ${1
/ \sqrt{6}}$ \\
\hline 6.c&  $ \sqrt{ {2 / 3}}$ & $  {1 / \sqrt{6}}$ & ${1
/ \sqrt{18}}$\\
\hline
\end{tabular}
\end{center}
The different values for $\td \mu'$ correspond to distinct admissible
values for $A_{32}$. Finally, for the values of $\td \mu''$, we again
meet two cases depending on which of $A_{34}$ or $A_{43}$ is equal
to $-1$. In each case, one has still to check the subdominant
conditions.

1. The two possibilities lead to a Cartan matrix: either $\td \mu'' =
2\sqrt{2}$ or $\td \mu'' = \sqrt{2}$. The former case is ruled out
because the subdominant conditions cannot be satisfied. The Dynkin
diagram of the remaining case describes the twisted overextension
$D_3^{(2)\wedge}$
\begin{center}
\scalebox{.5}{
\begin{picture}(180,60)
%nom
\put(-45,10){4-6} \thicklines
\multiput(10,10)(40,0){4}{\circle{10}} \put(15,10){\line(1,0){30}}
%double line (la deuxieme)
\put(55,12.5){\line(1,0){30}} \put(55,7.5){\line(1,0){30}}
%fleche vers la gauche
\put(65,10){\line(1,1){10}} \put(65,10){\line(1,-1){10}}
%double ligne (la troisieme) plus fleche vers la droite
\put(95,12.5){\line(1,0){30}} \put(95,7.5){\line(1,0){30}}
\put(105,0){\line(1,1){10}} \put(105,20){\line(1,-1){10}}
\end{picture}
}
\end{center}

2. a. Either $\td \mu'' = \sqrt{2/3}$ or $\td \mu'' =
\sqrt{6}$; both lead to hyperbolic algebras which correspond
respectively to the overextension $G_2^{\wedge\wedge}$

\begin{center}
\scalebox{.5}{
\begin{picture}(180,60)
%nom
\put(-45,10){4-7 } \thicklines
\multiput(10,10)(40,0){4}{\circle{10}}
\multiput(15,10)(40,0){2}{\line(1,0){30}}
%triple line (deuxieme ligne)
\put(95,12.5){\line(1,0){30}} \put(95,7.5){\line(1,0){30}}
\put(95,10){\line(1,0){30}}
%fleche
\put(105,0){\line(1,1){10}} \put(105,20){\line(1,-1){10}}
\end{picture}
}
\end{center}
and to the twisted overextension $D_4^{(3)\wedge}$
\begin{center}
\scalebox{.5}{
\begin{picture}(180,60)
%nom
\put(-45,10){4-8 } \thicklines
\multiput(10,10)(40,0){4}{\circle{10}}
\multiput(15,10)(40,0){2}{\line(1,0){30}}
\put(95,12.5){\line(1,0){30}} \put(95,10){\line(1,0){30}}
\put(95,7.5){\line(1,0){30}} \put(105,10){\line(1,1){10}}
\put(105,10){\line(1,-1){10}}
\end{picture}
}
\end{center}

2. b. Either $\td \mu'' = \sqrt{1/2}$ or $\td \mu'' =
\sqrt{2}$; the Dynkin diagrams correspond respectively to
the twisted overextension $A_4^{(2)\wedge}$
\begin{center}
\scalebox{.5}{
\begin{picture}(180,60)
%nom
\put(-45,10){4-9} \thicklines
\multiput(10,10)(40,0){4}{\circle{10}} \put(15,10){\line(1,0){30}}
%double line (la deuxieme)
\put(55,12.5){\line(1,0){30}} \put(55,7.5){\line(1,0){30}}
%fleche vers la droite
\put(65,0){\line(1,1){10}} \put(65,20){\line(1,-1){10}}
%double ligne (la troisieme) plus fleche vers la droite
\put(95,12.5){\line(1,0){30}} \put(95,7.5){\line(1,0){30}}
\put(105,0){\line(1,1){10}} \put(105,20){\line(1,-1){10}}
\end{picture}
}
\end{center}
and to the overextension $C_2^{\wedge\wedge}$
\begin{center}
\scalebox{.5}{
\begin{picture}(180,60)
%nom
\put(-45,10){4-10}\thicklines
\multiput(10,10)(40,0){4}{\circle{10}} \put(15,10){\line(1,0){30}}
%double line (la deuxieme)
\put(55,12.5){\line(1,0){30}} \put(55,7.5){\line(1,0){30}}
%fleche vers la droite
\put(65,0){\line(1,1){10}} \put(65,20){\line(1,-1){10}}
%double ligne (la troisieme) plus fleche vers la gauche
\put(95,12.5){\line(1,0){30}} \put(95,7.5){\line(1,0){30}}
\put(105,10){\line(1,1){10}} \put(105,10){\line(1,-1){10}}
\end{picture}
}
\end{center}

2. c. Only the value $\td \mu''= 2/\sqrt{6}$ is
compatible with the subdominant conditions. The corresponding algebra is
given by
\begin{center}
\scalebox{.5}{
\begin{picture}(180,60)
%nom
\put(-45,10){4-11}
%trois premiers vertex + 1 ligne entre chaque
\thicklines \multiput(10,10)(40,0){4}{\circle{10}}
\multiput(15,10)(40,0){3}{\line(1,0){30}}
% deux lignes supplementaires (deuxieme)
\put(55,12.5){\line(1,0){30}} \put(55,7.5){\line(1,0){30}}
%fleche vers la droite
\put(65,0){\line(1,1){10}} \put(65,20){\line(1,-1){10}}
\end{picture}
}
\end{center}

3. Here again, only the value $\td \mu''=1$ can be retained on account of the
subdominant conditions. This leads to
\begin{center}
\scalebox{.5}{
\begin{picture}(180,60)
%nom
\put(-45,10){4-12}
%double ligne + fleche vers la droite
\thicklines \multiput(10,10)(40,0){4}{\circle{10}}
\put(15,12){\line(1,0){30}} \put(15,8){\line(1,0){30}}
\put(25,0){\line(1,1){10}} \put(25,20){\line(1,-1){10}}
%double line (la deuxieme)
\put(55,12.5){\line(1,0){30}} \put(55,7.5){\line(1,0){30}}
%fleche vers la gauche
\put(65,10){\line(1,1){10}} \put(65,10){\line(1,-1){10}}
%double ligne (la troisieme) plus fleche vers la droite
\put(95,12.5){\line(1,0){30}} \put(95,7.5){\line(1,0){30}}
\put(105,0){\line(1,1){10}} \put(105,20){\line(1,-1){10}}
\end{picture}
}
\end{center}

4. a. Either $\td \mu'' = \sqrt{3}$ or $\td \mu'' = 1/
\sqrt{3}$; both values are admissible. They lead to
\begin{center}
\scalebox{.5}{
\begin{picture}(180,60)
%nom
\put(-45,10){4-13}
%double ligne + fleche vers la droite
\thicklines \multiput(10,10)(40,0){4}{\circle{10}}
\put(15,12){\line(1,0){30}} \put(15,8){\line(1,0){30}}
\put(25,0){\line(1,1){10}} \put(25,20){\line(1,-1){10}}
%simple line (la deuxieme)
\put(55,10){\line(1,0){30}}
%triple ligne (la troisieme) plus fleche vers la droite
\put(95,12.5){\line(1,0){30}} \put(95,7.5){\line(1,0){30}}
\put(95,10){\line(1,0){30}} \put(105,10){\line(1,1){10}}
\put(105,10){\line(1,-1){10}}
\end{picture}
}
\end{center}

\begin{center}
\scalebox{.5}{
\begin{picture}(180,60)
%nom
\put(-45,10){4-14}
%double ligne + fle\`{A}che vers la droite
\thicklines \multiput(10,10)(40,0){4}{\circle{10}}
\put(15,12){\line(1,0){30}} \put(15,8){\line(1,0){30}}
\put(25,0){\line(1,1){10}} \put(25,20){\line(1,-1){10}}
%simple line (la deuxieme)
\put(55,10){\line(1,0){30}}
%triple ligne (la troisieme) plus fleche vers la gauche
\put(95,12.5){\line(1,0){30}} \put(95,7.5){\line(1,0){30}}
\put(95,10){\line(1,0){30}} \put(105,0){\line(1,1){10}}
\put(105,20){\line(1,-1){10}}
\end{picture}
}
\end{center}

4. b. Either $\td \mu'' = 1/2$ or $\td \mu'' = 1$; both values are allowed and
they give respectively
\begin{center}
\scalebox{.5}{
\begin{picture}(180,60)
%nom
\put(-45,10){4-15}
%double ligne + fleche vers la droite
\thicklines \multiput(10,10)(40,0){4}{\circle{10}}
\put(15,12){\line(1,0){30}} \put(15,8){\line(1,0){30}}
\put(25,0){\line(1,1){10}} \put(25,20){\line(1,-1){10}}
%double line (la deuxieme)
\put(55,12.5){\line(1,0){30}} \put(55,7.5){\line(1,0){30}}
%fleche vers la droite
\put(65,0){\line(1,1){10}} \put(65,20){\line(1,-1){10}}
%double ligne (la troisieme) plus fle\`{A}che vers la droite
\put(95,12.5){\line(1,0){30}} \put(95,7.5){\line(1,0){30}}
\put(105,0){\line(1,1){10}} \put(105,20){\line(1,-1){10}}
\end{picture}
}
\end{center}

\begin{center}
\scalebox{.5}{
\begin{picture}(180,60)
%nom
\put(-45,10){4-16}
%double ligne + fleche vers la droite
\thicklines \multiput(10,10)(40,0){4}{\circle{10}}
\put(15,12.5){\line(1,0){30}} \put(15,7.5){\line(1,0){30}}
\put(25,0){\line(1,1){10}} \put(25,20){\line(1,-1){10}}
%double line (la deuxieme)
\put(55,12.5){\line(1,0){30}} \put(55,7.5){\line(1,0){30}}
%fleche vers la droite
\put(65,0){\line(1,1){10}} \put(65,20){\line(1,-1){10}}
%double ligne (la troisieme) plus fleche vers la gauche
\put(95,12.5){\line(1,0){30}} \put(95,7.5){\line(1,0){30}}
\put(105,10){\line(1,1){10}} \put(105,10){\line(1,-1){10}}
\end{picture}
}
\end{center}

4. c. Does not correspond to any hyperbolic algebra. 

5. Only the value $\td \mu''= \sqrt{2/3}$ is compatible with the
subdominant conditions but again there is no corresponding
hyperbolic algebra.

6. a. Only the first of the 2 values $\td \mu''=\sqrt{2}$ and
$\td \mu''=\sqrt{2}/3$ leads to an hyperbolic algebra, which is
\begin{center}
\scalebox{.5}{
\begin{picture}(180,60)
%nom
\put(-45,10){4-17}
%triple ligne + fleche vers la droite
\thicklines \multiput(10,10)(40,0){4}{\circle{10}}
\put(15,10){\line(1,0){30}} \put(15.5,12.5){\line(1,0){30}}
\put(15,7.5){\line(1,0){30}} \put(25,0){\line(1,1){10}}
\put(25,20){\line(1,-1){10}}
%simple line (la deuxieme)
\put(55,10){\line(1,0){30}}
%triple ligne (la troisieme) plus fleche vers la gauche
\put(95,12.5){\line(1,0){30}} \put(95,7.5){\line(1,0){30}}
\put(95,10){\line(1,0){30}} \put(105,10){\line(1,1){10}}
\put(105,10){\line(1,-1){10}}
\end{picture}
}
\end{center}

6. b. and 6. c. do not give an hyperbolic algebra.

\subsubsection*{One electric wall and two magnetic ones \label{EMM}}

This case can be eliminated on account of the subdominant
conditions. Indeed, without loss of generality, one may choose the
parametrization of the dominant walls such that the electric wall
takes a simple form, \ie, such that
\begin{eqnarray}
\alpha_1 &=& \beta^2 - \beta^1 \nonumber\\
\alpha_2&=&\beta^1 -\td \lambda\,\phi - \td \mu\,\varphi \nonumber\\
\alpha_3 &=&\beta^1 -\td \lambda'\,\phi +\td \mu'\,\varphi \nonumber\\
\alpha_4 &=& \td \lambda''\,\phi.\label{428}\end{eqnarray} Being
assumed subdominant, the electric walls associated to $\alpha_2$
and $\alpha_3$, namely $\tilde \alpha_2 = \td \lambda\,\phi +
\td \mu\,\varphi$ and $\tilde \alpha_3 = \td \lambda'\,\phi
-\td \mu'\,\varphi$ need be proportional to $\alpha_4$; this happens
only when $\td \mu= \td \mu'=0$,  but then (\ref{428}) does no longer
define a rank four root system.

\subsection*{Rank 4 Hyperbolic algebras: Oxidation}

Our aim is now to determine which of the 17 algebras selected in
the previous section admit Lagrangians in higher spacetime
dimensions and to provide the maximal oxidation dimension and the
$p$--forms content with its characteristic features. By considering
each Dynkin diagram and looking at the length of the A--chain
starting from the symmetry root $\alpha_1$, we establish the
following "empirical" oxidation rule:  if the A--chain has length
$k$ one can oxidise the spatial dimension up to (i) $d=k+1$ if the
norm squared of the next connected root is smaller than $2$ and up
(ii) to $d=k$ if the norm squared of the next connected root is
greater than $2$. In particular, the subdominant conditions are
always satisfied. Explicitly,

\begin{enumerate}
\item{Diagram $(4-1)$} is the overextension
$A_2^{\wedge\wedge}$. We know from \cite{Damour:2002fz} that it
corresponds to pure gravity in
$D_{max}=5$
\item{Diagrams $(4-2)$ and $(4-3)$} have an A--chain of length 1; the
3--dimensional theory cannot be oxidised.
\item{Diagram $(4-4)$} : $D_{max}=4$. The walls are given by
\begin{eqnarray}\alpha_1 &=&
\beta^3-\beta^2,
\quad\quad\quad\quad
\alpha_2=
\beta^2-\beta^1,\\ \alpha_3 &=& \beta^1 - \phi/ \sqrt{2},\quad\quad
\alpha_4=\beta^1 + \phi/ \sqrt{2} .\end{eqnarray} The last two are the
electric and magnetic dominant walls of a one--form coupled to the dilaton.
One sees immediately that $\tilde\alpha_3 = \alpha_4$ and
$\tilde\alpha_4=\alpha_3$.
\item{Diagram $(4-5)$} : the 3--dimensional Lagrangian has
no parent in
$D>3$.
\item{Diagram $(4-6)$} is the twisted overextension $D_3^{(2)\wedge}$.
The 3-D Lagrangian cannot be oxidised the reason being that
$\Vert\alpha_3\Vert^2>2$.
\item{Diagram $(4-7)$} is the overextension $G_2^{\wedge \wedge}$. We
know from \cite{Damour:2002fz} that the theory can be oxidised up to 
$D_{max}= 5$
where the Lagrangian is that of the Einstein-Maxwell system.
\item{Diagram $(4-8)$} describes $D_4^{(3)\wedge}$. The A--chain has length
$k=3$ and the next connected root is longer than $\sqrt{2}$. The maximal
oxidation dimension is
$D_{max}=4$ and the dominant walls are given by
\begin{eqnarray}
\alpha_1 &=&
\beta^3-\beta^2,\quad\quad\quad\,\,\,\,\, \alpha_2= \beta^2-\beta^1,\\
\alpha_3 &=&
\beta^1 - \sqrt{3/2} \phi, \quad\quad \alpha_4= \sqrt{6} \phi
\end{eqnarray} 
The root $\alpha_3$ is the electric wall of a
$1$--form,
$\alpha_4$ is the electric wall of a $0$--form. One easily checks that
the subdominant magnetic walls satisfy 
\begin{eqnarray}
\tilde\alpha_3 &=&
\beta^1+
\sqrt{3/2}
\phi\,\, =\,\, \alpha_3+\alpha_4 \\
\tilde\alpha_4 &=&
\beta^1+\beta^2-\sqrt{6} \phi\,\, =\,\, 2\alpha_3+\alpha_2.
\end{eqnarray}
\item{Diagram $(4-9)$} represents $A_4^{(2)\wedge}$. $D_{max}=4$.
Its billiard realisation requires
\begin{eqnarray} 
\alpha_1 &=&
\beta^3-\beta^2,\quad\quad\quad\quad \alpha_2= \beta^2-\beta^1,\\
\alpha_3 &=&
\beta^1 - \sqrt{1/2} \phi,\quad\quad\alpha_4 = \sqrt{1/2} \phi.
\end{eqnarray} 
The last two are again the electric walls of a $1$--form and
a zero--form; only the dilaton couplings differ from the previous
ones. The subdominant conditions are fulfilled: indeed, one obtains
$\tilde\alpha_3=\alpha_3+2\alpha_4$ and
$\tilde\alpha_4=2\alpha_3+\alpha_4+\alpha_2$.
\item{Diagram $(4-10)$} is the overextension $C_2^{\wedge\wedge}$. We
know from \cite{Damour:2002fz} that the theory can be oxidised up to 
$D_{max}=4$.
\item{Diagram $(4-11)$} has $D_{max}=4$ and
\begin{eqnarray}
\alpha_1 &=&
\beta^3-\beta^2,\quad\quad\quad\quad \alpha_2= \beta^2-\beta^1,\\
\alpha_3 &=& \beta^1 - \sqrt{1/6} \phi, \quad\quad\alpha_4=
\sqrt{2/3} \phi .\end{eqnarray} It has the same form content as
$(4-8)$ and $(4-9)$ but the dilaton couplings are still different.
Again, the subdominant conditions are
fulfilled: $\tilde\alpha_3=\alpha_3+\alpha_4$ and
$\tilde\alpha_4=2\alpha_3+\alpha_2$.
\item{Diagrams $(4-12)$ to $(4-17)$} : their 3-D Lagrangians cannot be
oxidised because there is a unique root of norm squared equal to 2.
\end{enumerate}

\textbf{Comments}

a) The subdominant conditions are indeed always fulfilled in $D>3$ and
only positive integer coefficients enter the linear combinations.

b) In case $(4-4)$, $\alpha_3$ and $\alpha_4$ are the electric and
magnetic walls of the same one--form. In the other cases, they are
respectively assigned to a single one--form and a single zero--form.
The root multiplicity being one, there is no room for various
$p$--forms with identical couplings.

\subsection*{Rank 5 Hyperbolic Algebras: $D=3$}

The 3--dimensional Lagrangians need $N=r-d=3$ dilatons
($\phi^1=\phi, \phi^2=\varphi, \phi^3=\psi$); there are two scale
factors and one symmetry wall $\alpha_1= \beta^2-\beta^1$. In
order to reproduce the other four simple roots of the algebra in
terms of dominant walls, one has a priori three different cases to
consider: indeed, the set of dominant walls can comprise (i) one
magnetic wall and three electric ones, (ii) two electric walls and
two magnetic ones and (iii) one electric wall and three magnetic
ones. Only the first possibility will survive because as soon as
the set of dominant walls contains more than one magnetic wall,
one can show that the corresponding electric walls cannot fulfill
the subdominant conditions. Although the proof is a
straightforward generalisation of the one given in subsection
(\ref{EMM}), we will provide it at the end of this section.

\subsubsection*{One magnetic wall and three electric ones}

As in the previous sections, we use the
freedom to redefine dilatons through an orthogonal transformation and
choose the parametrization of the dominant walls such that
\begin{eqnarray}
\alpha_1 &=& \beta^2-\beta^1 \\
\alpha_2 &=& \beta^1 - \td \lambda \phi\\\alpha_3 &=&\td \lambda'\phi - \td \mu'
\varphi\\ \alpha_4 &=&\td \lambda'' \phi + \td \mu'' \varphi - \td \nu''
\psi\\ \alpha_5 &=& \td \lambda''' \phi + \td \mu''' \varphi + \td \nu'''
\psi.
\end{eqnarray}

One sees immediately that the symmetry root $\alpha_1$ is only linked to
the magnetic root $\alpha_2$ while
$\alpha_2$ can further be connected to one, two or three roots.  According
to
\cite{S}, five different structures for the Dynkin diagrams can be
encountered; we classify them below according to the total number of roots
connected to
$\alpha_2$; this number is 4 in case A, 3 in cases B and C, 2 in cases D
and E.

\begin{center} \scalebox{.5}{
\begin{picture}(180,60)
%nom
\put(-45,10){A}
%trois vertex + une ligne entre chaque
\thicklines \multiput(10,10)(40,0){3}{\circle{10}}
\multiput(15,10)(40,0){2}{\line(1,0){30}}
%vertex au dessus
\put(50,50){\circle{10}} \put(50,15){\line(0,1){30}}
%vertex du dessous
\put(50,-30){\circle{10}} \put(50,5){\line(0,-1){30}}
%nom des racines
\put(5,-5){$\alpha_1$} \put(30,-5){$\alpha_2$}
 \put(30,-30){$\alpha_4$}  \put(30,45){$\alpha_5$}\put(85,-5){$\alpha_3$}
\end{picture}
}
\end{center}

\begin{center}
\scalebox{.5}{
\begin{picture}(180,60)
%nom
\put(-45,10){B}
%deux vertex + une ligne entre
\thicklines \multiput(10,10)(40,0){2}{\circle{10}}
\put(15,10){\line(1,0){30}}
%vertex au dessus un peu plus loin
\put(70,30){\circle{10}} \put(55,15){\line(1,1){10}}
%vertex du dessous un peu plus loin
\put(70,-10){\circle{10}} \put(55,5){\line(1,-1){10}}
%dernier vertex + lignes
\put(90,10){\circle{10}} \put(85,15){\line(-1,1){10}}
\put(85,5){\line(-1,-1){10}}
%nom des racines
\put(5,-5){$\alpha_1$} \put(30,-5){$\alpha_2$}
 \put(80,-15){$\alpha_4$}  \put(80,30){$\alpha_5$}\put(95,0){$\alpha_3$}
\end{picture}
}
\end{center}

\begin{center}
\scalebox{.5}{
\begin{picture}(180,60)
%nom
\put(-45,10){C}
%quatre vertex + lignes
\thicklines \multiput(10,10)(40,0){4}{\circle{10}}
\multiput(15,10)(40,0){3}{\line(1,0){30}}
%un vertex vers le haut
\put(50,50){\circle{10}} \put(50,15){\line(0,1){30}}
%nom des racines
\put(5,-5){$\alpha_1$} \put(45,-5){$\alpha_2$}
 \put(125,-5){$\alpha_5$}  \put(30,45){$\alpha_3$}\put(85,-5){$\alpha_4$}
\end{picture}
} \end{center}

\begin{center}
\scalebox{.5}{
\begin{picture}(180,60)
%nom
\put(-45,10){D}
%quatre vertex + lignes
\thicklines \multiput(10,10)(40,0){4}{\circle{10}}
\multiput(15,10)(40,0){3}{\line(1,0){30}}
%un vertex vers le haut
\put(50,50){\circle{10}} \put(50,15){\line(0,1){30}}
%nom des racines
\put(5,-5){$\alpha_4$} \put(45,-5){$\alpha_3$}
 \put(125,-5){$\alpha_1$}  \put(30,45){$\alpha_5$}\put(85,-5){$\alpha_2$}
\end{picture}
} \end{center}
Note that C and D simply differ by the assignment of the symmetry
root.

\begin{center}
\scalebox{.5}{
\begin{picture}(180,60)
%nom
\put(-45,10){E}
%cinq vertex + lignes
\thicklines \multiput(10,10)(40,0){5}{\circle{10}}
\multiput(15,10)(40,0){4}{\line(1,0){30}}
%nom des racines
\put(5,-5){$\alpha_1$} \put(45,-5){$\alpha_2$}
 \put(125,-5){$\alpha_4$}  \put(165,-5){$\alpha_5$}\put(85,-5){$\alpha_3$}
\end{picture}
} \end{center}

\textbf{Case A} - This case may be discarded. Indeed, there are in
fact two hyperbolic algebras with a Dynkin diagram of that shape:
one of them has a long and four short roots, while the other one
has one short and four long roots. Either one cannot find
couplings that reproduce their Cartan matrix or it is the
subdominant condition that is violated. More concretely:

A.1. Consider first the case for which
$\alpha_1,\alpha_2,\alpha_3,\alpha_4$ correspond to the short roots and
$\alpha_5$ is the long root. Then, according to \cite{S}, one needs
\beq
\Vert\alpha_1\Vert^2=\Vert\alpha_2\Vert^2 =\Vert\alpha_3\Vert^2 =
\Vert\alpha_4\Vert^2= 2 \quad\mbox{and}\quad \Vert\alpha_5\Vert^2=4 \ .
\nn
\eeq
These conditions are immediately translated into 
\beq
\td \lambda^2=2\quad,\quad
\td \lambda^{\prime 2}+ \td \mu^{\prime 2}=2\quad,\quad \td \lambda^{\prime\prime
2}+\td \mu^{\prime\prime 2} + \td \nu^{\prime\prime 2}=2\quad,\quad
\td \lambda^{\prime\prime\prime 2}+\td \mu^{\prime\prime\prime 2} +
\td \nu^{\prime\prime\prime 2}=4 \ ;
\nn
\eeq Hence $\td \lambda=\sqrt{2}$. From the shape of
the diagram or equivalently from the elements of the Cartan matrix, one
infers successively
\begin{enumerate}
\item{$A_{23}=-1=-\td \lambda\td \lambda'$} which gives $\td \lambda' = 1/\sqrt{2}$ and
$\td \mu'=\sqrt{3/2}$;
\item{$A_{24}=-1=-\td \lambda\td \lambda''$ and
$A_{34}=0=\td \lambda'\td \lambda''- \td \mu' \td \mu''$} which gives
$\td \lambda''= 1/\sqrt{2}$, $ \td \mu''=1/\sqrt{6}$ and $\td \nu''= 2/\sqrt{3}$
\item{$A_{25}=-2=-\td \lambda\td \lambda'''$} which gives $\td \lambda'''=\sqrt{2}$
\item{$A_{35}=0=\td \lambda'\td \lambda'''- \td \mu' \td \mu'''$} which gives
$\td \mu'''=\sqrt{2/3}$ and, using the norm of $\alpha_5$,
$\td \nu'''=2/\sqrt{3}$.
\end{enumerate}
Notice that the condition $A_{45}=0=\td \lambda''\td \lambda'''+\td \mu'' \td \mu'''
-\td \nu'' \td \nu'''$ is identically satisfied.

In summary, in order to fit the Dynkin diagram displayed in A
(with simple lines between $\alpha_2$ and
$\alpha_1,\alpha_3,\alpha_4$ and a double line between $\alpha_2$
and $\alpha_5$ oriented towards $\alpha_2$), besides the symmetry
wall, we need the following set of dominant walls
\begin{eqnarray}
\alpha_2 = \beta^1-\sqrt{2}\phi\quad\quad &,&\quad \alpha_4 =
\frac{\phi}{\sqrt{2}} + \frac{\varphi}{\sqrt{6}}
-\frac{2\,\psi}{\sqrt{3}} \\ \alpha_3 =
\frac{\phi}{\sqrt{2}}-\sqrt{\frac{3}{2}}\,\varphi\quad &,&\quad
\alpha_5 = \sqrt{2}\phi + \sqrt{\frac{2}{3}}\,\varphi
+\frac{2\,\psi}{\sqrt{3}} \end{eqnarray} It is now easy to verify,
for instance, that $\tilde\alpha_3 =
\beta^1-\frac{\phi}{\sqrt{2}}+\sqrt{\frac{3}{2}}\,\varphi$ cannot
be written as a positive linear combination of the $\alpha_i,
i=2,...,5$. Accordingly, on account of the subdominant conditions,
this case has to be rejected.

A.2.  There is another possibility producing the same diagram as
in A.1. above where the symmetry wall $\alpha_1$ now plays the
r\^ole of the long root: their norms are 
\beq
\Vert\alpha_1\Vert^2=2\quad\mbox{and}\quad
\Vert\alpha_2\Vert^2=\Vert\alpha_3\Vert^2 =\Vert\alpha_4\Vert^2 =
\Vert\alpha_5\Vert^2= 1 
\nn
\eeq 
but the equations giving the couplings
analogous to eq.1. to eq.4. above have no solution.

A.3. In the third case, there a short and four long roots with
norms 
\beq 
\Vert\alpha_1\Vert^2=\Vert\alpha_2\Vert^2
=\Vert\alpha_3\Vert^2 = \Vert\alpha_4\Vert^2= 2
\quad\mbox{and}\quad \Vert\alpha_5\Vert^2=1.
\nn
\eeq 
One can solve the
equations for the couplings and write the following set of
billiard walls: the symmetry wall $\alpha_1=\beta^2-\beta^1$ and
\begin{eqnarray}
\alpha_2 =
\beta^1-\sqrt{2}\phi\quad\quad &,&\quad \alpha_4 = \frac{\phi}{\sqrt{2}} +
\frac{\varphi}{\sqrt{6}} -\frac{2\,\psi}{\sqrt{3}} \\ \alpha_3 =
\frac{\phi}{\sqrt{2}}-\sqrt{\frac{3}{2}}\,\varphi\quad &,&\quad \alpha_5 =
\frac{\phi}{\sqrt{2}} +
\frac{\varphi}{\sqrt{6}} +\frac{\psi}{\sqrt{3}}. 
\nn
\end{eqnarray}
However, like in case A.1. above, one sees immediately that
$\tilde\alpha_3 = \beta^1 -
\frac{\phi}{\sqrt{2}} +
\frac{3\varphi}{\sqrt{6}}$, for instance, is not
subdominant; that is the reason why we discard this possibility.

\textbf{Cases B} - There are three hyperbolic algebras with a Dynkin
diagram of this shape.

B.1. The first one admits the following
couplings
\begin{eqnarray}
\td \lambda &=& \sqrt{2} \ ; \ \td \lambda' ={1\over \sqrt{2}} \ ; \ \td \mu' = \sqrt{{3
\over 2}} \nonumber
\nonumber\\
\td \lambda''&=& 0\quad \ ; \ \td \mu'' = \sqrt{{2 \over 3}}  \ ; \ \td \nu'' = {2
\over
\sqrt{3}}\ ; \
\td \lambda''' ={1\over \sqrt{2}} \ ; \ \td \mu''' = {1 \over \sqrt{6}}
\ ; \ \td \nu''' = {2 \over \sqrt{3} }
\end{eqnarray}
and is the overextension
$A_3^{\wedge\wedge}$
\begin{center}
\scalebox{.5}{
\begin{picture}(180,60)
%nom
\put(-45,10){5-1}
%deux vertex + une ligne entre
\thicklines \multiput(10,10)(40,0){2}{\circle{10}}
\put(15,10){\line(1,0){30}}
%vertex au dessus un peu plus loin
\put(70,30){\circle{10}} \put(55,15){\line(1,1){10}}
%vertex du dessous un peu plus loin
\put(70,-10){\circle{10}} \put(55,5){\line(1,-1){10}}
%dernier vertex + lignes
\put(90,10){\circle{10}} \put(85,15){\line(-1,1){10}}
\put(85,5){\line(-1,-1){10}}
\end{picture}
}
\end{center}

B.2. The second one has the
following Dynkin diagram
\begin{center}
\scalebox{.5}{
\begin{picture}(180,60)
%nom
\put(-45,10){5-2}
%deux vertex + deux ligne entre + fleche
 \thicklines
 \put(15,8){\line(1,0){30}} \put(15,12){\line(1,0){30}}
\multiput(10,10)(40,0){2}{\circle{10}} \put(25,0){\line(1,1){10}}
\put(25,20){\line(1,-1){10}}
%vertex au dessus un peu plus loin
\put(70,30){\circle{10}} \put(55,15){\line(1,1){10}}
%vertex du dessous un peu plus loin
\put(70,-10){\circle{10}} \put(55,5){\line(1,-1){10}}
%dernier vertex + lignes
\put(90,10){\circle{10}} \put(85,15){\line(-1,1){10}}
\put(85,5){\line(-1,-1){10}}
\end{picture}
}
\end{center}
and the following set of dilaton couplings:
\begin{eqnarray}
\td \lambda &=& 1\ ; \
\td \lambda' = {1\over 2} \ ; \ \td   \mu' = {\sqrt{3} \over 2} \nonumber \\
\td \lambda''&=& 0 \ ; \ \td   \mu'' = {1 \over \sqrt{3}}  \ ; \ \td \nu'' =
{\sqrt{2}
 \over \sqrt{3}} \ ; \
\td \lambda''' = {1\over 2} \ ; \  \td \mu''' = {1 \over 2\sqrt{3}} \ ; \
\td \nu''' = { \sqrt{2} \over \sqrt{3} }.
\end{eqnarray}

B.3. The diagram of the third one is the same as $(5-2)$ but with
the reversed arrow: this is impossible since in the present
context the norms are required to satisfy
$\Vert\alpha_1\Vert^2\geq \Vert\alpha_2\Vert^2$.

\textbf{Case C} - In order to generate this kind of structure, one
needs $\td \lambda'''= \td \mu''' =0$ and $\td \lambda' \td \lambda'' =  \td \mu'
 \td \mu''$. Next, from the subdominant condition for $\tilde\alpha_3$,
we deduce that $A_{32}$ can be $-2$ or $-3$ but since we want
hyperbolic algebras, only the value $A_{32} = -2$ can be retained.
Therefore $A_{23} = -1$ and $\td \lambda=\sqrt{2}$, $\td \lambda'=
\td \lambda''= 1 / \sqrt{2}$, $ \td \mu'=1 / \sqrt{2}$, $ \td \mu''= 1 /
\sqrt{2}$ and $\td \nu''= 1$.  A priori, one might still have
$\td \nu'''=2,1,\sqrt{2}$ but only one value is compatible with the
magnetic wall $\tilde\alpha_5$ being subdominant, namely $\td \nu''' =
1$. Accordingly, the couplings need to be defined as
\begin{eqnarray}
\td \lambda &=& \sqrt{2} \ ; \
\td \lambda' = {1\over \sqrt{2}} \ ; \  \td \mu' = {1\over \sqrt{2}} \nonumber
\\
\td \lambda''&=& {1\over \sqrt{2}} \ ; \  \td \mu'' = {1\over \sqrt{2}}  \ ; \
\td \nu'' =1 \ ; \
\td \lambda''' = 0\quad \ ; \  \td \mu''' = 0\quad \ ; \ \td \nu''' = 1 .
\nn
\end{eqnarray}
and the Dynkin diagram is the following,
\begin{center}
\scalebox{.5}{
\begin{picture}(180,60)
\put(-45,10){5-3}
%quatre vertex + lignes
\thicklines  \multiput(10,10)(40,0){4}{\circle{10}}
 \multiput(15,10)(40,0){2}{\line(1,0){30}}
%double derniere ligne + fleche vers la droite
\put(95,7.5){\line(1,0){30}} \put(95,12.5){\line(1,0){30}}
\put(105,0){\line(1,1){10}} \put(105,20){\line(1,-1){10}}
%un vertex vers le haut
\put(50,50){\circle{10}} \put(52.5,15){\line(0,1){30}}
\put(47.7,15){\line(0,1){30}}
%fleche vers le haut(dernier vertex)
\put(40,25){\line(1,1){10}} \put(50,35){\line(1,-1){10}}
\end{picture}
} \end{center}

\textbf{Cases D} - Dynkin diagrams of this shape can only be
recovered with
\be \td \lambda'' =
\td \lambda''' = 0\quad\mbox{and either}\quad \td \lambda =
\sqrt{2}\quad\mbox{or}\quad
\td \lambda = 1.\ee

\textbf{D.1.
$\td \lambda =
\sqrt{2}$}

All hyperbolic diagrams of that type have in their Cartan matrix
$A_{34} = A_{43} = -1$ which means that
$\Vert\alpha_3\Vert^2=\Vert\alpha_4\Vert^2$. Two additional cases
must be considered depending on which of $\alpha_2$ or $\alpha_5$
has a norm equal to the norm of $\alpha_3$:
\begin{enumerate}
\item in case D.1.1. we assume that the norms of
$\alpha_3$,
$\alpha_4$ and
$\alpha_5$ are equal
\item in case D.1.2. we assume that the norms of $\alpha_2$,
$\alpha_3$ and
$\alpha_4$ are equal.
\end{enumerate}
The subdominant conditions here simply reduce to
$A_{23}=-1$.

D.1.1. Again two hyperbolic algebras correspond to this case. For the
first one, the billiard walls are built out of the following
couplings
\begin{eqnarray}
\td \lambda &=& \sqrt{2}  \ ; \
\td \lambda'= {1\over \sqrt{2}} \ ; \  \td \mu' = \sqrt{{3\over 2}} \nonumber
\\
\td \lambda''&=& 0 \quad \ ; \  \td \mu'' = \sqrt{{2\over 3}}  \ ; \ \td \nu''
={1\over\sqrt{3}}  \ ; \
\td \lambda''' = 0\quad \ ; \  \td \mu''' = \sqrt{{2\over 3}} \ ; \
\td \nu''' = {2 \over \sqrt{3 }}
\label{435}
\end{eqnarray}
and the Dynkin diagram is that of the overextension $B_3^{\wedge\wedge}$
\begin{center}
\scalebox{.5}{
\begin{picture}(180,60)
%nom
\put(-45,10){5-4}
\thicklines  \multiput(10,10)(40,0){4}{\circle{10}}
 \multiput(55,10)(40,0){2}{\line(1,0){30}}
\put(15,8){\line(1,0){30}} \put(15,12){\line(1,0){30}}
\put(25,10){\line(1,1){10}} \put(25,10){\line(1,-1){10}}
%un vertex vers le haut
\put(50,50){\circle{10}} \put(50,15){\line(0,1){30}}
\end{picture}
} \end{center}
One can produce a billiard for the second one using the same
couplings as in (\ref{435}) except for
\begin{eqnarray}
\td \lambda''&=& 0 \ ; \  \td \mu'' = 2 \sqrt{{2\over 3}}  \ ; \ \td \nu''
={2\over\sqrt{3}} \ . 
\nn 
\end{eqnarray}
The Dynkin diagram here describes the twisted overextension
$A_5^{(2)\wedge}$
\begin{center}
\scalebox{.5}{
\begin{picture}(180,60)
%nom
\put(-45,10){5-5} \thicklines
 \multiput(10,10)(40,0){4}{\circle{10}}
 \multiput(55,10)(40,0){2}{\line(1,0){30}}
\put(15,8){\line(1,0){30}} \put(15,12){\line(1,0){30}}
\put(25,0){\line(1,1){10}} \put(25,20){\line(1,-1){10}}
%un vertex vers le haut
\put(50,50){\circle{10}} \put(50,15){\line(0,1){30}}
\end{picture}
} \end{center}

D.1.2.  Here also, two hyperbolic algebras correspond to this case
but one is eliminated on account of the subdominant conditions.
For the remaining one, the couplings are
\begin{eqnarray}
\td \lambda &=& \sqrt{2}  \ ; \
\td \lambda' ={1\over \sqrt{2}} \ ; \  \td \mu' = {1\over \sqrt{2}} \nonumber
\\
\td \lambda''&=& 0 \ ; \  \td \mu'' = {1\over \sqrt{2}}  \ ; \ \td \nu'' ={1\over
\sqrt{2}} \ ; \
\td \lambda''' = 0 \ ; \  \td \mu''' ={1\over \sqrt{2}} \ ; \ \td \nu'''
= {1\over \sqrt{2}}
\nn
\end{eqnarray}
and the Dynkin diagram is
\begin{center}
\scalebox{.5}{
\begin{picture}(180,60)
%nom
\put(-45,10){5-6 }
%quatre vertex + lignes simples
\thicklines  \multiput(10,10)(40,0){4}{\circle{10}}
\put(15,10){\line(1,0){30}} \put(95,10){\line(1,0){30}}
%double line (la deuxieme)
\put(55,12.5){\line(1,0){30}} \put(55,7.5){\line(1,0){30}}
%fleche vers la gauche
\put(65,10){\line(1,1){10}} \put(65,10){\line(1,-1){10}}
%un vertex vers le haut
\put(50,50){\circle{10}} \put(50,15){\line(0,1){30}}
\end{picture}
} \end{center}

\textbf{D.2. $\td \lambda = 1$}

\noindent In table 2 of reference \cite{S}, there are
two hyperbolic algebras with a Dynkin diagram of this shape. Both are
admissible for our present purpose:

D.2.1. The first one has couplings given by
\begin{eqnarray}
\td \lambda &=& 1  \ ; \
\td \lambda' = {1\over 2} \ ; \  \td \mu' = { \sqrt{3}\over 2} \nonumber \\
\td \lambda'' &=& 0 \ ; \  \td \mu'' ={2\over \sqrt{3}} \ ; \ \td \nu'' =
\sqrt{{2\over 3}} \ ; \
\td \lambda'''= 0 \ ; \  \td \mu''' = {1\over \sqrt{3}}  \ ; \
\td \nu''' =  \sqrt{{2\over3}}
\nn
\end{eqnarray}
and corresponds to
\begin{center}
\scalebox{.5}{
\begin{picture}(180,60)
%nom
\put(-45,10){5-7} \thicklines
 \multiput(10,10)(40,0){4}{\circle{10}} \put(55,10){\line(1,0){30}}
\put(15,8){\line(1,0){30}} \put(15,12){\line(1,0){30}}
\put(25,0){\line(1,1){10}} \put(25,20){\line(1,-1){10}}
%un vertex vers le haut
\put(50,50){\circle{10}} \put(50,15){\line(0,1){30}}
%double line (la troisieme)
\put(95,12.5){\line(1,0){30}} \put(95,7.5){\line(1,0){30}}
%fleche vers la gauche
\put(105,10){\line(1,1){10}} \put(105,10){\line(1,-1){10}}
\end{picture}
} \end{center}

D.2.2. The second one requires
\begin{eqnarray}
\td \lambda &=& 1  \ ; \
\td \lambda' = {1\over 2} \ ; \  \td \mu' = { \sqrt{3}\over 2} \nn \\
\td \lambda'' &=& 0 \ ; \  \td \mu'' ={1\over \sqrt{3}} \ ; \ \td \nu'' =
{1\over \sqrt{6}} \ ; \
\td \lambda'''= 0 \ ; \  \td \mu''' = {1\over \sqrt{3}}  \ ; \
\td \nu''' =  \sqrt{{2\over3}}
\nn
\end{eqnarray}
and has the following diagram
\begin{center}
\scalebox{.5}{
\begin{picture}(180,60)
%nom
\put(-45,10){5-8} \thicklines
 \multiput(10,10)(40,0){4}{\circle{10}} \put(55,10){\line(1,0){30}}
\put(15,8){\line(1,0){30}} \put(15,12){\line(1,0){30}}
\put(25,10){\line(1,1){10}} \put(25,10){\line(1,-1){10}}
%un vertex vers le haut
\put(50,50){\circle{10}} \put(50,15){\line(0,1){30}}
%double line (la troisieme)
\put(95,12.5){\line(1,0){30}} \put(95,7.5){\line(1,0){30}}
%fleche vers la gauche
\put(105,10){\line(1,1){10}} \put(105,10){\line(1,-1){10}}
\end{picture}
} \end{center}

\textbf{Cases E} - Table 2 of reference \cite{S} displays two
hyperbolic algebras of rank 5 which are duals of each other and
have linear diagrams. Only one of these two can be associated to a
billiard the walls of which correspond to

E.1. $\td \lambda =
\sqrt{2}$, $\td \lambda'' =
\td \lambda'''=
 \td \mu''' =0$  and all other dilaton couplings equal
to
$1/\sqrt{2}$.

Its Dynkin diagram is the twisted overextension
$A_{6}^{(2)\wedge}$ and is given by

\begin{center}
\scalebox{.5}{
\begin{picture}(180,60)
%nom
\put(-45,10){5-9}
%cinq vertex + lignes simples
\thicklines  \multiput(10,10)(40,0){5}{\circle{10}}
\put(15,10){\line(1,0){30}} \put(95,10){\line(1,0){30}}
%double ligne (deuxieme)
\put(55,8){\line(1,0){30}} \put(55,12){\line(1,0){30}}
%fleche vers la droite
\put(65,0){\line(1,1){10}} \put(65,20){\line(1,-1){10}}
%double ligne (quatrieme)
\put(135,8){\line(1,0){30}} \put(135,12){\line(1,0){30}}
%fleche vers la droite
\put(145,0){\line(1,1){10}} \put(145,20){\line(1,-1){10}}
\end{picture}
} \end{center}

There are however two more hyperbolic algebras with such linear Dynkin
diagrams; they are missing in \cite{S} but perfectly relevant in
the present context:

E.2. the first one is the overextension
$C_3^{\wedge\wedge}$
\begin{center}
\scalebox{.5}{
\begin{picture}(180,60)
%nom
\put(-45,10){5-10}
%cinq vertex + lignes simples
\thicklines  \multiput(10,10)(40,0){5}{\circle{10}}
\put(15,10){\line(1,0){30}} \put(95,10){\line(1,0){30}}
%double ligne (deuxieme)
\put(55,8){\line(1,0){30}} \put(55,12){\line(1,0){30}}
%fleche vers la droite
\put(65,0){\line(1,1){10}} \put(65,20){\line(1,-1){10}}
%double ligne (quatrieme)
\put(135,8){\line(1,0){30}} \put(135,12){\line(1,0){30}}
%fleche vers la gauche
\put(145,10){\line(1,1){10}} \put(145,10){\line(1,-1){10}}
\end{picture}
} \end{center}
whose couplings are equal to the previous ones except $\td \nu'''=\sqrt{2}$.

E.3. The second one is the dual of $C_3^{\wedge\wedge}$ known as
the twisted overextension $D_4^{(2)\wedge}$; its Dynkin diagram
corresponds to the previous one with reversed arrows

\begin{center}
\scalebox{.5}{
\begin{picture}(180,60)
%nom
\put(-45,10){5-11}
%cinq vertex + lignes simples
\thicklines  \multiput(10,10)(40,0){5}{\circle{10}}
\put(15,10){\line(1,0){30}} \put(95,10){\line(1,0){30}}
%double ligne (deuxieme)
\put(55,8){\line(1,0){30}} \put(55,12){\line(1,0){30}}
%fleche vers la gauche
\put(65,10){\line(1,1){10}} \put(65,10){\line(1,-1){10}}
%double ligne (quatrieme)
\put(135,8){\line(1,0){30}} \put(135,12){\line(1,0){30}}
%fleche vers la droite
\put(145,0){\line(1,1){10}} \put(145,20){\line(1,-1){10}}
\end{picture}
} \end{center}
and the dilaton couplings are such that $\td \lambda = \td \lambda'= \td \mu' =  \td \mu''
= \td \nu'' = \td \nu''' = \sqrt{2}$ while $\td \lambda'' = \td \lambda''' =  \td \mu'''=0$.

\subsubsection*{Two or more magnetic walls}

That these cases may be discarded will be proved on a particular
case but the argument can be easily generalised. Suppose the
dominant set comprises two magnetic walls and two electric ones,
we can always choose the parametrization such that
\begin{eqnarray} \alpha_1 &=& \beta^2-\beta^1\nonumber\\ \alpha_2 &=&
\beta^1 -\td \lambda\,\phi - \td \mu\,\varphi -\td \nu\,\psi\nonumber\\ \alpha_3 &=&
\beta^1 -\td \lambda'\,\phi - \td \mu'\,\varphi +\td \nu'\,\psi \label{550}\\ \alpha_4
&=&
\td \lambda''\,\phi + \td \mu''\,\varphi \nonumber\\ \alpha_5 &=&
\td \lambda'''\,\phi.\nonumber
\end{eqnarray}
Being assumed subdominant, the electric walls $\tilde\alpha_2 =
\td \lambda\,\phi + \td \mu\,\varphi +\td \nu\,\psi$ and $\tilde\alpha_3 =
\td \lambda'\,\phi + \td \mu'\,\varphi -\td \nu'\,\psi$, independent of
$\beta^1$, must be written as positive linear combinations of
$\alpha_4$ and $\alpha_5$ only; this requires $\td \nu=\td \nu'=0$ but
then (\ref{550}) can no longer describe a rank five root system.

\noindent The same argument remains of course valid for more magnetic walls.

\subsection*{Rank 5 Hyperbolic Algebras: Oxidation}

The empirical oxidation rule set up in the previous sections also
holds for the rank 5 algebras:
\begin{enumerate}
\item{Diagram $(5-1)$} is the Dynkin diagram of the overextension
$A_3^{\wedge\wedge}$. The Lagrangian is that of pure gravity in
$D_{max}=6$.
\item{Diagram $(5-2)$} : the 3--dimensional Lagrangian
cannot be oxidised because of the norm of $\alpha_2$.
\item{Diagram $(5-3)$} : The Lagrangian can be oxidised twice,
up to $D_{max}=5$. The dominant walls are the three symmetry walls
$\alpha_1 =
\beta^4-\beta^3$,
$\alpha_2=\beta^3-\beta^2$, $\alpha_3=\beta^2-\beta^1$ and the
electric wall of a $1$--form 
\beq 
\alpha_4 = \beta^1 - \sqrt{1/3}  \phi
\nn
\eeq 
and its magnetic wall
\beq 
\alpha_5 =
\beta^1 +
\beta^2 +
\sqrt{1/3}  \phi \ .
\nn
\eeq 
Obviously, $\tilde\alpha_4=\alpha_5$ and
$\tilde\alpha_5=\alpha_4$.
\item{Diagram $(5-4)$} represents $B_3^{\wedge\wedge}$. The
maximally oxidised Lagrangian is six--dimensional. The dominant walls
are here the four symmetry walls
$\alpha_1 =
\beta^5-\beta^4$, $\alpha_2=\beta^4-\beta^3$,
$\alpha_3=\beta^3-\beta^2$, $\alpha_4 = \beta^2-\beta^1$ and
\beq
\alpha_5 = \beta^1 + \beta^2
\nn
\eeq 
which is the electric
or magnetic wall of a self-dual $2-$form:
obviously $\tilde\alpha_5=\alpha_5$.
\item{Diagram $(5-5)$} is the twisted overextension
$A_5^{(2)\wedge}$. Here, $D_{max}=4$ and besides the symmetry
walls $\alpha_1=\beta^3-\beta^2$ and $\alpha_2= \beta^2-\beta^1$,
one finds
\begin{eqnarray} 
\alpha_3 &=& \beta^1-\sqrt{3/2} \phi \ , \nn \\ 
\alpha_4 &=& \sqrt{2/3}\phi - 2/ \sqrt{3} \varphi \ ,  \nn \\ 
\alpha_5 &=& 2 \sqrt{2/3} \varphi + 2/ \sqrt{3} \psi \ , 
\nn 
\end{eqnarray} 
which are the electric walls respectively of a one--form
and two
$0$--forms. One easily checks that $\tilde\alpha_3 =
\alpha_3+\alpha_4+\alpha_5$, $\tilde\alpha_4=
\alpha_2+2\alpha_3+\alpha_5$ and
$\tilde\alpha_5=\alpha_2+2\alpha_3+\alpha_4.$
\item{Diagram $(5-6)$} : $D_{max}=4$ and one needs
$\alpha_1=\beta^3-\beta^2$, $\alpha_2= \beta^2-\beta^1$ and
\begin{eqnarray}
\alpha_3 &=& \beta^1-\sqrt{1/2} \phi \ , \nn \\ 
\alpha_4 &=&
\sqrt{1/2}\phi - \sqrt{1/2} \varphi \ , \nn \\
\alpha_5 &=& \sqrt{1/2} \phi+ \sqrt{1/2} \varphi \ .
\nn 
\end{eqnarray} 
The form-field content is the
same as the previous one but the dilaton couplings are different.
Moreover: $\tilde\alpha_3=\alpha_3+\alpha_4+\alpha_5$,
$\tilde\alpha_4= \alpha_2+2\alpha_3+\alpha_5$ and $\tilde\alpha_5=
\alpha_2+2\alpha_3+\alpha_4$.
\item{Diagrams $(5-7)$ and $(5-8)$} : their 3-D Lagrangians cannot be
further oxidised because of the norm of $\alpha_2$.
\item{Diagram $(5-9)$} describes $A_{6}^{(2)\wedge}$. Here, $D_{max}=4$.
One obtains the billiard with $\alpha_1=\beta^3-\beta^2$,
$\alpha_2= \beta^2-\beta^1$ and
\begin{eqnarray}\alpha_3 &=&
\beta^1-\sqrt{1/2} \phi \ , \nn \\ 
\alpha_4 &=& \sqrt{1/2}\phi - \sqrt{1/2}
\varphi  \ , \nn  \\
\alpha_5 &=& \sqrt{1/2} \varphi \ . 
\nn
\end{eqnarray} 
One draws
the same conclusion as for $(5-6)$ and $(5-9)$ above. Here again:
$\tilde\alpha_3=\alpha_3+\alpha_4+\alpha_5$, $\tilde\alpha_4=
\alpha_2+2\alpha_3+\alpha_4+2\alpha_5$ and $\tilde\alpha_5=
\alpha_2+2\alpha_3+2\alpha_4+\alpha_5$.
\item{Diagram $(5-10)$} is the overextension
$C_3^{\wedge\wedge}$; the maximal oxidation dimension is $D_{max}=
4$ and the corresponding Lagrangian can be found in \cite{Damour:2002fz}.
\item{Diagram $(5-11)$} is the twisted overextension $D_4^{(2)\wedge}$. No
Lagrangian exists in higher dimensions.
\end{enumerate}

\textbf{Comment}

\noindent The results of this section show again that the subdominant
conditions play an important r\^ole in three dimensions where they
effectively contribute to the elimination of several Dynkin
diagrams. However, once they are satisfied in three dimensions,
they are always fulfilled in all dimensions where a Lagrangian
exists and only integers enter the linear combinations.

\subsection*{Rank 6 Hyperbolic Algebras: $D=3$}

The number of dilatons in the 3--dimensional Lagrangian is equal to
$n=4$: we denote them by $\phi^1=\phi, \phi^2=\varphi,
\phi^3=\psi, \phi^4=\chi$. A straightforward generalisation of the
argument used in the previous sections implies that a single
configuration for the set of dominant walls has to be considered.
It comprises one magnetic wall and four electric ones.

\noindent 
After allowed simplifications, the dominant walls are parametrized
according to:
\begin{eqnarray}
\alpha_1 &=&\beta^2-\beta^1 \ , \nn \\
\alpha_2 &=& \beta^1 - \td \lambda \phi  \ , \nn \\ 
\alpha_3 &=&\td \lambda'\phi -  \td \mu' \varphi \ , \nn \\ 
\alpha_4 &=&\td \lambda'' \phi +  \td \mu''\varphi - \td \nu'' \psi \ , \nn \\ 
\alpha_5 &=& \td \lambda''' \phi +  \td \mu''' \varphi + \td \nu'''
\psi - \rho''' \chi \ , \nn \\
\alpha_6 &=& \td \lambda'''' \phi +
 \td \mu'''' \varphi + \td \nu'''' \psi + \rho''''\chi \ . \nn 
\end{eqnarray}
The structure of the Dynkin diagrams is therefore displayed in one
of the cases labelled $A$ to $E$ below, depending on the number of
vertices connected to $\alpha_2$. When necessary, further
subclasses are introduced according to the number of vertices
linked to $\alpha_3$,...

\textbf{Case A} - The central vertex is labelled $\alpha_2$ and is
connected to the five other vertices:
\begin{center}
\scalebox{.5}{
\begin{picture}(180,60)
%nom
\put(-45,10){A}
%trois vertex + lignes simples
\thicklines  \multiput(10,10)(40,0){3}{\circle{10}}
 \multiput(15,10)(40,0){2}{\line(1,0){30}}
%vertex du dessus
\put(50,40){\circle{10}} \put(50,15){\line(0,2){20}}
\put(30,15){$\alpha_2$}
%vertex's du dessous
\put(35,-20){\circle{10}} \put(65,-20){\circle{10}}
\put(65,-15){\line(-1,2){11}} \put(35,-15){\line(1,2){11}}
\end{picture}
} 
\end{center}
There is a single hyperbolic algebra of this type in \cite{S}; one can
solve the equations for the dilaton couplings, but the
subdominant walls are not expressible as positive linear combinations of
the dominant ones.

\textbf{Case B} - The root $\alpha_2$ is connected to four vertices:
\begin{center}
\scalebox{.5}{
\begin{picture}(180,60)
%nom
\put(-45,10){B}
%quatre vertex + lignes simples
\thicklines  \multiput(10,10)(40,0){4}{\circle{10}}
 \multiput(15,10)(40,0){3}{\line(1,0){30}}
%vertex du dessous
\put(50,-30){\circle{10}} \put(50,5){\line(0,-1){30}}
%vertex du dessus
\put(50,50){\circle{10}} \put(50,15){\line(0,1){30}}
%nom des racines
\put(5,-5){$\alpha_1$} \put(30,-5){$\alpha_2$}
 \put(125,-5){$\alpha_6$}  \put(30,45){$\alpha_3$}\put(85,-5){$\alpha_5$}
 \put(30,-30){$\alpha_4$}
\end{picture}
} 
\end{center}
There are three hyperbolic algebras with that kind of Dynkin
diagram but none of them can be retained: indeed, couplings exist
but the subdominant conditions cannot be fulfilled.

\textbf{Cases C} - $\alpha_2$ has three links.
One has first the loop diagram
\begin{center}
\scalebox{.5}{
\begin{picture}(180,60)
%nom
\put(-45,10){C.1}
%quatre vertex + lignes simples
\thicklines  \multiput(10,10)(40,0){4}{\circle{10}}
 \multiput(15,10)(40,0){3}{\line(1,0){30}}
%deux vertex du dessus
 \multiput(90,50)(40,0){2}{\circle{10}}
\put(130,15){\line(0,1){30}} \put(50,15){\line(1,1){35}}
\put(95,50){\line(1,0){30}}
%nom des racines
\put(5,-5){$\alpha_1$} \put(45,-5){$\alpha_2$}
 \put(125,-5){$\alpha_4$}  \put(65,45){$\alpha_6$}\put(85,-5){$\alpha_3$}
  \put(140,45){$\alpha_5$}
\end{picture}
} 
\end{center}
One hyperbolic algebra has such a Dynkin diagram, namely, the
overextension $A_4^{\wedge\wedge}$. As we already know from
\cite{Damour:2002et}, the searched for $3$--dimensional Lagrangian coincides
with the toroidal dimensional reduction of the seven--dimensional
Einstein-Hilbert Lagrangian. The dilaton couplings are given by
\begin{eqnarray}
\td \lambda &=& \sqrt{2} \ ; \
\td \lambda' = {1\over \sqrt{2}} \ ; \  \td \mu' =  \sqrt{{3\over 2}} \ ; \
\td \lambda''= 0 \ ; \  \td \mu'' = \sqrt{{2\over 3}}  \ ; \ \td \nu'' =  { 2
\over
\sqrt{3}} \ ; \
\td \lambda''' = 0 \ ; \  \td \mu''' =0;  \nn \\ 
\td \nu''' &=& { \sqrt{3} \over
2} \ ; \ \rho'''={ \sqrt{5} \over 2} \ ; \
\td \lambda''''={1\over \sqrt{2}}  \ ; \  \td \mu''''={1\over
\sqrt{6}} \ ; \ \td \nu''''={1\over 2 \sqrt{3}}  \ ; \
\rho''''={ \sqrt{5} \over 2} \nn
\end{eqnarray}
and its Dynkin diagram is
\begin{center} \scalebox{.5}{
\begin{picture}(180,60)
\put(-50,10){6-1}
%quatre vertex + lignes simples
\thicklines  \multiput(10,10)(40,0){4}{\circle{10}}
 \multiput(15,10)(40,0){3}{\line(1,0){30}}
%deux vertex du dessus
 \multiput(90,50)(40,0){2}{\circle{10}}
\put(130,15){\line(0,1){30}} \put(50,15){\line(1,1){35}}
\put(95,50){\line(1,0){30}}
\end{picture}
} 
\end{center}
Next comes the tree diagram
\begin{center}
\scalebox{.5}{
\begin{picture}(180,60)
%nom
\put(-45,10){C.2}
%cinq vertex + lignes simples
\thicklines  \multiput(10,10)(40,0){5}{\circle{10}}
 \multiput(15,10)(40,0){4}{\line(1,0){30}}
%vertex du dessus
\put(50,50){\circle{10}} \put(50,15){\line(0,1){30}}
%nom des racines
\put(5,-5){$\alpha_1$} \put(45,-5){$\alpha_2$}
 \put(125,-5){$\alpha_5$}  \put(30,45){$\alpha_3$}\put(85,-5){$\alpha_4$}
  \put(165,-5){$\alpha_6$}
\end{picture}
} 
\end{center}
Two hyperbolic algebras have a Dynkin diagram of this shape;  but
they cannot be associated to billiards again because of the
impossibility to satisfy the subdominant conditions.

\noindent One also has to allow a relabelling of the vertices according to
\begin{center}
\scalebox{.5}{
\begin{picture}(180,60)
%nom
\put(-45,10){C.3}
%cinq vertex + lignes simples
\thicklines  \multiput(10,10)(40,0){5}{\circle{10}}
 \multiput(15,10)(40,0){4}{\line(1,0){30}}
%un vertex du dessus
\put(90,50){\circle{10}} \put(90,15){\line(0,1){30}}
%nom des racines
\put(5,-5){$\alpha_4$} \put(45,-5){$\alpha_3$}
 \put(125,-5){$\alpha_5$}  \put(70,45){$\alpha_1$}\put(85,-5){$\alpha_2$}
  \put(165,-5){$\alpha_6$}
\end{picture}
} \end{center}
There are five hyperbolic algebras of that type; but for only one of
them can one fulfill all conditions. The dilaton couplings are given by
\begin{eqnarray}
\td \lambda &=& \sqrt{2} \ ; \
\td \lambda' = {1\over \sqrt{2}} \ ; \  \td \mu' =  \sqrt{{3\over 2}} \ ; \
\td \lambda''= 0 \ ; \  \td \mu'' = \sqrt{{2\over 3}}  \ ; \ \td \nu'' =  { 1 \over
\sqrt{3}}  \ ; \
\td \lambda''' = {1\over \sqrt{2}};\nonumber \\  \td \mu''' &=&{1\over
\sqrt{6}} \ ; \ \td \nu''' = {1\over \sqrt{3}}  \ ; \ \rho'''=1 \ ; \
\td \lambda''''= 0 \ ; \ \alpha''''=0 \ ; \ \beta''''=0 \ ; \
\rho''''=1
\nn
\end{eqnarray}
and its Dynkin diagram is the following
\begin{center}
\scalebox{.5}{
\begin{picture}(180,60)
%nom
\put(-45,10){6-2}
%cinq vertex + lignes simples
\thicklines  \multiput(10,10)(40,0){5}{\circle{10}}
 \multiput(55,10)(40,0){2}{\line(1,0){30}}
%double premiere ligne
\put(15,7.5){\line(1,0){30}}\put(15,12.5){\line(1,0){30}}
%double derniere ligne
\put(135,7.5){\line(1,0){30}}\put(135,12.5){\line(1,0){30}}
%fleche vers la droite
\put(145,0){\line(1,1){10}} \put(145,20){\line(1,-1){10}}
%un vertex du dessus
\put(90,50){\circle{10}} \put(90,15){\line(0,1){30}}
%fleche vers la gauche
\put(25,10){\line(1,1){10}} \put(25,10){\line(1,-1){10}}
\end{picture}
} \end{center}

\textbf{Cases D} - are characterized by the fact that $\alpha_2$ has
two links:

\noindent D.1. corresponds further to $\alpha_3$ having four links
\begin{center}
\scalebox{.5}{
\begin{picture}(180,60)
%nom
\put(-45,10){D.1}
%4 vertex + lignes simples
\thicklines  \multiput(10,10)(40,0){4}{\circle{10}}
 \multiput(15,10)(40,0){3}{\line(1,0){30}}
%un vertex du dessus
\put(90,50){\circle{10}} \put(90,15){\line(0,1){30}}
%un vertex en dessous
\put(90,-30){\circle{10}} \put(90,5){\line(0,-1){30}}
%nom des racines
\put(5,-5){$\alpha_1$} \put(45,-5){$\alpha_2$}
 \put(125,-5){$\alpha_5$}  \put(70,45){$\alpha_6$}\put(70,-5){$\alpha_3$}
 \put(70,-30){$\alpha_4$}
\end{picture}
} \end{center}
Three diagrams of \cite{S} fit in this shape; only two of them are
realised through billiards. The couplings of the first one are given by
\begin{eqnarray}
\td \lambda &=& \sqrt{2} \ ; \
\td \lambda' = {1\over \sqrt{2}} \ ; \  \td \mu' =  \sqrt{{3\over 2}} \ ; \
\td \lambda''= 0 \ ; \  \td \mu'' = \sqrt{{2\over 3}}=  \td \mu''' \ ; \ \td \nu'' =  { 2 \over
\sqrt{3}} ;\quad
\td \lambda''' = 0;\nonumber \\ \td \nu'''
&=& {1\over
\sqrt{3}};\quad\rho'''=1;\quad
\td \lambda'''' = 0;\quad \td \mu''''=\sqrt{{ 2 \over 3}};\quad
\td \nu''''={1\over \sqrt{3}};\quad
\rho''''=1 
\label{560}
\end{eqnarray}
they provide the Dynkin diagram which is $D_4^{\wedge\wedge}$:
\begin{center}
\scalebox{.5}{
\begin{picture}(180,60)
%nom
\put(-45,10){6-3}
%4 vertex + lignes simples
\thicklines  \multiput(10,10)(40,0){4}{\circle{10}}
 \multiput(15,10)(40,0){3}{\line(1,0){30}}
%un vertex du dessus
\put(90,50){\circle{10}} \put(90,15){\line(0,1){30}}
%un vertex en dessous
\put(90,-30){\circle{10}} \put(90,5){\line(0,-1){30}}
\end{picture}
} 
\end{center}
For the second one, $\td \lambda = 1$ and all the other couplings are
those given in (\ref{560}) divided by $\sqrt{2}$. They lead to the
following diagram
\begin{center}
\scalebox{.5}{
\begin{picture}(180,60)
%nom
\put(-45,10){6-4}
%4 vertex + lignes simples
\thicklines  \multiput(10,10)(40,0){4}{\circle{10}}
 \multiput(55,10)(40,0){2}{\line(1,0){30}}
%double premiere ligne
\put(15,7.5){\line(1,0){30}}\put(15,12.5){\line(1,0){30}}
%fl\`{A}che vers la droite
\put(25,0){\line(1,1){10}} \put(25,20){\line(1,-1){10}}
%un vertex du dessus
\put(90,50){\circle{10}} \put(90,15){\line(0,1){30}}
%un vertex en dessous
\put(90,-30){\circle{10}} \put(90,5){\line(0,-1){30}}
\end{picture}
} 
\end{center}
D.2. corresponds to $\alpha_3$ having three connections
\begin{center}
\scalebox{.5}{
\begin{picture}(180,60)
%nom
\put(-45,10){D.2}
%5 vertex + lignes simples
\thicklines  \multiput(10,10)(40,0){5}{\circle{10}}
 \multiput(15,10)(40,0){4}{\line(1,0){30}}
%un vertex du dessus
\put(90,50){\circle{10}} \put(90,15){\line(0,1){30}}
%nom des racines
\put(5,-5){$\alpha_1$} \put(45,-5){$\alpha_2$}
 \put(125,-5){$\alpha_5$}  \put(70,45){$\alpha_4$}\put(85,-5){$\alpha_3$}
  \put(165,-5){$\alpha_6$}
\end{picture}
} 
\end{center}
and differs from C.3. above by the assignment of the
symmetry root. There are 4 Dynkin diagrams representing hyperbolic algebras
of this type and they all admit a billiard. 

\noindent (D.2.1.) The couplings are
\begin{eqnarray}
\td \lambda &=& 1 \ ; \
\td \lambda' = {1\over 2} \ ; \  \td \mu' =  {\sqrt{3}\over 2} \ ; \
\td \lambda'' = 0\ ; \  \td \mu'' = {1\over\sqrt{ 3}}  \ ; \ \td \nu'' =
\sqrt{{2\over3}} \ ; \
\td \lambda''' = 0 \ ; \  \td \mu''' = {1 \over \sqrt{3} } \nonumber \\
\td \nu''' &=& {1\over \sqrt{6}} \ ; \
\rho'''= {1 \over \sqrt{2}} \ ; \
\td \lambda'''' = 0 \ ; \  \td \mu''''=0  \ ; \ \td \nu''''=0 \ ; \
\rho''''=\sqrt{2}.\label{xx}
\end{eqnarray}
and the Dynkin diagram corresponds to
\begin{center}
\scalebox{.5}{
\begin{picture}(180,60)
%nom
\put(-45,10){6-5}
%5 vertex + lignes simples
\thicklines  \multiput(10,10)(40,0){5}{\circle{10}}
 \multiput(55,10)(40,0){2}{\line(1,0){30}}
%double premiere ligne
\put(15,7.5){\line(1,0){30}}\put(15,12.5){\line(1,0){30}}
%fleche vers la droite
\put(25,0){\line(1,1){10}} \put(25,20){\line(1,-1){10}}
%double derniere ligne
\put(135,7.5){\line(1,0){30}}\put(135,12.5){\line(1,0){30}}
%fleche vers la gauche
\put(145,10){\line(1,1){10}} \put(145,10){\line(1,-1){10}}
%un vertex du dessus
\put(90,50){\circle{10}} \put(90,15){\line(0,1){30}}
\end{picture}
} 
\end{center}
(D.2.2.) The couplings are the same as in (\ref{xx}) above except
$\rho''''$ which reads 
\beq \rho'''' = 1/ \sqrt{2} \ .
\nn
\eeq
The Dynkin diagram is
\begin{center}
\scalebox{.5}{
\begin{picture}(180,60)
%nom
\put(-45,10){6-6}
%5 vertex + lignes simples
\thicklines  \multiput(10,10)(40,0){5}{\circle{10}}
 \multiput(55,10)(40,0){2}{\line(1,0){30}}
%double premiere ligne
\put(15,7.5){\line(1,0){30}}\put(15,12.5){\line(1,0){30}}
%fleche vers la droite
\put(25,0){\line(1,1){10}} \put(25,20){\line(1,-1){10}}
%double derniere ligne
\put(135,7.5){\line(1,0){30}}\put(135,12.5){\line(1,0){30}}
%fleche vers la droite
\put(145,0){\line(1,1){10}} \put(145,20){\line(1,-1){10}}
%un vertex du dessus
\put(90,50){\circle{10}} \put(90,15){\line(0,1){30}}
\end{picture}
} 
\end{center}
(D.2.3.) The dilaton couplings are given by
\begin{eqnarray}
\td \lambda &=& \sqrt{2} \ ; \
\td \lambda' = {1\over \sqrt{2}} \ ; \  \td \mu' =  \sqrt{{3\over 2}} \ ; \
\td \lambda''= 0= \ ; \  \td \mu'' = \sqrt{{2\over 3}}  \ ; \ \td \nu'' =  { 2
\over
\sqrt{3}} \ ; \
\td \lambda''' = 0;\nonumber \\  \td \mu''' &=&\sqrt{{ 2 \over 3}}\ ; \ \td \nu''' =
{1\over
\sqrt{3}}  \ ; \ \rho'''=1 \ ; \
\td \lambda'''' = 0 \ ; \  \td \mu''''=0  \ ; \ \td \nu''''=0 \ ; \
\rho''''=1
 \label{563}
\end{eqnarray}
they provide the Dynkin diagram of $B_4^{\wedge\wedge}$
\begin{center}
\scalebox{.5}{
\begin{picture}(180,60)
%nom
\put(-45,10){6-7}
%5 vertex + lignes simples
\thicklines  \multiput(10,10)(40,0){5}{\circle{10}}
 \multiput(15,10)(40,0){3}{\line(1,0){30}}
%double derniere ligne
\put(135,7.5){\line(1,0){30}}\put(135,12.5){\line(1,0){30}}
%fleche vers la droite
\put(145,0){\line(1,1){10}} \put(145,20){\line(1,-1){10}}
%un vertex du dessus
\put(90,50){\circle{10}} \put(90,15){\line(0,1){30}}
\end{picture}
} 
\end{center}
(D.2.4.) The couplings are the same as in (\ref{563}) except \be
\rho''''=2\ee and
the algebra is $A_{7}^{(2)\wedge}$
\begin{center}
\scalebox{.5}{
\begin{picture}(180,60)
%nom
\put(-45,10){6-8}
%5 vertex + lignes simples
\thicklines  \multiput(10,10)(40,0){5}{\circle{10}}
 \multiput(15,10)(40,0){3}{\line(1,0){30}}
%double derniere ligne
\put(135,7.5){\line(1,0){30}}\put(135,12.5){\line(1,0){30}}
%fleche vers la gauche
\put(145,10){\line(1,1){10}} \put(145,10){\line(1,-1){10}}
%un vertex du dessus
\put(90,50){\circle{10}} \put(90,15){\line(0,1){30}}
\end{picture}
} 
\end{center}
D.3. describes the general structure below in which $\alpha_2$
and $\alpha_3$ have two links while $\alpha_4$ is connected three times
\begin{center}
\scalebox{.5}{
\begin{picture}(180,60)
%nom
\put(-45,10){D.3}
%5 vertex + lignes simples
\thicklines  \multiput(10,10)(40,0){5}{\circle{10}}
 \multiput(15,10)(40,0){4}{\line(1,0){30}}
%nom des racines
\put(5,-5){$\alpha_1$} \put(45,-5){$\alpha_2$}
 \put(125,-5){$\alpha_4$}  \put(110,45){$\alpha_5$}\put(85,-5){$\alpha_3$}
  \put(165,-5){$\alpha_6$}
%un vertex du dessus
\put(130,50){\circle{10}} \put(130,15){\line(0,1){30}}
\end{picture}
} 
\end{center}
There are two hyperbolic algebras of that type but only one satisfies all
billiard conditions. Its non zero couplings are
\beq
\td \lambda = \sqrt{2}\quad\mbox{and}\quad\td \lambda' =
 \td \mu' =  \td \mu'' = \td \nu'' =  \td \nu''' = \rho''' = \rho''''=
1/  \sqrt{2} \ . 
\nn
\eeq
The Dynkin
diagram is
\begin{center}
\scalebox{.5}{
\begin{picture}(180,60)
%nom
\put(-45,10){6-9}
%5 vertex + lignes simples
\thicklines  \multiput(10,10)(40,0){5}{\circle{10}}
 \multiput(95,10)(40,0){2}{\line(1,0){30}}
\put(15,10){\line(1,0){30}}
%double ligne (deuxieme)
\put(55,7.5){\line(1,0){30}}\put(55,12.5){\line(1,0){30}}
%fleche vers la droite
\put(65,0){\line(1,1){10}} \put(65,20){\line(1,-1){10}}
%un vertex du dessus
\put(130,50){\circle{10}} \put(130,15){\line(0,1){30}}
\end{picture}
} 
\end{center}

\textbf{Cases E.} - This set provides all linear diagrams. There
are seven hyperbolic algebras of this kind and all of them are
admissible
\begin{center}
\scalebox{.5}{
\begin{picture}(180,60)
%nom
\put(-45,10){E}
%6 vertex + lignes simples
\thicklines  \multiput(10,10)(40,0){6}{\circle{10}}
 \multiput(15,10)(40,0){5}{\line(1,0){30}}
%nom des racines
\put(5,-5){$\alpha_1$} \put(45,-5){$\alpha_2$}
 \put(125,-5){$\alpha_4$}  \put(205,-5){$\alpha_6$}\put(85,-5){$\alpha_3$}
  \put(165,-5){$\alpha_5$}
\end{picture}
} 
\end{center}
E.1. has the following couplings
\begin{eqnarray}
\td \lambda &=& \sqrt{2} \ ; \
\td \lambda' = {1 \over \sqrt{ 2}}  \ ; \  \td \mu' =   \sqrt{ { 3 \over 2}}
\ ; \
\td \lambda''= 0 \ ; \  \td \mu'' = \sqrt{ { 2 \over 3}} \ ; \ \td \nu'' = {2
\over
\sqrt{ 3}} \ ; \
\td \lambda''' = 0; \nonumber \\  \td \mu''' &=& 0\ ; \ \td \nu''' =\sqrt{ 3}  \ ;
\
\rho'''=1 \ ; \
\td \lambda'''' = 0  \ ; \  \td \mu''''=0   \ ; \ \td \nu''''=0\  \ ; \
\rho''''=2
\nn
\end{eqnarray} and its Dynkin diagram belongs to $E_6^{(2)\wedge}$
\begin{center}
\scalebox{.5}{
\begin{picture}(180,60)
%nom
\put(-45,10){6-10}
%6 vertex + lignes simples
\thicklines  \multiput(10,10)(40,0){6}{\circle{10}}
 \multiput(15,10)(40,0){3}{\line(1,0){30}}
\put(175,10){\line(1,0){30}}
%double ligne (quatrieme)
\put(135,7.5){\line(1,0){30}}\put(135,12.5){\line(1,0){30}}
%fleche vers la gauche
\put(145,10){\line(1,1){10}} \put(145,10){\line(1,-1){10}}
\end{picture}
} 
\end{center}
E.2. corresponds to
\begin{eqnarray}
\td \lambda &=&  \sqrt{2} \ ; \
\td \lambda' = {1 \over \sqrt{ 2}}  \ ; \  \td \mu' =  \sqrt{{3\over 2}} \ ; \
\td \lambda''= 0=\ ; \  \td \mu'' = \sqrt{{2\over 3}}  \ ; \ \td \nu'' =  { 1 \over
\sqrt{3}} \ ; \
\td \lambda''' = 0 \nonumber\\  \td \mu''' &=&0 \ ; \ \td \nu''' = { \sqrt{3} \over
2}
\ ;
\ \rho'''={1 \over 2} \ ; \
\td \lambda'''' = 0 \ ; \  \td \mu''''=0  \ ; \ \td \nu''''=0 \ ; \
\rho''''=1
\nn
\end{eqnarray}
and its algebra is associated to
\begin{center}
\scalebox{.5}{
\begin{picture}(180,60)
%nom
\put(-45,10){6-11}
%6 vertex + lignes simples
\thicklines  \multiput(10,10)(40,0){6}{\circle{10}}
 \multiput(15,10)(40,0){2}{\line(1,0){30}}
 \multiput(135,10)(40,0){2}{\line(1,0){30}}
%double ligne (troisieme)
\put(95,7.5){\line(1,0){30}}\put(95,12.5){\line(1,0){30}}
%fleche vers la droite
\put(105,0){\line(1,1){10}} \put(105,20){\line(1,-1){10}}
\end{picture}
} 
\end{center}
E.3. The walls are defined through the following set of parameters
\begin{eqnarray}
\td \lambda &=&  \sqrt{2} \ ; \
\td \lambda' = {1 \over \sqrt{ 2}}  \ ; \  \td \mu' =  \sqrt{{3\over 2}} \ ; \
\td \lambda''= 0\quad \ ; \  \td \mu'' = \sqrt{{2\over 3}}  \ ; \ \td \nu'' =  { 2 \over
\sqrt{3}} \ ; \
\td \lambda'''= 0 \nonumber\\  \td \mu''' &=&0 \ ; \ \td \nu''' = { \sqrt{3} \over 2}
\ ;
\ \rho'''={1 \over 2} \ ; \
\td \lambda'''' = 0 \ ; \  \td \mu''''=0 \ ; \ \td \nu''''=0 \ ; \
\rho''''=1
\nn
\end{eqnarray} 
the algebra is $F_4^{\wedge \wedge}$
\begin{center}
\scalebox{.5}{
\begin{picture}(180,60)
%nom
\put(-45,10){6-12}
%6 vertex + lignes simples
\thicklines  \multiput(10,10)(40,0){6}{\circle{10}}
 \multiput(15,10)(40,0){3}{\line(1,0){30}}
\put(175,10){\line(1,0){30}}
%double ligne (quatrieme)
\put(135,7.5){\line(1,0){30}}\put(135,12.5){\line(1,0){30}}
%fleche vers la droite
\put(145,0){\line(1,1){10}} \put(145,20){\line(1,-1){10}}
\end{picture}
} 
\end{center}
E.4. has the following couplings
\begin{eqnarray}
\td \lambda &=&  \sqrt{2} \ ; \
\td \lambda' = {1 \over \sqrt{ 2}} \ ; \  \td \mu' = {1 \over \sqrt{ 2}} \ ; \
\td \lambda''= 0 =\ ; \  \td \mu'' = {1 \over \sqrt{ 2}}   \ ; \ \td \nu'' =  {1
\over \sqrt{ 2}} \ ; \
\td \lambda''' = 0 \nonumber \\  \td \mu''' &=& 0 \ ; \ \td \nu''' = {1 \over \sqrt{
2}}
\ ;
\ \rho'''={1 \over \sqrt{ 2}}\ ; \
\td \lambda'''' = 0 \ ; \  \td \mu''''=0  \ ; \ \td \nu''''=0 \ ; \
\rho''''=\sqrt{2}  
\label{574}
\end{eqnarray}
and its diagram corresponds to $C_4^{\wedge\wedge}$
\begin{center}
\scalebox{.5}{
\begin{picture}(180,60)
%nom
\put(-45,10){6-13}
%6 vertex + lignes simples
\thicklines  \multiput(10,10)(40,0){6}{\circle{10}}
 \multiput(95,10)(40,0){2}{\line(1,0){30}}
\put(15,10){\line(1,0){30}}
%double ligne (2)
\put(55,7.5){\line(1,0){30}}\put(55,12.5){\line(1,0){30}}
%fleche vers la droite
\put(65,0){\line(1,1){10}} \put(65,20){\line(1,-1){10}}
%double ligne (derniere)
\put(175,7.5){\line(1,0){30}}\put(175,12.5){\line(1,0){30}}
%fleche vers la gauche
\put(185,10){\line(1,1){10}} \put(185,10){\line(1,-1){10}}
\end{picture}
} 
\end{center}
E.5. has the same couplings as those given in (\ref{574}) except
\beq 
\rho'''' = {1\over \sqrt{2}} \ .
\nn
\eeq 
Its diagram corresponds to
$A_{8}^{(2)\wedge}$
\begin{center}
\scalebox{.5}{
\begin{picture}(180,60)
%nom
\put(-45,10){6-14}
%6 vertex + lignes simples
\thicklines  \multiput(10,10)(40,0){6}{\circle{10}}
 \multiput(95,10)(40,0){2}{\line(1,0){30}}
\put(15,10){\line(1,0){30}}
%double ligne (2)
\put(55,7.5){\line(1,0){30}}\put(55,12.5){\line(1,0){30}}
%fleche vers la droite
\put(65,0){\line(1,1){10}} \put(65,20){\line(1,-1){10}}
%double ligne (derniere)
\put(175,7.5){\line(1,0){30}}\put(175,12.5){\line(1,0){30}}
%fleche vers la droite
\put(185,0){\line(1,1){10}} \put(185,20){\line(1,-1){10}}
\end{picture}
} 
\end{center}
E.6. is characterized by
\begin{eqnarray}
\td \lambda &=&  \sqrt{2} \ ; \
\td \lambda' = \sqrt{2} \ ; \  \td \mu' = \sqrt{2} \ ; \
\td \lambda''= 0 \ ; \  \td \mu'' = \sqrt{2} \,  \ ; \ \td \nu'' =  \sqrt{2}
\ ; \
\td \lambda''' = 0; \nonumber\\  \td \mu''' &=& 0 \ ; \ \td \nu''' = \sqrt{2}   \ ; \
\rho'''=\sqrt{2} \ ; \
\td \lambda'''' = 0 \ ; \  \td \mu''''=0  \ ; \ \td \nu''''=0 \ ; \
\rho''''=\sqrt{2}
\nn
\end{eqnarray}
and its diagram describes $D_5^{(2)\wedge}$
\begin{center}
\scalebox{.5}{
\begin{picture}(180,60)
%nom
\put(-45,10){6-15}
%6 vertex + lignes simples
\thicklines  \multiput(10,10)(40,0){6}{\circle{10}}
 \multiput(95,10)(40,0){2}{\line(1,0){30}}
\put(15,10){\line(1,0){30}}
%double ligne (2)
\put(55,7.5){\line(1,0){30}}\put(55,12.5){\line(1,0){30}}
%fleche vers la gauche
\put(65,10){\line(1,1){10}} \put(65,10){\line(1,-1){10}}
%double ligne (derniere)
\put(175,7.5){\line(1,0){30}}\put(175,12.5){\line(1,0){30}}
%fleche vers la droite
\put(185,0){\line(1,1){10}} \put(185,20){\line(1,-1){10}}
\end{picture}
} 
\end{center}
E.7. is the last one of this rank; its couplings are
\begin{eqnarray}
\td \lambda &=&  1 \ ; \
\td \lambda' = {1 \over 2}  \ ; \  \td \mu' =  {\sqrt{3} \over 2} \ ; \
\td \lambda''= 0 \ ; \  \td \mu'' = {1 \over \sqrt{ 3}} \,\, \ ; \ \td \nu'' =
\sqrt{ {2 \over 3}} \ ; \
\td \lambda''' = 0; \nonumber \\  \td \mu''' &=&0 \ ; \ \td \nu''' = { \sqrt{3} \over 2
\sqrt{2 }}  \ ; \ \rho'''={1 \over 2 \sqrt{ 2}} \ ; \
\td \lambda'''' = 0 \ ; \  \td \mu''''=0 \ ; \ \td \nu''''=0 \ ; \
\rho''''={1 \over \sqrt{ 2}}
\nn
\end{eqnarray}
and its diagram gives $A_{8}^{(2)\prime\wedge}$
\begin{center}
\scalebox{.5}{
\begin{picture}(180,60)
%nom
\put(-45,10){6-16}
%6 vertex + lignes simples
\thicklines  \multiput(10,10)(40,0){6}{\circle{10}}
 \multiput(55,10)(40,0){2}{\line(1,0){30}}
\put(175,10){\line(1,0){30}} \put(15,12.5){\line(1,0){30}}
\put(15,7.5){\line(1,0){30}} \put(25,0){\line(1,1){10}}
\put(25,20){\line(1,-1){10}}
%double ligne (quatrieme)
\put(135,7.5){\line(1,0){30}}\put(135,12.5){\line(1,0){30}}
%fleche vers la droite
\put(145,0){\line(1,1){10}} \put(145,20){\line(1,-1){10}}
\end{picture}
} 
\end{center}

\subsection*{Rank 6 Hyperbolic Algebras: Oxidation}

Our next task is again to study which of the 16 algebras admitting
a three--dimensional billiard model allow in addition a higher
dimensional Lagrangian description.
\begin{enumerate}
\item{Diagram $(6-1)$} is the overextension $A_4^{\wedge\wedge}$. The
maximal oxidation dimension is $D_{max}=7$ where the Lagrangian
describes pure gravity \cite{Damour:2001sa}.
\item{Diagram $(6-2)$} : Here, $D_{max}=5$. The dominant
walls are the symmetry walls
$\alpha_1=\beta^4-\beta^3$,
$\alpha_2=\beta^3-\beta^2$, $\alpha_3=\beta^2-\beta^1$ and 
\begin{eqnarray}
\alpha_4 &=&\beta^1 - 1/ \sqrt{3} \phi \ , \nn \\ 
\alpha_5 &=&\beta^1 + \beta^2
+ 1/ \sqrt{3} \phi - \psi \ , \nn \\ 
\alpha_6 &=&\psi \ . 
\nn 
\end{eqnarray} 
These are
respectively the electric walls of a one--form, a two--form and a
zero--form. One easily checks that $\tilde\alpha_4=\alpha_5+\alpha_6$,
$\tilde\alpha_5=\alpha_4+\alpha_6$ and $\tilde\alpha_6=
\alpha_2+\alpha_3+\alpha_4+\alpha_5$.
\item{Diagram $(6-3)$} is the overextension $D_4^{\wedge\wedge}$; its 3-D
version can be oxidised up to $D_{max}=6$ and the Lagrangian is
written in references \cite{Cremmer:1999du} and \cite{Damour:2002fz}.
\item{Diagrams $(6-4)$, $(6-5)$ and $(6-6)$} : their Lagrangians have no
higher dimensional parent.
\item{Diagram $(6-7)$} is the overextension $B_4^{\wedge\wedge}$. Remark
that since the diagram has a fork one can oxidise in two different
ways: both lead to $D_{max}=6$. The Lagrangians can again be found
in references \cite{Cremmer:1999du} and \cite{Damour:2002fz}.
\item{Diagram $(6-8)$} is the twisted overextension
$A_{7}^{(2)\wedge}$.
$D_{max}=6$. The dominant walls are the symmetry walls
$\alpha_1 = \beta^5-\beta^4$, $\alpha_2 = \beta^4-\beta^3$,
$\alpha_3=\beta^3-\beta^2$ and $\alpha_4 =\beta^2-\beta^1$ and
\begin{eqnarray} 
\alpha_5 &=&\beta^1+\beta^2 -\phi \ , \nn \\ 
\alpha_6 &=&2\phi 
\nn 
\end{eqnarray} 
which are the electric walls of a $2$--form
and a $0$--form. Their respective magnetic walls are
subdominant: indeed one finds 
\begin{eqnarray}
\tilde\alpha_5 &=&\beta^1+\beta^2+\phi = \alpha_5+\alpha_6 \ , \nn \\ 
\tilde\alpha_6 &=&
\beta^1+\beta^2+\beta^3+\beta^4-2\phi = 2\alpha_5 +
\alpha_4 +2\alpha_3+\alpha_2 \ . 
\nn
\end{eqnarray}
\item{Diagram $(6-9)$} : $D_{max}=4$. The
wall system reads $\alpha_1 = \beta^3-\beta^2,
\alpha_2 =\beta^2-\beta^1$ and
\begin{eqnarray} 
\alpha_3 &=& \beta^1- 1/ \sqrt{2} \phi \ , \nn \\
\alpha_4 &=&1/ \sqrt{2} (\phi-\psi) \ ,  \nn \\ 
\alpha_5 &=&1/ \sqrt{2} (\psi-\chi) \ , \nn \\ 
\alpha_6 &=&1/ \sqrt{2} (\psi+\chi) \ ; 
\nn
\end{eqnarray} 
the last
four are the electric walls of a $1$--form and three $0$--forms. The
subdominant condition is fulfilled: indeed, one finds $\tilde\alpha_3 =
\alpha_3+2\alpha_4+\alpha_5+\alpha_6$, $\tilde\alpha_4 =
\alpha_2+2\alpha_3+\alpha_4+\alpha_5+\alpha_6$, $\tilde\alpha_5 =
\alpha_2+2\alpha_3+2\alpha_4+\alpha_6$, $\tilde\alpha_6 =
\alpha_2+2\alpha_3+2\alpha_4+\alpha_5$.
\item{Diagram $(6-10)$} is the twisted overextension $E_6^{(2)\wedge}$.
The oxidation rule gives the maximal dimension $D_{max}=5$. The walls
other than the symmetry ones are
\begin{eqnarray}
\alpha_4 &=& \beta^1-2/ \sqrt{3} \phi \ , \nn \\ 
\alpha_5 &=& \sqrt{3}\phi - \varphi \ , \nn \\ 
\alpha_6 &=& 2  \varphi \ .
\nn 
\end{eqnarray} 
One checks that $\tilde\alpha_4 =\alpha_3+
2\alpha_4+2\alpha_5+\alpha_6$, $\tilde\alpha_5 = \alpha_2+2\alpha_3+
3\alpha_4+\alpha_5+\alpha_6$, $\tilde\alpha_6 = \alpha_2+2\alpha_3+
3\alpha_4+2\alpha_5$.
\item{Diagram $(6-11)$} : a Lagrangian exists in $D_{max}=5 $ which
produces besides the symmetry walls
\begin{eqnarray}
\alpha_4 &=& \beta^1-1/ \sqrt{3} \phi \ , \nn \\ 
\alpha_5 &=& \sqrt{3} /2 \phi - 1/2 \varphi \ , \nn \\
\alpha_6 &=&  \varphi \ . 
\nn
\end{eqnarray} 
The subdominant conditions read
$\tilde\alpha_4= \alpha_3+2\alpha_4+2\alpha_5+\alpha_6$, $\tilde\alpha_5=
\alpha_2+2\alpha_3+3\alpha_4+\alpha_5+\alpha_6$ and $\tilde\alpha_6=
\alpha_2+2\alpha_3+3\alpha_4+2\alpha_5$.
\item{Diagram $(6-12)$} is the overextension $F_4^{\wedge\wedge}$;
the maximally oxidised theory is 6 dimensional and contains the
metric, one dilaton, one zero--form, two one--forms, a two--form and
a self- dual three--form field strength \cite{Cremmer:1999du, Damour:2002fz}.
\item{Diagram $(6-13)$} is the overextension $C_4^{\wedge\wedge}$
\cite{Cremmer:1999du}. This is the last one of its series: remember that the
$C_n^{\wedge\wedge}$ algebras are hyperbolic only for $n\leq 4$.
The maximal oxidation dimension is $D_{max}=4$; besides the
symmetry walls, the other dominant ones are
\begin{eqnarray}
\alpha_3 &=& \beta^1-1/ \sqrt{2} \phi \ , \nn \\ 
\alpha_4 &=& 1/ \sqrt{2}( \phi - \varphi) \ , \nn \\ 
\alpha_5 &=& 1/ \sqrt{2}( \varphi-\psi) \ , \nn \\ 
\alpha_6 &=& \sqrt{2} \psi \ . 
\nn 
\end{eqnarray} 
The subdominant conditions are satisfied, they read
$\tilde\alpha_3 =
\alpha_3+2\alpha_4+2\alpha_5+\alpha_6$, $\tilde\alpha_4 =\alpha_2+
2\alpha_3+\alpha_4+2\alpha_5+\alpha_6$, $\tilde\alpha_5 =
\alpha_2+2\alpha_3+2\alpha_4+\alpha_5+\alpha_6$, $\tilde\alpha_6 =
\alpha_2+2\alpha_3+2\alpha_4+2\alpha_5$.
\item{Diagram $(6-14)$} is the twisted overextension
$A_{8}^{(2)\wedge}$. There is no higher dimensional theory.
\item{Diagram $(6-15)$} represents $D_5^{(2)\wedge}$. In $D_{max}=4$, the
dominant walls other than the symmetry ones are
given by 
\begin{eqnarray}
\alpha_3 &=& \beta^1-1/ \sqrt{2} \phi \ , \nn \\ 
\alpha_4 &=& 1/ \sqrt{2}( \phi - \varphi)  \ ,  \nn \\ 
\alpha_5 &=& 1/ \sqrt{2}( \varphi-\psi) \ , \nn \\
\alpha_6 &=& 1/ \sqrt{2} \psi \ .
\nn
\end{eqnarray} 
One obtains easily the following expressions $\tilde\alpha_3 =
\alpha_3+2\alpha_4+2\alpha_5+\alpha_6$, $\tilde\alpha_4 =\alpha_2+
2\alpha_3+\alpha_4+2\alpha_5+2\alpha_6$, $\tilde\alpha_5 =
\alpha_2+2\alpha_3+2\alpha_4+\alpha_5+2\alpha_6$, $\tilde\alpha_6 =
\alpha_2+2\alpha_3+2\alpha_4+2\alpha_5+\alpha_6$.
\item{Diagram $(6-16)$} describes $A_{8}^{(2)\prime\wedge}$; it
cannot be associated to a billiard in
$D>3$.
\end{enumerate}

\textbf{Comment}

\noindent Here again, in $D>3$, the subdominant conditions are
always satisfied; it is only in $D=3$ that their r\^ole is crucial in
the selection of the admissible algebras. Hence, they do not add any
constraint in the oxidation construction.

\section{Rank 7, 8, 9 and 10 Hyperbolic algebras}

These hyperbolic algebras fall into two classes: the first one
comprises all algebras of rank $7 \leq r\leq 10$ that are overextensions of the
following finite simple Lie algebras $A_n, \, B_n, \,D_n, E_6, E_7, E_8$. They
are

$A_n^{\wedge\wedge},\quad (n=5,6,7)$

\begin{center}
\scalebox{.5} {
\begin{picture}(180,60)
%nom des racines
\put(5,-5){$\alpha_{-1}$} \put(45,-5){$\alpha_0$}
 \put(125,-5){$\alpha_2$}  \put(65,45){$\alpha_n$}\put(85,-5){$\alpha_1$}
  \put(140,45){$\alpha_3$}
%quatre vertex + lignes simples
\thicklines  \multiput(10,10)(40,0){4}{\circle{10}}
 \multiput(15,10)(40,0){3}{\line(1,0){30}}
%deux vertex du dessus
 \multiput(90,50)(40,0){2}{\circle{10}}
\put(130,15){\line(0,1){30}} \put(50,15){\line(1,1){35}}
\dashline[0]{2}(95,50)(105,50)(115,50)(125,50)
\end{picture}
}
\end{center}

$B_n^{\wedge\wedge},\quad (n= 5,6,7,8)$

\begin{center}
\scalebox{.5}{
\begin{picture}(180,60)
%nom des racines
\put(5,-5){$\alpha_{-1}$} \put(45,-5){$\alpha_0$}
 \put(125,-5){$\alpha_{n-1}$}  \put(70,45){$\alpha_2$}\put(85,-5){$\alpha_1$}
  \put(165,-5){$\alpha_n$}
%5 vertex + lignes simples
\thicklines  \multiput(10,10)(40,0){5}{\circle{10}}
 \multiput(15,10)(40,0){2}{\line(1,0){30}}
\dashline[0]{2}(95,10)(105,10)(115,10)(125,10)
%double derni\`{A}re ligne
\put(135,7.5){\line(1,0){30}}\put(135,12.5){\line(1,0){30}}
%fl\`{A}che vers la droite
\put(145,0){\line(1,1){10}} \put(145,20){\line(1,-1){10}}
%un vertex du dessus
\put(90,50){\circle{10}} \put(90,15){\line(0,1){30}}
\end{picture}
} \end{center}

$D_n^{\wedge\wedge}, \quad (n= 5,6,7,8)$

\begin{center}
\scalebox{.5}{
\begin{picture}(180,60)
%nom des racines
\put(5,-5){$\alpha_{-1}$} \put(45,-5){$\alpha_0$}
 \put(125,-5){$\alpha_{n-2}$}  \put(70,45){$\alpha_2$}\put(85,-5){$\alpha_1$}
  \put(165,-5){$\alpha_n$} \put(140,45){$\alpha_{n-1}$}
%4 vertex + lignes simples
\thicklines  \multiput(10,10)(40,0){4}{\circle{10}}
 \multiput(15,10)(40,0){2}{\line(1,0){30}}
\dashline[0]{2}(95,10)(105,10)(115,10)(125,10)
%deux vertex du dessus
\put(90,50){\circle{10}} \put(90,15){\line(0,1){30}}
\put(130,50){\circle{10}} \put(130,15){\line(0,1){30}}
%deux dernier vertex
 \multiput(130,10)(40,0){2}{\circle{10}}
\put(135,10){\line(1,0){30}}
\end{picture}
} \end{center}

$E_6^{\wedge \wedge}$

\begin{center}
\scalebox{.5}{
\begin{picture}(180,60)
%nom des racines
\put(5,-5){$\alpha_{-1}$} \put(45,-5){$\alpha_0$}
\put(85,-5){$\alpha_1$}
 \put(125,-5){$\alpha_2$}
  \put(165,-5){$\alpha_3$} \put(205,-5){$\alpha_4$}
  \put(140,45){$\alpha_5$}   \put(140,85){$\alpha_6$}
%6 vertex + lignes simples
\thicklines  \multiput(10,10)(40,0){6}{\circle{10}}
 \multiput(15,10)(40,0){5}{\line(1,0){30}}
%deux vertex du dessus
\put(130,50){\circle{10}} \put(130,15){\line(0,1){30}}
\put(130,90){\circle{10}} \put(130,55){\line(0,1){30}}
\end{picture}
} \end{center}

$E_7^{\wedge \wedge}$

\begin{center}
\scalebox{.5}{
\begin{picture}(180,60)
%nom des racines
\put(5,-5){$\alpha_{-1}$} \put(45,-5){$\alpha_0$}
\put(85,-5){$\alpha_1$}
 \put(125,-5){$\alpha_2$}
  \put(165,-5){$\alpha_3$} \put(205,-5){$\alpha_4$}
  \put(245,-5){$\alpha_5$}   \put(285,-5){$\alpha_6$}
  \put(180,45){$\alpha_7$}
%8 vertex + lignes simples
\thicklines  \multiput(10,10)(40,0){8}{\circle{10}}
 \multiput(15,10)(40,0){7}{\line(1,0){30}}
%1 vertex du dessus
\put(170,50){\circle{10}} \put(170,15){\line(0,1){30}}
\end{picture}
} \end{center}

$E_8^{\wedge \wedge}$

\begin{center}
\scalebox{.5}{
\begin{picture}(180,60)
%nom des racines
\put(5,-5){$\alpha_{-1}$} \put(45,-5){$\alpha_0$}
\put(85,-5){$\alpha_1$}
 \put(125,-5){$\alpha_2$}
  \put(165,-5){$\alpha_3$} \put(205,-5){$\alpha_4$}
  \put(245,-5){$\alpha_5$}   \put(285,-5){$\alpha_6$}
  \put(325,-5){$\alpha_7$}
  \put(260,45){$\alpha_8$}
%9 vertex + lignes simples
\thicklines  \multiput(10,10)(40,0){9}{\circle{10}}
 \multiput(15,10)(40,0){8}{\line(1,0){30}}
%1 vertex du dessus
\put(250,50){\circle{10}} \put(250,15){\line(0,1){30}}
\end{picture}
} 
\end{center}
In the second class, one finds the
four duals of
the
$B_n^{\wedge\wedge}, (n= 5,6,7,8)$, \ie the algebras known as $CE_{n+2} =
A_{2n-1}^{(2)\wedge}$

$CE_{n+2} = A_{2n-1}^{(2)\wedge}$

\begin{center}
\scalebox{.5}{
\begin{picture}(180,60)
%nom des racines
\put(5,-5){$\alpha_{-1}$} \put(45,-5){$\alpha_0$}
 \put(125,-5){$\alpha_{n-1}$}  \put(70,45){$\alpha_2$}\put(85,-5){$\alpha_1$}
  \put(165,-5){$\alpha_n$}
%5 vertex + lignes simples
\thicklines  \multiput(10,10)(40,0){5}{\circle{10}}
 \multiput(15,10)(40,0){2}{\line(1,0){30}}
\dashline[0]{2}(95,10)(105,10)(115,10)(125,10)
%double derni\`{A}re ligne
\put(135,7.5){\line(1,0){30}}\put(135,12.5){\line(1,0){30}}
%fl\`{A}che vers la gauche
\put(145,10){\line(1,1){10}} \put(145,10){\line(1,-1){10}}
%un vertex du dessus
\put(90,50){\circle{10}} \put(90,15){\line(0,1){30}}
\end{picture}
} \end{center}

\subsection{Overextensions of finite simple Lie algebras}
The algebras of the first class have already been encountered as
billiards of some three--dimensional $\cG/\cK$ coset theories, see Table \ref{kmth}.
Those of rank $10$, $E_{10}, BE_{10}$ and $DE_{10}$ have been
found \cite{Damour:2000hv} to describe the billiards of the seven string
theories, $M, IIA, IIB, I, HO, HE$ and the closed bosonic string
in 10 dimensions. More precisely, these theories split into three
separate blocks which correspond to three distinct billiards:
namely, ${\cal B}_2 = \{M, IIA, IIB\}$ leads to $E_{10}$, ${\cal
B}_1 = \{I, HO, HE\}$ corresponds to $BE_{10}$ and ${\cal B}_0 =
\{D=10\, \mbox{closed bosonic string}\}$ gives $DE_{10}$, see Table 
\ref{cordes}.

\noindent For sake of completeness, we here simply recall the maximal
spacetime dimensions and the specific
$p$--forms menus producing
the billiards.

\begin{enumerate}
\item{$A_n^{\wedge \wedge},\quad (n=5,6,7$)} : the Lagrangian is that of pure
gravity in
$D_{max}=n+3$.

\item{$B_n^{\wedge\wedge},\quad (n=5,6,7,8$)} : the maximally oxidised
Lagrangian lives in
$D_{max}= n+2$ where it comprises a dilaton, a $1$--form coupled to the
dilaton with coupling equal to $\td \lambda^{(1)}(\phi)=\phi/\sqrt{d-1}$ and a
$2$--form coupled to the dilaton with coupling equal to
$\td \lambda^{(2)}(\phi)=2\phi/ \sqrt{d-1}$.

\item{$D_n^{\wedge \wedge}, \quad (n=5,6,7,8$)} : a Lagrangian exists in
$D_{max}= n+2$ and comprises a dilaton and a $2$--form coupled to
the dilaton with coupling equal to $\td \lambda^{(2)}(\phi)= 2\phi/ \sqrt{d-1}$.

\item{$E_6^{\wedge \wedge}$} : the maximal oxidation dimension is $D_{max}=8$.
The Lagrangian has a dilaton, a $0$--form with coupling
$\td \lambda^{(0)}(\phi)=\phi\,\sqrt{2}$ and a
$3$--form with coupling
$\td \lambda^{(3)}(\phi)=-\phi/
\sqrt{2}$.

\item{$E_7^{\wedge \wedge}$} :  the maximal spacetime
dimension is $D_{max}=10$. The Lagrangian describes gravity and a
$4$--form: it is a truncation of type IIB supergravity.

\item{$E_8^{\wedge \wedge}$} : $D_{max}=11$. The
Lagrangian describes gravity coupled to a $3$--form; it is the bosonic sector
of eleven dimensional supergravity.

\end{enumerate}

\subsection{The algebras $CE_{q+2} = A_{2q-1}^{(2)\wedge}$}

The Weyl chamber of the algebras $CE_{q+2}$ ($q=5,6,7,8$), which
are dual to $B_q^{\wedge\wedge}$, allows a billiard realisation in
maximal dimension $D_{max} = q+1 = d+1$. The field content of the
theory is the following: there are two dilatons, $\phi$ and
$\varphi$, a $0$--form coupled to the dilatons through 
\beq
\td \lambda^{(0)}(\phi) = 2 \sqrt{\frac{(d-1)}{d}}\,\phi - \frac{2}{
\sqrt{d}}\,\varphi \ , 
\nn 
\eeq 
a one form with dilaton couplings 
\beq
\td \lambda^{(1)}(\phi) = -\sqrt{\frac{d}{(d-1)}}\,\phi \ , 
\nn
\eeq 
and a 2--form with the following couplings 
\beq 
\td \lambda^{(2)}(\phi) = -
\frac{2}{ \sqrt{d (d-1)}} \,\phi - \frac{2}{
\sqrt{d}}\,\varphi \ .
\nn
\eeq
 In particular, the Lagrangian in $D_{max}=9$ producing the
billiard identifiable as the fundamental Weyl chamber of $CE_{10}$
corresponds to $q=8=d$ and is explicitly given by \cite{MHBJ}
\begin{eqnarray} {\cal L}_{9} &=& ^{(9)}R\star\unity - \star
d\phi\wedge d\phi - \star d\varphi \wedge d\varphi
-\frac{1}{2}\,e^{(2\phi\sqrt{\frac{7}{2}} - \varphi\sqrt{2} \,)}\star
F^{(1)}\wedge F^{(1)}\nonumber
\\ & \, & -\frac{1}{2}\,e^{-4\phi\sqrt{\frac{2}{7}}}\star
F^{(2)}\wedge
F^{(2)}-\frac{1}{2}\,e^{-(\phi\sqrt{\frac{2}{7}}+\varphi\sqrt{2}\,)}\star
F^{(3)}\wedge F^{(3)}.
\label{Lagr}\end{eqnarray}
This Lagrangian is obtained as the \emph{minimal }one that exhibit the hyperbolic Kac--Moody algebra $CE_{10}$ \emph{\`a la limite }BKL. $CE_{10}$ is the fourth hyperbolic algebra of rank 10; contrary to
the other three cited above, which belong to the class of the
overtextensions of finite simple Lie algebras, its Lagrangian
(\ref{Lagr}) does not stem from string theories.

\section{Conclusions}

In this chapter we have presented all Lagrangian systems in which
gravity, dilatons and $p$--forms combine in such a way as to
produce a billiard that can be identified with the Weyl chamber of
a given hyperbolic Kac--Moody algebra. Exhaustive results have been
systematically obtained by first constructing Lagrangians in three
spacetime dimensions, at least for the algebras of rank $r\leq 6$.
We insist on the fact that our three--dimensional Lagrangians are
not assumed to realise a coset theory. We also have solved the
oxidation problem and provided the Lagrangians in the maximal
spacetime dimension with their $p$--forms content and specific
dilaton couplings. It turns out that the subdominant conditions
play no r\^ole in the oxidation analysis. The positive integer
coefficients that appear when expressing the subdominant walls in
terms of the dominant ones in the maximal oxidation dimension have
been systematically worked out.
The search for Lagrangians 
invariant under these algebras [which are not over--extended algebras] has not been considered. Non--linear realisations based on these algebras might give a hint on how to handle this 
problem.

\section{More hyperbolic algebras}

For completeness, we draw hereafter the Dynkin diagrams of 6
hyperbolic algebras missing in reference \cite{S}. This raises their
total number to $142$. Reference \cite{WZX} lists all hyperbolic algebras. 

\begin{center}
\begin{tabular}{l c c}
Rank 3 & Rank 4 & Rank 5 \\
& & \\
\scalebox{.5}{
\begin{picture}(180,60)
%trois vertex
\thicklines  \multiput(10,10)(40,0){3}{\circle{10}}
%trois premieres lignes
\put(15,7.5){\line(1,0){30}} \put(15,12.5){\line(1,0){30}}
\put(15,10){\line(1,0){30}}
%fleche vers la droite
\put(25,0){\line(1,1){10}} \put(25,20){\line(1,-1){10}}
%trois lignes
\put(55,7.5){\line(1,0){30}} \put(55,12.5){\line(1,0){30}}
\put(55,10){\line(1,0){30}}
%fleche vers la droite
\put(65,0){\line(1,1){10}} \put(65,20){\line(1,-1){10}}
\end{picture}
}
&
\scalebox{.5}{
\begin{picture}(180,60)
%trois premiers vertex + 1 ligne entre deux premiers
\thicklines  \multiput(50,10)(40,0){2}{\circle{10}}
%deux lignes
\put(55,8){\line(1,0){30}} \put(55,12){\line(1,0){30}}
% dernier vertex
\put(50,50){\circle{10}}
%deux lignes(haut)
\put(55,48){\line(1,0){30}} \put(55,52){\line(1,0){30}}
% dernier vertex(haut)
\put(90,50){\circle{10}}
%2 lignes vers le haut
\put(48,15){\line(0,1){30}} \put(52,15){\line(0,1){30}}
%2 lignes vers le haut (deuxiemes)
\put(88,15){\line(0,1){30}} \put(92,15){\line(0,1){30}}
%fleche vers la droite
\put(65,0){\line(1,1){10}} \put(65,20){\line(1,-1){10}}
%fleche vers la droite (haut)
\put(65,40){\line(1,1){10}} \put(65,60){\line(1,-1){10}}
%fleche vers le haut(dernier vertex)
\put(40,25){\line(1,1){10}} \put(50,35){\line(1,-1){10}}
%fleche vers le haut(dernier vertex) (deuxieme)
\put(80,25){\line(1,1){10}} \put(90,35){\line(1,-1){10}}
\end{picture}
}
&
\scalebox{.5}{
\begin{picture}(180,60)
%cinq vertex + lignes simples
\thicklines  \multiput(10,10)(40,0){5}{\circle{10}}
\put(15,10){\line(1,0){30}} \put(95,10){\line(1,0){30}}
%double ligne (deuxieme)
\put(55,8){\line(1,0){30}} \put(55,12){\line(1,0){30}}
%fleche vers la droite
\put(65,0){\line(1,1){10}} \put(65,20){\line(1,-1){10}}
%double ligne (quatrieme)
\put(135,8){\line(1,0){30}} \put(135,12){\line(1,0){30}}
%fleche vers la gauche
\put(145,10){\line(1,1){10}} \put(145,10){\line(1,-1){10}}
\end{picture}
}
  \\
\scalebox{.5}{
\begin{picture}(180,60)
%trois vertex
\thicklines  \multiput(10,10)(40,0){3}{\circle{10}}
%quatre premieres lignes
\put(15,8.75){\line(1,0){30}} \put(15,11.25){\line(1,0){30}}
\put(15,6.25){\line(1,0){30}} \put(15,13.75){\line(1,0){30}}
%fleche vers la droite
\put(25,0){\line(1,1){10}} \put(25,20){\line(1,-1){10}}
%quatre lignes
\put(55,8.75){\line(1,0){30}} \put(55,11.25){\line(1,0){30}}
\put(55,6.25){\line(1,0){30}} \put(55,13.75){\line(1,0){30}}
%fleche vers la droite (deuxieme ligne)
\put(65,0){\line(1,1){10}} \put(65,20){\line(1,-1){10}}
\end{picture}
} &

\scalebox{.5}{
\begin{picture}(180,60)
%triple ligne + fleche vers la droite
\thicklines  \multiput(10,10)(40,0){4}{\circle{10}}
\put(15,10){\line(1,0){30}} \put(15.5,12.5){\line(1,0){30}}
\put(15,7.5){\line(1,0){30}} \put(25,0){\line(1,1){10}}
\put(25,20){\line(1,-1){10}}
%simple line (la deuxieme)
\put(55,10){\line(1,0){30}}
%triple ligne (la troisieme) plus fleche vers la droite
\put(95,12.5){\line(1,0){30}} \put(95,7.5){\line(1,0){30}}
\put(95,10){\line(1,0){30}} \put(105,0){\line(1,1){10}}
\put(105,20){\line(1,-1){10}}
\end{picture}
}
&
\scalebox{.5}{
\begin{picture}(180,60)
%cinq vertex + lignes simples
\thicklines  \multiput(10,10)(40,0){5}{\circle{10}}
\put(15,10){\line(1,0){30}} \put(95,10){\line(1,0){30}}
%double ligne (deuxieme)
\put(55,8){\line(1,0){30}} \put(55,12){\line(1,0){30}}
%fleche vers la gauche
\put(65,10){\line(1,1){10}} \put(65,10){\line(1,-1){10}}
%double ligne (quatrieme)
\put(135,8){\line(1,0){30}} \put(135,12){\line(1,0){30}}
%fleche vers la droite
\put(145,0){\line(1,1){10}} \put(145,20){\line(1,-1){10}}
\end{picture}}
\\
\end{tabular}
\end{center}
%%%%%%%%%%%%%%%%%%%%%%%%%%%%%%%%%%
%%%%%%%%%%%%%%%%%%%%%%%%%%%%%%%%%%%
\cleardoublepage
\part{Non-linear realisations of Kac--Moody Algebras}
\cleardoublepage
%%%%%%%%%%%%%%%%%%%%%%%%%%%%%%%%%%%%
%%%%%%%%%%%%%%%%%%%%%%%%%%%%%%%%%%%%
%%\include{intro_nlr}
\chapter{General Framework}
 \markboth{GENERAL  {F}RAMEWORK}{}
 
In the first part of this thesis, we focussed on the
study of the asymptotic behaviour of the gravitational 
field for theories containing 
gravity suitably coupled to forms and dilatons, see Table 
\ref{kmth}. This study revealed some Kac--Moody algebras
through the ``Cosmological Billiards'' analysis. 
The questions addressed in this first part 
were, directly or indirectly, related to the more fundamental
questions: 

\centerline{\emph{Are these Kac-Moody algebra symmetries of these theories?}}

\centerline{\emph{If yes, how is the symmetry realised?}}

\noindent The point of view is changed in the second part, which is based on  the conjecture made in reference \cite{West:2001as}, to answer the preceding 
questions. This conjecture states that 

\centerline{\emph{$E_8^{+++}$ is the symmetry of M-theory.}}

\noindent  $E_8^{+++}$ ($\cG^{+++}$) is the Kac--Moody group having as Lie algebra  
the very--extended algebra $\mf{e}_8^{+++}$ ($\mf{g}^{+++}$) obtained from $\mf{e}_8^{++}$ 
($\mf{g}^{++}$) by adding one
node connected to the overextended node by a single line \cite{Gaberdiel:2002db}.
This conjecture is motivated in \cite{West:2001as} by the 
reformulation of the eleven--dimensional supergravity, at the level of the equations of motion, as a
simultaneous non--linear realisation of the group $E_8^{+++}$ and the conformal
group. This conjecture has been generalised to all \emph{maximally oxidised 
theories} $\cG$ in \cite{Englert:2003zs}:  the
assertion is that the maximally oxidised theories $\cG$ --- or 
some extensions of them --- possess the much larger very-extended 
Kac-Moody symmetry
$\cG^{+++}$.  Reference \cite{Englert:2003py} proposed to build actions explicitly invariant under 
$\cG^{+++}$  as non--linear $\s$--models based 
on the coset spaces $\cG^{+++}/\td{\cK}^{+++}$ where $\td{\cK}^{+++}$ is the subgroup of $\cG^{+++}$ invariant under the \emph{temporal involution }defined later. The hope is that these
actions encode the corresponding maximally oxidised theories $\cG$ with perhaps additional degrees of freedom. Some progress has been achieved in this direction but 
most of the questions are still open today. 

This chapter explains in detail the construction of these actions
invariant under very extended Kac--Moody groups $\cG^{+++}$. 
 A \emph{level decomposition} of the Kac--Moody algebra $\mf{g}^{+++}$ ($\mf{g}^{+++}$ is the Lie algebra 
 of the group $\cG^{+++}$) is introduced in order to handle with
 the infinite number of fields that parametrise the coset space $\cG^{+++}/\td{\cK}^{+++}$. The 
 level decomposition rests on the selection of an $\mf{sl}(D)$ subalgebra of
 $\mf{g}^{+++}$ which brings a handle in order to interpret the fields of the non--linear $\s$--model as 
the  fields of the maximally oxidised theories. The generators of 
$\mf{g}^{+++}$ are organised within
 finite dimensional  irreducible representations of the $\mf{sl}(D)$ subalgebra. The  subgroup 
 $\td{\cK}^{+++}$ is then 
 specified via the introduction of a \emph{temporal involution} $\Omega_1$, 
 $\td \cK^{+++}Ê= \Omega_1(\td \cK^{+++})$. 
Once these ingredients have been collected,  the searched for actions can be written in a recursive way. 
  Two 
 non equivalent  truncations of these actions are considered. They are  a useful guide 
 to make contact with the maximally oxidised theories $\cG$. 

\subsubsection{Invariant Actions: Geodesic Motion}

\begin{figure}[h]
  \centering
\begin{picture}(0,0)%
\includegraphics{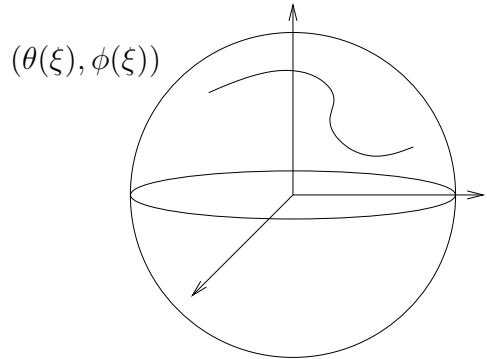}%
\end{picture}%
\setlength{\unitlength}{3315sp}%
\begingroup\makeatletter\ifx\SetFigFont\undefined%
\gdef\SetFigFont#1#2#3#4#5{%
  \reset@font\fontsize{#1}{#2pt}%
  \fontfamily{#3}\fontseries{#4}\fontshape{#5}%
  \selectfont}%
\fi\endgroup%
\begin{picture}(3567,2675)(1531,-3894)
\put(1531,-1726){\makebox(0,0)[lb]{\smash{{\SetFigFont{12}{14.4}{\familydefault}{\mddefault}{\updefault}{\color[rgb]{0,0,0}$(\th(\xi), \phi(\xi))$}%
}}}}
\end{picture}%
  \caption{\small Geodesic motion on a sphere $SO(3)/SO(2)$: 
    the coordinates $\{\th, \phi\}$
    of the sphere are promoted to be fields depending on an evolution 
    parameter $\xi$, the motion is determined by the minimisation of the
    length of the world line  $(\th(\xi), \phi(\xi))$.}
  \label{manifold}
\end{figure}

\noindent A natural way of writing down a theory explicitly invariant under a
Lie Group $G$ is to consider a geodesic motion on the group manifold 
$G /H$, where $H$ is a subgroup of $G$. The coordinates of the group manifold 
are promoted to be 
fields depending on an evolution parameter and the action $S_G$ is taken to be
the length of the world line spanned by the parameter. 
For an illustration see Figure \ref{manifold}. 

\noindent 
This simple construction is applied here to the poorly known Kac--Moody groups. Before 
going into details, let us sketch the general structure of the actions. 
An  action $S_{\cG^{+++}}$ invariant under the very extended Kac--Moody group $\cG^{+++}$ is 
defined on a world-line in terms of fields $\phi(\xi)$ parametrising the
 coset space $\cG^{+++}/\tilde{\cK}^{+++}$, where $\xi$ spans the world-line. The coset space 
 $\cG^{+++}/ \td{\cK}^{+++}$ is the space of elements $x$ of $\cG^{+++}$ endowed with the 
 equivalence relation $x \sim x k $ where $k $ belongs to $\td{\cK}^{+++}$. 
 Here $\td{\cK}^{+++}$
 is the subgroup of $\cG^{+++}$ chosen such that its Lie algebra 
 $\td{\mf{k}}^{+++}$ is the subalgebra of $\mf{g}^{+++}$ invariant under 
 the \emph{temporal involution} 
 $\Omega_1$ (defined later).  
This action is formulated in terms of a one-parameter 
dependent group element 
\beq 
\cv=\cv(\xi)\in \cG^{+++} / \tilde{\cK}^{+++}
\label{cosetrep}
\eeq
and its Lie algebra valued
derivative 
\beq 
v(\xi) :=\frac{d\cv}{d\xi}\cv^{-1}(\xi) \in \mf{g}^{+++} 
\ .
\label{vvv}
\eeq 
The action is 
\beq 
S_{\cG^{+++}} = \int {d\xi \over n(\xi)} 
\langle v_{sym}(\xi)\vert v_{sym}(\xi) \rangle 
, 
\label{invaction}
\eeq 
where
\begin{itemize}
\item The symmetric projection 
\beq 
v_{sym} := \frac{1}{2} (v+v^T)
\label{vsym}
\eeq  
is introduced in order to define an evolution on the coset space. 
  $v^T$ is the 
``transposed" of $v$ defined with respect to the  \emph{temporal involution }$\Omega_1$  as $v^T = -\Omega_1(v)$. 
\item $\langle . \vert . \rangle$
is the standard invariant bilinear form on $\cG^{+++}$.
\item The lapse function $n(\xi)$  ensures that the trajectory is a null geodesic though its equations of motion, which imply $\langle v \vert v \rangle = 0$.  
\item $\xi$ is a priori not related to spacetime.
\end{itemize}

\noindent The main point of this construction is that this action is
invariant under \emph{global }transformations of $\cG^{+++}$ and 
\emph{local }transformations of $\td \cK^{+++}$:
\beq 
\cv(\xi) \rightarrow k(\xi) \cv(\xi) g \hspace{1cm} \mbox{where}\hspace{1cm} k(\xi) 
\in \tilde{\cK}^{+++} \, \mathrm{and} \,
g\in \cG^{+++} 
\nn
\ .
\eeq 
The role of $k(\xi)$ is to restore the transformed $\cv(\xi)$ to the chosen gauge. 

\subsubsection{Level Decomposition}

\noindent The 
number of fields $\phi(\xi)$ is infinite since such is  the dimension of 
the coset space $\cG^{+++} / \tilde{\cK}^{+++}$. 
An organisation of these 
fields seems useful. The method here chosen is to select a \emph{finite} 
dimensional subalgebra\footnote{ Level expansions
of very-extended algebras in terms of the subalgebra $A_{D-1}$
have been considered in \cite{West:2002jj, Nicolai:2003fw, Kleinschmidt:2003mf}.} $\mf{sl}(D)$ of $\mf{g}^{+++}$. This is achieved 
by selecting successive nodes ({\it i.e.} simple roots) which form an 
$\mf{sl}(D)$
sub--diagram 
of the Dynkin diagram of $\mf{g}^{+++}$.  Let $\{\a_i\}_{i=1}^r$ be the 
simple roots of $\mf{g}^{+++}$ and suppose that the roots $\{\a_j\}_{j=1}^s$ 
span an $\mf{sl}(s+1)$ subalgebra. Each root $\a = \sum_{i=1}^r a_i \a_i$ of $\mf{g}^{+++}$ can be assigned a level $\ell_k$   with respect to $\a_k$ ($k= s+1,...,r$) in 
the following way:
    the levels $\ell_k$ with respect to $\a_k$ are defined to be $\ (\ell_s,\ell_{s+1},
\dots,\ell_r) = (a_s,a_{s+1},\dots,a_r)$. The sum of the levels $\sum_{k=s+1}^r
\ell_k$ is denoted $\ell$. 
Two cases are possible, either $\a$ is a positive root and $\ell_k \geq 0$  $\forall k Ê\in \{ s+1,...,r \}$ or $\a$ is a negative
root and $\ell_k \leq 0$ $\forall k Ê\in \{ s+1,...,r \}$. 
 It can be shown that at each level, there is only a finite number of roots.

\begin{quote} \textit{Level of a root}: Consider
the algebra $E_6^{+++}$, with simple roots $\a_i$
($i=1,\ldots,9$) labelled according to the Dynkin diagram:

\vspace{.45cm}
\begin{center}
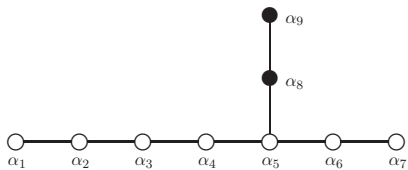
\begin{figure}[h]
%\begin{center}
\centering
\scalebox{.6}{
\begin{picture}(180,60)
%nom des racines
\put(5,-5){$\a_{1}$} \put(45,-5){$\a_2$}
\put(85,-5){$\a_3$}
 \put(125,-5){$\a_4$}
  \put(165,-5){$\a_5$} \put(205,-5){$\a_6$}
  \put(245,-5){$\a_7$}
  \put(180,45){$\a_8$}   \put(180,85){$\a_9$}
%6 vertex + lignes simples
\thicklines \multiput(10,10)(40,0){7}{\circle{10}}
\multiput(15,10)(40,0){6}{\line(1,0){30}}
%deux vertex du dessus
\put(170,50){\circle*{10}} \put(170,15){\line(0,1){30}}
\put(170,90){\circle*{10}} \put(170,55){\line(0,1){30}}
\end{picture}
}
%\end{center}
\label{dynkindiagrame6+++}
 \caption{{\ft Dynkin diagram of $\mf{e}_6^{+++}$. The black nodes are the ones which  are not
is the Dynkin diagram of the chosen $\mf{sl}(8)$ subalgebra. }} 
\end{figure}
\end{center}
One defines a level decomposition of $\mf{e}_6^{+++}$
under its $\mf{sl}(8)=A_7$ subalgebra admitting the sub-Dynkin
diagram with  nodes from 1 to 7. This singles out the nodes $8$ and
$9$ which do not belong to the $A_7$ subdiagram. Any positive root
$\a$ of $E_6^{+++}$ can be written as $\a = \sum_{i=1}^9 m_i \a_i =
\sum_{s=1}^7 m_s \a_s+ \sum_{g= 8,9} \ell_g \a_g$ with $m_s$ and
$\ell_g$ non-negative integers. Here, $\ell_8$ and $\ell_9$ are
called respectively the $\a_8$  level and the $\a_9$ level of
$\a$.
\end{quote}
\noindent One can also  organise the $\mf{g}^{+++}$ algebra 
generators\footnote{The generators of an algebra are understood, depending on the context, as the generators of the algebra \emph{or} generators of the underlying vector space.} according
to the level decomposition. Remember that the  $\mf{g}^{+++}$ generators can be chosen to
be the Chevalley--Serre generators
$h_i, \, e_i, \, f_i \ Ê(i = 1, ... , r) $ --- obeying the commutation relations given by Eqs (\ref{serrerelationskm}) --- and 
their multicommutators subject to the Serre relations (\ref{serrerel}). According to the triangular 
decomposition, any of these multicommutators can be written either as $  [e_i, [e_j, ....,[e_k,e_m]...]] $ or $[f_i, [f_j, ....,[f_k,f_m]...]] $. The $h_i$'s generate the Cartan subalgebra, they are also called Cartan generators.  
The $\mf{sl}(D)$ subalgebra generators and the Cartan generators of $\mf{g}^{+++}$ are of level $\ell = 0$. The
ladder generators, \ie the $  [e_i, [e_j, ....,[e_k,e_m]...]] $'s and  $[f_i, [f_j, ....,[f_k,f_m]...]] $'s,  inherit their level from their corresponding root. There are only
\emph{finitely} many generators of a given level.
% {\color{red} level \cite{} Kac?}.
\begin{quote}
{\it Level of a generator}:
The algebra $\mf{e}_6^{+++}$ decomposes according to the 
triangular decomposition (\ref{triangkm}), $\mf{e}_6^{+++} = n_-  \oplus
 \mf{g}_0 \oplus n_+$. The generators of level 0 
are $h_i, \ [e_i, [e_j, ....,[e_k,e_m]...]] $,  $ [f_i,[f_j,...,[f_k,f_m]...]]$ 
with $i = 1...7$, {\it i.e.} the $\mf{sl}(8)$ generators that also fit 
the triangular decomposition, together with $h_8,  \ h_9$.  The 
generators of a given level $(\ell_8,\ell_9)$ are in $n_+$ and can 
be written as a multicommutator $[e_i,...[e_8,...[e_9,...[e_j,e_k]...]]]$ with $e_8$ appearing
$\ell_8$ times and $e_9$ appearing $\ell_9$ times.
\end{quote}

\subsubsection{Generators of $\mf{g}^{+++}$ and Representations of $\mf{sl}(D)$}
 
\noindent The set of  $\mf{g}^{+++}$ generators  splits into

\noindent \emph{\textbf{$\ell =  0$ generators}}: The choice of a $\mf{sl}(D)$ subalgebra (rather than other 
 subalgebra) is made to provide an interpretation of the fields $\phi(\xi)$ as spacetime fields, though their $\mf{sl}(D)$ covariance properties.  
The $\ell= 0$ generators are the $\mf{sl}(D)$ generators and the Cartan 
generators not belonging to the $\mf{sl}(D)$ subalgebra. The $\mf{sl}(D)$ generators
plus one of the remaining Cartan generators form a $\mf{gl}(D)$ algebra. 
Even if there is more than one  generator out of the 
$\mf{sl}(D)$ subalgebra, only one contributes to 
the enhancement of $\mf{sl}(D)$ to $\mf{gl}(D)$ (the others are later
associated with dilatons).
The involution $\Omega_1$ defining the subalgebra $\mf{\tilde{k}}^{+++}$ though $\Omega_1(
\mf{\td{k}}^{+++}) = \mf{\td{k}}^{+++}$ is chosen 
such that the restriction of $\mf{k}^{+++}$ to the level $\ell = 0$ is $\mf{so}(D-1,1)$.
Therefore, at level $\ell =0$, the coset reduces to $GL(D) / 
SO(D-1,1)$ and
its coordinates are 
interpreted as the gravitational vielbein matrix,
 $D$ 
becoming the spacetime dimension.
 Indeed, the gravitational vielbein can be seen as an element of $GL(D)$ (invertible
matrices), and two gravitational vielbeins differing by a  Lorentz
group rotation are equivalent since they reproduce the  same metric.
The $\mf{sl}(D)$ sub--Dynkin diagram is called 
the \emph{gravity line}. To summarise: the $\ell =0$ generators are either (i) $\mf{gl}(D)$ 
generators or  (ii) the "remaining" Cartan generators. 

\noindent (i) The definition representation $K^a_{~b} \ (a,b=1,2,\ldots ,D)$ of the $\mf{gl}(D)$ subalgebra is,
\beq 
\ [ K^a{}_b, K^c{}_d ]  = \d^c_b K^a{}_d - \d^a_d K^c{}_b 
\,  .
\label{comm00}
 \eeq
 
\noindent (ii) The "remaining" Cartan generators are denoted $R_u \,( u=1 \dots q)$, where
$q = r-D+1$ ($r$ is the rank of $\mf{g}^{+++}$). These Cartan generators commute with the 
$K^a{}_b$'s for all $\mf{g}^{+++}$ .
%except $\mf{a}_n^{+++}$, \ie in the pure gravity case).
From the point of view of the corresponding maximally oxidised action
$S_{\cal G}$, these generators are associated with the $q$ dilatons\footnote{
All the maximally oxidised  theories have at most one dilaton except
the $C_{q+1}$-series characterised by $q$ dilatons. 
}. 

\noindent \emph{\textbf{ $\ell > 0$ generators}}: The generators of level $\ell>0$ are the ladder generators
of level $\ell >0$. The $\mf{sl}(D)$ generators act  
on the generators of $\mf{g}^{+++}$ through the adjoint action,
\beq
\G(x) \cdot g = [ x,g ]  \in \mf{g}^{+++} \, ,  Ê\hspace{2.5cm}   g \in \mf{g}^{+++}, \ x \in \mf{sl}(D) \, .
\nn
\eeq
It is clear from the structure of Kac--Moody algebras and from the definition 
of the level that the set of generators of a given level is invariant under the action of $\mf{sl}(D)$. 
Therefore 
generators of a given level $\ell$ form a \emph{finite dimensional 
representation} of $\mf{sl}(D)$.
These representation spaces are tensors of the
$\mf{sl}(D)$ subalgebra $R^{ d_1\dots d_s}$, 
the symmetry of which is encoded in a \emph{Young
tableau} or equivalently in its \emph{Dynkin labels} (see appendix  \ref{lie}).  
In principle it is possible to determine all the 
irreducible representations
present at a given level \cite{Nicolai:2003fw,Kleinschmidt:2003mf}.
These representations have a natural interpretation in terms of spacetime fields as $\mf{sl}(D)$ representations. For instance, the lowest levels generally contain completely antisymmetric tensors $R^{a_1...a_r}$ that have a natural interpretation as 
spacetime fields electric and magnetic
potentials of the maximally oxidised action $S_{\cal
G}$. These completely antisymmetric $R^{a_1\dots a_r}$, which appear at the lowest levels,  satisfy the tensor  relations
\beq
\label{comm01} 
\  [K^a{}_b ,R^{a_1\dots a_r} ]   = \d^{a_1}_b R^{a a_2 \dots a_r} +\dots +
\d^{a_r}_b R^{a_1 \dots a_{r-1}a} \, , 
\eeq
and the $R_u$ satisfy the scaling relations, 
\beq 
\label{commphi1} 
\ [R_u, R^{a_1\dots a_r} ] =   -\frac{\varepsilon_p \l_p^u}{2}\,  R^{a_1\dots a_r}\, ,
\eeq
where $\l_p^u$ is the coupling constant of the dilaton $\phi^u$ to the
 field strength form (\ref{keyaction})  and
$\varepsilon_p$ is $+1\, (-1)$ for an electric (magnetic) root. 

\begin{quote}  {\it Decomposition of $\mf{g}^{+++}$  into $\mf{sl}(D)$ representations}:
Consider the level $(1,0)$ generators of $\mf{e}^{+++}_6$, they belong to a 
representation $\G$ of  $\mf{sl}(8)$. Each representation is completely 
characterised by its lowest (or equivalently highest) weight (see appendix \ref{lie}).
The generator $e_8$ is of level $(1,0)$ and
\beq
\G(f_i) \cdot e_8 = [f_i,e_8] = 0 \  \ i = 1,...,7 \ .
\nn
\eeq
Therefore, $e_8$ is a lowest weight of the representation. The Dynkin labels
of this representation are defined through
\beq
\G(h_i) \cdot e_8 &=& [h_i,e_8] = 0  \ \  i = 1,2,3,4,6,7 \ \nn \\
\G(h_5) \cdot e_8 &=& [h_5,e_8] = -e_8 \ .
\nn 
\eeq
The Dynkin labels are therefore $[0,0,0,0,1,0,0]$ and the corresponding 
Young tableau is 
\begin{center}
\scalebox{.9}{
\begin{picture}(30,30)(0,-10)
\multiframe(0,10)(10.5,0){1}(10,10){$ $}
\multiframe(0,-0.5)(10.5,0){1}(10,10){$ $}
\multiframe(0,-11)(10.5,0){1}(10,10){$ $} 
\end{picture}
}
\end{center}
which is associated with a completely antisymmetric tensor with three indices (one per box). 
One interprets this tensor as a 3-form
potential $A_{a_1a_2a_3}$ ($a_i \in \{ 1,...,8=D\}$). The representations appearing at the lowest levels 
are given in Table \ref{e6+++rep} \cite{Kleinschmidt:2003mf}, they can be interpreted as the field content
of the maximally oxidised theory $E_6$.
\begin{table}[h]
\begin{center}
\scalebox{0.7}{
\begin{tabular}{|c|c| l|}
\hline
$(\ell_8,\ell_9)$ & Dynkin Labels &  Interpretation \\
\hline
(0,0) & $[1, \ 0, \ 0, \ 0, \ 0 , \ 0 , \ 1] $ & $h_a^{\ b}$  (vielbeins) \\
(0,0) & $[0, \ 0, \ 0, \ 0, \ 0  , \ 0, \ 0]$ & $\phi$ (dilaton)\\
(0,1) & $[0, \ 0, \ 0, \ 0, \ 0 , \ 0 , \ 0]$ & $\chi$ (scalar field)\\
(1,0)& $[0, \ 0, \ 0, \ 0, \ 1 , \ 0 , \  0]$ & $A_{a_1 a_2 a_3}$ (3-form potential)\\
(1,1) &$[0, \ 0, \ 0, \ 0, \ 1 , \ 0 , \  0]$ & $\tilde{A}_{a_1 a_2 a_3}$ (3-form potential)\\
(2,1) & $[0, \ 0, \ 0, \ 1, \ 0 , \ 0 , \  1]$ & $A_{b \arrowvert a_1 \dots a_5 }$ ("dual of the graviton")\\
(2,0) &$[0, \ 1, \ 0, \ 0, \ 0 , \ 0 , \  0]$ & $\tilde{A}^\chi_{a_1 \dots a_6}$ (6-form potential)\\
(2,1) &$[0, \ 1, \ 0, \ 0, \ 0 , \ 0 , \  0]$ & $\tilde{A}^\phi_{a_1 \dots a_6}$ (6-form potential)\\
... & & \\
\hline
\end{tabular}}
\end{center}
\begin{center}
\caption{ {\ft First level representations appearing in the decomposition of $\mf{e}_6^{+++}$ as 
representations of the $\mf{sl}(8)$ subalgebra depicted in Figure \ref{dynkindiagrame6+++}. Their interpretation in terms
of spacetime fields correspond to the field content of the maximally oxidised theory $E_6$. 
One should note that the fields and their dual appear independently (a field with tilde is the dual of another one).}  }
\end{center}
\label{e6+++rep}
\end{table}
\end{quote}
\noindent \emph{\textbf{$\ell<0$ generators}}: if at a given level $\ell >0$ the representation 
$ R^{\quad
d_1\dots   d_r}$  appears then at level $- \ell$ one finds its dual  
representation  $\bar  R_{ d_1\dots   d_r}$. 
They satisfy the tensor and scaling relations
\beq
\label{comm0-l} 
\  [K^a{}_b , \bar R_{a_1\dots a_r} ]   &=& -\d^{a_1}_b \bar R_{a a_2 \dots a_r} -\dots -
\d^{a_r}_b \bar R_{a_1 \dots a_{r-1}a} \, , \\
\label{commphi-l} 
\ [R_u, \bar R_{a_1\dots a_r} ] &=&   -\frac{\varepsilon_p \l_p^u}{2}\,  \bar R_{a_1\dots a_r}\, .
\eeq

\subsubsection{$\mf{\tilde{k}}^{+++}$ Subalgebra}

\noindent The \emph{temporal involution} $\Omega_1$ 
generalises the Chevalley involution. It is  defined by
\beq
 \Omega_1(K^a_{~b}) &=&
-\epsilon_a\epsilon_b K^b_{~a} \nn \ , \\
\Omega_1(R_u)&=& -R_u \nn \ , \\
 \Omega_1( R_{ d_1\dots d_s}{}^{ c_1\dots c_r})&=&
-\epsilon_{c_1}\dots\epsilon_{c_r}\epsilon_{d_1}\dots\epsilon_{d_s}
  \bar R_{ c_1\dots c_r}^{\quad d_1\dots d_s} \ , 
\label{tempinv}
\eeq
with $\epsilon_a =-1$ if $a=1$ and
$\epsilon_a=+1$ otherwise. Remember that for the Chevalley involution
$\epsilon_a = 1$ for all $a$.  The  subalgebra 
$\mf{\tilde{k}}^{+++}$ is defined to be the subalgebra of
$\mf{g}^{+++}$ invariant under $\Omega_1$. The restriction of $\td{\mf{k}}^{+++}$ to the 
level $\ell=0$ is $\mf{sl}(D-1,1)$.

\subsubsection{Coset Representative}

\noindent The fields $\phi(\xi)$,   which are the coordinates on the coset space 
${\cG}^{+++}/\td{\cK}^{+++}$,  are denoted $A_{c_1\dots c_r}{}^{d_1\dots d_s}$ 
if they are  associated with the $\ell >0$
ladder generators $R_{ d_1\dots d_s}{}^{ c_1\dots c_r} $; $h^b{}_a$ if they are associated with the
Cartan generators or with the positive\footnote{A generator is said to be positive if  the corresponding root is a positive root.}  ladder generators $K^a{}_b$ ($b \ge a$)
of the $\mf{gl}(D)$ 
subalgebra; $\phi^u$ if associated with $R_u$; in short 
\beq 
\phi(\xi) = \{h^b{}_a (\xi) \, (\, b \ge a),  A_{c_1\dots c_r}{}^{d_1\dots d_s} (\xi), \phi^u(\xi) \} \, .
\nn
\eeq
The  coset representative $\cV(\xi) $ of (\ref{cosetrep}) is taken to be
\beq
{\cal V(\xi)} &=&   \cV_\circ(\xi) \cV_+(\xi) \, , \label{borel} \\
\cV_\circ(\xi) &=& \exp (\sum_{a\ge b}
h_b^{~a}(\xi)K^b_{~a} -\sum_{u=1}^q
\phi^u(\xi) R_u)  \, , \nn \\
\cV_+(\xi)&=& \exp (\sum
\frac{1}{r!s!} A_{ b_1\dots b_s}{}^{ a_1\dots a_r}(\xi) R_{
a_1\dots   a_r}{}^{b_1\dots b_s} +\cdots)\, ,
\nn 
\eeq 
where the exponential defining $\cV_\circ$ and $\cV_+$ contain respectively only the level 0 operators and only the positive operators. 
Although this parametrisation of the coset space  $\cG^{+++}/\cK^{+++}$ provides a very useful \emph{local }parametrisation, it does not cover the coset globally (since $\cK^{+++}$ is not the maximal compact subgroup of $\cG^{+++}$) \cite{Keurentjes:2005jw}.

\subsubsection{Lie Algebra Valued Derivative}

\noindent To get the action (\ref{invaction}), one needs first to compute
$v$ (\ref{vvv}). 
Inserting (\ref{borel}) into (\ref{vvv}), one obtains
\beq 
v(\xi) &=& v_\circ(\xi)  + v_+ (\xi) \nn \, , \\
v_\circ (\xi) &=& {d\cV_\circ (\xi) \over d\xi} \cV_\circ(\xi) ^{-1} \, , \nn \\
v_+(\xi) &=& \cV_\circ(\xi) {d\cV_+(\xi)  \over d\xi}\cV_+(\xi) ^{-1}  \cV_\circ(\xi) ^{-1} \, . \nn \eeq
Using  the commutation relations  (\ref{comm00}) and  the
Campbell-Hausdorff formulas, 
\beq 
d e^X e^{-X} &=& dX + \frac{1}{2} [X,dX] + \frac{1}{3!} [X,[X,dX]] + \dots 
\, ,
\nn \\
 e^X Y e^{-X} &=& Y + [X,Y] + \frac{1}{2}  [X,[X,Y]] + \dots  
\, ,
\nn
\eeq 
twice the first formula to obtain $v_\circ$ and once the second  to obtain $v_+$, one gets 
 \beq
 v_\circ &=& -\sum_{a \geq b} [e^h ({de^{-h} \over d\xi})]_b{}^a K^b{}_a -\sum_{u=1}^q {d\phi^u  \over d\xi}R_u 
 \, ,  
 \nn \\
 v_+ &=&  {1\over r!s!} {dA_{m_1\dots m_r}{}^{n_1\dots n_s} \over d \xi} e^{-\sum_{u=1}^q \l_u\phi^u}
 e_{a_1}{}^{m_1}\dots e_{a_r}{}^{m_r}e_{n_1}{}^{b_1}\dots e_{n_s}{}^{b_s} R_{b_1\dots b_s}{}^{a_1\dots a_r} + \dots
 \, , 
 \nn
 \eeq  
 where 
 \beq
 e_\mm{}^\n = (e^{h})_\mm{}^\n \hspace{1cm} \th_\n{}^\mm = (e^{-h})_\n{}^\mm \, .
 \label{vielbeins}
 \eeq 
 The  two types of indices have been introduced to indicate their $GL(D)$ or $SO(D-1,1)$ nature. $\cV_\circ$ belongs
to the \emph{left} coset space $GL(D) /SO(D-1,1)$,  therefore the line index of $\cV_\circ$ 
is an $SO(D-1,1)$ index while the column index is a $GL(D)$ index. The fields $h$ inherit this 
index structure. The choice of Greek indices with and without brackets refers to the  
interpretation of $e_\mm{}^\n$ as spacetime vielbeins and $A_{\m_1...\m_r}{}^{\n_1...\n_s}$ as
spacetime fields. The particular index 1 is  interpreted as a temporal index and the others  as 
space--like indices. One should emphasise that  the expression of 
 $v_+$ written above is obtained by assuming that the $R_{b_1\dots b_s}{}^{a_1\dots a_r}$'s 
 commute; the $\dots$ stand for the commutators we should have computed. These commutators are formally taken into 
 account via the replacement of the derivatives 
 $dA_{\m_1\dots \m_r}{}^{\n_1\dots \n_s}$ by covariant derivatives 
  $DA_{\m_1\dots \m_r}{}^{\n_1\dots \n_s}$  (see \cite{Englert:2003py} for an explicit example). 
  In principle, the covariant
 derivatives can be computed 
 recursively when the algebra $\cG$ is given \cite{Kleinschmidt:2003mf}. 
 
Finally $v_{sym}$ is obtained using the temporal involution $\Omega_1$, 
\beq v_{sym} &=& v_{\circ \, sym} + v_{+ \, sym} \nn \, , \\
v_{\circ \, sym}Ê&= & 
-{1\over 2} \sum_{a \geq b} [e^h ({de^{-h}  \over d\xi})]_\nnn{}^\mm (K^\nnn{}_\mm +\epsilon_\mm \epsilon_\nnn
K^\mm{}_\nnn)-
\sum_{u=1}^q {d\phi^u \over d\xi} R_u 
 \, ,  
 \nn \\
 v_{+ \, sym} &=&  {1\over 2r!s!} {dA_{\m_1\dots \m_r}{}^{\n_1\dots \n_s} \over d \xi} e^{-\sum_{u=1}^q \l_u\phi^u}
 e_{\sst{(\m_1)}}{}^{\m_1}\dots e_{\sst{(\m_r)}}{}^{\m_r}e_{\n_1}{}^{\sst{(\n_1)}}\dots e_{\n_s}{}^{\sst{(\n_s)}} 
\nn \\
& &  (R_{\sst{(\n_1)}\dots \sst{(\n_s)}}{}^{\sst{(\m_1)}\dots \sst{(\m_r)}} +\epsilon_{\sst{(\m_1)}}\dots\epsilon_{\sst{(\m_r)}}\epsilon_{\sst{(\n_1)}}\dots\epsilon_{\sst{(\n_s)}}
  \bar R_{ \sst{(\m_1)}\dots \sst{(\m_r)}}{}^{\sst{(\n_1)}\dots \sst{(\n_s)}})+ \dots
 \, , 
 \nn
 \eeq

\subsubsection{Invariant Action}

\noindent The invariant scalar product relations are 
\begin{eqnarray}
\label{scalarproducts}
&&\langle K_{~a}^a,K_{~b}^b\rangle =G_{ab}\, ,\quad \langle
K^b_{~a},K_{~c}^d\rangle=
\d_c^b\d_a^d \ a\neq b\, , \quad\langle R, R\rangle
=\frac{1}{2}
\,,\\
\nn
&&\langle
R^{\quad a_1\dots a_r}_{ b_1\dots b_s} , \bar  R_{ d_1\dots   d_r}^{\quad
c_1\dots c_s}\rangle
=\d^{c_1}_{b_1}\dots\d^{c_s}_{b_s}\d^{a_1}_{d_1}\dots\d^
{a_r}_{d_r}\,.
\end{eqnarray}
Here $G= I_D -
\frac{1}{2}\Xi_D$ where $\Xi_D$ is the $D\times D$ matrix with all
entries  equal to unity. Decomposing $S_{\cG^{+++}}$ as
\begin{equation}
\label{full} S_{{\cal G}^{+++}} =S_{{\cal G}^{+++}}^{(0)}+\sum_A
S_{{\cal G}^{+++}}^{(A)}\, ,
\end{equation} where $S_{{\cal G}^{+++}}^{(0)}$ contains all level
zero contributions, one obtains
\begin{equation}
\label{fullzero} S_{{\cal G}^{+++}}^{(0)} =\frac{1}{2}\int d\xi
\frac{1}{n(\xi)}\left[\frac{1}{2}(g^{\m\n}g^{\sigma\tau}-
\frac{1}{2}g^{\m\sigma}g^{\n\tau})\frac{dg_{\m\sigma}}{d\xi}
\frac{dg_{\n\tau}}{d\xi}+\sum_{u=1}^q
\frac{d\phi^u}{d\xi}\frac{d\phi^u}{d\xi}\right],
\end{equation}
\begin{equation}
\label{fulla} S_{{\cal G}^{+++}}^{(A)}=\frac{1}{2 r! s!}\int d\xi
\frac{ e^{- 2\l
\phi}}{n(\xi)}\left[
\frac{DA_{\m_1\dots \m_r}^{\quad \n_1\dots
\n_s}}{d\xi} g^{\m_1\s_1}...\,
g^{\m_r\s_r}g_{\n_1\rho_1}...\,
g_{\n_s\rho_s}
\frac{DA_{\s_1\dots \s_r}^{\quad
\rho_1\dots \rho_s}}{d\xi}\right].
\end{equation} 
  The $\xi$-dependent fields $g_{\m\n}$ are defined as
$g_{\m\n} =e_\m{}^{\rrr}e_\n{}^{\sss}\eta_{\rrr \sss}$, where $e_\m{}^{\rrr}=(e^{-h(\xi)})_\m{}^{\rrr}$. The appearance of the
Lorentz metric $\eta_{\mm \nnn}$ with $\eta_{11}=-1$ is a consequence of the
temporal involution $\Omega_1$.  $\l$ is the
generalisation of the scale parameter
$-\varepsilon_A \l_p^u/2$ to all roots.

\subsubsection{How to Make Contact with the Original Theories?} 

\noindent The field content of the original maximally oxidised theories can be 
identified in the way  sketched  above, but
there are \emph{infinitely} many more fields in the sigma model. 
Some of these 
fields may be auxiliary fields. Others might be related to new degrees of freedom, 
such as those 
describing the perturbative string spectrum. Moreover, the fields of the sigma model depend on $\xi $ and they are not
spacetime fields. 
In \cite{Damour:2002cu} it was conjectured that the derivatives 
at all orders of the 
fields (identified  as those) of the maximally oxidised theories  are present
among  the infinite number of fields of the coset model. It was shown that the
tensors, with the symmetry appropriate to encode these derivatives, were present in the coset model.

To make contact with maximally oxidised theories, a truncation of the 
coset model is very instructive \cite{Englert:2004ph}. Two different truncations and their 
physical relevance are 
presented in the next sections. 
The first consists in putting to zero all the fields $\phi(\xi)$ associated with 
the very extended node and the second follows the same procedure but after 
a Weyl reflection. These procedures provide actions invariant under 
$\cG^{++}$. The first one is called $\cG_B^{++}$ 
and the second one $\cG_C^{++}$, where the subscript ${}_C$ stands for Cosmological. 
Such truncations  of the $\cG^{+++}$ theory are consistent in the sense that  all the solutions of the equations of
motion of $S_{{\cal G}_{C/B}^{++}}$ are also solutions of the
equations of motion of $S_{\cG^{+++}}$. Note that the $\cG_{C/B}^{++}$ action was already written in \cite{Damour:2002cu}, 
before the $\cG^{+++}$ invariant actions.

\section{ The $\cG^{++}_C$-theory}

Consider the overextended algebra $\cG^{++}_C$ obtained from the very-extended
algebra $\cG^{+++}$  by deleting the node labelled 1 from the Dynkin
diagrams of $\cG^{+++}$ depicted in Fig.1.
The action $S_{\cG^{++}_C}$ describing the $\cG^{++}_C$--theory is obtained from $\cG^{+++}$ by performing the following
consistent truncation. One sets to zero in  the coset representative (\ref{borel}) the
field multiplying the Chevalley generator
$h_1=  K_{~1}^1- K_{~2}^2$ and all the
fields multiplying the positive ladder operators associated with  a root $\a$ such that its decomposition in terms of simple roots contains $\a_1$. As shown in reference \cite{Englert:2004ph}, this is equivalent 
\begin{itemize}
\item[$\cdot$] to put to zero all the fields $ A_{\m_1\dots \m_r}{}^{ \n_1\dots
\n_s}$  with a least one index equal to  1. To avoid confusion, the remaining
fields will be denoted by $B_{m_1\dots m_r}{}^{n_1\dots
n_s}$ where the indices are now \emph{latin} indices $ a = (2,...,D)$ and  these fields are  interpreted 
as the space--like components of spacetime fields;
\item[$\cdot$] to put to zero the $g_{1a}$ fields;
\item[$\cdot$] to replace $g_{11}$ by $ g$ where $g= $det$(g_{ab})$: this condition amounts to cancel the field in front of the Cartan generator $h_1$. 
\end{itemize}
Performing this truncation one obtains the following action
\beq
S_{\cG^{++}_C} = S_{\cG^{++}_C}^{(0)}+\sum_B
S_{\cG^{++}_C}^{(B)} \, ,
\eeq 
where
\beq
\label{fullop} 
S_{\cG^{++}_C}^{(0)} &=&
\frac{1}{2} \int dt
\frac{1}{n(t)} ( \frac{1}{2} (g^{ab}g^{cd}-
g^{ac}g^{bd}) \frac{dg_{ac}}{dt}
\frac{dg_{bd}}{dt}+
\frac{d\phi}{dt} \frac{d\phi}{dt} ) \, , 
\nn
\\
S_{\cG^{++}_C}^{(B)} &=&
\frac{1}{2 r! s!} \int dt
\frac{ e^{- 2\lambda\phi} }{n(t)}
\left[
\frac{DB_{m_1\dots m_r}{}^{n_1 \dots n_s} }{dt} g^{m_1p_1}...\,
g^{m_r p_r} g_{n_1q_1}
...\,    g_{n_s q_s}
\frac{DB_{p_1\dots p_r}{}^{q_1\dots q_s} }{dt} \right] .
\nn
\eeq
The
latin indices $a, \ b, ... $ run over $(2,\dots D)$;  $\xi$  has been renamed $t$ and  $\d g = - g^{-1} g^{ab}\d g_{ab}$ has been used.
This theory describes a motion on the coset
$\cG^{++}_C/\cK^{++}$ where  $\cK^{++}$ is the subalgebra of $\cG^{++}$
invariant under the Chevalley involution. The restriction to the level $\ell = 0$ of $\cK^{++}$ is $SO(D)$. This fact and the interpretation of $g_{\m\n}$ as the spacetime metric lead  to the interpretation of the indices $a=(2,...,D)$ as space--like indices, $g_{ab}$ as the spatial metric and the evolution parameter $\xi =t  $ as the remaining time coordinate. These \emph{cosmological} actions $S_{\cG^{++}_C}$ generalise to
all $\cG^{++}$  the $E_{10}\, (\equiv E_8^{++})$ action of reference \cite{Damour:2002cu, Damour:2004zy} proposed in the context of M--theory and cosmological billiards. The rest of this 
section summarises the results of reference \cite{Damour:2002cu}, which will 
be generalised in chapter \ref{gravitino} by including the fermionic sector. 

\subsubsection{"Small Tension Expansion" of $D=11$ Supergravity}

\noindent The billiard shape is only governed by  the 
simple roots, \ie height one roots, of $\mf{g}^{++}$, 
the dominant walls being in one-to-one correspondence with them.  
Some of the subdominant walls also have an algebraic interpretation in terms
of higher-height positive roots. This enables one to go beyond the billiard picture. 
The idea developed in \cite{Damour:2002cu}  is to reformulate the eleven dimensional supergravity action as the non--linear $\s$--model
$S_{E_8^{++}}$\footnote{
This $\s$-model is constructed along the same lines shown above for
very extended groups (\ref{full}) $S_{\cG^{+++}}$ except that the evolution parameter is taken to 
be the time and the chosen subgroup $\cK^{++}$ is the maximally compact one: accordingly, the
Iwasawa decomposition can be fully used.
}. 
The relevance of this $\s$--model  stems from the fact that there exists a group
theoretical interpretation of the billiard motion, namely:  \emph{the asymptotic
BKL dynamics is equivalent,
at each spatial point, to the  asymptotic dynamics of the
one--dimensional nonlinear $\sigma$-model based on the
infinite--dimensional coset space $\cG^{++}_C/\cK^{++}$}. 
More precisely, a "gradient expansion" of the equations of motion of the eleven--dimensional supergravity
perfectly matches a level expansion of the equations of motion of the $\s$--model. The 
matching is established up to height 30 \cite{Damour:2002cu} (and \cite{Damour:2004zy} for more details). 
 Reference \cite{Damour:2002cu} conjectures that the spatial derivatives
of the fields can be identified with coset elements. This conjecture is based on the fact that
there 
is ``enough room'' in $\mf{e}_8^{++}$ for all the spatial gradients: 
the three infinite sets of admissible $\mf{sl}(10)$ Dynkin 
labels $(00100000k)$, $(00000100k)$ and $(10000001k)$ with 
highest weights obeying $\Lambda^2 = 2$, at levels $\ell = 3k+1, \, 3k+2$ 
and $3k + 3$, respectively appear in the decomposition of $\mf{e}_8^{++}$ as $\mf{sl}(10)$ representations. These correspond to three infinite towers of  $\mf{e}_8^{++}$ elements 
\beq 
R_{a_1 ...a_n}{}^{ b_1 b_2 b_3}, \, \,
R_{a1 ...an}{}^{b_1 ...b_6}, \, \, 
R_{a1 ...an}{}^{ b_0 | b_1 ...b_8} \, , 
\nn
\eeq  
which possess the right symmetry to be interpreted as the gradients of $A_{b_1 b_2 b_3}, 
\, A_{b_1 ...b_6}$ and $A_{b_0 | b_1 ...b_8}$.

\section{ The $\cG^{++}_B$-theory}

The action $S_{\cG^{+++}}$ is invariant under the Weyl reflections of $\cG^{+++}$ by construction. The truncated action
 $S_{\cG^{++}_C}$ is no 
longer invariant under the Weyl reflections $W_{\a_1}$ with respect to the deleted root $\a_1$. 
This fact and the non commutativity of the temporal involution with $W_{\a_1}$ implies the existence of an action $S_{{\cal G}_B^{++}}$ 
that is not equivalent to $S_{\cG^{++}_C}$. 
The $\cG^{++}_B$-theory is obtained by performing the same truncation
as the one presented above (namely  all the fields in the $\cG^{+++}$--invariant action (\ref{full}) that 
multiply the generators  involving the root $\alpha_1$ are equated to zero) but this
truncation is performed {\em after} the $\cG^{+++}$ Weyl transformation Eq.(\ref{permute})
which
transmutes the time index 1 to a space index. This yields an action $S_{\cG^{++}_B}$ which is
formally identical to the one given by Eq. (\ref{fullop}),
 but with a Lorentz signature for the
metric, which in the flat coordinates amounts to a negative sign for the
Lorentz metric component $\eta_{22}$, and with $\xi$ identified with the missing
space coordinate (instead of $t$).
The action $S_{{\cal G}_B^{++}}$, and hence $S_{\cG^{+++}}$, contains
{\em exact} intersecting extremal  brane solutions of space-time
covariant theories reduced on all dimensions but one \cite{Englert:2003py, Englert:2004it, Englert:2004ph}. These solutions are 
important because they provide a
laboratory for analysing the significance of at least some subset of the
infinitely many fields describing the Kac--Moody invariant theories.
Extremal branes in more non--compact dimensions differ from the
one dimensional ones only by the dependence
of a harmonic function on the number of non--compact dimensions. For
such  decompactified solutions to exist in
the Kac--Moody theory,   higher level fields must provide the
derivatives necessary to obtain  higher dimensional harmonic
functions \cite{Englert:2003py}. This test is crucial to settle the issue of whether or not
the Kac--Moody theories discussed here can really describe uncompactified
space-time covariant theories.
In the rest of this section, we review in detail the effects of 
Weyl reflections with respect to a root, belonging to the gravity line or not, on the signature of the 
theory. The last reflections may affect the signs of the $p$--form kinetic terms.
Finally exotic signatures are discussed.

\subsubsection{Weyl Reflections w.r.t. roots belonging to the gravity line}

\noindent A  Weyl transformation $W$  can be expressed as a conjugation by a group
element
$U_W$ of $\cG^{+++}$. The involution $\Omega^\prime$ operating on
the conjugate elements is defined by
\begin{equation}
\label{newinvolve}
U_W\, \Omega T\, U^{-1}_W=\Omega^\prime \, U_W  T U^{-1}_W\, ,
\end{equation}
where $T$ is any generator of $\cG^{+++}$.
The effect of the Weyl reflection $W_{\alpha_1}$
generated by the simple root $\a_1$ (the very extended root) is, 
\begin{eqnarray}
\label{permute}
&&U_1\, \Omega K^2_{\ 1} \, U^{-1}_1= \rho K^2_{\ 1}= \rho\Omega^\prime
 K^1_{\ 2}\nonumber\, ,\\
&&U_1\, \Omega K^1_{\ 3} \, U^{-1}_1= \sigma K^3_{\ 2}=
\sigma\Omega^\prime
 K^2_{\ 3} \, ,\\
&&U_1\, \Omega K^i_{\ i +1} \, U^{-1}_1= -\tau K^{i+1}_{\ \, i}=
\tau\Omega^\prime  K^i_{\ i +1}\quad i >2\, .\nonumber
\end{eqnarray}
Here $\rho,\sigma,\tau$ are plus or minus signs which may arise since
ladder operators are representations of the Weyl group
up to signs. Eqs.(\ref{permute}) illustrate the general result
that such signs always cancel in the determination of
$\Omega^\prime$. The content of Eqs.(\ref{permute}) is
represented in Table \ref{involutionK}. The signs below the generators of the gravity
line indicate the sign in front of the
 negative ladder operator obtained by the involution: a
minus sign is in agreement with the conventional Chevalley involution
and indicates that the indices in
$K^a_{\ a +1}$ are either both space--like or time--like indices, while a plus sign
indicates that one index must be time--like and the other  space--like.
\begin{table}[h]
\label{involutionK}
\caption{\small Involution switch from $\Omega$ to
$\Omega^\prime$ due to the Weyl reflection $W_{\alpha_1}$}
\begin{center}
\begin{tabular}{|c|ccccc|c|}
\hline
gravity line&$K^1_{\ 2}$&$K^2_{\ 3}$&$K^3_{\ 4}$&$\cdots$&$K^{D-1}_{\
D}$&time coordinate\\
\hline\hline
$\Omega$&$+$&$-$&$-$&$-$&$-$&1\\
\hline$\,\Omega^\prime$&$+$&$+$&$-$&$-$&$-$&2\\
\hline
\end{tabular}
\end{center}
\end{table}

%%%%%%%%%%%%%%%%%%
If we choose the description that leaves
unaffected coordinates attached to planes invariant under the Weyl
transformation, Table \ref{involutionK} shows that
the  time coordinate must be identified with 2.
The generic Weyl reflection $W_{\a_a}$ generated by  the simple root $\a_a$ of the gravity line
exchanges the indices $a$ and $a+1$ together with the spacetime nature of the corresponding coordinates.

\subsubsection{Weyl Reflections w.r.t. roots not belonging to the gravity line}

\noindent Weyl reflections generated by  simple roots not belonging to the gravity line
relate ladder operators of different levels. As a consequence \cite{Keurentjes:2004bv,Keurentjes:2004xx}, these
 may potentially induce changes of signature far less trivial than the simple change of the index
 identifying the time coordinate. These changes have been studied from the algebraic point
 of view in great details for $E_{11}$  (and more generally for $E_n$) in
 \cite{Keurentjes:2004bv,Keurentjes:2004xx}\footnote{Some algebraic considerations in this context for others groups $\cal G$ are presented in the Appendix of  reference \cite{Gaberdiel:2002db}.}.
 
 In order to tackle the question of how the  involution acts on a generic ladder operator
 $R^{a_1 \dots a_r}$ of level greater than zero in a given irrreducible representation of $A_{D-1}$, let us introduce some notations.
 First, given an involution $\tilde \Omega$, one defines
 $\mathrm{sign}(\tilde \Omega X)$ for any given positive ladder operator  $X$  in  the following way
 \begin{equation}
 \tilde \Omega \, X \equiv \mathrm{sign}(\tilde \Omega X) \, {\bar X},
 \label{sidef}
 \end{equation}
 where $\bar X$ denotes the negative ladder operator conjugate to $X$.
Second, we also introduce a sign associated with a given  positive ladder operator of level greater than zero\footnote{As far as the action of the involution is concerned the symmetry properties of a ladder operator given by its Dynkin labels do not play any role.}
 $R^{a_1 \dots a_r}$
in the following way
 \begin{eqnarray}
\label{signa}
&+ \quad : &\mathrm{sign}(\tilde \Omega R^{a_1 \dots a_r})= -\epsilon_{a_1}\dots\epsilon_{a_r}  \\
\label{signb}
&- \quad : &\mathrm{sign}(\tilde \Omega R^{a_1 \dots a_r})=+\epsilon_{a_1}\dots\epsilon_{a_r},
\end{eqnarray}
where $\epsilon_a=-1$ if $a$ is a timelike index, and
$\epsilon_a=+1$ if $a$ is a space-like index, the space-time
nature of the coordinate labelled by the index $a$ being defined
by the action of $\tilde \Omega$ on the generators $K^a_{\, b}$. The $+$ sign
defined in Eq.(\ref{signa}) leads to a positive kinetic
energy term for the corresponding field in the action while
the $-$ sign defined in Eq.(\ref{signb}) leads to a negative
kinetic energy term.
Finally, if we perform a Weyl reflection $W_Y$ generated by a
simple root not belonging to the gravity line and associated with a
ladder operator $Y$, then Eq.(\ref{newinvolve}) gives
\begin{equation}
\label{kisi}
\mathrm{sign}(\Omega Y) = \mathrm{sign}(\Omega^\prime Y),
\end{equation}
because $\mathrm{sign}(\tilde \Omega Y)= \mathrm{sign}(\tilde \Omega {\bar Y})$ where
$\bar Y$ is the negative ladder operator conjugate to $Y$.

\subsubsection{Exotic Signatures}

\noindent The $S_{\cG^{++}_B}$ action is characterised by a signature $(1,D-2,+)$
where  the $+$ sign means that Eq.(\ref{signa}) is fulfilled for all the simple positive ladder operators, which implies
 that all the kinetic energy terms in the action are positive.
As this involution does not generically commute with Weyl reflections, the same coset can be
described by actions $S_{{\cal G}{}_{(i{}_1i{}_2\dots i{}_t)}^{++}}^{(T,S,\varepsilon)}$, where
the global signature is $(T,S,\varepsilon)$ with
$\varepsilon$ denoting  a set of  signs, one for each
simple ladder operator that does not belong to the gravity line, defined by Eqs.(\ref{signa}) and (\ref{signb}), and $i{}_1i{}_2\dots i{}_t$ are the time indices. The equivalence of the different actions has been
shown by deriving differential equations relating the fields parametrising the different coset
representatives \cite{Englert:2004ph} and in the special case of $\cG^{++}_B=E_8^{++}$ all the signatures in the orbits of
$(1,D-2,+)$ have been found to agree with \cite{Keurentjes:2004bv,Keurentjes:2004xx} and \cite{Hull:1998vg,Hull:1998fh,Hull:1998ym}.
In the last chapter we derive all the signatures in the orbit of $(1,D-2,\{ \epsilon =+\})$ for all $\cG^{++}_B$--theories.
%%%%%%%%%%%%%%%%%%%%%%
%%%%%%%%%%%%%%%%%%%%%%%%
\cleardoublepage
%%%%%%%%%%%%%%%%%%%%%%
%%%%%%%%%%%%%%%%%%%%%%
%%\include{chap_dirac_fermions}
\chapter{Dirac Fermions}
\markboth{DIRAC FERMIONS}{}
\label{dirac}

One intriguing feature of the hidden symmetries is the fact that
when the coupled Einstein--$\cG/K(\cG)$ system is the bosonic sector of
a supergravity theory, then, important properties of
supergravities which are usually derived on the grounds of
supersymmetry may alternatively be obtained by invoking the hidden
symmetries. This is for instance the case of the Chern--Simons term
and of the precise value of its coefficient in eleven dimensional
supergravity, which is required by supersymmetry
\cite{Cremmer:1978km}, but which also follows from the $E$ ($E_{8(8)}$
or $E_{8(8)}^{++}$) symmetry of the Lagrangian \cite{Cremmer:1978ds,Cremmer:1979up,Damour:2002cu}. Quite
generally, the spacetime dimension $11$ is quite special for the
Einstein--$3$--form system, both from the point of view of
supersymmetry and from the existence of hidden symmetries. Another
example will be provided below (subsection \ref{e8}). One might
thus be inclined to think that there is a deep connection between
hidden symmetries and supergravity. However, hidden symmetries
exist even for bosonic theories that are not the bosonic sectors
of supersymmetric theories. For this reason, they appear to have a
wider scope.

In order to further elucidate hidden symmetries, we have
investigated how fermions enter the picture. Although the
supersymmetric case is most likely ultimately the most
interesting, we have initially considered only spin--$1/2$ fermions, for
two reasons. First, this case is technically simpler. Second, in
the light of the above comments, we want to deepen the
understanding of the connection ---~or the absence of connection~---
between hidden symmetries and supersymmetry. The spin--$3/2$ is 
envisaged in chapter \ref{gravitino}. 

The Einstein--($p$--form)--Dirac system by itself is not
supersymmetric and yet we find that the Dirac fermions are
compatible with the $\cG$--symmetry, for all (split) real simple Lie
groups. Indeed, one may arrange for the fermions to form
representations of the compact subgroup $K(\cG)$. This is automatic
for the pure Einstein--Dirac system. When $p$--forms are present,
the hidden symmetry invariance requirement fixes the Pauli
couplings of the Dirac fermions with the $p$--forms, a feature
familiar from supersymmetry. In particular, $E_8$--invariance of
the coupling of a Dirac fermion to the Einstein--$3$--form system
reproduces the supersymmetric covariant Dirac operator of
$11$--dimensional supergravity \cite{Cremmer:1978km,Julia:1986qq}.
A similar feature holds for ${\cal N}=1$ supergravity in $5$
dimensions \cite{Chamseddine:1980sp}.  Thus, we see again that hidden
symmetries of gravitational theories appear to have a wider scope
than supersymmetry but yet, have the puzzling feature of
predicting similar structures when supersymmetry is available.

We formulate the theories both in $3$ spacetime dimensions, where
the symmetries are manifest, and in the maximum oxidation dimension,
where the Lagrangian is simpler. To a large extent, one may thus
view this work as an extension of the oxidation analysis
\cite{Julia:1980gr,Julia:1981wc,J'',Cremmer:1999du,Damour:2002fz,Henneaux:2003kk,Henry-Labordere:2002dk,Henry-Labordere:2002xh,Keurentjes:2002xc,Keurentjes:2002rc,deBuyl:2003ub} to
include Dirac fermions. Indeed the symmetric Lagrangian is known in
$3$ dimensions and one may ask the question of how high it oxidises.
It turns out that in most cases, the Dirac fermions do not bring new
obstructions to oxidation in addition to the ones found in the
bosonic sectors. If the bosonic Lagrangian lifts up to $n$ dimensions,
then the coupled bosonic--Dirac Lagrangian (with the Dirac fields
transforming in appropriate representations of the maximal compact
subgroup $K(\cG)$) also lifts up to $n$ dimensions. This absence of
new obstructions coming from the fermions is in line with the
results of Keurentjes \cite{Keurentjes:2003yu}, who has shown that the topology
of the compact subgroup $K(\cG)$ is always appropriate to allow for
fermions in higher dimensions when the bosonic sector can be
oxidised.

We then investigate how the fermions fit in the conjectured
infinite--dimensional symmetry $\cG^{++}$ and find indications that
the fermions form representations of $K(\cG^{++})$ up to the level
where the matching works for the bosonic sector. We study next the
BKL limit \cite{Belinsky:1970ew,Belinsky:1982pk,Damour:2002et} of the systems with fermions. We extend
to all dimensions the results of \cite{Belinsky:1988mc}, where it was found
that the inclusion of Dirac spinors (with a non--vanishing
expectation value for fermionic currents) eliminates chaos in four
dimensions. Our analysis provides furthermore a group theoretical
interpretation of this result: elimination of chaos follows from
the fact that the geodesic motion on the symmetric space
$\cG^{++}/K(\cG^{++})$, which is lightlike in the pure bosonic case
\cite{Damour:2002et}, becomes timelike when spin--$1/2$ fields are included
(and their currents acquire non--vanishing values) --- the mass term
being given by the Casimir of the maximal compact subgroup
$K(\cG^{++})$ in the fermionic representation.

This chapter is organized as follows. In the first section, we recall the construction of 
non--linear realisation based on the coset spaces $\cG/\cK$. In the section
\ref{gravity},  
we consider the dimensional
reduction to three dimensions of the pure Einstein--Dirac system in
$D$ spacetime dimensions and show that the fermions transform in
the spinorial representation of the maximal compact subgroup
$SO(n+1)$ in three dimensions, as they should. 
The $SO(n,n)$ case is treated in section \ref{dncase}, by relying on
the pure gravitational case. The maximal compact subgroup is now
$SO(n) \times SO(n)$. We show that one can choose the Pauli
couplings so that the fermions transform in a representation of
$SO(n) \times SO(n)$ (in fact, one can adjust the Pauli couplings
so that different representations arise).
In section \ref{e8case}, we turn to the $E_n$--family. We show that
again, the Pauli couplings can be adjusted so that the spin--$1/2$
fields transform in a representation of $SO(16)$, $SU(8)$ or
$Sp(4)$ when one oxidises the theory along the standard lines. 
in section
\ref{nonsimplylaced},  we show that the
$G_2$--case admits also covariant fermions in 5 dimensions in section
\ref{g2case} and treat next all the other non simply laced groups
from their embeddings in simply laced ones.

\noindent In section \ref{G++}, we show that the Dirac fields fit (up to the
same level as the bosons) into the conjectured $\cG^{++}$ symmetry,
by considering the coupling of Dirac fermions to the $(1+0)$ non
linear sigma model of \cite{Damour:2002cu}. In section \ref{BBKKLL}, we
analyse the BKL limit and argue that chaos is eliminated by the
Dirac field because the Casimir of the $K(\cG^{++})$ currents in the
fermionic representation provides a mass term for the geodesic
motion on the symmetric space $\cG^{++}/K(\cG^{++})$. Finally, we
close this chapter with some conclusions.

In the analysis of the models, we rely very much on the papers
\cite{Cremmer:1997ct,Cremmer:1998px,Cremmer:1999du}, where the
maximally oxidised theories have been worked out in detail and
where the patterns of dimensional reduction that we shall need
have been established.

We shall exclusively deal in this chapter with the split real forms
of the Lie algebras, defined as above in terms of the same
Chevalley--Serre presentation but with coefficients that are
restricted to be real numbers. Remarks on the non--split case are
given in the conclusions.

\section{Non--linear realisation based on the coset space $\cG/\cK$ }

The coset space $\cG/\cK$ (with the gauge subgroup
$\cK$ acting from the left) is parametrised by taking the group elements $\cV$ in
the upper--triangular ``Borel gauge''. 
The differential 
\be v =
\ud \cV \cV^{-1} \rlap{} \ee is in the Borel subalgebra of $\mf{g}$, \ie is a
linear combination of the $h_i$'s and the $e_\alpha$'s (notations explained in appendix \ref{lie}).
The differential $v$  is
invariant under right multiplication. One defines $\cP$ as its
symmetric part and $\cQ$ as its antisymmetric part, \be
\cP=\frac{1}{2}(v+v^T), \; \; \; \; \; \;
\cQ=\frac{1}{2}(v-v^T) \, ,  \ee where $v^T = \tau(v) $ and $\tau$ is the 
Cartan involution. $\cQ$ is in the compact
subalgebra. Under a gauge transformation $\cP$ is covariant
whereas the antisymmetric part $\cQ$ transforms as a gauge
connection, 
\be \cV \ \longrightarrow \ H \cV \rlap{\ ,} \; \; \;
\; \; \;  \cP \ \longrightarrow \ H \cP H^{-1}\rlap{\ ,} \; \; \;
\; \; \; \cQ \ \longrightarrow \ \ud H H^{-1}+ H \cQ H^{-1}  \ee
(with $H \in \cK$).
One may parametrise $\cV$ as $\cV = \cV_1 \, \cV_2$ where $\cV_1$
is in the Cartan torus 
$\cV_1 = e^{\frac{1}{2} \phi^i H_i} $
and $\cV_2$ is in the nilpotent subgroup generated by the
$e_\alpha$'s (the $H_i$ are as in the orthonormal basis, see Eq.(\ref{13.56}) and the $e_\a$ are as in the 
Chevalley basis "up to signs") \cite{Cremmer:1997ct}\footnote{The
 explicit choice for $\cV_2$ for the maximally oxidised theories which comprise at most one $p$--form 
$F^{\sst{(p)}} = dA^{\sst{(p-1)}}$ and possibly Chern--Simons terms --- \ie for $\mf{a}_n$, $\mf{d}_n$, 
$\mf{e}_8$ and $\mf{g}_2$ ---   is $\cV_2=\cV_3 \cV_4 \cV_5 \cV_6$, 
\beq 
\cV_3 &=& \Pi_{i<j} U_{ij} \nn \, , \\
\cV_4 &=&    e^{\sum_{i_1,... , i_p} A^\0_{i_1... i_{p-1}} E^{i_1... i_{p-1}}} \nn \, , \\
\cV_5 &=&  e^{\sum_{ij} \psi^{i_1...i_{p-2}} D_{i_1...i_{p-2}}} \nn \, ,  \\
\cV_6 &=&   e^{\sum_j \chi^j D_j}  \nn \, ,
\eeq
where $U_{ij} = e^{\cA^i_{\0 \, j} E^i{}_j}$ (without sum), the $\cA^i_{\0 \, j}$ are as in Eqs.(\ref{redmetric}), the $A^\0_{i_1... i_p}$ are as in Eq.(\ref{reda}), the $\chi^j$ and $\psi^{i_1...i_{p-2}} $ as is Eqs.(\ref{d3genkkmod}). The $e_\a$'s are $\{ E^i{}_j, \, E^{i_1... i_{p-1}}, \, D_{i_1...i_{p-2}}, \,  D_j \}$}. In terms of this parametrisation, one finds \be v
= \frac{1}{2} \, \ud \phi^i H_i + \sum_{\alpha \in \Delta_+}
e^{\frac{1}{2} \vec{\a} . \vec{\phi}} \cF_\alpha e_\alpha
\label{formofG}\ee where $\Delta_+$ is the set of positive roots.
Here, the one-forms $\cF_\alpha$ are defined through \be \ud \cV_2
\cV_2^{-1} = \sum_{\alpha \in \Delta_+} \cF_\alpha e_\alpha. \ee
In the infinite-dimensional case, there is also a sum over the
multiplicity index. Thus we get \be \cP = \frac{1}{2}\, \ud \phi^i
H_i + \frac{1}{2} \sum_{\alpha \in \Delta_+} e^{\frac{1}{2}
\vec{\a} . \vec{\phi}} \cF_\alpha \left(e_\alpha- f_\alpha \right)
\label{formofP}\ee and \be \cQ = \frac{1}{2}\sum_{\alpha \in
\Delta_+} e^{\frac{1}{2} \vec{\a} . \vec{\phi}} \cF_\alpha \,
k_\alpha  \equiv \sum_{\alpha \in \Delta_+} \cQ_{(\a)} \, k_\a
\label{formofQ}\ee with \be \cQ_{ (\a)} = \frac{1}{2}e^{\frac{1}{2}
\vec{\a} . \vec{\phi}} \cF_{ \alpha}. \ee We see therefore that
the one-forms $(1/2) e^{\frac{1}{2} \vec{\a} . \vec{\phi}}
\cF_{\alpha }$ appear as the components of the connection one-form
of the compact subgroup $K(G)$ in the basis of the $k_\alpha$'s.

The Lagrangian for the coset model $\cG/\cK$ reads \be \cL_{\cG/\cK}
= - K\left( \cP ,\w
* \cP \right) \ee If we expand the Lagrangian according to
(\ref{formofP}), we get 
\be {\cal L}_{\cG/\cK} = - \fft12
{*d\vec{\phi}}\wedge d\vec{\phi} - \frac{1}{2}\sum_{\alpha \in
\Delta_+} N_\a \, e^{ \vec{\a} . \vec{\phi}}\, {*\cF_\alpha}\wedge
\cF_\alpha \label{generalform}\ee where the factor $N_\a$ is
defined in (\ref{norma}).

The coupling of a field $\psi$ transforming in a representation
$J$ of the ``unbroken" subgroup $\cK$ is straightforward. One
replaces ordinary derivatives $\partial_\m$ by covariant
derivatives $D_\m$ where \be D_\m \psi = \partial_\m \psi-
\sum_{\alpha \in \Delta_+}\cQ_{\mu  (\alpha)} J_{\alpha} \psi
\label{cova}\ee with  $\cQ_{(\alpha)} = \cQ_{\m (\alpha) }dx^\m $.  In
(\ref{cova}), $J_{\alpha}$ are the generators of the representation
$J$ of $\cK$ in which $\psi$ transforms, $J_{\alpha} = J(k_\alpha)$
(the generators $J_{\alpha}$ obeys the same commutation relations as
$k_\alpha$). This guarantees $\cK$ -- and hence $\cG$ --
invariance. The three-dimensional Dirac Lagrangian is thus (in
flat space) \be \bar{\psi} \gamma^\m \left( \partial_\m
-\sum_{\alpha \in \Delta_+}\cQ_{\mu  (\alpha)} J_{\alpha} \right)
\psi \ee

\section{Dimensional reduction of the Einstein--Dirac System}
\label{gravity}

We start with the simplest case, namely, that of the coupled
Einstein--Dirac system without extra fields.  To show that the Dirac field is
compatible with the hidden symmetry is rather direct in this case.

\subsection{Reduction of gravity}

Upon dimensional reduction down to $d=3$, gravity in $D=3+n$ gets
a symmetry group $SL(n+1)$, beyond the $SL(n)$
symmetry of the reduced dimensions. Moreover, in three dimensions
the scalars describe a $SL(n+1)/SO(n+1)$
$\sigma$--model. A 
pedestrian presentation
 of pure gravity dimensional reduction is given in appendix \ref{app_dim_reduction}, here 
 details are skipped.  Following (\ref{metricreduction}), we parametrize
the $D=3+n$ vielbein in a triangular gauge as 
\be 
\hat{e} = \left(
\begin{array}{cc} 
e^{\frac{1}{2} \vec{s}.\vec{\phi}} e_\mu{}^\nnn
&
e^{\vec{\gamma_i}.\vec{\phi}} \cA_\1{}_\mu{}^j \\
0 & M_i{}^j
\end{array} \right)
\label{vb} 
\ee 
with $\mu, (\n) = 0,D-2,D-1$ and $i,j = 1, ..., n$. We
choose the non--compactified dimensions to be $0$, $D-2$ and $D-1$
so that indices remain simple in formulas. The vielbein in three
spacetime dimensions is $e_\mu{}^\nnn$. We denote by $M$ the
upper--triangular matrix\footnote{The indices would have been  at the same ``floor''
if $h^i = dz^i + \cA_\0{}^i {}_j dz^j+ \cA_\1{}^i$ has been defined by 
$h^i = dz^i + dz^j\cA_\0{}_j{}^i + \cA_\1{}^i$. }
\be M_i{}^j =
e^{\vec{\gamma_i}.\vec{\phi}} (\d_i^j + \cA_\0{}^i{}_j) =
e^{\vec{\gamma_i}.\vec{\phi}} \td{\g}^i{}_j 
\nn
\ee 
One can
check that $\det(M) = e^{-\frac{1}{2}\vec{s}.\vec{\phi}}$.
After dualising the Kaluza--Klein vectors $\cA_\1{}^i$ into scalars
$\chi_j$ as
\be 
e^{\vec{b}_i.\vec{\phi}} * \! \left(\td\g^i{}_j \ud
(\g^j{}_m {\cA}_\1^m )\right) = \gamma^j{}_i \left( \ud \chi_j \right)
\label{dual-a1} 
\ee 
one can form the upper--triangular $(n+1)
\times (n+1)$ matrix 
\be 
\cV^{-1} = \left(
\begin{array}{cc}
M_i{}^j & \chi_i e^{\frac{1}{2}\vec{s}.\vec{\phi}} \\
0 & e^{\frac{1}{2} \vec{s}.\vec{\phi}}
\end{array} \right)
\label{V-An} \ee which parametrizes a
$SL(n+1)/SO(n+1)$ symmetric space. With this
parametrization, the three--dimensional reduced Lagrangian becomes
\be 
\cL_E^\3 = R *1 - \frac{1}{2} \mathrm{Tr}\left( \cP \w *
\cP \right) 
\nn
\ee 
where one finds explicitly from (\ref{V-An}) 
\be 
\cG =
\ud \cV \cV^{-1}= - \cV (\ud \cV^{-1}) = - \left(
\begin{array}{cc}
M^{-1} \ud M & M^{-1} \ud \chi e^{\frac{1}{2}\vec{s}.\vec{\phi}} \\
0 & \frac{1}{2} \vec{s}.\ud \vec{\phi}
\end{array} \right)
\nn 
\ee 
and 
\bea \cP &=& - \left(
\begin{array}{cc} \frac{1}{2} \left( M^{-1} \ud M + (M^{-1} \ud M)^T
\right)
& \frac{1}{2} M^{-1} \ud \chi e^{\frac{1}{2}\vec{s}.\vec{\phi}} \nn \\
\frac{1}{2} \left(M^{-1} \ud \chi e^{\frac{1}{2}\vec{s}.\vec{\phi}}
\right)^T & \frac{1}{2} \vec{s}.\ud \vec{\phi}
\end{array} \right) \\
\cQ &=& - \left( \begin{array}{cc} \frac{1}{2} \left( M^{-1} \ud M -
(M^{-1} \ud M)^T \right)
& \frac{1}{2} M^{-1} \ud \chi e^{\frac{1}{2}\vec{s}.\vec{\phi}} \\
-\frac{1}{2} \left(M^{-1} \ud \chi e^{\frac{1}{2}\vec{s}.\vec{\phi}}
\right)^T & 0
\end{array} \right)
\label{Q-An} \rlap{\ .} 
\eea

\subsection{Adding spinors}

We now couple a Dirac spinor to gravity in $D=3+n$ dimensions: 
\be
\hat \cL = \hat \cL_E +  \hat \cL_D \nn
\ee 
with $ \hat \cL_E$ the Einstein Lagrangian and
$\hat \cL_D$ the Dirac Lagrangian, 
\be 
\hat \cL_D = \he \bar{\hat  \psi} \Sh{ \hat D} \hat \psi = \hat e
\bhpsi
\g^{\hat \m} \left( \partial_{\hat \m} - \frac{1}{4} \hat \omega_{\hat \m, {\sst{(\hat{\n})}} {\sst{(\hat{\rho})}}} \gamma^{{\sst{(\hat{\n})}} {\sst{(\hat{\rho})}}}
\right) \hpsi 
\label{einst-dirac}
\ee 
$\he$ is the determinant of the vielbein and $\hat \m,  (\hat \n) =
0,... ,\Dm - 1$. The indices $(\hat \m)$, $(\hat \n)$, ... are internal indices
and $\hat \m$, $\hat \n$ are spacetime indices. The spin connection $\hat \omega$  can 
be computed from the vielbein, 
\bea
\hat \omega_{\hat{\m}, {\sst{(\hat{\n})}} {\sst{(\hat{\rho})}}} 
&=& 
\frac{1}{2} e_{\sst{(\hat{\n})}}{}^{\hat \s} 
(\partial_{\hat \s}  e_{\hat{\m}}{}_{\sst{(\hat{\rho})}} - \partial_{\hat{\m}} e_{\hat \s} {}_{\sst{(\hat{\rho})}}) 
- \frac{1}{2} e_{\sst{(\hat{\rho})}}{}^{\hat \s} 
(\partial_{\hat \s}  e_{\hat \m} {}_{\sst(\hat{\n})} - \partial_{\hat{\m}} e_{\hat \s} {}_{\sst{(\hat{\n})}})  
\nn \\ && \hspace{3cm} 
- \frac{1}{2} e_{\sst{(\hat{\n})}}{}^{\hat \lambda} e_{\sst{(\hat{\rho})}}{}^{\hat \s} 
(\partial_{\hat \s}  e_{\hat \lambda}{}_ {\sst{(\hat \d)}}  - \partial_{\hat \lambda} e_{\hat \s} {}_ {\sst{(\hat \d)}}) e_{\hat \m}{}^ {\sst{(\hat \d)}} \rlap{\ .}  
\nn
\eea
Note that the
numerical matrices $\gamma^{\sst{(\hat \m)}}$ (with $(\hat \m)$ an internal index) are
left unchanged in the reduction process, but this is not the case
for $\gamma^{\hat \m}$ (with $\hat \m $ a spacetime index). Indeed, $\g^{\hat \m }$
 has to be understood as $\hat e_{\sst{(\hat \n)}}{}^{\hat \mu} \g^{\sst{(\hat \n)}}$ in $D$
dimensions, and as $ e_{\sst{(\n)}}{}^\mu \g^{\sst{(\n)}} $ in 3 dimensions.
Nevertheless, we do not put hats on three--dimensional
$\gamma$--matrices with a spatial index as no confusion should
arise. Remember that the Dirac spinor  belongs to a representation of $Spin(D-1,1)$ which is 
the universal covering group of the 
orthochronous Lorentz group $SO(D-1,1)^\uparrow_+$.  At the Lie algebra level, the 
dirac spinor belongs to a representation of the Lie algebra $\mf{so}(D-1,1)$. The $\mf{so}(D-1,1)$ 
generators $J^{\hat \m \hat \nu}$ are represented by gamma matrices $\fft12 \g^{\sst{(\hat \m)(\hat \n)}}$,
see Lagrangian (\ref{einst-dirac}).

\noindent We perform a dimensional reduction, with the vielbein parametrised
by (\ref{vb}), by imposing the vanishing of derivatives $\partial_{\hat \m}$
for $\hat \m  \geq 3$. We also rescale $\hpsi$ by a power $f$ of the
determinant of the reduced part of the vielbein $M$: 
\be 
\psi =
e^{- \frac{1}{2} f \vec{s}.\vec{\phi}} \hpsi \rlap{\ .} \nn
\ee
After reassembling the various terms, we find that the reduced
Dirac Lagrangian can be written as 
\beq
\cL_D^\3 &=& \  e^{(\frac{1}{2}+f)\vec{s}.\vec{\phi}} e \bpsi
\Sh{D} \psi + \left( \frac{1}{2} f + \frac{1}{4} \right)
e^{(\frac{1}{2}+f)\vec{s}.\vec{\phi}}  
e \, \partial_\mu(\vec{s}.\vec{\phi}) \, \bpsi \gamma^\mu \psi \nn \\
&  &+ \frac{1}{8} e^{(\frac{1}{2}+f)\vec{s}.\vec{\phi}} e \left(
M^{-1}{}_j{}^k \partial_\mu M_k{}^i - M^{-1}{}_i{}^k \partial_\mu M_k{}^j \right)
\bpsi \gamma^\mu \gamma^{ij} \psi \nn \\
& &+ \frac{1}{8} e^{f\vec{s}.\vec{\phi}} e \,  e_\rrr{}^\mu
e_\sss{}^\nu \! \left( \partial_\mu(\cA_\1{}_\nu{}^j M^{-1}{}_j{}^k) -
\partial_\nu(\cA_\1{}_\mu{}^j M^{-1}{}_j{}^k) \right) \! M_k{}^i \, \bpsi
\gamma^i \gamma^{\rrr \sss} \psi 
\nn
\eeq 
where $e$ is the determinant of the dreibein.
\noindent dualising $\cA_\1$ according to (\ref{dual-a1}), we get 
\beq
\cL_D^\3 &=& \  e^{(\frac{1}{2}+f)\vec{s}.\vec{\phi}} e \bpsi
\Sh{D} \psi  + \left( \frac{1}{2} f + \frac{1}{4} \right)
e^{(\frac{1}{2}+f)\vec{s}.\vec{\phi}}
e \, \partial_\mu(\vec{s}.\vec{\phi}) \, \bpsi \gamma^\mu \psi \nn \\
&& + \frac{1}{8} e^{(\frac{1}{2}+f)\vec{s}.\vec{\phi}} e \left(
M^{-1}{}_j{}^k \partial_\mu M_k{}^i - M^{-1}{}_i{}^k \partial_\mu M_k{}^j \right)
\bpsi \gamma^\mu \gamma^{ij} \psi \nn \\
& &+ \frac{1}{8} e^{(1 + f) \vec{s}.\vec{\phi}} e \,
\epsilon_{\rrr \sss \sst{(\d)}}{}^{\sst{(\d)}} e_{\sst{(\d)}}^\mu M^{-1}{}_i{}^j \partial_\mu \chi_j \bpsi
\gamma^i \gamma^{\rrr \sss} \psi \rlap{\ .} 
\nn
\eeq
If we choose the scaling power of the spinor as 
\be f =
-\frac{1}{2} \rlap{\ ,} 
\nn
\ee 
the Lagrangian simplifies to 
\beq
\cL_D^\3 &= &\  e \bpsi \Sh{D} \psi  + \frac{1}{8} e
\left( M^{-1}{}_j{}^k \partial_\mu M_k{}^i - M^{-1}{}_i{}^k \partial_\mu M_k{}^j
\right)
\bpsi \gamma^\mu \gamma^{ij} \psi  \nn \\
& &+ \frac{1}{4} e \, e^{\frac{1}{2} \vec{s}.\vec{\phi}}
M^{-1}{}_i{}^j \partial_\mu \chi_j \bpsi \gamma^\mu \hgamma \gamma^i
\psi
\label{cld} 
\eeq 
where we have used the notation $\hat{\gamma} =
\gamma^0 \gamma^{D-2} \gamma^{D-1}$.

\noindent In fact, the three--dimensional Lagrangian can be rewritten using a
covariant derivative including a connection with respect to the
gauge group $SO(n+1)$: 
\be 
\cL_D^\3 = e \bpsi
\Sh{\nabla} \psi \nn  
\ee 
with 
\be 
\nabla_\mu = \partial_\mu - \frac{1}{4}
{\omega}_{\mu, \rrr \sss} \gamma^{\rrr\sss} - \frac{1}{2} \cQ_{\mu,ij}
J^{ij} 
\label{3.18}
\ee 
where $\cQ$ is the $SO(n+1)$
connection (\ref{Q-An}), acting on Dirac spinors through 
\be
J_{ij} = \frac{1}{2} \gamma^{ij}\, , \; \; \; \; \; \; \; \;
J_{i(n+1)} = \frac{1}{2} \hgamma \gamma^i \,  \; \; \; \; \; \;\;
\;  (i,j=1..n) \rlap{\ .} 
\nn
\ee
These matrices define a spinorial representation of
$\mf{so}(n+1)$. The commutations relations are indeed 
\be
\begin{array}{ccl}
\left[ \frac{1}{2} \gamma^{ij}, \frac{1}{2} \gamma^{kl} \right] &=& 0 \\
\left[ \frac{1}{2} \gamma^{ij}, \frac{1}{2} \gamma^{ik} \right] &=&
-\frac{1}{2}
\gamma^{jk} \\
\left[ \frac{1}{2} \hat{\gamma} \gamma^i, \frac{1}{2} \gamma^{jk} \right] &=& 0 \\
\left[ \frac{1}{2} \hat{\gamma} \gamma^i, \frac{1}{2} \gamma^{ij} \right]
&=&
\frac{1}{2} \hat{\gamma} \gamma^j \\
\left[ \frac{1}{2} \hgamma \gamma^i, \frac{1}{2} \hgamma \gamma^j \right]
&=& -\frac{1}{2} \gamma^{ij}
\end{array}
\nn
\ee 
where different indices are supposed to be distinct.
Equivalently, we can remark that $\hat{\gamma}$ commutes with
$\gamma^m$ for $m=0,D-2,D-1$ and anticommutes with $\gamma^i$ for
$i=1,...,n$. As we have also $\hat{\gamma}^2 = 1$, it follows that
$\gamma^a$'s for $a=1,...,n$ and $\hat{\gamma}$ generate an
internal $n+1$ dimensional  Clifford algebra, commuting with the
spacetime Clifford algebra generated by $\gamma^m, m=0,D-2,D-1$.

\noindent In other words, the $Spin(n+2,1)$ representation of Dirac fermions
in dimension $D=3+n$ is reduced to a $Spin(2,1) \times  Spin(n+1)$
representation in dimension 3, ensuring that the Dirac fermions
are compatible with the hidden symmetry. Note that if $D$ is even,
one can impose chirality conditions on the spin $1/2$ field in $D$
dimensions. One gets in this way a chiral spinor of $Spin(n+1)$
after dimensional reduction.

\subsection{Explicit Borel decomposition}

One may write the Lagrangian in the form (\ref{generalform}) by
making a full Borel parametrization of the matrix $M$. The algebra
element $\cG$ reads 
\be 
\cG = \frac{1}{2} \ud \vec{\phi} . \vec{H} +
\sum_{i<j} e^{ \frac{1}{2} \vec{b}_{ij}.\vec{\phi}} \cF_\1{}^i{}_j
e_{b_{ij}} + \sum_i e^{-\frac{1}{2} \vec{b}_i.\vec{\phi}} \cG_\1{}^i
e_{b_i} 
\nn
\ee 
where the dilaton vectors and the ``field strengths'' $\cF^i_{\1 j}$ and $\cG_\1{}^i$
(which are also the $SO(n+1)$ connections) are given in the appendix \ref{app_dim_reduction}. 
The positive
roots of $\mf{sl}(n+1)$ are $(\vec b_{ij}, -\vec b_i)$ and the
corresponding root vectors $e_{b_{ij}}$ and $e_{b_i}$ are the
multiple commutators of the generators $e_i$ not involving $e_1$
(for $e_{b_{ij}}$) or involving $e_1$ (for $e_{b_i}$), i.e.,
$e_{b_{ii+1}} = e_{i+1}$ ($i = 1, \cdots, n-1$), $e_{b_{ij}} = [e_i,
[e_{i+1},[ ... [e_{j-2}, e_{j}] \cdots ] $ ($i, j = 2, \cdots, n$,
$i+1 < j$), $e_{b_i} = [e_1, e_{b_{2i}}]$ ($i\geq 3$), $e_{b_n} =
e_n$. These root vectors are such that the normalization factors
$N_\a$ are all equal to one.
The Lagrangian $ \cL^{(3)} = \cL_E^{(3)} + \cL_D^\3$ reads 
\bea 
{\cal L}^{(3)} &=& R\,
{*\oneone} - \fft12 {*d\vec\phi}\wedge d\vec\phi - \fft12 \sum_i
e^{-\vec b_i\cdot\vec\phi}\, {*\cG_{\1 i}}\wedge \cG_{\1 i} -\fft12
\sum_{i<j} e^{\vec b_{ij}\cdot\vec\phi}\, {*\cF^i_{\1 j}} \wedge
\cF^i_{\1 j} \nn\\
&&+ \he \bhpsi \gamma^\mu \left( \partial_\mu - \frac{1}{4}
\hat{\omega}_{\mu,\rrr \sss} \gamma^{\rrr \sss} - \fft14 \sum_{i<j} e^{\fft12 \vec
b_{ij}\cdot\vec\phi}\, {\cF^i_{\1 j}} \gamma^{ij} - \fft14 \sum_i
e^{-\fft12 \vec b_i \cdot\vec\phi}\, \cG_{\1 i} \hat{\gamma}
\gamma^{i}\right) \psi \hspace{.5cm} .
\nn
\eea

\section{$D_n$ case}
\label{dncase}
\subsection{Bosonic sector}

Following \cite{Cremmer:1999du}, we consider the gravitational
lagrangian ${\cal L}_E$ with an added three form field strength
$F_\3$ coupled to a dilaton field $\varphi$,
%%%%%
\be 
{\cal L} = R \, {*\oneone} - \fft12 {*d\varphi} \wedge d\varphi -
\fft12 e^{a\varphi} \, {*F_\3} \wedge F_\3 
\label{dnlag} 
\ee
%%%%%
in the dimension $\Dm = n+2$, where the coupling constant $a$ is
given by $a^2=8/(\Dm-2)$. Upon toroidal reduction to $D=3$, this
yields the Lagrangian
%%%%%
\bea 
{\cal L}^{(3)} &=& R\, {*\oneone} - \fft12 {*d\vec\phi}\wedge
d\vec\phi - \fft12 \sum_i e^{\vec b_i\cdot\vec\phi}\,
{*\cF_\2^i}\wedge \cF_\2^i -\fft12 \sum_{i<j} e^{\vec
b_{ij}\cdot\vec\phi}\, {*\cF^i_{\1 j}} \wedge
\cF^i_{\1 j}\nn\\
&&-\fft12 \sum_i e^{\vec a_i\cdot\vec\phi}\, {*F_{\2 i}}\wedge
F_{\2 i} -\fft12 \sum_{i<j} e^{\vec a_{ij}\cdot\vec\phi}\, {*F_{\1
ij}}\wedge F_{\1 ij}\ .\label{dnlag1} \eea
%%%%%
Note that here $\vec\phi$ denotes now the set of dilatons
$(\phi_1,\phi_2,\ldots, \phi_{{\sst \Dm}-3} )$, augmented by
$\varphi$ (the dilaton in $\Dm$ dimensions) as a zeroth component;
$\vec\phi=(\varphi, \phi_1,\phi_2,\ldots, \phi_{{\sst \Dm}-3})$.
The dilaton vectors entering the exponentials in the Lagrangian
are given by $\vec a_i, \,  \vec
a_{ij}, \,   \vec b_i$ and $\vec b_{ij}$ (see appendix \ref{app_dim_reduction}) augmented by a zeroth component that is equal to
the constant $a$ in the case of $\vec a_i$ and $\vec a_{ij}$, and
is equal to zero in the case of $\vec b_i$ and $\vec b_{ij}$. The
field strengths are given in appendix \ref{app_dim_reduction}.
After dualising the 1--form potentials $\cA^i_\1$ and $A_{\1 i}$ to
axions $\chi_i$ and $\psi^i$ respectively, the three--dimensional
Lagrangian (\ref{dnlag1}) can written as the purely scalar
Lagrangian
%%%%%
\bea {\cal L}_{D_n }^{(3)} &=& R\, {*\oneone} - \fft12
{*d\vec\phi}\wedge d\vec\phi - \fft12 \sum_i e^{-\vec
b_i\cdot\vec\phi}\, {*\cG_{\1 i}}\wedge \cG_{\1 i} -\fft12
\sum_{i<j} e^{\vec b_{ij}\cdot\vec\phi}\, {*\cF^i_{\1 j}} \wedge
\cF^i_{\1 j}\nn\\
&&-\fft12 \sum_i e^{-\vec a_i\cdot\vec\phi}\, {* G_\1^i}\wedge
G_\1^i -\fft12 \sum_{i<j} e^{\vec a_{ij}\cdot\vec\phi}\, {*F_{\1
ij}}\wedge F_{\1 ij}\ ,\label{dnlag2} \eea
%%%%%
where the dualised field strengths $\cG_{\1 i}$ and $G_\1^i$ are  given in appendix \ref{app_dim_reduction}.
%%%%%
The positive roots of $D_n$
are given by $(\vec b_{ij}, -\vec b_i, \vec a_{ij}, -\vec a_i)$ ,
the simple roots being $\vec a_{12}$, $\vec b_{i,i+1}$ ($i\le
n-1$) and $-\vec a_n$ \cite{Cremmer:1999du}. The three--dimensional
Lagrangian (\ref{dnlag2}) describes a $SO(n,n) / (SO(n)\times
SO(n) ) $ $\sigma$--model in the Borel gauge coupled to gravity.
The field strength of this $\sigma$--model is 
\beq
\cG &=& \frac{1}{2} \ud \vec{\phi} . \vec{H} + \sum_{i<j} e^{
\frac{1}{2} \vec{b}_{ij}.\vec{\phi}} \cF_\1{}^i{}_j e_{b_{ij}} +
\sum_i e^{-\frac{1}{2} \vec{b}_i.\vec{\phi}} \cG_\1{}^i e_{b_i}
\nn \\
& &+ \sum_{i<j} e^{\frac{1}{2} \vec{a}_{ij}.\vec{\phi}} F_{\1 ij}
e_{a_{ij}} + \sum_{i} e^{-\frac{1}{2} \vec{a}_{i}.\vec{\phi}}
G_\1{}^{i} e_{a_i} \rlap{\ .}
\label{g-dn} 
\eeq 
$\vec{H}$ is the vector of Cartan generators. We
 express the positive generators of $D_{n}$ as follow, 
\bea
e_{b_{ij}} & \equiv & [e_i,[...,[e_{j-2},e_{j-1}]...] \ \ i<j \nn
\\ e_{b_i} & \equiv & [\tilde{e}_{n-1},e_{b_{1i}}] \nn
\\
e_{a_i} & \equiv & [e_{b_{in-1}},e_{n-1}] \\
e_{a_{ij}} & \equiv & [[e_n,e_{b_{2j}}],e_{b_{1i}}] \ i<j 
\eea
where $\tilde{e}_{n-1} = [e_{n},[e_{b_{2n-1}},e_{n-1}]]$ and where
$ i = 1,... ,n-1$ and $e_{b_{ii}}$ must be understood as being
absent. The Chevalley--Serre generators of $D_n$, namely $\{e_m \
\arrowvert \ m = 1,...,n \}$, are given by $e_i = e_{b_{i i+1}}$
($i=1,...,n-2$), $e_{n-1} = e_{a_{n-1}}$ and $e_n = e_{a_{12}}$.
These generators are associated to the vertices numbered as shown
in the following Dynkin diagram,
\begin{center}
\scalebox{.5}{
\begin{picture}(180,60)
%nom des racines
\put(5,-5){$n-1$} \put(45,-5){$n-2$} \put(85,-5){$3$}
\put(125,-5){$2$} \put(165,-5){$1$} \put(140,45){$n$}
%4 vertex + lignes simples
\thicklines 
\multiput(10,10)(40,0){4}{\circle{10}}
\multiput(95,10)(40,0){2}{\line(1,0){30}}
\dashline[0]{2}(55,10)(65,10)(75,10)(85,10)
%deux vertex du dessus
%\put(90,50){\circle{10}} \put(90,15){\line(0,1){30}}
\put(130,50){\circle{10}} \put(130,15){\line(0,1){30}}
%deux dernier vertex
\multiput(130,10)(40,0){2}{\circle{10}}
\put(15,10){\line(1,0){30}}
\end{picture}
} 
\end{center} 
Their non vanishing commutation relations are \bea
\ [e_{b_{ij}},e_{b_{mn}}] &=& \d_{jm} e_{b_{in}} - \d_{in}e_{b_{mj}} \nn \\
\ [e_{b_{ij}},e_{b_m} ] &=& - \d_{im} e_{b_j}\nn \\
\ [e_{b_{ij}},e_{a_{mn}}] &=& -\d_{im} e_{a_{jn}} - \d_{in}e_{a_{mj}} + \d_{im} e_{a_{nj}} \nn\\
\ [e_{b_{ij}},e_{a_m}] &=& \d_{jm} e_{a_i} \nn \\
\ [e_{a_{ij}},e_{a_m} ]& = &- \d_{im} e_{b_j} + \d_{jm} e_{b_i}
\eea Notations are similar for the negative generators (with $f$'s
instead of $e$'s). One easily verifies that the normalization
factors $N_\a$ are all equal to one, $K(e_\a, f_\b) = - \d_{\a
\b}$.

The gravitational subalgebra $A_{n-1}$ is generated by $h_1,
\cdots, h_{n-2}, \tilde{h}_{n-1}$ (Cartan generators), $e_1,
\cdots, e_{n-2}, \tilde{e}_{n-1}$ (raising operators) and $f_1,
\cdots, f_{n-2}, \tilde{f}_{n-1}$ (lowering operators), with
$\tilde{h}_{n-1} = - h_n - h_2 - h_3 - \cdots - h_{n-1}$. The
simple root $\tilde{\a}_{n-1}$ is connected to $\a_1$ only, with a
single link. Note that although it is a simple root for the
gravitational subalgebra $A_{n-1}$, it is in fact the highest root
of the $A_{n-1}$ subalgebra associated with the Dynkin subdiagram
$n, 2, 3 , \cdots, n-1$.

\subsection{Fermions}

We want to add Dirac fermion in $D_{max}$, with a coupling which
reduces to $SO(2,1) \times (SO(n)\times SO(n))$, or more precisely 
$Spin(2,1) \times (Spin(n)\times Spin(n))$. The coupling to
gravitational degrees of freedom is already fixed to the spin
connection by invariance under reparametrization; we know from the
first section that it reduces to the  $Spin(2,1)\times Spin(n)$
connection in $D=3$. From the structure of the theory, we know
that the fermions must have linear couplings with the $3$--form
$F_3$. Indeed, the $D=3$ couplings must be of the following form
\be 
e \bpsi \gamma^{\mu} \left( \partial_\mu - \frac{1}{4} \hat{\omega}
_{\mu,\mm \nnn} \gamma^{\mm \nnn} - \cQ_{\mu (\alpha)} J^{(\alpha)} \right) \psi
\nn 
\ee 
where $\cQ$ can be read off from (\ref{g-dn}) above and the
$J^{(\a)}$'s are a representation of $SO(n) \times SO(n)$. The
possible Lorentz--covariant coupling of this kind are the Pauli
coupling and its dual, 
\be 
\ -\sqrt{- \hat g} e^{\frac{1}{2} a
\varphi}\bar{\hpsi} { 1 \over 3!}(\a \g^{\sst{(\hat \m )(\hat \n)(\hat \rho)}} + \b \g^{\sst{(\hat \m )(\hat \n)(\hat \rho)}} \g)
F_\3{}_{ \, \sst{(\hat \m )(\hat \n)(\hat \rho)}} 
\label{dnlagf0} \hpsi 
\ee 
where $\a$ and $\b$ are
arbitrary constants, which will be determined below. The dilaton
dependence is fixed so as to reproduce the roots $\vec{a}_i$ and
$\vec{a}_{ij}$ in the exponentials in front of the fermions in the
expressions below. The matrix $\g$ is the product of all gamma
matrices $\g = \g^0 \g^1 ... \g^{D-1}$. One has $\gamma^2 =
-(-1)^{[ {D \over 2} ]}$. Notice that in odd dimensions this
matrix is proportional to the identity and therefore we can put
$\b=0$ without loss of generality. Thus we add to the bosonic
lagrangian (\ref{dnlag}) the following term,
%%%%%
\be {\cal L}_{\hpsi} = \ \sqrt{- \hat g} \bar{\hpsi} (
\g^{\hat \m}\partial_{\hat \m} - {1 \over 4} \hat \o_{\hat \m, \sst{(\hat \m) (\hat \n)}} \g^{\sst{(\hat \m) (\hat \n)}} - { 1 \over
3!}(\a \g^{\sst{(\hat \m )(\hat \n)(\hat \rho)}} + \b \g^{\sst{(\hat \m )(\hat \n)(\hat \rho)}} \g)  e^{\frac{1}{2} a
\varphi} F_\3{}_{ \sst{(\hat \m )(\hat \n)(\hat \rho)}} \label{dnlagf} )
\hpsi \, .\ee
%%%%%
Upon toroidal reduction to $D=3$, the last term of (\ref{dnlagf})
becomes, 
\beq 
&-& { 1 \over 3!} \sqrt{-{g}} \bar{\psi} \Big(
e^{\fft12 \vec a_i\cdot\vec\phi } 3 ( \a \g^{\sst{(\m)(\n)}} \g^{i}+ \b \g^{\sst{(\m)(\n)}
i} \g ) F_\2{}_{i \ \sst{(\m)(\n)}} \nn \\
&+& e^{\fft12 \vec a_{ij}\cdot\vec\phi } 3!( \a
\g^{\sst{(\m)}} \g^{j} \g^{i} + \b \g^{\sst{(\m)}} \g^{j}\g^{i} \g ) F_{\sst{(1)}ji \ {\sst{(\m)}}} \Big)
\psi \, .
\nn
\eeq
By using the
relation $\e_{\sst{(\rho)(\m)(\n)}} \g^{\sst{(\m)(\n)}}= 2 \g_{\sst{(\rho)}} \hg $
and dualising the 2 form field strengths, we get for the
dimensional reduction of the Lagrangian (\ref{dnlagf}), 
\bea {\cal L}_{\psi}^{(3)} = \sqrt{-g}
\bar{\psi} \g_c(\g^{\m}\partial_{\m} &-& {1 \over 4}
\o_{\m,\mm \nnn} \g^{\mm \nnn} \nn \\&-& {1 \over 2} e^{-\fft12 \vec
a_i\cdot\vec\phi } \Gamma_{\vec a_i} G^c_i - {1 \over 2}e^{\fft12
\vec a_{ij}\cdot\vec\phi } \Gamma_{\vec a_{ij}} F^c_{\sst{(1)} ij} \nn \\
&-& {1 \over 2}e^{-{1 \over 2}\vec b_i . \vec \phi} \Gamma_{b_i}
\cG^c_i - {1 \over 2}e^{{1 \over 2}\vec b_{ij} . \vec \phi }
\Gamma_{b_{ij}} \cF^c_{\sst{(1)} ij}) \psi \label{lagdn3}
\eea 
where
\bea \label{rep-dn}&& \Gamma_{\vec a_i} = 2(\a \hg \g^i + \b \hg
\g^i \g) , \; \; \; \; \;  \Gamma_{\vec a_{ij}} = 2(\a \g^i \g^j +
\b \g^i \g^j \g ),\nn \\ && \Gamma_{\vec b_i} = \fft12 \hg \g^i ,
\; \; \; \; \;  \; \; \; \; \Gamma_{\vec b_{ij}} = \fft12 \g^i
\g^j\eea
We have to compare this expression with 
\be 
e
\bar{\psi} \g^{\m}(\partial_{\m} - {1 \over 4}
{\o}_{\m,\rrr \sss} \g^{\rrr \sss} - \cQ ^{ ( \a)}_\mu J^{( \a)})\psi 
\nn
\ee
where $\cQ^{(\a)}_{ \mu}$ are the coefficients of the $K(SO(n,n)) =
SO(n) \times SO(n)$ gauge field. From (\ref{g-dn}), we find that
\beq
\cQ& = & \frac{1}{2} \sum_{i<j} e^{ \frac{1}{2}
\vec{b}_{ij}.\vec{\phi}} \cF_\1{}^i{}_j (e_{b_{ij}} + f_{b_{ij}})
+ \frac{1}{2} \sum_i e^{-\frac{1}{2} \vec{b}_i.\vec{\phi}}
\cG_\1{}^i (e_{b_i}+f_{b_i})
\nn \\
& &+ \frac{1}{2} \sum_{i<j} e^{\frac{1}{2} \vec{a}_{ij}.\vec{\phi}}
F_{\1 ijk} (e_{a_{ij}} + f_{a_{ij}}) + \frac{1}{2} \sum_{i}
e^{-\frac{1}{2} \vec{a}_{i}.\vec{\phi}} G_\1{}^{i} (e_{a_i} +
f_{a_i}) \rlap{\ . }
\nn
\eeq 
To check the correspondence, we need the commutation relations of the maximally 
compact subalgebra of $D_n$. Remember that  the Cartan involution $\tau$ is such that $\tau(h_i)=
-h_i$, $\tau(e_{\a})= f_{\a}$ and $\tau(f_{\a})=e_{\a}$ so that a
basis of the maximally compact subalgebra of $D_n$ reads $k_{\a} =
e_{\a} + f_{\a}$ where $ \a = \{a_{ij},a_i,b_{ij},b_i \}$ and
$i<j= 1,..., n-1$. The commutation relations of the $k_\a$'s are
\bea \ [k_{b_{ij}},k_{b_{mn}}] &=& \d_{jm} k_{b_{in}} -
\d_{in}k_{b_{mj}} +
\d_{im}(k_{b_{nj}}-k_{b_{jn}}) + \d_{jn}(k_{b_{mi}} - k_{b_{im}}) \nn \\
\ [k_{b_{ij}},k_{b_m} ] &=& - \d_{im} k_{b_j} + \d_{jm}k_{b_{i}} \nn \\
\ [k_{b_{ij}},k_{a_{mn}}] &=& -\d_{im} k_{a_{jn}} -
\d_{in}k_{a_{mj}} +
\d_{im} k_{a_{nj}} + \d_{jm}k_{b_{in}} + \d_{jn}(k_{b_{mi}}-k_{b_{im}}) \nn\\
\ [k_{b_{ij}},k_{a_m}] &=& \d_{jm} k_{a_i} - \d_{im} k_{a_{j}} \nn \\
\ [k_{b_{ij}},k_{a_m} ]& = &- \d_{im} k_{a_j} + \d_{jm} k_{a_i}\nn \\
\ [k_{b_i},k_{b_j}] &=& -k_{b_{ij}}+k_{b_{ji}} \nn \\
\ [k_{b_{i}},k_{a_{mn}} ]& = &- \d_{in} k_{a_m} + \d_{im} k_{a_n}
\nn \\
\ [k_{b_i},k_{a_j}] &=& -k_{a_{ij}}+k_{a_{ji}} \nn \\
\ [k_{a_{ij}},k_{a_{mn}}] &=& \d_{jm} k_{b_{in}} -
\d_{in}k_{b_{mj}} +
\d_{im}(k_{b_{nj}}-k_{b_{jn}}) + \d_{jn}(k_{b_{mi}} - k_{b_{im}}) \nn \\
\ [k_{a_{ij}},k_{a_m} ]& = &- \d_{im} k_{b_j} + \d_{jm} k_{b_i}
\label{compdn}
\eea Notice that by going to the new basis $\{k_b + k_a, k_b - k_a \}$, one
easily recognizes the algebra $\mf{so}(n) \oplus \mf{so}(n)$.

 One has to fix the values of
$\a$ and $\b$ in Eq. (\ref{dnlagf0}) such that the generators $\Gamma_{\vec a_i}$,
$\Gamma_{\vec b_i}, \ \Gamma_{\vec a_{ij}}$ and $\Gamma_{\vec
b_{ij}} $ obey the commutation relations (\ref{compdn}). The conditions we
found are  
\be 
\label{cond-dn} \a^2 + \b^2 \g^2 = {1 \over 16} \,
, \; \; \; \; \; \;  \;  \a \b = 0 \, . 
\ee
In odd dimension, we have set $\b=0$. This implies $\a = \pm
\frac{1}{4}$. We get for each choice of $\a$ a representation
which is trivial for either the left or the right $\mf{so}(n)$
factor of the compact gauge group. With $\b=0$, (\ref{rep-dn})
generates indeed $\mf{so}(n)$, as our analysis of the
gravitational sector has already indicated.\newline
In even dimension, the choices $ \b = 0, \a = \pm \frac{1}{4}$ are
still solutions to (\ref{cond-dn}), but in addition one can have
$\a =0$, $\b = \pm \frac{\iota}{4} $, where the constant $\iota$
is $1$ or $i$ such that $(\iota \g)^2 = 1$. In this case,
(\ref{rep-dn}) combines with the gravitational $\mf{so}(n)$ to
give a $\mf{so}(n)\times\mf{so}(n)$ representation which
is nontrivial on both factors. The two factors
$\mf{so}(n)_\pm$ are generated in the spinorial space by the
matrices 
\be
\begin{array}{c}
\frac{1}{4} (1 \pm \iota \gamma) \gamma^{ij} \\
\frac{1}{4} (1 \pm \iota \gamma) \hgamma \gamma^i
\end{array}
\nn
\ee 
The (reduced) gravitational sector is given by the diagonal
$\mf{so}(n)$. If one imposes a chirality condition in $\Dm$
dimensions, the solution with $\b = 0$ and the solution with $\a =
0$ are of course equivalent and the representation is
trivial on one of the $\mf{so}(n)$.

This completes the proof that the Dirac spinors are compatible
with the $D_n$ hidden symmetry.

\section{$E_n$ sequence}
\label{e8case}

\subsection{$E_8$ -- bosonic}

We consider now the bosonic part of 11---dimensional supergravity,
\ie gravity coupled to a 3-form in 11 dimensions with the specific
value of the Chern--Simons term dictated by supersymmetry.  We
denote the 3--form $A_\3$ and its field strength $F_\4 = \ud A_\3$.
The Lagrangian is \cite{Cremmer:1978km} 
\be 
\cL = R\, {*\oneone} -
\frac{1}{2}  *\! F_\4 \w F_\4 - \frac{1}{3!} F_\4 \w F_\4 \w A_\3
\rlap{\ .} 
\label{e8-bos} 
\ee 
Prior to dualisation, the 3--form
term of the Lagrangian reduces in three dimensions to 
\bea
{\cL}^{(3)} &=& -\frac{1}{2} \sum_{i<j<k}
e^{\vec{a}_{ijk}.\vec{\phi}}  * F_\1{}_{ijk} \w F_\1{}_{ijk} - \frac{1}{2} 
\sum_{i<j} e^{\vec{a}_{ij}.\vec{\phi}} *  F_\2{}_{ij}\w F_\2{}_{ij} \nn
\\ && \hspace{2cm} - \frac{1}{144} \ud A_\0{}_{ijk} \w \ud A_\0{}_{lmn} \w
A_\1{}_{pq} \epsilon^{ijklmnpq} 
\label{sugra-3} 
\eea 
In addition
to the gravitational degrees of freedom described in section
\ref{gravity}, we have 56 scalars $A_\0{}_{ijk}$ and 28 1--forms
$A_\1{}_{ij} = A_{\mu (i+2)(j+2)} \ud x^\mu$, with $i,j,k = 1..8$.
The reduced field strength are defined as \bea
F_\1{}_{ijk} &=& \gamma^l{}_i \gamma^m{}_j \gamma^n{}_k \, \ud A_\0{}_{lmn} \\
F_\2{}_{ij} &=& \gamma^k{}_i \gamma^l{}_j ( \ud A_\1{}_{kl} -
\gamma^m{}_n \, \ud A_\0{}_{klm} \w \cA_\1{}^n ) \label{F2}
\rlap{\ .} \eea The 1--forms $A_\1{}_{ij}$ are then dualised into
scalars $\lambda^{kl}$: \be e^{\vec{a_{ij}}.\vec{\phi}} * \!
F_\2{}_{ij} = G_\1{}^{ij} = (\gamma^{-1})^i{}_k
(\gamma^{-1})^j{}_l ( \ud \lambda^{kl} +\frac{1}{72} \ud A_{\0
mnp} A_{\0 qrs} \epsilon^{klmnpqrs} ) \rlap{\ .} \label{dual-f2}
\ee Moreover, the gravitational duality relation (\ref{dual-a1})
has to be modified to take into account the 3--form degrees of
freedom 
\be 
e^{\vec{b}_i.\vec{\phi}} * \! \cF_\2{}^i = \cG_\1{}^i
= \gamma^j{}_i \left( \ud \chi_j - \frac{1}{2} A_{\0 jkl} \ud
\lambda^{kl} - \frac{1}{432} \ud A_{\0 klm} A_{\0 npq} A_{\0 rsj}
\epsilon^{klmnpqrs} \right) \rlap{\ .} 
\nn
\ee
Taking all this into account, the full 3--dimensional Lagrangian
can be written as 
\beq
{\cL} &=& R\, {*\oneone} - \frac{1}{2} \!*\! \ud \vec{\phi} \w
 \ud \vec{\phi} - \frac{1}{2}  \sum_{i<j}
e^{\vec{b}_{ij}.\vec{\phi}} \!*\!  \cF_\1{}^i{}_j \w  \cF_\1{}^i{}_j
-\frac{1}{2}  \sum_i e^{-\vec{b}_i.\vec{\phi}} \!*\!  \cG_\1{}^i \w
 \cG_\1{}^i
\nn
\\
& &-\frac{1}{2}  \sum_{i<j<k} e^{\vec{a}_{ijk}.\vec{\phi}} \!*\!  F_{\1
ijk} \w F_{\1 ijk} -\frac{1}{2}  \sum_{i<j}
e^{-\vec{a}_{ij}.\vec{\phi}} \!*\! G_\1{}^{ij} \w  G_\1{}^{ij}
\nn
\eeq 
which describes a $E_{8}/SO(16)$ $\sigma$--model coupled to
gravity \cite{Cremmer:1978ds,Cremmer:1979up,Marcus:1983hb,Cremmer:1997ct}, in the Borel
gauge, with field strength 
\beq
\cG &=& \frac{1}{2} \ud \vec{\phi} . \vec{H} + \sum_{i<j} e^{
\frac{1}{2} \vec{b}_{ij}.\vec{\phi}} \cF_\1{}^i{}_j e_{ij} + \sum_i
e^{-\frac{1}{2} \vec{b}_i.\vec{\phi}} \cG_\1{}^i e_i
\nn \\
& &+ \sum_{i<j<k} e^{\frac{1}{2} \vec{a}_{ijk}.\vec{\phi}} F_{\1 ijk}
\tilde{e}_{ijk} + \sum_{i<j} e^{-\frac{1}{2}
\vec{a}_{ij}.\vec{\phi}} G_\1{}^{ij} \tilde{e}_{ij} \rlap{\ .}
\label{g-e8} 
\eeq
The explicit expressions for the couplings $\vec{a}_{ijk}$ and
$\vec{a}_{ij}$ are (see \cite{Cremmer:1997ct}) 
\be 
\vec{a}_{ijk} =
-2(\vec{\gamma}_i + \vec{\gamma}_j + \vec{\gamma}_k), \; \; \;
\vec{a}_{ij} = -2 (\vec{\gamma}_i + \vec{\gamma}_j) -\vec{s}. 
\nn
\ee
The positive roots are $\vec{b}_{ij}$, $- \vec{b}_{i}$,
$\vec{a}_{ijk}$ and $-\vec{a}_{ij}$.  The elements $e_{ij}$
($i<j$), $e_i$, $\te_{ijk}$ and $\te_{ij}$ (with antisymmetry over
the indices for the two last cases) are the raising operators.
Note that $e_{ij}$ and $e_i$ generate the $\mf{sl}(9,\RR)$ subalgebra
coming from the gravitational sector. In addition, there are
lowering operators $f_{ij}$, $f_i$, $\tf_{ijk}$ and $\tf_{ij}$. 
We
give all the commutation relations in that basis of $E_8$ here under.
We take a basis of the Cartan subalgebra $(h_i)$, such that \be
\begin{array}{rcl}
\left[ f_{ij} , e_{ij} \right] &=& h_i - h_j \\
\left[ f_{i} , e_{i} \right] &=& -h_i \\
\left[ \tilde{f}_{ijk} , \te_{ijk} \right] &=& \frac{1}{3} (h_1 + h_2 +
\ldots +
h_8) -h_i - h_j -h_k \\
\left[ \tilde{f}_{ij} , \te_{ij} \right] &=& - \frac{1}{3} (h_1 + h_2 +
\ldots + h_8) + h_i + h_j
\end{array}
\ee \be
\begin{array}{rclrcl}
\left[ h_i , e_{ij} \right] &=& e_{ij} & \qquad \left[ h_i , f_{ij} \right] &=& -f_{ij} \\
\left[ h_j , e_{ij} \right] &=& -e_{ij} & \left[ h_j , f_{ij} \right] &=& f_{ij} \\
\left[ h_i , e_i \right] &=& - e_i & \left[ h_i , f_i \right] &=& f_i \\
\left[ h_i , \te_{ijk} \right] &=& -\te_{ijk} & \left[ h_i , \tf_{ijk} \right]
&=&
\tf_{ijk} \\
\left[ h_i , \te_{jk} \right] &=& -\te_{jk} & \left[ h_i , \tf_{jk} \right] &=&
\tf_{jk}
\end{array}
\ee where distinct indices are supposed to have different values.
The vectors associated with the simple roots are $e_{i\,i+1}$,
$\te_{123}$. Other non vanishing commutations relations are the following, with
the same convention on indices. \be
\begin{array}{rclrcl}
\left[ e_{ij} , e_{jk} \right] &=& e_{ik}
& \qquad \left[ f_{ij} , f_{jk} \right] &=& f_{ik} \\
\left[ \te_{ijk}, e_{kl} \right] &=& e_{ijl}
& \left[ \tf_{ijk}, f_{kl} \right] &=& f_{ijl} \\
\left[ \te_{ijk} , \te_{lmn} \right] &=& \frac{1}{2} \epsilon^{ijklmnpq}
\te_{pq} & \left[ \tf_{ijk} , \tf_{lmn} \right] &=& \frac{1}{2}
\epsilon^{ijklmnpq} \tf_{pq}
\\
\left[ e_{ij} , \te_{jk} \right] &=& \te_{ik}
& \left[ f_{ij} , \tf_{jk} \right] &=& \tf_{ik} \\
\left[ \te_{ijk} , \te_{jk} \right] &=& e_i
& \left[ \tf_{ijk} , \tf_{jk} \right] &=& f_i \\
\left[ e_i , e_{ij} \right] &=& e_j & \left[ f_i , f_{ij} \right] &=& f_j
\end{array}
\ee \be
\begin{array}{rcllrcll}
\left[ f_{ij} , e_{kj} \right] &=& e_{ki} & \textrm{ if } i>k & \qquad
\left[ f_{ij} , e_{kj} \right] &=& -f_{ik} & \textrm{ if } i<k \\
\left[ f_{ji} , e_{jk} \right] &=& -e_{ik} & \textrm{ if } i<k &
\left[ f_{ji} , e_{jk} \right] &=& f_{ki} & \textrm{ if } i>k \\
\left[ f_{ij} , \te_{klj} \right] &=& e_{kli} &&
\left[ e_{ij} , \tf_{klj} \right] &=& f_{kli} \\
\left[ \tf_{ijk} , \te_{ijl} \right] &=& -e_{kl} & \textrm{ if } k<l &
\left[ \tf_{ijk} , \te_{ijl} \right] &=& f_{lk} & \textrm{ if } k>l \\
\left[ f_{ji} , \te_{jk} \right] &=& -\te_{ik} &&
\left[ e_{ji} , \tf_{jk} \right] &=& -\tf_{ik} \\
\left[ \tf_{ijk} , \te_{lm} \right] &=& \multicolumn{2}{l}{-\frac{1}{3!}
\epsilon^{ijklmnpq} \te_{npq}} & \left[ \te_{ijk} , \tf_{lm} \right] &=&
\multicolumn{2}{l}{-\frac{1}{3!} \epsilon^{ijklmnpq} \tf_{npq}}
\\
\left[ f_{ij} , e_j \right] &=& e_i &&
\left[ e_{ij} , f_j \right] &=& f_i \\
\left[ \tf_{ij} , \te_{ik} \right] &=& e_{kj} & \textrm{ if } j>k &
\left[ \tf_{ij} , \te_{ik} \right] &=& -f_{jk} & \textrm{ if }\ j<k \\
\left[ \tf_{ijk} , e_k \right] &=& -\te_{ij} &&
\left[ \te_{ijk} , f_k \right] &=& -\tf_{ij} \\
\left[ \tf_{ij} , e_k \right] &=& \te_{ijk} &&
\left[ \te_{ij} , f_k \right] &=& \tf_{ijk} \\
\left[ f_{i} , e_j \right] &=& -e_{ij} & \textrm{ if } i<j & \left[ f_{i} ,
e_j \right] &=& f_{ji} & \textrm{ if } i>j
\end{array}
\ee
The Chevalley--Serre generators are $h_i-h_{i+1}$, $h_{123} \equiv
\frac{1}{3}(h_1 + \cdots + \h_8) - h_1 - h_2 - h_3$, $e_{i i+1}$,
$\te_{123}$,  $f_{i i+1}$ and $\tf_{123}$.  The scalar products of
the $h_i$'s are $K(h_i, h_j) = \d_{ij} + 1$ and the factors
$N_{\alpha}$ are equal to unity.

\subsection{$E_8$ -- fermions}
\label{e8}

The maximal compact subgroup of $E_8$ is $Spin(16)/\ZZ_2$. 
We want to add Dirac
fermions in $D=11$, with a coupling which reduces to a
$Spin(2,1) \times Spin(16)$--covariant derivative in
three dimensions. The coupling to gravitational degrees of freedom
is already fixed to the spin connection by invariance under
reparametrisations; we know from the first section that it reduces
to the relevant 
$Spin(2,1) \times Spin(9)$ connection
in $D=3$.

{}From the structure of the reduced theory, we know that the
fermion must have a linear coupling to the 4--form $F_\4$. The only
Lorentz--covariant coupling of this kind for a single Dirac fermion
in $D=11$ is a Pauli coupling 
\be 
\hat e a \frac{1}{4!} \bhpsi F_{\hat \mu
\hat \nu \hat \rho \hat \sigma} \gamma^{\hat \mu \hat  \nu \hat \rho \hat \sigma} \hpsi \ee where $a$
is a constant. Indeed in odd dimensions, the product of all
$\gamma$ matrices is proportional to the identity, so the dual
coupling is not different: 
\be 
\frac{1}{7!} (*F)_{\hat \mu_1 \ldots
\hat \mu_7} \gamma^{\hat \mu_1 \ldots \hat \mu_7} = \frac{1}{4!} F_{\hat \mu_1 \ldots
\hat \mu_4} \gamma^{\hat \mu_1 \ldots \hat \mu_4} \rlap{\ .} 
\nn
\ee 
Thus we add to
the bosonic Lagrangian (\ref{e8-bos}) the fermionic term 
\be
\hat{\cL}_{\hpsi} = \hat e \bhpsi ( \gamma^{\hat{\mu}} \partial_{\hat{\mu}}  - \frac{1}{4}
\hat \omega_{\hat \mu}{}^{\sst{(\hat \rho)(\hat \s)}} \gamma^{\hat \mu} \gamma^{\sst{(\hat \rho)(\hat \s)}} - \frac{1}{4!} a
F_{\hat \mu \hat \nu \hat \rho \hat \sigma} \gamma^{\hat \mu \hat \nu \hat \rho \hat \sigma} )  \hpsi 
\nn
\ee 
where
$\gamma$ matrices with indices $\hat \m$ must be understood as
$\gamma^{\hat \mu} = \hat e_{\sst{(\hat \n)}}{}^{\hat \mu} \gamma^{\sst{(\hat \n)}}$.

\noindent Dimensional reduction to $D=3$ leads to 
\be 
\cL_\psi^\3 = \cL_D^\3
- e a \frac{1}{3!} e^{\frac{1}{2} \vec{a}_{ijk}.\vec{\phi}}
F_{\1 \mu ijk} \bpsi \gamma^\mu \gamma^{ijk} \psi - e a
\frac{1}{2.2} e^{\frac{1}{2} \vec{a}_{ij}.\vec{\phi}} F_{\2 \mu\nu
ij} \bpsi \gamma^{\mu\nu} \gamma^{ij} \psi 
\nn
\ee 
where $\cL_D^\3$
is part not containing the 3--form computed previously in
(\ref{cld}), and with the same rescaling of $\hpsi$ into $\psi$.
Dualisation (\ref{dual-f2}) of $F_{\2 ij}$ can be written as 
\be
\frac{1}{2} e^{\frac{1}{2} \vec{a}_{ij}.\vec{\phi}} F_{\2 \mu\nu
ij} \gamma^{\mu\nu} = e^{-\frac{1}{2} \vec{a}_{ij}.\vec{\phi}}
G_{\1 \mu ij} \gamma^\mu \hgamma \rlap{\ .} 
\nn
\ee 
It gives the fully
dualised fermionic term 
\be 
\cL_\psi^\3 = \cL_D^\3 - e a
\frac{1}{3!} e^{\frac{1}{2} \vec{a}_{ijk}.\vec{\phi}} F_{\1 \mu
ijk} \bpsi \gamma^\mu \gamma^{ijk} \psi - e a \frac{1}{2}
e^{-\frac{1}{2} \vec{a}_{ij}.\vec{\phi}} G_{\1 \mu ij} \bpsi
\gamma^\mu \hgamma \gamma^{ij} \psi \rlap{\ .} 
\label{l3-e8} 
\ee

We have to compare this expression to 
\be 
e \bpsi \gamma^\mu
\left( \partial_\mu - \frac{1}{4} {\omega} _{\mu,\mm \nnn} \gamma^{\mm \nnn} -
\cQ_{\mu}^{ (\alpha)} J^{(\alpha)} \right) \psi . 
\nn
\ee 
{}From
(\ref{g-e8}), we have 
\beq
\cQ &= & \frac{1}{2} \sum_{i<j} e^{ \frac{1}{2}
\vec{b}_{ij}.\vec{\phi}} \cF_\1{}^i{}_j (e_{ij} + f_{ij}) +
\frac{1}{2} \sum_i e^{-\frac{1}{2} \vec{b}_i.\vec{\phi}}
\cG_\1{}^i (e_i+f_i)
\nn
\\
& &+ \frac{1}{2} \sum_{i<j<k} e^{\frac{1}{2}
\vec{a}_{ijk}.\vec{\phi}} F_{\1 ijk} (\te_{ijk} + \tf_{ijk}) +
\frac{1}{2} \sum_{i<j} e^{-\frac{1}{2} \vec{a}_{ij}.\vec{\phi}}
G_\1{}^{ij} (\te_{ij} + \tf_{ij}) \rlap{\ .}
\nn
\eeq 
In fact, we have precisely the correct gauge connection that
appears in (\ref{l3-e8}). We have only to check that the products
of gamma matrices that multiply the connection in (\ref{l3-e8})
satisfy the correct commutation relations. The generators of the compact subalgebra $\mf{so}(16)$ are 
\be
\begin{array}{rcll}
k_{ij} &=& e_{ij} + f_{ij} & \mathrm{for} \ i<j \nn\\
k_i &=& e_i+f_i \nn\\
\tk_{ijk} &=& \te_{ijk} + \tf_{ijk} \nn\\
\tk_{ij} &=& \te_{ij} + \tf_{ij} \rlap{\ .}
\end{array}
\ee It is convenient to define $k_{ij} = - k_{ji} = -e_{ji} -
f_{ji}$ for $i>j$. Their non vanishing commutators are \be
\begin{array}{rclrcl}
\left[ k_{ij} , k_{jk}\right] &=& k_{ik} &
\left[ \tk_{ijk} , k_{kl}\right] &=& \tk_{ijl} \\
\left[ \tk_{ijk} , \tk_{lmn}\right] &=& \frac{1}{2} \epsilon^{ijklmnpq}
\tk_{pq} &
\left[ \tk_{ijk} , \tk_{ijl}\right] &=& - k_{kl} \\
\left[ k_{ij} , \tk_{jk}\right] &=& \tk_{ik} &
\left[ \tk_{ijk} , \tk_{jk}\right] &=& k_{l} \\
\left[ \tk_{ijk} , \tk_{lm}\right] &=& - \frac{1}{3!} \epsilon^{ijklmnpq}
\tk_{npq} &
\left[ \tk_{ij} , \tk_{ik}\right] &=& - k_{jk} \\
\left[ k_{ij} , k_j \right] &=& k_i &
\left[ \tk_{ijk} , k_k \right] &=& -\tk_{ij} \\
\left[ \tk_{ij} , k_k \right] &=& \tk_{ijk} & \left[ k_{i} , k_j \right] &=&
-k_{ij}
\end{array}
\label{so16} \ee where it is assumed that distinct indices have
different values.

We
find that the coupling constant must be $a=-\frac{1}{2}$. The
spinorial generators are then given by 
\be
\begin{array}{ll}
k_{ij}: & \frac{1}{2} \gamma^{ij} \\
k_i: & \frac{1}{2} \hgamma \gamma^i \\
\tk_{ijk}: & -\frac{1}{2} \gamma^{ijk} \\
\tk_{ij}: & -\frac{1}{2} \hgamma \gamma^{ij}
\end{array}
\nn
\ee 
(we define $k_{ij} = - k_{ji} = -e_{ji} - f_{ji} $ for  $
i>j$). We have  recovered the well known feature that the spinorial
representation of $\mf{so}(9)$ is the vector representation of
$\mf{so}(16)$ (see \cite{Keurentjes:2003yu} for more on this).

We see also that $E_{8}$--invariance forces one to
introduce the covariant Dirac operator
\be 
\gamma^{\hat{\m}} \hat D_{\hat{\m}} \hpsi =
\gamma^{\hat{\m}} (\partial_{\hat{\m}} - \frac{1}{4} \omega_{\hat{\m}}{}^{\nnn \rrr} \gamma_{\nnn \rrr} )
\hpsi + \frac{1}{2 . 4!} F_{\hat{\m} \hat \nu \hat \rho \hat \sigma} \gamma^{\hat{\m}
\hat \nu \hat \rho \hat \sigma} \hpsi \nn
\ee
for the Dirac field. This is exactly the
\emph{same} which appears in $D=11$ supergravity for the supersymmetry 
transformation parameter $\epsilon$ of reference \cite{Cremmer:1978km} but it is
obtained in that context from supersymmetry.

\subsection{IIB}

The oxidation of the $E_{8}/SO(16)$ coset
theory has another endpoint, in $D=10$: the bosonic sector of type
\emph{IIB} supergravity. There is no manifestly covariant
Lagrangian attached to this theory.  Indeed, the theory contains a
selfdual 4--form, which has no simple (quadratic) manifestly
covariant Lagrangian (although it does admit a quadratic non
manifestly covariant Lagrangian \cite{Henneaux:1988gg}, or a non polynomial
manifestly covariant Lagrangian \cite{Pasti:1996vs}). In spite of
the absence of a  covariant Lagrangian, the equations of motion
are covariant and one may address the following question: is there
a ``covariant Dirac operator'' for fermions in $D=10$ which reduces to
the same $SO(16)$ covariant derivative in $D=3$?

Following the notations of \cite{Cremmer:1998px}, we have for this
theory,  in addition to the metric, a dilaton $\phi$, an other
scalar $\chi$, two 2--forms $A_\2^1$ and $A_\2^2$ with field strength
$F_\3^1$ and $F_\3^2$, and a 4--form $B_\4$ with selfdual field
strength $H_\5$.
If it exists, the $D=10$ ``covariant Dirac operator'' would have the
form 
\beq
\gamma^{\hat{\m}} \nabla_{\hat{\m}} &=&  \gamma^{\hat{\m}} \partial_{\hat{\m}} - \frac{1}{4} \omega_{\hat{\m}}{}^{\sst{(\hat \d)(\hat \lambda)}}
\gamma^{\hat{\m}} \gamma^{\sst{(\hat \d)(\hat \lambda)}} -
e^\phi \partial_{\hat{\m}} \chi (a + \tilde{a} \gamma) \gamma^{\hat{\m}} - \frac{1}{3!}
e^{\frac{1}{2}\phi} F^1_{{\hat{\m}} {\hat{\n}}{\hat{\rho}}}
(b + \tilde{b} \gamma) \gamma^{{\hat{\m}}{\hat{\n}}{\hat{\rho}}} \nn \\
& &- \frac{1}{3!} e^{-\frac{1}{2}\phi} F^2_{{\hat{\m}} {\hat{\n}}{\hat{\rho}}} (c + \tilde{c}
\gamma) \gamma^{{\hat{\m}}{\hat{\n}}{\hat{\rho}}} - \frac{1}{5!} H_{{\hat{\m}}{\hat{\n}}{\hat{\rho}}{\hat{\s}}{\hat{\tau}}} f
\gamma^{{\hat{\m}}{\hat{\n}}{\hat{\rho}}{\hat{\s}}{\hat{\tau}}} \rlap{\ .} 
\label{cd-iib}
\eeq 
$\gamma = \gamma^{11}$ is the product of the ten $\gamma^i$
matrices. As $H_\5$ is selfdual, the dual term 
\be
H_{{\hat{\m}}{\hat{\n}}{\hat{\rho}}{\hat{\s}}{\hat{\tau}}} \gamma\gamma^{{\hat{\m}}{\hat{\n}}{\hat{\rho}}{\hat{\s}}{\hat{\tau}}} =
(*H)_{{\hat{\m}}{\hat{\n}}{\hat{\rho}}{\hat{\s}}{\hat{\tau}}} \gamma^{{\hat{\m}}{\hat{\n}}{\hat{\rho}}{\hat{\s}}{\hat{\tau}}}  \nn
\ee 
is
already taken into account. The powers of the dilaton are fixed so
that the field strength give the expected fields in $D=3$.

Now, the axion term $e^\phi \partial_{\hat{\m}} \chi$ is the connection
for the $SO(2)$--subgroup of the $SL(2)$ symmetry
present in 10 dimensions.  Under $SO(2)$--duality,  the
two two--forms rotate into each other. So, the commutator of the
generator $(a + \tilde{a} \gamma)$ multiplying the connection
$e^\phi \partial_{\hat{\m}} \chi$ with the generators $(b + \tilde{b}
\gamma) \gamma^{{\hat{\n}}{\hat{\rho}}}$ multiplying the connection
$e^{\frac{1}{2}\phi} F^1_{{\hat{\m}} {\hat{\n}}{\hat{\rho}}}$ should reproduce the
generator $(c + \tilde{c} \gamma) \gamma^{{\hat{\n}}{\hat{\rho}}}$ multiplying
the connection $e^{\frac{1}{2}\phi} F^2_{{\hat{\m}} {\hat{\n}}{\hat{\rho}}}$. But one
has $[(a + \tilde{a} \gamma), (b + \tilde{b} \gamma)
\gamma^{{\hat{\n}}{\hat{\rho}}}] = 0$, leading to a contradiction.

 The problem just described comes from the fact that we have taken a
single Dirac fermion.  Had we taken instead two Weyl fermions, as it
is actually the case for type IIB supergravity, and assumed that
they transformed appropriately into each other under the
$SO(2)$--subgroup of the $SL(2)$ symmetry, we could have constructed
an appropriate covariant derivative.  This covariant derivative is
in fact given in \cite{Schwarz:1983wa,Howe:1983sr,Mizoguchi:1998wv,Mizoguchi:1999fu}, to which we refer the reader.
The $SO(2)$ transformations rules of the spinors ---~as well as the
fact that they must have same chirality in order to transform indeed
non trivially into each other ---~follow from $E_8$--covariance in 3
dimensions.

\subsection{$E_7$ case}

The $E_{7}$ exceptional group is a subgroup of $E_{8}$. As a
consequence, the $D=3$ coset $E_{7}/SU(8)$ can be seen as a
truncation of the $E_{8}/SO(16)$ coset theory. In fact, this
truncation can be made in higher dimension \cite{Cremmer:1999du}.
One can truncate the $D=9$ reduction of the gravity + 3--form
theory considered in the last section. If one does not worry about
Lagrangian, one can go one dimension higher and view the theory as
the truncation of the bosonic sector of type \emph{IIB}
supergravity in which one keeps only the vielbein and the chiral
4--form.

In $D=9$, the coupling to fermions obtained in section \ref{e8} is
truncated in a natural way: the components of the covariant
Dirac operator acting on fermions are the various fields of the
theory, so some of them just disappear with the truncation. The
symmetry of the reduced $D=3$ theory is thus preserved: the
fermions are coupled to the bosonic fields through a
$SU(8)$ covariant derivative, the truncation of the
$SO(16)$ covariant derivative of the $E_8$ case.

The question is about oxidation to $D=10$. Can we obtain this
truncated covariant Dirac operator from a covariant Dirac operator of the
$D=10$ theory? For the reasons already exposed, if it exists, this operator
would act on Dirac fermions as 
\be \gamma^{ \hat \mu}
\nabla_{ \hat \mu} =\gamma^{ \hat \mu} \partial_{ \hat \mu} - \frac{1}{4} \hat \omega_{ \hat \mu}{}^{\sst{( \hat \d)( \hat \lambda)}}
\gamma^{ \hat \mu} \gamma^{\sst{( \hat \d)( \hat  \lambda)}} - a \frac{1}{5!} H_{\hat \mu \hat \nu \hat \rho \hat \sigma \hat \tau}
\gamma^{\hat \mu \hat \nu \hat \rho \hat \sigma \hat \tau} \nn
\ee 
where we have denoted by $H$ the
selfdual field strength.
With notations analogous to the $E_8$ case, we can write
the $D=3$ reduction of the covariant Dirac operator as 
\beq
\gamma^\mu \nabla_\mu &=& \gamma^\mu \partial_\mu - \frac{1}{4}
{\omega}_\mu{}^{\sst{(\d)( \lambda)}} \gamma^\mu \gamma^{\sst{(\d)( \lambda)}} \nn \\
& &- \frac{1}{4}
e^{\frac{1}{2}\vec{b}_{i}.\vec{\phi}} \cF_{\2 \mu\nu i}
\gamma^{\mu\nu} \gamma^i - \frac{1}{4}
e^{\frac{1}{2}\vec{b}_{ij}.\vec{\phi}}
\cF_{\1 \mu ij} \gamma^\mu \gamma^{ij} \nn \\
& &-a  \frac{1}{2.3!} e^{\frac{1}{2}\vec{a}_{ijk}.\vec{\phi}} H_{\2
\mu\nu ijk} \gamma^{\mu\nu} \gamma^{ijk} - a \frac{1}{4!}
e^{\frac{1}{2}\vec{a}_{ijkl}.\vec{\phi}} H_{\1 \mu ijkl}
\gamma^{\mu} \gamma^{ijkl} \rlap{\ .}
\nn
\eeq 
Because of the selfduality of $H$, the 2-forms $H_\2$ and the
1-forms $H_\1$ are in fact dual. Using $\gamma = \gamma^0 \gamma^1
\ldots \gamma^9$, the covariant Dirac operator turns into 
\beq
\gamma^\mu \nabla_\mu &=& \gamma^\mu \partial_\mu + \frac{1}{4}
{\omega}_\mu{}^{ab} \gamma^\mu \gamma^{ab} \nn \\
& &+ \frac{1}{2}
e^{-\frac{1}{2}\vec{b}_{i}.\vec{\phi}} \cG_{\1 \mu i} \gamma^{\mu}
\hgamma \gamma^i + \frac{1}{2.2}
e^{\frac{1}{2}\vec{b}_{ij}.\vec{\phi}}
\cF_{\1 \mu ij} \gamma^\mu \gamma^{ij} \nn \\
& & + a \frac{1}{4!} e^{\frac{1}{2}\vec{a}_{ijkl}.\vec{\phi}} H_{\1
\mu ijkl} \gamma^{\mu} (1+\gamma) \gamma^{ijkl} \rlap{\ .}
\label{cd-su8} 
\eeq
The embedding of $E_{7}/SU(8)$ in
$E_{8}/SO(16)$ gives the following
identifications: \bea
H_{\1 1ijk} &=& F_{\1 (i+1)(j+1)(k+1)} \nn\\
H_{\1 ijkl} &=& -\frac{1}{2} \epsilon^{12(i+1)(j+1)(k+1)(l+1)mn}
G_{\1 mn}
\nn\\
\cF_{\1 1i} &=& F_{\1 12(i+1)} \nn\\
\cG_{\1 1} &=& - G_{\2 12} \nn\\
\cF_{\1 ij} &=& \cF_{\1 (i+1)(j+1)} \nn\\
\cG_{\1 i} &=& \cG_{\1 (i+1)} \nn
\eea 
with $2 \leq i,j,k,l \leq 7$.
We thus have to check that the matrices in (\ref{cd-su8}) form the
following representation: \be
\begin{array}{rcl}
a (1+\gamma) \gamma^{1ijk} &\sim& \tk_{(i+1)(j+1)(k+1)} \\
a (1+\gamma) \gamma^{ijkl} &\sim&\frac{1}{2}
\epsilon^{12(i+1)(j+1)(k+1)(l+1)mn} \tk_{mn}\\
\frac{1}{2}\gamma^{1i} &\sim& \tk_{12(i+1)} \\
\frac{1}{2}\hgamma\gamma^1 &\sim& -\tk_{12} \\
\frac{1}{2}\gamma^{ij} &\sim& k_{(i+1)(j+1)} \\
\frac{1}{2}\hgamma\gamma^i &\sim& k_{(i+1)} \rlap{\ .}
\end{array}
\label{alg-su8} \ee 
This is true if and only if 
\be -4 a^2
(1+\gamma) = \frac{1}{2} \rlap{\ .} \nn
\ee 
This has to be understood
as an identity between operators acting on fermions. In fact, this
means that we must restrict to Weyl spinors, with $\gamma= +1$
when acting on them. Due to the even number of $\gamma$ matrices
involved in all generators in (\ref{alg-su8}), the
$\mf{su}(8)$ algebra preserves the chirality of spinors. We
get in addition the value of the coupling constant: 
\be a = \pm
\frac{i}{4} \rlap{\ .} \ee

\subsection{$E_6$  case}

The $E_6$ case is more simple. One has a Lagrangian in all
dimensions. In dimension 3, the scalar coset is $E_6/Sp(4)$.
Maximal oxidation is a $D=8$ theory with a 3--form, a
dilaton and an axion (scalar) \cite{Cremmer:1999du}. It can be
seen as a truncation of the $E_8$ case in all dimensions. In the
compact subalgebra of $\mf{so}(16)$ given in (\ref{so16}), one
should remove generators with one or two indices in $ \{1,2,3\} $
while keeping $\tk_{123}$.

In fact, all the matrices involved in the Dirac representation can
be expressed in terms of a $D=8$ Clifford algebra. For most
generators, it is trivial to check that they involve only
$\gamma^i$ matrices with $i\neq 1,2,3$. The single nontrivial case
is $\tk_{123}$ which is represented by $-\frac{1}{2} \gamma^{123}$
in the eleven--dimensional Clifford algebra. But from the fact that
$\gamma^{(10)} = \gamma^0 \gamma^1 \ldots \gamma^9$, we can write
$\gamma^{123}$ as the product of all other $\gamma$ matrices:
$\gamma^{123} = \gamma^{0456789(10)}$. As a consequence, the $D=8$
Clifford algebra is sufficient to couple a Dirac fermion to this
model: we can couple a single $D=8$ Dirac fermion.

\section{$G_2$ case}
\label{g2case}

\subsection{Bosonic sector}

Let us consider the Einstein--Maxwell system in $D=5$, with the
$FFA$ Chern--Simons term prescribed by supersymmetry \cite{Chamseddine:1980sp},
%%%%%
\be {\cal L}_5 = R\, {*\oneone} - \fft12 {*F_\2}\wedge F_\2
+\fft1{3\sqrt3} F_\2\wedge F_\2\wedge A_\1\ .\label{g2lag} \ee
%%%%%
This action is known to be relevant to $G_2$
\cite{Mizoguchi:1998wv,Mizoguchi:1999fu,Cremmer:1999du}. Upon reduction to $D=3$, the
Lagrangian is \cite{Cremmer:1999du}
%%%%%
\bea {\cal L} &=& R\, {*\oneone} -\fft12 {*d\vec\phi}\wedge
d\vec\phi - \fft12 e^{\phi_2 -\sqrt3\phi_1}\, {*\cF^1_{\1 2}}\wedge
\cF^1_{\1 2}
-\fft12 e^{\fft2{\sqrt3}\phi_1}\, {*F_{\1 1}}\wedge F_{\1 1} \nn\\
&&-\fft12 e^{\phi_2 -\fft1{\sqrt3}\phi_1}\, {*F_{\1 2}}\wedge F_{\1
2} - \fft12 e^{-\phi_2 -\sqrt3\phi_1}\, {*\cF_\2^1}\wedge \cF_\2^1
\label{d5einstmax}\\
&&- \fft12 e^{-2\phi_2}\, {*\cF_\2^2}\wedge \cF_\2^2 -\fft12
e^{-\phi_2 -\fft1{\sqrt3}\phi_1}\, {*F_\2}\wedge F_\2 +
\ft{2}{\sqrt3} \ud A_{\0 1}\wedge dA_{\0 2}\wedge A_\1\ .\nn \eea
%%%%%
After dualising the vector potentials to give axions, there will
be six axions, together with the two dilatons. The dilaton vectors
$\vec\a_1=(-\sqrt3,1)$ and $\vec\a_2=(\ft2{\sqrt3}, 0)$,
corresponding to the axions $\cA^1_{\0 2}$ and $A_{\0 1}$, are the
simple roots of $G_2$, with the remaining dilaton vectors
expressed in terms of these as
%%%%%
\be (-\fft1{\sqrt3}, 1)=\vec\a_1+\vec \a_2\ ,\quad
(\fft1{\sqrt3},1)= \vec\a_1 + 2\vec\a_2\ ,\quad (\sqrt3,1)=
\vec\a_1+3\vec\a_2\ ,\quad (0,2)=2\vec\a_1+3\vec a_3\ . \ee
%%%%%
The resulting $D=3$ lagrangian is a $G_2 / SO(4)$ $\sigma$--model
coupled to gravity. The field strength of this $\sigma$--model is
\beq
\cG &=& \frac{1}{2} \ud \vec{\phi} . \vec{H} + e^{ \frac{1}{2}
\vec{\a}_{1}.\vec{\phi}} \cF_\1{}^1{}_2 \e_{1} + e^{\frac{1}{2}
(\vec \a_1 + 3 \vec \a_2).\vec{\phi}} \cG_\1{}^1 \e_{5}+
e^{\frac{1}{2} (2 \vec \a_1 + 3 \vec \a_2).\vec{\phi}} \cG_\1{}^2
\e_{6}
\nn \\
& &+ e^{\frac{1}{2} \vec{\a}_{2}.\vec{\phi}} F_{\1 1} \e_{2}+
e^{\frac{1}{2} (\vec{\a}_{1} + \vec \a_2).\vec{\phi}} F_{\1 2}
\e_{3} + e^{-\frac{1}{2} (\vec{\a}_1 + 2 \vec \a_2).\vec{\phi}}
G_\1{} \e_{4} \rlap{\ .}
\label{g-dn2} 
\eeq
where $G_\1 $ is the dual of $F_2$. The $\e_i$ generators are the 
Chevalley generators of $G_2$ up to normalisation. 
Let $e_1$ and $e_2$ be the positive Chevalley generators of $G_2$
corresponding to the two simple roots $\a_1$ and $\a_2$. The other
positive generators are \bea
e_3 &=& [e_2,e_1] \hspace{1cm} e_4 = [e_2,[e_2,e_1]] \nn \\
e_5 &=& [e_2,[e_2,[e_2,e_1]]] \hspace{1cm} e_6 =
[[e_2,[e_2,[e_2,e_1]]], e_1] .\eea Their non vanishing commutation
relations are,
\bea \ [e_1, e_2 ] &=& -e_3 \hspace{1cm} [e_1, e_5 ] = - e_6 \nn \\
\ [e_2, e_3 ] &=& e_4 \hspace{1cm} [e_2, e_4 ] = e_5 \nn \\
\ [e_3,e_4] &=& e_6\hspace{1cm} \eea
The normalizing factors $N_\a$ for the simple roots are given by
$N_1 = 1$ and $N_2 = 3$ since $(\a_1 \vert \a_1) = 2$ and $(\a_2
\vert \a_2) = \frac{2}{3}$.  It follows that $N_3 = 3$, $N_4 =
12$, $N_5 = 36$ and $N_6 = 36$.  We define the vectors $\e_i$ in
order to absorb these factors, i.e., $\e_1 = e_1$, $\e_2 =
\frac{1}{\sqrt{3}} e_2$, $\e_3 = \frac{1}{\sqrt{3}} e_3$, $\e_4 =
\frac{1}{2\sqrt{3}} e_4$, $\e_5 = \frac{1}{6} e_5$, $\e_6 =
\frac{1}{6} e_6$. This implies $K(\e_i, \tau(\e_i)) = -1$.

\subsection{Fermions}

We want to add Dirac fermion in $D=5$, with a coupling which in
the $D=3$ reduction is covariant with respect to  $Spin(1,2) \times
Spin(4)$. From what we have already learned, this
should be possible,  with a representation which is trivial on one
of the two $SU(2)$ factors of $SO(4) \simeq (SU(2) \times SU(2))/\ZZ_2$,
since we have already seen in the
analysis of the gravitational sector that the Clifford algebra
contains $Spin(1,2) \times Spin(3)$ representations.

To check if we can indeed derive such a representation from a
consistent $D=5$ coupling, we add to the lagrangian (\ref{g2lag}) a
Dirac fermion with a Pauli coupling,
%%%%%
\be \hat  {\cal L}_{\hpsi} = \ \he \bhpsi ( 
\g^{\hat \m}\partial_{\hat \m} - {1 \over 4} \hat \o_{\hat \m, \sst{(\hat \rho)(\hat \s)}} \g^{ \sst{(\hat \rho)(\hat \s)}} - { 1 \over
2}\a \g^{\hat \m \hat \n } F_\2{}_{ \ \hat \m \hat \n} \label{g2lagf} ) \hpsi \ee
%%%%%
where $\a$ is a coupling constant which will be determined below.
Upon toroidal reduction to $D=3$, the last term of (\ref{g2lagf})
becomes, 
\bea { \a \over 2} e \bar{\psi} (e^{-\fft12
(\vec \a_1 + 2 \vec \a_2)\cdot\vec\phi } \g^{\m \n} F_\2{}_{ \ \m\n} + 2
e^{\fft12 \vec \a_2 \cdot\vec\phi } \g^\m \g^{1} F_{\sst{(1)}1 \ \m} + 2
e^{\fft12 (\vec \a_1 + \vec \a_2 )\cdot\vec\phi } \g^\m \g^{2}
F_{\sst{(1)}2 \ \m}) \psi \eea 
Let us dualise the 2 form field
strengths.  We get for the dimensional reduction of the whole
lagrangian (\ref{g2lagf}), 
\beq 
{\cal L}_{\psi}^{(3)} &=&
e \bar{\psi} \g^{\m} ( \partial_{\m} - {1
\over 4} {\o}_{\m,\rrr \sss} \g^{\rrr \sss} \nn
\\ && \hspace{.5cm} -{1 \over 2} e^{\fft12 (\vec \a_1 + 2 \vec \a_2)\cdot\vec\phi
} \Gamma_{4} G_\m - {1 \over 2}e^{\fft12 \vec \a_2\cdot\vec\phi }
\Gamma_{2} F_{\sst{(1)} 1}{}_\m
- {1 \over 2}e^{\fft12 (\vec \a_1 + \vec \a_2)\cdot\vec\phi }
\Gamma_{3} F_{\sst{(1)} 2}{}_\m \nn \\
&& \hspace{.5cm} - {1 \over 2}e^{{1 \over 2}(\vec \a_1 + 3 \vec
\a_2). \vec \phi} \Gamma_{5} \cG_1{}_\m -{1 \over 2}e^{{1 \over 2}(2
\vec \a_1 + 3 \vec \a_2). \vec \phi} \Gamma_{6} \cG_2{}_\m - {1 \over
2}e^{{1 \over 2}\vec \a_1 . \vec \phi } \Gamma_{1} \cF^{
1}_{\sst{(1)} \ 2}{}_\m) \psi 
\nn
\eeq 
where $\Gamma_1 = {1\over 2} \g^{12}, \
\Gamma_2 = 2 \a \g^1, \ \Gamma_3 = 2 \a \g^2, \ \Gamma_4 = 2 \a
\hat{\g}, \ \Gamma_5 = {1\over 2 } \hat{\g} \g^1 $ and $\Gamma_6 =
{1\over 2} \hat{\g} \g^2$. Notice that $\hat{\g} = -i \g^{12}$
because the product of all gamma matrices $\g^0\g^1\g^2\g^3 \g^4 =
\hat{\g} \g^1\g^2$ in $D=5$ can be equated to $i$.

As in the case of the other algebras encountered above, we need to
check that the $\Gamma_i$'s obey the commutation relations of the
maximally compact subalgebra of $G_2$, i.e., obey the same
commutation relations as the compact generators $k_i = \e_i + \tau(\e_i)$.  The
commutators of the compact subalgebra are \bea && [k_1, k_2] = -
k_3, \qquad [k_1, k_3] = k_2, \qquad [k_1, k_4] = 0, \nn \\
&& [k_1, k_5] = - k_6, \qquad [k_1, k_6] = k_5 , \qquad [k_2, k_3]
= \frac{2}{\sqrt{3}} k_4 - k_1 \nn \\ && [k_2, k_4] = k_5 -
\frac{2}{\sqrt{3}} k_3 , \qquad [k_2, k_5] = - k_4 , \qquad [k_2,
k_6] = 0, \nn \\ &&  [k_3, k_4] =  k_6 + \frac{2}{\sqrt{3}}
k_2 , \qquad [k_3, k_5] =  0 , \qquad [k_3, k_6] = - k_4 ,
\nn
\\ && [k_4, k_5] = k_2 , \qquad [k_4, k_6] =  k_3, \qquad [k_5,
k_6] = - k_1. \eea  In the basis \bea && \xi_1 = \frac{1}{4}(3 k_1
+ \sqrt{3} k_4),\qquad \xi_2 = \frac{1}{4}(\sqrt{3} k_2 - 3
k_6),\qquad \xi_3 = - \frac{1}{4} (\sqrt{3} k_3 + 3 k_5) \nn \\ &&
X_1 = \frac{1}{4}(k_1 - \sqrt{3} k_4),\qquad X_2 =
\frac{1}{4}(\sqrt{3} k_2 + k_6),\qquad X_3 = - \frac{1}{4}
(\sqrt{3} k_3 - k_5),\eea the commutation relations read \be
[\xi_i, \xi_j] = \varepsilon_{ijk} \xi_k, \qquad [\xi_i, X_j] = 0,
\qquad [X_i, X_j] = \varepsilon_{ijk} X_k \ee and reveal the
$\mf{su}(2) \oplus \mf{su}(2)$ structure of the algebra.

We find that the commutation relations are
indeed fulfilled provided we take $\a = i a$, with $a$ solution of
the quadratic equation $16 a^2 + \frac{8}{\sqrt{3}} a - 1 = 0$,
which implies $ \a = - i \frac{\sqrt{3}}{4}$ or $\a = \frac{i}{4
\sqrt{3}}$.  The two different solutions correspond to a non
trivial representations for either the left or the right factor
$\mf{su}(2)$. Thus, we see again that the fermions are compatible with
$G_2$--invariance and we are led to introduce the covariant Dirac
operator 
\be 
\gamma^{\hat \mu} D_{\hat \mu} \hat \psi = \gamma^{\hat \mu} (\partial_{\hat \m} -
{1 \over 4} \hat \o_{\hat \m, \sst{(\hat \rho) (\hat \s)}} \g^{\sst{(\hat \rho) (\hat \s)}}) \hpsi - { 1 \over 2}\a \g^{ \hat \m \hat \r
\hat \s } F_\2{}_{ \ \hat \r \hat \s} \hpsi 
\nn
\ee 
(with $\a$ equal to one of the above
values) for the spin--$1/2$ field. This is the \emph{same} expression as
the one that followed from supersymmetry \cite{Chamseddine:1980sp}.

Another approach of this problem is to remember that $G_2$ can be
embedded in $D_4 = SO(4,4)$ \cite{Cremmer:1999du}. The maximal
oxidation is $D=6$ and contains a 2--form in addition to gravity.
After reduction on a circle, we get two dilatons and three
1-forms: the original 2-form and its Hodge dual both reduce to
1--forms, and we have also the Kaluza--Klein 1--form. The model we
are dealing with is obtained by equating these three 1--forms, and
setting the dilatons to zero \cite{Cremmer:1999du}. It is clear
that this projection do respect the covariance of the fermionic
coupling obtained by reduction of (\ref{dnlagf0}). In $D=3$, the
compact gauge group is projected from $SO(4) \times SO(4)$ onto
$SO(4)$, in addition to the unbroken
$SO(1,2)$. All other terms in the connection are indeed set to
zero by the embedding. The $D=6$ spinor can be chosen to have a
definite chirality. Each chirality corresponds to a different
choice of $\a$ after dimensional reduction.

\section{Non--simply laced algebras $B_n$, $C_n$, $F_4$}
\label{nonsimplylaced}

All the non-simply laced algebras can be embedded in simply laced
algebras \cite{Cremmer:1999du}. Therefore, we can find the
appropriate coupling by taking the one obtained for the simply
laced algebras and by performing the same identifications as for
the bosonic sector.

$B_n = SO(n,n+1)$ (with maximal compact subgroup $SO(n) \times
SO(n+1)$) can be obtained from $D_{n+1}$ by modding out the
$\ZZ_2$ symmetry of the diagram. As the $D_{n+1}$ coset can be
oxidised up to $D=n+3$, we must consider a $D=n+3$ Clifford
algebra. The $B_n$ coset has its maximal oxidation in one
dimension lower. \newline
If $n$ is even, $D=n+3$ and $D=n+2$ Dirac spinors
are the same: the embedding gives a coupling to a single $D=n+2$
Dirac spinor. \newline
If $n$ is odd, this argument is no longer
sufficient. However, due to the fact that all elements of the
compact subalgebra of $SO(n+1,n+1)$ are represented by a product
of an even number of gamma matrices, we can take a Weyl spinor in
$D=n+3$: it gives a single Dirac spinor in $D=n+2$. It is thus
possible couple the maximal oxidation of the $B_n$ coset to a
single Dirac spinor, such that it reduces to a Dirac coupling in
$D=3$, covariant with respect to $SO(1,2) \times SO(n) \times
SO(n+1)$.  We leave the details to the reader.

For $C_n = Sp(n)$ (with maximal compact subgroup $U(n)$),
the maximal oxidation lives in $D=4$. The embedding in
$A_{2n-1}$ couples the bosonic degrees of freedom to a $D=2n+2$
spinor. As it is an even dimension, the Weyl condition can be
again imposed, so that we get a $D=2n$ Dirac spinor. It is not
possible to reduce further the number of components: the
representation involves product of odd numbers of gamma matrices.
In $D=4$, this gives a coupling to $2^{n-2}$ Dirac spinors.

The situation for $F_4$ is similar, when considering the embedding
in $E_6$. The $E_6$ coset can be oxidised up to $D=8$, with a
consistent fermionic coupling to a Dirac spinor. As the coupling
to the 3--form involves the product of 3 gamma matrices, it is not
possible to impose the Weyl condition. The maximal oxidation of
the $F_4/(SU(2) \times Sp(3))$ coset,
which lives in dimension 6, is thus coupled to a
pair of Dirac spinors.

For $C_n$ and $F_4$, the embeddings just described give a coupling
to respectively $2^{n-2}$ and $2$ Dirac spinors in the maximally
oxidised theory. We have not investigated in detail whether one
could construct invariant theories with a smaller number of
spinors.

\section{$\cG^{++}$ Symmetry}
\label{G++} 

The somewhat magic emergence of unexpected symmetries
in the dimensional reduction of gravitational theories has raised
the question of whether these symmetries, described by the algebra
$\cG$ in three dimensions, are present prior to reduction or are
instead related to toroidal compactification. It has been argued
recently that the symmetries are, in fact, already present in the
maximally oxidised version of the theory (see \cite{deWit:1985iy,deWit:1986mz,Nicolai:1986jk,deWit:2000wu} for
early work on the $E_8$--case) and are part of a much bigger,
infinite--dimensional symmetry, which could be the overextended
algebra $\cG^{++}$ \cite{Julia:1980gr,Julia:1981wc,Nicolai:1991kx,Damour:2000hv}, the very extended
algebra $\cG^{+++}$ \cite{West:2001as,Gaberdiel:2002db,Englert:2003zs,Englert:2003pd,
Miemiec:2004iv,Kleinschmidt:2003jf,West:2004kb,West:2004wk,Schnakenburg:2004vd,West:2004iz,Englert:2003py,Englert:2004it,Englert:2004ug,Englert:2004ph,deBuyl:2005it}, or a Borcherds superalgebra
related to it \cite{Henry-Labordere:2002xh,Henry-Labordere:2002dk}. There are different
indications that this should be the case, including a study of the
BKL limit of the dynamics \cite{Belinsky:1970ew,Belinsky:1982pk}, which leads to
``cosmological billiards'' \cite{Damour:2002et}.

In \cite{Damour:2002cu}, an attempt was made to make the symmetry manifest
in the maximal oxidation dimension by reformulating the system as
a $(1+0)$--non linear sigma model $\cG^{++}/K(\cG^{++})$.  The explicit
case of $E_{10}$ was considered.  It was shown that at low levels,
the equations of motion of the bosonic sector of 11--dimensional
supergravity can be mapped on the equations of motion of the non
linear sigma model $E_{10}/ K(E_{10})$.  The matching works for
fields associated with roots of $E_{10}$ whose height does not
exceed 30  (see also \cite{Damour:2004zy}).

We now show that this matching works also for Dirac spinors.  We
consider again the explicit case of $E_{10} = E_8^{++}$ for definiteness.  We
show that the Dirac Lagrangian for a Dirac spinor in eleven
dimensions, coupled to the supergravity three--form as in section
\ref{e8}, is covariant under $K(E_{10})$, at least up to the level
where the bosonic matching is successful.  [For related work on
including fermions in these infinite--dimensional algebras, see
\cite{Nicolai:2004nv,Kleinschmidt:2004dy}.]

Our starting point is the action for the non linear sigma model
$E_{10}/ K(E_{10})$ in $1+0$ dimension coupled to Dirac fermions
transforming in a representation of $K(E_{10})$.  We follow the
notations of \cite{Damour:2002et}.  The Lagrangian reads 
\be 
\cL =
\frac{1}{2} n^{-1} <\cP \vert \cP> + i \Psi^{\dagger} D_t \Psi
\label{basic1+0}
\ee 
where we have introduced a lapse function $n$
to take into account reparametrisation invariance in time. The
$K(E_{10})$ connection is 
\be \cQ = \sum_{\a \in \d_+}
\sum_{s=1}^{mult(\a)} \cQ_{\a,s} K_{\a, s} \nn
\ee 
while the covariant
derivative is 
\be  D_t \Psi = \dot{\Psi} - \sum_{\a,s}\cQ_{\a,s}
T_{\a,s} \Psi \nn
\ee 
where the $T_{\a,s}$ are the generators of the
representation in which $\Psi$ transforms (there is an infinity of
components for $\Psi$).

In the Borel gauge, the fermionic part of the Lagrangian becomes
\be 
i \Psi^{\dagger} \dot{\Psi} - \frac{i}{2} \sum_{\a,s}
e^{\a(\b)} j_{\a,s} \Psi^{\dagger} T_{\a,s} \Psi
\label{sigmafermionic}
\ee 
where $\beta^\m$ are now the Cartan
subalgebra variables (i.e., we parametrize the elements of the
Cartan subgoup as $\exp (\b^\m h_\m$)) and $\a(\b)$ the positives
roots. The ``currents'' $j_{\a,s}$ (denoted by $\cF_{\a}$ in 
(\ref{generalform}) in finite--dimensional case, the addition of the index $s$ is introduce to take into
account  the
multiplicity of $\a$) are, as before, the coefficients of the
positive generators in the expansion of the algebra element
$\dot{\cV} \cV^{-1}$, 
\be \dot{\cV} \cV^{-1} = \dot{\b}^\m h_\m +
\sum_{\a \in \Delta_+} \sum_{s=1}^{mult(\a)} \exp
{\left(\a(\b)\right)} j_{\a,s} E_{\a,s} 
\nn
\ee  
We must compare
(\ref{sigmafermionic}) with the Dirac Lagrangian in 11 dimensions
with coupling to the 3--form requested by $E_8$ invariance, 
\be e
\bpsi \left(\gamma^\mu \partial_\mu - \frac{1}{4} \omega_{\mu \ \rrr \sss}
\gamma^\mu \gamma^{\rrr \sss} - \frac{1}{2.4!} F_{\mu\nu\rho\sigma}
\gamma^{\mu\nu\rho\sigma} \right) \psi
\label{Dirac11}
\ee  
where $e$
is the determinant of the space-time vielbein; we are no more 
dealing with dimensional reduction therefore the hats are suppressed;  $\bpsi = i \psi^\dagger \gamma^{\sst{(0)}}$ and we denote $\partial \psi / \partial x^0$ by $\dot{\psi}$. To make the
comparison easier, we first take the lapse $n$ equal to one
standard lapse $N$ equal to $e$) since both
(\ref{sigmafermionic}) and (\ref{Dirac11}) are reparametrization
invariant in time.  We further split the Dirac Lagrangian
(\ref{Dirac11}) into space and time using a zero shift ($N^k = 0$)
and taking the so--called time gauge for the vielbeins $e_\m{}^\nnn$,
namely no mixed space--time component. This yields 
\bea  i
\chi^{\dagger} ( \dot{\chi} &-& \frac{1}{4} \omega_{\aaa\bb}^{R}
\gamma^{\aaa\bb} \chi - \frac{1}{2.3!} F_{0\aaa\bb\ccc}\gamma^{\aaa\bb\ccc} \chi \nn \\
&-&
\frac{e}{2 .4! \, 6!}\varepsilon_{\aaa\bb\ccc\ddd \sst{(p_1)
(p_2)}  \cdots \sst{(p_6)}} F^{\aaa\bb\ccc\ddd} \gamma^{\sst{(p_1)} \cdots \sst{(p_6)}}\chi )\nn \\
 +\,  i \chi^{\dagger} ( &-& \frac{e}{4 .8!}
\omega_{k}^{\; \; \aaa\bb} \varepsilon_{\aaa\bb \sst{(p_1)} \cdots \sst{(p_8)}} \gamma^k
\gamma^{\sst{(p_1)} \cdots \sst{(p_8)}} \chi  + \frac{e}{10!} \varepsilon_{\sst{(p_1)} \cdots \sst{(p_{10})}}\gamma^k \gamma^{\sst{(p_1)} \cdots \sst{(p_{10})}}
\partial_k \chi ) 
\label{keykey}
\eea 
where $e$ is now the determinant of
the spatial vielbein and where the Dirac field is taken to be
Majorana (although this is not crucial) and has been rescaled as
$\chi = e^{1/2} \psi$. In (\ref{keykey}), the term
$\omega_{ab}^{R}$ stands for 
\be \omega_{\aaa\bb}^{R} = -{1 \over
2}(e_{\aaa}{}^k \dot{e}_{k \bb} - e_{\bb}{}^k \dot{e}_{k \aaa}) \nn 
\ee

A major difference between (\ref{sigmafermionic}) and
(\ref{keykey}) is that $\Psi$ has an infinite number of components
while $\chi$ has only $32$ components. But $\Psi$ depends only on
$t$, while $\chi$ is a spacetime field.  We shall thus assume, in
the spirit of \cite{Damour:2002cu}, that $\Psi$ collects the values of
$\chi$ and its successive spatial derivatives at a given spatial
point, 
\be \Psi^{\dagger} = (\chi^{\dagger}, \partial_k
\chi^{\dagger} , \cdots) \nn
\ee 
[The dictionary between $\Psi$ on the
one hand and $\chi$ and its successive derivatives on the other
hand might be more involved (the derivatives might have to be
taken in privileged frames and augmented by appropriate
corrections) but this will not be of direct concern for us here.
We shall loosely refer hereafter to the ``spatial derivatives of
$\chi$'' for the appropriate required modifications.] We are thus
making the strong assumption that by collecting $\chi$ and its
derivatives in a single infinite dimensional object, one gets a
representation of $K(E_{10})$.  It is of course intricate to check
this assertion, partly because $K(E_{10})$ is poorly understood
\cite{Nicolai:2004nv}. Our only justification is that it makes sense
at low levels.

Indeed, by using the bosonic, low level, dictionary of \cite{Damour:2002cu},
we do see the correct connection terms appearing in (\ref{keykey})
at levels $0$ ($\omega_{\aaa\bb}^{R}$ term), $1$ (electric field term)
and $2$ (magnetic field term).  The corresponding generators
$\gamma^{\sst{(a_1)( a_2)}}$, $\gamma^{\sst{(a_1)( a_2)( a_3)}}$ and $\gamma^{\sst{(a_1)} \cdots
\sst{(a_6)}}$ do reproduce the low level commutation relations of
$K(E_{10})$.

The matching between the supergravity bosonic equations of motion
and the nonlinear sigma model equations of motion described in
\cite{Damour:2002cu} goes slightly beyond level $2$ and works also for some
roots at level $3$.  We shall refer to this as ``level $3^-$''. To
gain insight into the matching at level $3^-$ for the fermions, we
proceed as in \cite{Damour:2002cu} and consider the equations in the
homogeneous context of Bianchi cosmologies \cite{Demaret:1985js,Henneaux:1980ft,Henneaux:1981vr}
(see also \cite{Saha:2003xv}). The derivative term $\partial_k \chi$ then
drops out --- we shall have anyway nothing to say about it here,
where we want to focus on the spin connection term $\omega_k^{\;
\; \aaa\bb}$.  In the homogeneous context, the spin connection term
becomes 
\be \omega_{\aaa\bb\ccc} = e_{\aaa}{}^k \o_{k\bb\ccc}  = \frac{1}{2}\left(C_{\ccc\aaa\bb} + C_{\bb\ccc\aaa} -
C_{\aaa\bb\ccc} \right) \nn
\ee 
in terms of the structure constants $C^{\aaa}_{\; \;
\bb\ccc} = - C^{\aaa}_{\; \; \ccc\bb}$ of the homogeneity group expressed in
homogeneous orthonormal frames (the $C^{\aaa}_{\; \; \bb\ccc}$ may depend on
time). We assume that the traces $C^{\aaa}_{\; \; \aaa\ccc}$ vanish since
these correspond to higher height and go beyond the matching of
\cite{Damour:2002cu}, i.e., beyond level $3^-$. In that case, one may
replace $\omega_{\aaa\bb\ccc}$ by $(1/2)C_{\aaa\bb\ccc}$ in (\ref{keykey}) as can
be seen by using the relation
$$\varepsilon_{\aaa\bb\sst{(p_1)} \cdots \sst{(p_8)}}\gamma^k \gamma^{\sst{(p_1)} \cdots \sst{(p_8)}} =
\varepsilon_{\aaa\bb\sst{(p_1)} \cdots \sst{( p_8)}} \gamma^{k \sst{(p_1)} \cdots \sst{(p_8)}} +
\varepsilon_{\aaa\bb k\sst{(p_2)} \cdots \sst{(p_8)}} \gamma^{\sst{(p_2)} \cdots \sst{(p_8)}}. $$ The
first term drops from (\ref{keykey}) because $\omega^{\aaa}_{\; \; \bb\aaa} =
0$, while the second term is completely antisymmetric in $a$, $b$,
$k$. Once $\omega_{\aaa\bb\ccc}$ is replaced by $(1/2)C_{\aaa\bb\ccc}$, one sees
that the remaining connection terms in (\ref{keykey}), i.e., the one
involving a product of nine $\gamma$--matrices, agree with the
dictionary of \cite{Damour:2002cu}. {}Furthermore, the corresponding level
three generators $\gamma^{\ccc} \gamma^{\sst{(p_1)} \cdots \sst{(p_8)}}$ also fulfill the
correct commutation relations of $K(E_{10})$ up to the requested
order.

To a large extent, the $E_{10}$ compatibility of the Dirac
fermions up to level $3^-$ exhibited here is not too surprising,
since it can be viewed as a consequence of $SL_{10}$ covariance
(which is manifest) and the hidden $E_8$ symmetry, which has been
exhibited in previous sections. The real challenge is to go beyond
level $3^-$ and see the higher positive roots emerge on the
supergravity side. These higher roots might be connected, in fact,
to quantum corrections \cite{Damour:2005zb} or higher spin degrees of
freedom.

\section{BKL limit}
\label{BBKKLL} We investigate in this final section how the Dirac
field modifies the BKL behaviour.  To that end, we first rewrite
the Lagrangian (\ref{basic1+0}) in Hamiltonian form.  The
fermionic part of the Lagrangian is already in first order form
(with $i \Psi^{\dagger}$ conjugate to $\Psi$), so we only need to
focus on the bosonic part. The conjugate momenta to the Cartan
fields $\b^\m$ are unchanged in the presence of the fermions since
the time derivatives $\dot{\b}^\m$ do not appear in the connection
$\cQ_{\a,s}$. However, the conjugate momenta to the off--diagonal
variables parameterizing the coset do get modified. How this
affects the Hamiltonian is easy to work out because the time
derivatives of these off-diagonal group variables occur linearly
in the Dirac Lagrangian, so the mere effect of the Dirac term is
to shift their original momenta.  Explicitly, in terms of the
(non--canonical) momentum--like variables 
\be \Pi_{\a,s} =
\frac{\d \cL}{\d j_{\a,s}} \nn
\ee 
introduced in
\cite{Matschull:1994vi,Damour:2002et}, one finds 
\be \Pi_{\a,s} = \Pi_{\a,s}^{old}  -
\frac{1}{2} \exp \left( \a(\b) \right) J^F_{\a,s} \nn
\ee 
where
$\Pi_{\a,s}^{old}$ is the bosonic contribution (in the absence of
fermions) and where $J^F_{\a,s}$ are the components of the
fermionic $K(\cG^{++})$--current, defined by 
\be J^F_{\a,s} = i
\Psi^\dagger T_{\a,s} \Psi \n . \nn
\ee 
The currents $J^F_{\a,s}$ are
real.

It follows that the Hamiltonian associated with (\ref{basic1+0})
takes the form 
\be 
\cH = n \left( \frac{1}{2} G^{\m \n} \pi_\m
\pi_\n + \sum_{\a \in \d_+} \sum_{s=1}^{mult(\a)} \exp
{\left(-2 \a(\b)\right)} \left(\Pi_{\a,s} + \frac{1}{2} \exp
\left( \a(\b) \right) J^F_{\a,s} \right)^2 \right)
\label{Hamiltonian}
\ee 
If one expands the Hamiltonian, one gets
\be 
\cH = n \left( \frac{1}{2} G^{\m \n} \pi_\m \pi_\n +
\sum_{\a,s} \exp {\left(-2 \a(\b)\right)} \Pi_{\a,s}^2  +
\sum_{\a,s} \exp \left(- \a(\b) \right) \Pi_{\a,s} J^F_{\a,s} +
\frac{1}{4} C \right) \nn
\ee 
where $C$ is (up to a numerical factor)
the quadratic Casimir of the fermionic representation, 
\be C =
\sum_{\a,s} (J^F_{\a,s})^2 \, . \nn
\ee  
We see that, just as in the pure
bosonic case, the exponentials involve only the positive roots
with negative coefficients. However, we obtain, in addition to the
bosonic walls, also their square roots. All the exponentials in
the Hamiltonian are of the form $\exp (-2 \rho(\b))$, where
$\rho(\b)$ are the positive roots or half the positive roots.

In order to investigate the asymptotic BKL limit $\b^\m
\rightarrow \infty$, we shall treat the $K(\cG^{++})$--currents as
classical real numbers and consider their equations of motion that
follow from the above Hamiltonian, noting that their Poisson
brackets $[J^F_{\a,s}, J^F_{\a',s'}]$ reproduce the
$K(\cG^{++})$--algebra. This is possible because the Hamiltonian in
the Borel gauge involves only the $\Psi$--field through the
currents. This is a rather remarkable property. [A ``classical''
treatment of fermions is well known to be rather delicate.  One
can regard the dynamical variables, bosonic and fermionic, as
living in a Grassmann algebra.  In that case, bilinear in fermions
are ``pure souls'' and do not influence the behaviour of the
``body'' parts of the group elements, which are thus trivially
governed by the same equations of motion as in the absence of
fermions. However, it is reasonable to expect that the currents
$J^F_{\a,s}$ have a non trivial classical limit (they may develop
non--vanishing expectation values) and one might treat them
therefore as non--vanishing real numbers.  This is technically
simple here because the currents obey closed equations of motion.
It leads to interesting consequences.]

Next we observe that $[J^F_{\a,s}, C]= 0$. It follows that the
quadratic Casimir $C$ of the fermionic representation is
conserved.  Furthermore, it does not contribute to the dynamical
Hamiltonian equations of motion for the group variables or the
currents. By the same reasoning as in \cite{Damour:2002et}, one can then
argue that the exponentials tend to infinite step theta functions
and that all variables except the Cartan ones, \ie the
off--diagonal group variables and the fermionic currents,
asymptotically freeze in the BKL limit.

Thus, we get the same billiard picture as in the bosonic case, with
same linear forms characterizing the walls (some of the exponential
walls are the square roots of the bosonic walls). But the free
motion is governed now by the Hamiltonian constraint 
\be G^{\m \n}
\pi_\m \pi_\n + M^2 \approx 0 \nn 
\ee  
with $M^2 = C/2>0$.   This implies
that the motion of the billiard ball is timelike instead of being
lightlike as in the pure bosonic case. This leads to a non--chaotic
behaviour, even in those cases where the bosonic theory is chaotic.
Indeed, a timelike motion can miss the walls, even in the hyperbolic
case. This is in perfect agreement with the results found in
\cite{BK} for the four--dimensional theory.

Our analysis has been carried out in the context of the sigma model
formulation, which is equivalent to the Einstein--Dirac--p--form system
only for low Kac--Moody levels. However, the low levels roots are
precisely the only relevant ones in the BKL limit (``dominant
walls'').  Thus, the analysis applies also in that case.  Note that
the spin $1/2$ field itself does not freeze in the BKL limit, even
after rescaling by the quartic root of the determinant of the
spatial metric, but asymptotically undergoes instead a constant
rotation in the compact subgroup, in the gauge $n=1$ (together with
the Borel gauge). Note also that the same behaviour holds if one
adds a mass term to the Dirac Lagrangian, since this term is
negligible in the BKL limit, being multiplied by $e$, which goes to
zero.

One might worry that the coefficients $\Pi_{\a,s} J^F_{\a,s}$ of the
square roots of the bosonic walls have no definite sign. This is
indeed true but generically of no concern for the following reason:
in the region $\a(\b) <0$ outside the billiard table where the
exponential terms are felt and in fact blow up with time at a given
configuration point ($\a(\b)\rightarrow - \infty$) \cite{Damour:2002et}, the
wall $\exp (- 2 \a(\b))$ dominates the wall $\exp (- \a(\b))$ coming
from the fermion and the total contribution is thus positive. The
ball is repelled towards the billiard table.

\section{Conclusions}

In this chapter, we have shown that the Dirac field is compatible
with the hidden symmetries that emerge upon toroidal dimensional
reduction to three dimensions, provided one appropriately fix its
Pauli couplings to the $p$--forms.  We have considered only the
split real form for the symmetry (duality) group in three
dimensions, but similar conclusions appear to apply to the
non--split forms (we have verified it for the four--dimensional
Einstein--Maxwell--Dirac system, which leads to the $SU(2,1)/S(U(2)
\times U(1))$ coset in three dimensions).  We have also indicated
that the symmetry considerations reproduce some well known
features of supersymmetry when supersymmetry is available.

We have also investigated the compatibility of the Dirac field
with the conjectured infinite--dimensional symmetry $\cG^{++}$ and
found perfect matching with the non--linear sigma model equations
minimally coupled to a $(1+0)$ Dirac field, up to the levels where
the bosonic matching works.

{}Finally, we have argued that the Dirac fermions destroy chaos
(when it is present in the bosonic theory), in agreement with the
findings of \cite{BK}.  This has a rather direct group theoretical
interpretation (motion in Cartan subalgebra becomes timelike) and
might have important implications for the pre--big--bang
cosmological scenario and the dynamical crossing of a cosmological
singularity \cite{Gasperini:2002bn,Veneziano:2003sz,Steinhardt:2001st,Wesley:2005bd}.

It would be of interest to extend these results to include the
spin $3/2$ fields, in the supersymmetric context. In particular,
11--dimensional supergravity should be treated.  To the extent that
$E_{10}$ invariance up to the level $3^-$ is a mere consequence
of $E_8$ invariance in three dimensions and $SL_{10}$ convariance,
one expects no new feature in that respect since the reduction to
three dimensions of full supergravity is indeed known to be $E_8$
invariant \cite{Marcus:1983hb}.  But perhaps additional structure
would emerge.  Understanding the BKL limit might be more
challenging since the spin $3/2$ fields might not freeze, even
after rescaling.
%%%%%%%%%%%%%%%%%%%%%
%%%%%%%%%%%%%%%%%%%%%%
\cleardoublepage
%%%%%%%%%%%%%%%%%%%%%%
%%%%%%%%%%%%%%%%%%%%%%%%
%%\include{chap_sugra}
\chapter{Eleven Dimensional Supergravity}
\markboth{ELEVEN DIMENSIONAL SUPERGRAVITY}{}
\label{gravitino}

\section{Introduction}

The hyperbolic Kac--Moody algebra $\mf{e}_{10(10)}$, whose Dynkin diagram is
given in Fig.1,  has repeatedly been argued to play a crucial role
in the symmetry structure of M--theory
\cite{Julia:1980gr,Julia:1981wc,Nicolai:1991kx,Mizoguchi:1997si}.
\newline
\begin{center}
\ \scalebox{.8}{
\begin{picture}(180,60)
%nom des racines
\put(-55,-5){$\alpha_{9}$} \put(-15,-5){$\alpha_8$}
\put(25,-5){$\alpha_7$} \put(65,-5){$\alpha_6$}
\put(105,-5){$\alpha_5$} \put(145,-5){$\alpha_4$}
\put(185,-5){$\alpha_3$} \put(225,-5){$\alpha_2$}
\put(265,-5){$\alpha_1$} \put(200,45){$\alpha_0$}
%9 vertex + lignes simples
\thicklines \multiput(-50,10)(40,0){9}{\circle{10}}
\multiput(-45,10)(40,0){8}{\line(1,0){30}}
%1 vertex du dessus
\put(190,50){\circle{10}} \put(190,15){\line(0,1){30}}
\end{picture}
}
\end{center}

\centerline{
{\footnotesize FIG. 1.}
{\small \emph{The Dynkin diagram of $E_{10}$} }}
$ \ $ 

This infinite--dimensional algebra has a complicated structure that
has not been deciphered yet. In order to analyse further its root
pattern, it was found convenient in \cite{Damour:2002cu} to introduce a
``level'' for any root $\a$, defined as the number of times the
simple root $\a_0$ occurs in the decomposition of $\a$.

The roots $\a_1$ through $\a_9$ define a subalgebra $\mf{sl}(10,\RR)$.
Reflections in these roots define the finite Weyl group $W_{A_9}$
($\simeq S_{10}$) of $A_9$, which acts naturally on the roots of
$\mf{e}_{10(10)}$. If we express the roots of $\mf{e}_{10(10)}$ in terms of the
spatial scale factors $\b^i$ appearing naturally in cosmology
\cite{Damour:2000wm,Damour:2000hv}, the action of $W_{A_9}$ is simply to permute the $\b$'s.
The level is invariant under $W_{A_9}$.  Consider the set $R_{E_8}$
of roots of the $E_{8(8)}$ subalgebra associated with the simple roots
$\a_0$ through $\a_7$. By acting with $W_{A_9}$ on $R_{E_8}$, one
generates a larger set $\tilde{R}_{E_8}$ of roots. This set will be
called the extended set of roots of $E_{8(8)}$. By construction, the
roots in $\tilde{R}_{E_8}$ are all real and have length squared
equal to $2$. There is an interesting description of the roots in
$\tilde{R}_{E_8}$ in terms of the level.  One can easily verify that
all the roots at level $0$, $\pm 1$ and $\pm 2$, as well as all the
{\it real} roots at level $\pm 3$ exhaust $\tilde{R}_{E_8}$.  This
includes, in particular, all the roots with $\vert$height$\vert$
$<30$.

Recently, following the analysis {\it \`a la} BKL \cite{Belinsky:1970ew,Belinsky:1982pk,Damour:2002et} of
the asymptotic behaviour of the supergravity fields near a
cosmological singularity, the question of the hidden symmetries of
eleven--dimensional supergravity has received a new impulse
\cite{Damour:2000wm,Damour:2000hv}. It has been argued that one way to exhibit the symmetry
was to rewrite the supergravity equations as the equations of motion
of the non--linear sigma model $E_{10(10)}/K(E_{10(10)})$ \cite{Damour:2002cu}.

The first attempt for rewriting the equations of motion of
eleven--dimensional supergravity as non--linear sigma model equations
of motion --- in line with the established result that the scalar
fields which appear in the toroidal compactification down to three
spacetime dimensions form the coset $E_{8(8)}
/(Spin(16)/\ZZ_2)$ \cite{Cremmer:1978ds,Cremmer:1979up} --- is
due to \cite{West:2001as}. In that approach, it is the larger
infinite--dimensional algebra $\mf{e}_{11(11)}$ which is priviledged. Various
evidence supporting $\mf{e}_{11(11)}$ was provided in \cite{West:2001as,Englert:2003zs,Englert:2003py}. Here,
we shall stick to (the subalgebra) $\mf{e}_{10(10)}$, for which the dynamical
formulation is clearer, although similar construction can be done for $\mf{e}_{11(11)}$. 

The idea of rewriting the equations of motion of eleven--dimensional
supergravity as equations of motion of $E_{10(10)}/K(E_{10(10)})$  was
verified in \cite{Damour:2002cu} for the first bosonic levels in a level
expansion of the theory. More precisely, it was verified that in the
coset model $E_{10(10)}/K(E_{10(10)})$, the fields corresponding to the
Cartan subalgebra and to the positive roots $\in \tilde{R}_{E_8}$
have an interpretation in terms of the (bosonic) supergravity fields
(``dictionary" of \cite{Damour:2002cu}). {}Furthermore, there is a perfect
match of the supergravity equations of motion and the coset model
equations of motion for the fields corresponding to these real
roots.  This extended $E_{8(8)}$--invariance, which combines the known
$E_{8(8)}$--invariance and the manifest $SL(10,\RR)$--invariance, is a first
necessary step in exhibiting the full $E_{10(10)}$ symmetry. Further
indication on the meaning of the fields associated with the higher
roots in terms of gradient expansions, using partly information from
$E_{9(9)}$, was also given in \cite{Damour:2002cu}.

The purpose of this chapter is to explicitly verify the extended
$E_{8(8)}$--invariance of the fermionic sector of $11$--dimensional
supergravity.  This amounts to showing that up to the requested
level, the fermionic part of the supergravity Lagrangian, which is
first order in the derivatives, can be written as \be i \Psi^T M D_t
\Psi \label{covariant}\ee where (i) $\Psi$ is an infinite object
that combines the spatial components of the gravitino field $\psi_a$
and its successive gradients \be \Psi = (\psi_a, \cdots) \ee in such
a way that $\Psi$ transforms in the spinorial representation of  $K(E_{10(10)})$
that reduces to the spin 3/2 representation of $SO(10)$; (ii) $M$ is
a $K(E_{10(10)})$--invariant (infinite) matrix; and (iii) $D_t$ is the
$K(E_{10(10)})$ covariant derivative. [We work in the gauge $\psi'_0=0$,
where $\psi'_0$ is the redefined temporal component of the gravitino
field familiar from dimensional reduction \cite{Cremmer:1978ds,Cremmer:1979up}, \be \psi'_0 =
\psi_0 - \g_0 \g^a \psi_a ,\label{redef}\ee so that the temporal
component of $\psi_\mu$ no longer appears.]

In fact, for the roots considered here, one can truncate $\Psi$ to
the undifferentiated components $\psi_a$.  The next components ---
and the precise dictionary yielding their relationship with the
gravitino field gradients --- are not needed. In view of the fact
that the undifferentiated components of the gravitino field form a spinorial
representation of the maximal compact subgroup $Spin(16)/\ZZ_2$ of $E_{8(8)}$
in the reduction to three dimensions, without the need to
introduce gradients or duals, this result is not unexpected.

The crux of the computation consists in constructing the spinorial
representation of $K(E_{10(10)})$ up to the required level. This is done
in the next section, where we compare and contrast the spin 1/2 and
spin 3/2 representations of $K(E_{10(10)})$.  The technically simpler
case of the spin 1/2 representation was investigated in the previous chapter, where it was shown that the Dirac Lagrangian was
compatible with extended $E_{8(8)}$ invariance provided one introduces an
appropriate Pauli coupling with the $3$--form.  Our work overlaps the
work \cite{West:2003fc} on the fermionic representations of $K(E_{11})$ as
well as the analyses of \cite{Keurentjes:2003yu,Keurentjes:2003hc} on the maximal compact
subgroups of $E_{n(n)}$ and of \cite{Nicolai:2004nv} on $K(E_{9})$.

We then investigate the conjectured infinite--dimensional symmetry
$E_{10(10)}$ of the Lagrangian of \cite{Cremmer:1978km}.   We find that
the fermionic part also takes the form dictated by extended
$E_{8(8)}$--invariance, with the correct covariant derivatives appearing
up to the appropriate level. As observed by previous authors and in
particular in \cite{deBuyl:2005zy}, there is an interesting interplay
between supersymmetry and the hidden symmetries.

\section{`Spin 3/2' Representation of $K(E_{10(10)})$}

\subsection{Level 0}

To construct the `spin 3/2' representation of $K(E_{10(10)})$, we have
to extend the level 0 part which is the usual $SO(10)$ `spin 3/2'
parametrised by a set of 10 spinors $\chi_m$, where $m=1\ldots 10$
is a space index.\footnote{We work at the level of the algebra that we denote here by $K(E_{10(10)})$ instead of our usual notation $k(\mf{e}_{10(10)})$.} [The level is not a grading for $K(E_{10(10)})$ but a
filtration, defined modulo lower order terms.]  The $\mf{so}(10)$
generators $k^{ij}$ act on $\chi_m$ as \be k^{ij}.\chi_m =
\frac{1}{2} \gamma^{ij} \chi_m + \delta_{m}^{\; \;i} \chi^j -
\delta_{m}^{\; \; j} \chi^i \rlap{\ .} \label{k-ab}\ee The aim is
to rewrite the Rarita--Schwinger term with all couplings of the
fermionic field, up to higher order fermionic terms, into the form
(\ref{covariant}).

\subsection{Level 1}

Beyond $SO(10)$, the first level couples to $F_{0abc}$. To
reproduce the supergravity Lagrangian, the level 1 generators must
contain products of $\gamma$ matrices where the number of matrices
is odd and at most five. Indeed, the matrix $M$ in
(\ref{covariant}) is proportional to the antisymmetric product
$\g^{ab}$ (as one sees by expanding the supergravity Lagrangian
${\cal L} \sim i \psi^T_a \g^{ab} \dot{\psi}_b + \cdots$), while
$F_{0abc}$ is coupled to fermions, in the supergravity Lagrangian,
through terms $\psi_m^T \g^{mabcn}\psi_n$ and $\psi_a^T \g^{bc}
\g^n\psi_n$. In addition, the generators must be covariant with
respect to $SO(10)$. This gives the general form \be
k^{abc}.\chi_m = A \gamma_m^{\; \;nabc} \chi_n + 3 B \delta_m^{[a}
\gamma^{bc]n} \chi_n + 3C \gamma_m^{\;\;[ab}\chi^{c]} + 6D
\delta_m^{[a} \gamma^b \chi^{c]} + E \gamma^{abc} \chi_m
\label{gen-kabc} \ee where $A,B,C,D,E$ are constants to be fixed.
In fact, it is well known that such generators do appear in the
dimensional reduction of supergravity. If $d$ dimensions are
reduced, the generators  mix only $\chi_m$ with $1 \leq m \leq d$.
Therefore we set the terms involving summation on $n$ on the right
hand side of (\ref{gen-kabc}) equal to zero: $A=B=0$.

To fix the coefficients $C$, $D$ and $E$, we must check the
commutations relations. Commutation with level 0 generators is
automatic, as (\ref{gen-kabc}) is covariant with respect to
$SO(10)$. What is non--trivial is commutation of the generators at
level 1 with themselves. The generators $K^{abc}$ of $K(E_{10(10)})$
at level 1 fulfill \be [K^{abc}, K^{def}] = K^{abcdef} - 18
\d^{ad} \d^{be} K^{cf} \ee (with antisymmetrisation in $(a,b,c)$
and $(b,c,d)$ in $\d_{ad} \d_{be} K_{cf}$) as it follows from the
$E_{10(10)}$ commutation relations. In order to have a representation
of $K(E_{10(10)})$, the $k^{abc}$ must obey the same algebra, \be
[k^{abc}, k^{def}] = k^{abcdef} - \d^{ad} \d^{be} k^{cf}.
\label{level1algebra}\ee When all the indices are distinct,
(\ref{level1algebra}) defines the generators at level 2. One gets
non trivial constraints when two or more indices are equal.
Namely, there are two relations which must be imposed: \beq
\left[ k^{abc} , k^{abd} \right] &=& -k^{cd} \label{rel1}\\
\left[ k^{abc} , k^{ade} \right] &=& 0 \label{rel2}\eeq where different
indices are supposed to be distinct. In fact, as we shall discuss in
the sequel, all other commutation relations which have to be checked
for higher levels can be derived from this ones using the Jacobi
identity. One can verify (\ref{rel1}) and (\ref{rel2}) directly or
using \emph{FORM} \cite{Vermaseren:2000nd}.  One finds that these two relations
are satisfied if and only if \be C=- \frac{1}{3} \epsilon , \; \; \;
\; D=\frac{2}{3} \epsilon, \; \; \; \;  E=\frac{1}{2} \epsilon \ee
with $\epsilon = \pm 1$. In fact one can change the sign of
$\epsilon$ by reversing the signs of all the generators at the odd
levels. This does not change the algebra. We shall use this freedom
to set $\epsilon =  1$ in order to match the conventions for the
supergravity Lagrangian. Putting everything together, the level 1
generator is \be k^{abc}.\chi_m =
 \frac{1}{2}  \gamma^{abc} \chi_m -
\gamma^{m[ab}\chi^{c]} +4  \delta_m^{[a} \gamma^b \chi^{c]}
\rlap{\ .} \label{k-abc} \ee

\subsection{Level 2}

The expression just obtained for the level 1 generators can be used
to compute the level 2 generator \be k^{abcdef} = \left[ k^{abc} ,
k^{def} \right] \ee which is totally antisymmetric in its indices, as it
can be shown using the Jacobi identity. Explicitly, Eq.(\ref{k-abc})
gives \be k^{abcdef}.\chi_m = \frac{1}{2} \gamma^{abcdef} \chi_m +4
\gamma_m^{\; \;[abcde} \chi^{f]} -10 \delta_m^{[a} \gamma^{bcde}
\chi^{f]} \rlap{\ .} \label{k-abcdef}\ee

\subsection{Level 3}

We now turn to level 3.  There are two types of roots. Real roots
have generators \be k^{a;abcdefgh} =  [ k^{abc} , k^{adefgh} ]
\ee (without summation on $a$ and all other indices distinct).
They are easily computed to act as \be k^{a;abcdefgh}.\chi_m =
\frac{1}{2}  \gamma^{bcdefgh} \chi_m + 2 \gamma_m^{\; \; abcdefgh}
\chi^a +16  \delta_m^a \gamma^{[abcdefg}\chi^{h]} -7 \gamma_m^{\;
\; [bcdefg}\chi^{h]} \rlap{\ .} \label{real}\ee In addition, there
are generators $k^{a;bcdefghi}$ with all indices distinct,
corresponding to null roots. {}From \be \ [ k^{abc} , k^{defghi}
] = 3 k^{[a;bc]defghi} \ee (with all indices distinct) one finds
\be k^{a;bcdefghi}.\chi_m = - 2  \left( \gamma_m^{\;
\;[abcdefgh}\chi^{i]} - \gamma_m^{\; \; bcdefghi}\chi^a \right) - 16
\left( \delta_m^{[a} \gamma^{bcdefgh} \chi^{i]} - \delta_m^a
\gamma^{[bcdefgh}\chi^{i]} \right) \rlap{\ .} \label{null}\ee
Combining these results, one finds that the level 3 generators can
be written as \be
\begin{split}
k^{a;bcdefghi}.\chi_m = -2 & \left( \gamma_m^{\; \;
[abcdefgh}\chi^{i]} - \gamma_m^{\; \; bcdefghi}\chi^a \right) - 16 \left(
\delta_m^{[a} \gamma^{bcdefgh} \chi^{i]} - \delta_m^a
\gamma^{[bcdefgh}\chi^{i]} \right)\\& + 4  \delta^{a[b}
\gamma^{cdefghi]} \chi_m - 56
\gamma^{m[bcdefg}\delta^{\hat{a}h}\chi^{i]} \rlap{\ .}
\end{split} \label{level3}
\ee (where the hat over $a$ means that it is not involved in the
antisymmetrization).  Note that if one multiplies the generator
(\ref{level3}) by a parameter $\m_{a;bcdefghi}$ with the symmetries
of the level 3 Young tableau (in particular, $\m_{[a;bcdefghi]} =
0$), the first terms in the two parentheses disappear.
{}Furthermore, the totally antisymmetric part of the full level 3
generator vanishes. The condition $\m_{[a;bcdefghi]}=0$ on $\m_{a;bcdefghi}$ is equivalent
to the tracelessness
of its dual.  
\subsection{Compatibility checks}
Having defined the generators of the `spin 3/2' representation up
to level 3, we must now check that they fulfill all the necessary
compatibility conditions expressing that they represent the
$K(E_{10(10)})$ algebra up to that level (encompassing the
compatibility conditions (\ref{rel1}) and (\ref{rel2}) found
above). This is actually a consequence of the Jacobi identity and
of the known $SO(16)$ invariance in 3 dimensions, as well as of
the manifest spatial $SL(10,\RR)$ covariance that makes all spatial
directions equivalent.

Consider for instance the commutators of level 1 generators with
level 2 generators.  The $K(E_{10(10)})$ algebra is
 \be [K^{abc}, K^{defghi}] = 3 K^{[a;bc]defghi} - 5! \d^{ad}
\d^{be} \d^{cf} K^{ghi} \label{1-2}\ee Thus, one must have \be
[k^{abc}, k^{defghi}] = 3 k^{[a;bc]defghi} - 5! \d^{ad} \d^{be}
\d^{cf} k^{ghi} \label{1-2bis}\ee These relations are constraints
on $k^{abc}$ and $k^{defghi}$ when the level 3 generators are
absent, which occurs when (at least) two pairs of indices are
equal.  But in that case, there are only (at most) 7 distinct
values taken by the indices and the relations are then part of the
known $SO(16)$ invariance emerging in 3 dimensions. In fact, the
relations (\ref{1-2bis}) are known to hold when the indices take
at most 8 distinct values, which allows $k^{a;acdefghi}$ with a
pair of repeated indices. These 8 values can be thought of as
parameterizing the 8 transverse dimensions of the dimensional
reduction. Note that since the index $m$ in (\ref{k-abc}) can be
distinct from the 8 ``transverse'' indices, we have both the `spin
1/2' and the `spin 3/2' (\ie the vector and the spinor)
representations of $SO(16)$, showing the relevance of the analysis
of chapter \ref{dirac} in the present context.

Similarly, the commutation of two level 2 generators read \be
[K^{abcdef},K^{ghijkl}] = - 6\cdot 6! \d^{ag} \d^{bh} \d^{ci}
\d^{dj} \d^{ek} K^{fl} + \hbox{ ``more"} \ee where ``more" denotes
level 4 generators.  Thus, one must have \be
[k^{abcdef},k^{ghijkl}] = - 6\cdot 6! \d^{ag} \d^{bh} \d^{ci}
\d^{dj} \d^{ek} k^{fl} + \hbox{ ``more"} \label{2-2}\ee  These
relations are constraints when the level 4 generators are
absent\footnote{When the level 4 generators are present, the
relations (\ref{2-2}) are consequences of the definition of the
level 4 generators -- usually defined through commutation of level
1 with level 3 --, as a result of the Jacobi identity.}. Now, the
level 4 generators are in the representation $(001000001)$
characterised by a Young tableau with one 9--box column and one
3--box column, and in the representation $(200000000)$
characterised by a Young tableau with one 10--box column and two
1--box columns \cite{Nicolai:2003fw,Fischbacher:2005fy}.  To get rid of these level 4
representations, one must again assume that the indices take at
most 8 distinct values to have sufficiently many repetitions.  [If
one allows 9 distinct values, one can fill the tableau
$(001000001)$ non trivially.] But then, $SO(16)$  ``takes over''
and guarantees that the constraints are fulfilled.  The same is
true for the commutation relations of the level 1 generators with
the level 3 generators, which also involve generically the level 4
generators unless the indices take only at most 8 distinct values
(which forces in particular the level-3 generators to have one
repetition, \ie to correspond to real roots).

{}Finally, the level 5 generators and the level 6 generators, which
occur in the commutation relations of level 2 with level 3, and
level 3 with itself,  involve also representations associated with
Young tableaux having a column with 9 or 10 boxes
\cite{Nicolai:2003fw,Fischbacher:2005fy}. For these to be absent, the indices must
again take on at most 8 distinct values. The commutation relations
reduce then to those of $SO(16)$, known to be valid.

\subsection{`Spin 1/2' representation}
We note that if one keeps in the above generators (\ref{k-ab}),
(\ref{k-abc}), (\ref{k-abcdef}) and (\ref{level3}) only the terms
in which the index $m$ does not transform, one gets the `spin 1/2'
representation investigated in \cite{deBuyl:2005zy}. A notable feature
of that representation is that it does not see the level 3
generators associated with imaginary roots, as one sees from
(\ref{null}).

It should be stressed that up to level 3, the commutation
relations of the $K(E_{n(n)})$ subgroups are all very similar for $n
\geq 8$ (\cite{West:2003fc,Keurentjes:2003yu,Keurentjes:2003hc,Nicolai:2004nv}). A more complete analysis
of the `spin 3/2' and `spin 1/2' representations of $K(E_{9(9)})$
are given in \cite{Paulot:2006zp}.

\section{Extended $E_{8(8)}$ Invariance of Supergravity Lagrangian}
\label{lagrangian}

The fermionic part of 11--dimensional supergravity is \be e^{(11)}(-\frac{1}{2}
\bpsi_\mu \gamma^{\mu \rho \nu} D_\rho \psi_\nu- \frac{1}{96}\bpsi_\mu \gamma^{\mu \nu \alpha \beta \gamma \delta }
\psi_\nu F_{\alpha \beta \gamma \delta}- \frac{1}{8} \bpsi^\a \gamma^{\gamma\delta} \psi^\b F_{\a\b \gamma \delta}
)\,, \label{sugra11}\ee where $e^{(11)}$ is the determinant of the
spacetime vielbein  and where we have dropped the terms with four
fermions. We want to compare this expression with the Lagrangian
(\ref{covariant}), where the $K(E_{10(10)})$ representation is the spin
3/2 one constructed in the previous section.  If we expand the
Lagrangian (\ref{covariant}) keeping only terms up to level 3 and
using the dictionary of \cite{Damour:2002cu} for the $K(E_{10(10)})$ connection,
we get (see Eq. (8.7) of \cite{deBuyl:2005zy}) \bea -\frac{i}{2} \psi^T_m
\gamma^{mn} (\dot{\psi}_n - \frac{1}{2}\omega_{ab}^{R} k^{ab}.\psi_n
- \frac{1}{3!}F_{0abc} k^{abc}.\psi_n -\frac{e}{4! \,
6!}\varepsilon_{abcdp_1p_2 \cdots p_6}
F^{abcd}k_{p_1p_2 \cdots p_6}.\psi_n \nn \\
- \frac{e}{2.2!\, 8!} C^a{}_{rs} \epsilon^{rsbcdefghi}
k_{a;bcdefghi}.\psi_n) \label{sigmaf}\eea where $M$ at this level is
given by $\gamma^{mn}$ and where $\omega_{ab}^{R} = -{1 \over
2}(e_a{}^\m \dot{e}_{\m b} - e_b{}^\m \dot{e}_{\m a}) $.  In
(\ref{sigmaf}), $e$ is the determinant of the spatial vielbein,
$e^{(11)} = N \, e$ with $N$ the lapse.

We have explicitly checked the matching between (\ref{sigmaf}) and
(\ref{sugra11}).  In order to make the comparison, we
\begin{itemize}
\item  take the standard lapse $N$ equal to $e$; \item split the
eleven dimensional supergravity Lagrangian (\ref{sugra11}) into
space and time using a zero shift ($N^k = 0$) and taking the
so-called time gauge for the vielbeins $e^a_\m$, namely no mixed
space-time component; \item rescale the fermions $\psi_n
\rightarrow e^{1/2} \psi_n$ as in the spin 1/2 case, so that
$\psi_n$ in (\ref{sigmaf}) is $e^{1/2} \psi_n$ in (\ref{sugra11});
\item take the gauge choice $\psi'_0=0$ (\ref{redef});
\item take the spatial gradient of the fermionic
fields equal to zero (these gradients would appear at higher
levels); \item assume that the spatial metric is (at that order)
spatially homogeneous (\ie neglect its spatial gradients in the
adapted frames) and that the structure constants $C^a_{\; \;bc} = -
C^a_{\; \; cb}$ of the homogeneity group are traceless (to match the
level 3 representation),
$$C^a_{\; \; ac}=0.$$
\end{itemize}
We have also verified that the matrix $M$ is indeed invariant up
to that level.

As for the spin 1/2 case, the matching between (\ref{sigmaf}) and
(\ref{sugra11}) fully covers level $3$ under the above condition of
tracelessness of $C^a_{\; \;bc}$, including the imaginary roots. For
the spin 1/2 case, this is rather direct since the null root part
vanishes, but this part does not vanish for the spin 3/2. However,
the dictionary of \cite{Damour:2002cu} is reliable only for extended $E_8$.

{}Finally, we recall that the covariant derivative of the
supersymmetry spin 1/2 parameter is also identical with the
$K(E_{10(10)})$ covariant derivative up to level 3 \cite{deBuyl:2005zy}, so
that the supersymmetry transformations are $K(E_{10(10)})$ covariant.
The $K(E_{10(10)})$ covariance of the supersymmetry transformations
might prove important for understanding the $K(E_{10(10)})$ covariance
of the diffeomorphisms, not addressed previously.  Information on
the diffeomorphisms would follow from the fact that the graded
commutator of supersymmetries yields diffeormorphisms
(alternatively, the supersymmetry constraints are the square roots
of the diffeomorphisms constraints \cite{Teitelboim:1977fs}).

\section{Conclusions}
In this chapter, we have shown that the gravitino field of
11--dimensional supergravity is compatible with the conjectured
hidden $E_{10(10)}$ symmetry up to the same level as in the bosonic
sector. More precisely, we have shown that the fermionic part of the
supergravity Lagrangian take the form (\ref{covariant}) with the
correct $K(E_{10(10)})$ covariant derivatives as long as one considers
only the connection terms associated with the roots in extended
$E_8$, for which the dictionary relating the bosonic supergravity
variables to the sigma--model variables has been established. The
computations are to some extent simpler than for the bosonic sector
because they involve no dualisation.  In sigma--model terms, the
supergravity action is given by the (first terms of the) action for
a spinning particle on the symmetric space $E_{10(10)}/K(E_{10(10)})$, with
the internal degrees of freedom in the `spin 3/2' representation of
$K(E_{10(10)})$ (modulo the 4--fermion terms).

This action takes the same form as the action for a Dirac spinor
with the appropriate Pauli couplings that make it $K(E_{10(10)})$
covariant \cite{deBuyl:2005zy}, where this time the internal degrees of
freedom are in the `spin 1/2' representation. We can thus analyse
its dynamics in terms of the conserved $K(E_{10(10)})$ currents along
the same lines as in chapter \ref{dirac} and conclude that the BKL limit
holds.

Although the work in this chapter is a necessary first step for
checking the conjectured $E_{10(10)}$ symmetry, much work remains to be
done to fully achieve this goal.  To some extent, the analysis
remains a bit frustrating because no really new light is shed on the
meaning of the higher levels.  Most of the computations are
controlled by $E_{8(8)}$ and manifest $sl_{10}$ covariance. In
particular, the imaginary roots, which go beyond $E_8$ and height
29,  still evade a precise dictionary. The works in \cite{Brown:2004jb} and in
\cite{Damour:2005zb} are to our knowledge the only ones where imaginary
roots are discussed and are thus particularly precious and important
in this perspective.

We have treated explicitly the case of maximal supergravity in this
chapter, but a similar analysis applies to the other supergravities,
described also by infinite--dimensional Kac--Moody algebras (sometimes
in non--split forms \cite{Henneaux:2003kk,Fre':2005sr}).

Similar results have been simultaneously derived in reference \cite{Damour:2005zs}.
%%%%%%%%%%%%%%%
%%%%%%%%%%%%%%%
\cleardoublepage
%%%%%%%%%%%%%%%%%
%%%%%%%%%%%%%%%%%%
%%\include{chap_signatures}
\chapter{Dualities and Signatures of $G^{++}$
Invariant Theories} 
\markboth{DUALITIES AND SIGNATURES OF $G^{++}$ INVARIANT THEORIES}{}
\label{signatures}

\begin{figure}[h]
\centering
\includegraphics[scale=0.6]{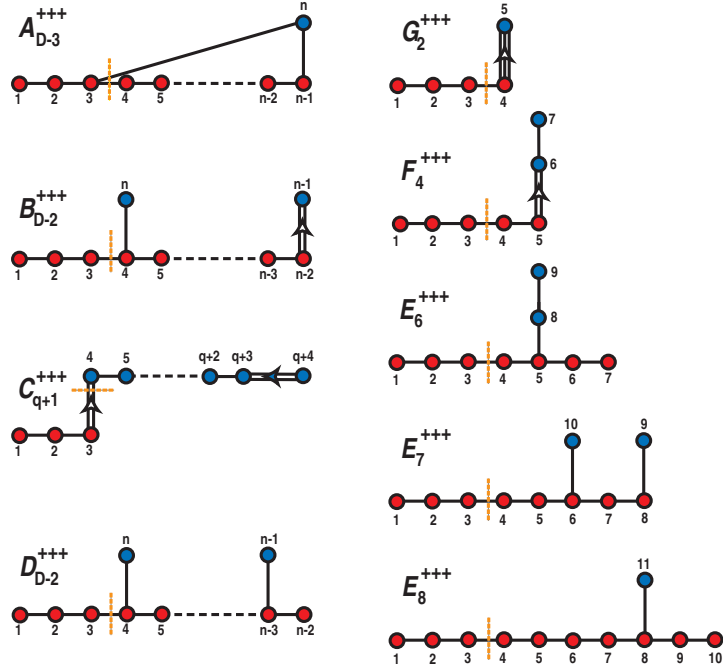}
\caption { \small The
nodes labelled 1,2,3 define the Kac--Moody extensions of the  Lie
algebras. The horizontal line
starting at 1 defines the `gravity line', which is the
Dynkin diagram  of a
$A_{D-1}$ subalgebra.}
\label{veryalg}
\end{figure}

The action of Weyl reflections generated by simple roots not
belonging to the gravity line on the exact extremal brane
solutions has been studied for all $\cG^{+++}$ --theory constructed with
the temporal involution selecting the index 1 as a time coordinate
\cite{Englert:2003py}. The existence of Weyl orbits of extremal brane
solutions similar to the U--duality orbits existing in M--theory
strongly suggests a general group--theoretical origin of
``dualities'' for all $\cG^{+++}$ --theories transcending string theories and
supersymmetry.
The precise analysis of the different possible signatures has been performed for $\cG^{++}_B=E_{8(8)}^{++}=E_{10(10)}$.
In this case the corresponding maximally oxidised theory is the bosonic sector of the  low effective action of M--theory.
For $E_{8(8)}^{++}$ the Weyl reflection $W_{\a_{11}}$ generated by the simple root $\a_{11}$ (see Figure \ref{veryalg}) corresponds to a double T--duality in the directions 9 and 10 followed by an exchange of the
two radii \cite{ Elitzur:1997zn,Obers:1998fb,Banks:1998vs,Englert:2003zs}. 
The signatures found in the analysis of references \cite{Keurentjes:2004bv,Keurentjes:2004xx} and in the context of $S_{\cG^{++}_B}$ in \cite{Englert:2004ph} match perfectly with the signature changing dualities and the exotic
phases of M--theories discussed in \cite{Hull:1998vg,Hull:1998fh,Hull:1998ym}.

In this context it is certainly interesting to extend to all
$\cG^{+++}$--theories the analysis of signature changing Weyl reflections.
This is the purpose of the present chapter.   We find for all the
$\cG^{++}_B$--theories all the possible signature $(T,S)$, where $T$
(resp. $S$) is the number of time--like (resp. space--like)
directions, related by Weyl reflections of $\cG^{++}$ to the signature
$(1,D-1)$ associated to the theory corresponding to the
traditional maximally oxidised theories. Along with the different
signatures the signs of the kinetic terms of the relevant fields
are also discussed. We start the analysis with $A_{D-3}^{++}$
corresponding to pure gravity in $D$ dimensions then we extend the
analysis to the other $\cG^{++}$, first to the simply laced ones and then to the
non-simply laced ones\footnote{
results have been independently obtained in \cite{Keurentjes:2005jw}.}. Each $\cG^{++}$ algebra contains a $A_{D-3}^{++}$
subalgebra, the signatures of $\cG^{++}$ should thus includes the one of
$A_{D-3}^{++}$. This is indeed the case, but some $\cG^{++}$ will
contain additional signatures. If one want to restrict our focus
on  string theory, the special cases of $D_{24}^{++}$ and
$B_{8}^{++}$ are interesting, the former being related to the
low-energy effective action of the bosonic string (without
tachyon) and the latter being related to the low-energy effective
action of the heterotic string (restricted to one gauge field).
The existence of signature changing dualities  are related to the
magnetic roots and suggests that these transformations correspond
to a generalisation of the S--duality existing in these two
theories \cite{Julia:1982gx,Sen:1994fa,Sen:1994wr}.

\section{The signatures of $\cG^{++}_B$--theories}
We will characterise the different signatures of $\cG^{++}_B$   in
terms of $\cG^{+++}$, namely  we will determine all the signatures in the
Weyl orbit of $(1,D-1,\{ \epsilon =+\})$ with the index 1 fixed to be  a
space--like coordinate. We first discuss in detail the
$\cG^{++}_B=A_{D-3}^{++}  \subset A_{D-3}^{+++}$ case. The other $\cG^{+++}$
contain as a subalgebra $A_{D-3}^{+++}$ (there is always a
graviphoton present at some level) consequently the signatures of
all $\cG^{++}_B$ include at least those of $A_{D-3}^{++}$.

\subsection{$A_{D-3}^{+++}$}

\subsubsection{$D>5$}

Our purpose is to determine all $S_{{A}{}_{(i{}_1i{}_2\dots
i{}_t)}^{++}}^{(T,S,\varepsilon)}$ equivalent to $S_{A^{++}_B}$,
i.e. all $\Omega^\prime$ related to $\Omega_2$ via a Weyl
reflection of $A_{D-3}^{++}$ (see Eq.(\ref{newinvolve})). As
explain above, the only Weyl reflections changing  the signature
in a non-trivial way  are the ones generated by simple roots not
belonging to the gravity line. Here there is only one such a
simple root , namely $\a_D$ (see Figure \ref{veryalg}). The  Weyl reflection
$W_{\a_D}$ exchanges the following roots,
\begin{eqnarray}
\label{ex1}
\alpha_{D-1} & \leftrightarrow & \alpha_{D-1} + \a_D \\
 \label{ex2}
 \a_{3} &  \leftrightarrow &  \a_3 + \a_{D}.
 \end{eqnarray}
One can express this Weyl reflection as a conjugaison by a group
element $U_{W_{\a_D}}$ of $A_{D-3}^{+++}$. The non-trivial action
of $U_{W_{\a_D}}$, on the step operators is given by,

\begin{eqnarray}
\label{exc1}
K^{D-1}{}_D & \leftrightarrow  & \sigma  R^{4...D,D-1} \\
\label{exc2}
K^3{}_4 &\leftrightarrow &  \rho R^{35...D,D},
 \end{eqnarray}
where $\sigma$ and $\rho$ are +1 or -1 and the tensor
$R^{a_4...a_{D},a_{D+1}}$ is in the representation\footnote{Here
we adopt the following convention : the   Dynkin labels of the
$A_{D-1}$ representations are labelled  from right to left when
compared with the labelling of the Dynkin diagram of Figure \ref{veryalg}.
For
instance the last label on the right refers to the fundamental
weight associated with the root   labelled 1 in Figure \ref{veryalg}.
In \cite{Kleinschmidt:2003mf}, the opposite convention is used.}
$[1,0,\dots 1,0,0]$
of $A_{D-1}$ that occurs at level one \cite{Kleinschmidt:2003mf}. 
$R^{4...D,D-1}$ is the generator of
$A_{D-3}^{+++}$ associated to the root $\a_{D-1} + \a_D$ and  $
R^{35...D,D}$ the one associated to $ \a_3 + \a_{D}$.

In order to obtain  a change of  signature, we need generically  $\Omega
K^3{}_4 \neq \Omega' K^3{}_4$ and/or $\Omega K^{D-1}{}_D \neq
\Omega' K^{D-1}{}_D$. Using Eqs.(\ref{exc1}), (\ref{exc2}) and Eq.(\ref{sidef}), 
these conditions are equivalent to
$\mathrm{sign} ( \Omega K^3{}_4 ) \neq \mathrm{sign} ( \Omega
R^{35...D,D} ) $ and/or $ \mathrm{sign} ( \Omega K^{D-1}{}_D )=
\mathrm{sign} ( \Omega R^{4...D,D-1} )$. The following
equalities,

\bea \label{signcond} \mathrm{sign} (\Omega  R^{4...D,D} ) &=&-
\mathrm{sign}(\Omega R^{35...D,D}).\mathrm{sign}(\Omega K^3{}_4) \nn \\
&=& -\mathrm{sign} (\Omega R^{4...D,D-1} ).\mathrm{sign}(\Omega
K^{D-1}{}_D),
\eea imply that we have only $\Omega
K^3{}_4 \neq \Omega' K^3{}_4$ \textit{and} $\Omega K^{D-1}{}_D
\neq \Omega' K^{D-1}{}_D$. These inequalities lead to four different possibilities
summarised in Table \ref{con1}.

\begin{table}[h]
\caption{\small  Conditions on $\Omega$'s leading to non-trivial signature
changes under the Weyl reflection $W_{\alpha_D}$}
\begin{center}
\begin{tabular}{|c|c|c|c|}
\hline  & $ \mathrm{sign}(\Omega K^3{}_4$) & $\mathrm{sign}(\Omega
R^{35...D,D})$ & $\mathrm{sign}(\Omega
K^{D-1}{}_D) $  \\
\hline
 a. & + & - & +   \\
 b. & + & - & -  \\
 c. & - & + & +   \\
 d. & - & + & -   \\
\hline
\end{tabular}
\end{center}
\label{con1}
\end{table}
Note that  by  Eq. (\ref{signcond}) the sign of $\Omega R^{4...D,D-1}$ is
deduced from the signs of $ \Omega K^3{}_4$, $\Omega R^{35...D,D}$
and $\Omega K^{D-1}{}_D $. All the signatures in the orbit of $(1,D-1,+)$, where $+$ is the sign
associated to the generatoræ $R^{a_4...a_{D},a_{D+1}}$ defined
by Eq.(\ref{signa}), are now derived.

\begin{itemize}
\item \textbf{Step 1}: Let us first consider a $\Omega$ characterised by a signature $(T,S,+)$  where $T$ is
odd. We consider this set of signatures because it contains our starting point
$(1,D-1,+)$ and also because other signatures of this type will be useful for the
recurrence, i.e. the signature obtained from $(1,D-1,+)$
will lead in some cases to new signatures of the general type
$(T,S,+)$ where $T$ is odd. We analyse the different possibilities of Table \ref{con1}.

\begin{itemize}
\item a. The coordinates $3$ and $4$ are of  different nature as well as the coordinates $D-1$ and $D$. Moreover they are an even
number of time coordinates in the set $\{ 3,5,...,D-1\}$ as a direct consequence of the sign $-$ in the second column of
Table \ref{con1} and the fact that $R^{a_4...a_{D},a_{D+1}}$ satisfy Eq.(\ref{signa}) . So they are an odd number of time coordinates in
the complementary subset $\{1,2,4,D\}$, i.e 1 or 3. These
conditions lead to the possibilities given in Table
\ref{gravity1}.

\begin{table}[h]
\caption{\small $\Omega$'s leading to  signature changes under the
Weyl reflection $W_{\alpha_D}$.  Space--like (resp. time--like)
coordinate are denoted by s (resp. t)}
\begin{center}
\begin{tabular}{|c|ccc|ccc|c|}
\hline &       1 & 2 & 3 & 4 & ... & D-1 & D \\
\hline a.1. &  s & t & t & s & ... & t & s \\
       a.2. &  s & s & s & t & ... & t & s \\
       a.3. &  s & s & t & s & ... & s & t \\
       a.4. &  s & t & s & t & ... & s & t \\
\hline
 \end{tabular}
\end{center}
 \label{gravity1}
\end{table}
 The nature of the coordinates 1, 2, 3 and $D$ (resp.
4,...,$D-1$) does not (resp. does) change under the action of the
Weyl reflection generated by $\alpha_{D}$ on $\Omega$'s satisfying
the conditions a  given in Table 3. We must also determine  which
sign Eq.(\ref{signa}) or Eq.(\ref{signb}) characterises
$R^{a_1...a_{D-3},a_{D-2}}$ under the action of the conjugated
involution $\Omega^\prime$ given by Eq.(\ref{newinvolve}). Using
Eq.(\ref{kisi}) one has $\mathrm{sign}(\Omega^\prime R^{4...D,D}) =
\mathrm{sign}(\Omega R^{4...D,D}) = -1$ because there is an odd number of time
coordinates in $\{4,...,D-1\}$ before Weyl reflection. The nature
of all these coordinates changes under the action of $W_{\a_D}$.
Therefore if $D$ is even, we have an odd number of time
coordinates in this subset and $\mathrm{sign}(\Omega^\prime
R^{4...D,D})$ satisfies Eq.(\ref{signa}). If $D$ is odd we get an
even number of time coordinates yielding to the other sign
Eq.(\ref{signb}). Therefore, the action of $W_{\alpha_D}$ on the
$\Omega$'s characterised by the signatures of Table \ref{gravity1}
yields $\Omega^\prime$'s characterised by the signatures given in
Table \ref{newsigna}.
\begin{table}[h]
\caption{\small  $\Omega^\prime$'s obtained by the Weyl reflection
$W_{\a_D}$ from $\Omega$'s given in Table \ref{gravity1}}
\begin{center}
\begin{tabular}{|c|l|c|}
\hline &        signature $\Omega'$ & conditions on $\Omega$ \\
\hline a.1. &
            $(S,T,(-)^D)$ & $T \geq 3 $ and $S\geq 3$ \\
       a.2. &
            $(S-4,T+4,(-)^D)$ & $T \geq 3 $ and $S\geq 4$ \\
       a.3. &
            $(S,T,(-)^D)$ & $T \geq 3 $ and $S\geq 4$ \\
       a.4.  &
            $(S,T,(-)^D)$ & $T \geq 3 $ and $S\geq 3$ \\
\hline \end{tabular}
\end{center}
\label{newsigna} \end{table}

\item b. We get the same possibilities as those of case a (see Table \ref{gravity1})  except for the
coordinate $D-1$ which is different. Therefore the new signatures
will be the same. Only the conditions on $S$ and $T$ can differ.
These conditions are given in Table \ref{newsignb}.
\begin{table}[h]
\caption{\small  $\Omega^\prime$'s obtained by the Weyl reflection
$W_{\a_D}$ from $\Omega$'s given in Table \ref{gravity1} with the
nature of the coordinate $D-1$ changed }
\begin{center}
\begin{tabular}{|c|c|c|}
\hline &          signature $\Omega'$ & conditions on $\Omega$ \\
\hline b.1. & $(S,T,(-)^D)$ & $T \geq 3 $ and $S\geq 4$ \\
       b.2. & $(S-4,T+4,(-)^D)$ & $T\geq 1 $ and $S\geq 5$ \\
       b.3. &  $(S,T,(-)^D)$ & $T \geq 3 $ and $S\geq 3$ \\
       b.4. &  $(S,T,(-)^D)$ & $T \geq 5$ and $S\geq 2$ \\
\hline \end{tabular}
\end{center}
\label{newsignb}
\end{table}

\item c. The coordinates $3$ and $4$ are of the same nature,  $D-1$ and $D$ are of different
nature.  Moreover there are an odd number of time coordinates in $\{ 3,5,...,D-1\}$ and therefore an even number of time
coordinates in the complementary subset $\{1,2,4,D\}$, i.e 0, 2 (1
being always space--like in $\cG^{++}_B$). These conditions lead to the
possibilities given in Table \ref{gravity3}. The new signatures
are given in Table \ref{newsignc} (the sign for
$R^{a_1...a_{D-3},a_{D-4}}$ is determined by applying a similar
reasoning as the one developed in case a).

\begin{table}[h]
\caption{\small $\Omega$'s leading to  signature changes under
the Weyl reflection $W_{\alpha_D}$ in case c. There is an odd number of time coordinates in
the subset $\{4,...,D-1\}$}
\begin{center}
\begin{tabular}{|c|ccc|ccc|c|}
\hline &       1 & 2 & 3 & 4 & ... & D-1 & D \\
\hline c.1. &  s & s & s & s & ... & t & s \\
       c.2. &  s & t & t & t & ... & t & s \\
       c.3. &  s & t & s & s & ... & s & t \\
       c.4. &  s & s & t & t & ... & s & t \\
\hline \end{tabular}
\end{center}
\label{gravity3} \end{table}

\begin{table}[h]
\caption{\small    $\Omega^\prime$'s obtained by the Weyl reflection
$W_{\a_D}$ from $\Omega$'s given in Table \ref{gravity3}}
\begin{center}
\begin{tabular}{|c|c|c|}
\hline &          signature $\Omega'$ & conditions on $\Omega$ \\
\hline c.1. & $(S-4,T+4,(-)^D)$ & $T \geq 1 $ and $S\geq 5$ \\
       c.2. & $(S,T,(-)^D)$ & $T\geq 5 $ and $S\geq 2$ \\
       c.3. &  $(S,T,(-)^D)$ & $T \geq 3 $ and $S\geq 4$ \\
       c.4. &  $(S,T,(-)^D)$ & $T\geq 3$ and $S\geq 3$ \\
\hline \end{tabular}
\end{center}
\label{newsignc}
\end{table}

\item d. We get the same signature as those of case c (see Table \ref{gravity3}) except for
the nature of the coordinate $D-1$. Again, only the conditions on
$T$ and $S$ can differ. The new signatures are given in Table
\ref{newsignd}.

\begin{table}[h]
\caption{\small  $\Omega^\prime$'s obtained by the Weyl reflection
$W_{\a_D}$ from $\Omega$'s given in Table \ref{gravity3} with the
nature of the coordinate $D-1$ changed }
\begin{center}
\begin{tabular}{|c|c|c|}
\hline &          signature $\Omega'$ & conditions on $\Omega$ \\
\hline d.1. & $(S-4,T+4,(-)^D)$ & $T \geq 1 $ and $S\geq 6$ \\
       d.2. & $(S,T,(-)^D)$ & $T \geq 3 $ and $S\geq 3$ \\
       d.3. &  $(S,T,(-)^D)$ & $T \geq 3 $ and $S\geq 3$ \\
       d.4. &  $(S,T,(-)^D)$ & $T \geq 5$ and $S\geq 2$ \\
\hline \end{tabular}
\end{center}
\label{newsignd}
\end{table}

\end{itemize}
\noindent \textbf{Summary}: From a signature $(T,S,+)$ with $T$
odd, we can
 reach the signatures
\bea (S-4,T+4,(-)^D) \ \ & & \mathrm{if} \ \ \{ \ T \geq 3 \
\mathrm{and} \ S \geq
4 \} \ \mathrm{or} \ \{T \geq 1 \ \mathrm{and} \ S \geq 5 \} \nn \\
\label{cong}
(S,T,(-)^D) \ \ & & \mathrm{if} \ \  \{ T\geq 5 \ \mathrm{and} \ S
\geq 2 \} \ \mathrm{or} \ \{ T \geq  3 \ \mathrm{and} \ S \geq 3
\}. \label{cond1} \eea
In order to find the Weyl orbits of $(1,D-1,+)$ we need to distinguish between  $D$ even and $D$ odd.

\textit{\textbf{$D$ even}}: The conditions given by Eq.
(\ref{cond1}) simplify to
\bea (S-4,T+4,+) \ \ & & \mathrm{if} \ \  \{T \geq 1 \ \mathrm{and} \ S \geq 5 \} \nn \\
\label{condsi}(S,T,+) \ \ & & \mathrm{if} \  \ \{ T \geq  3 \ \mathrm{and} \ S
\geq 3 \}. \label{cond2} \eea

If we start from $(1,D-1,+)$,  after the action of  $W_{\a_D}$ one
gets $(D-5,5,+)$ which is of the
generic type $(T,S,+)$ with $T$ odd furthermore all the other
signatures that we reach given by Eq.(\ref{condsi}) are of this
type. Therefore we can use the above analysis  and  taking into
account the conditions Eq.(\ref{condsi}) we conclude that when $D$
is even, the signatures in the $W_{\a_D}$ orbit of $(1,D-1,+)$ for
$\cG^{++}_B =A_{D-3}^{++}$  are given by ($n$ is an integer)

\bea (1,D-1,+) \nn \\
(1+4n , D-1-4n,+) \  && 3 \leq 4n+1 \leq D-3 \nn \\
\label{condev}
( D-1-4n, 1 + 4n , +) \ &&  5 \leq 4n+1 \leq D-1. \eea

\textit{\textbf{$D$ odd}}: After the action of  $W_{\a_D}$ the
signature $(D-5,5,-)$ is not of the type $(T,S,+)$ with $T$ odd.
To determine the orbit of $(1,D-1,+)$, we have thus to analyse the
signatures of the form $(T,S,-)$ with $T$ even and $S$ odd.

\item \textbf{Step 2}: Let us consider an  involution $\Omega$ characterised
by a signature $(T,S,-)$ where $T$ is
even and $S$ is odd ($D$ is odd). The discussion of the different
possible cases of signature change is similar to the ones
discussed in step 1. Indeed, the  even number
of time coordinates balances the   minus sign
 for the kinetic term of
$R^{a_1...a_{D-3},a_{D-4}}$ (see Eq.(\ref{signb})). We must be careful as far as the
conditions on $T$ and $S$ are concerned. The only difference with
the Tables \ref{newsigna}, \ref{newsignb}, \ref{newsignc},
\ref{newsignd} is that we have for $D$ odd the opposite parity for
the number of times in the coordinates $\{4,...,D-1\}$. Therefore,
starting with $(T,S,-)$ we will reach the signature $(S,T,+)$
under the following conditions, \bea (S,T,+) \ \ & & \mathrm{if} \
\  \{ T\geq 2 \ \mathrm{and} \ S \geq5 \} \ \mathrm{or} \ \{ T
\geq  4 \
\mathrm{and} \ S \geq 3 \}. \label{condd2} \eea
The sign + given by Eq.(\ref{signa})  for $\Omega^\prime
R^{a_1...a_{D-3},a_{D-4}}$ is obtained by a reasoning similar to the one
given below Table \ref{gravity1} taking into account the fact that
the number of dimensions is odd.

The new signature Eq.(\ref{condd2}) is of the type considered in
step 1. The conditions for getting new signatures by acting
again with $W_{\a_D}$ on this signature can thus be deduced from Eq.(\ref{cong}).
Starting from $(1,D-1,+)$ the step 1 gives us
$(D-5,5,-)$, then step 2 gives us $(5,D-5,+)$, step 1 can be used
again to obtain $(D-9,9,-)$. Repeating the argument, all the new
signatures are obtained using ``step 1'' or ``step 2''.

We conclude that when $D$ is odd, the following signatures can be
reach,
\bea (1+4n , D-1-4n,+) \  && 1 \leq 4n+1 \leq D-2 \nn \\
\label{condodd} ( D-1-4n, 1 + 4n ,-) \ &&  1< 4n+1 \leq D. \eea

\end{itemize}

\noindent We can rewrite Eq.(\ref{condev}) and Eq.(\ref{condodd})
in a more concise way and conclude that for {\it all} $D$ (odd and
even) the signatures of $\cG^{++}_B$ in the Weyl orbit of $(1,D-1,+)$
are given by

\bea (1+4n , D-1-4n,+) \  && 0 \leq n \leq [ \frac {D-3}{4} ] \nn \\
\label{confin}
( D-1-4n, 1 + 4n ,  (-)^D) \ &&  1< n \leq   [ \frac {D-1}{4} ]  ,
\eea where $n$ is an integer and $[x]$ is the integer part of $x$.
\subsubsection{$D=5$}

The reasoning of the previous section cannot apply for $D=5$
because the conditions on $\Omega$ to obtain new signatures
$\Omega^\prime$ given in Table \ref{con1} assumed $4< D-1$. These
conditions give in this case  $\mathrm{sign} (\Omega K^3{}_4 )\neq
\mathrm{sign}(\Omega R^{35,5})$ \textit{and} $\mathrm{sign}
(\Omega K^4{}_5) \neq \mathrm{sign}(\Omega R^{45,4})$. Starting
from a signature $(1,4,+)$, this implies that the coordinate 4
must be the time. We get by acting with the Weyl reflection
generated by $\alpha_5$ the signature $(0,5,-)$. Consequently all the
possible signatures for $D=5$ are

\bea (1,4,+) \ (0,5,-) .\eea

\subsubsection{$D=4$}

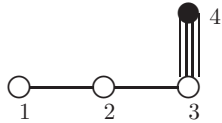
\begin{figure}[h]
\caption{ \small Dynkin diagram of $A_1^{+++}$}
\begin{center}
\scalebox{.8}{
\begin{picture}(180,60)
%nom des racines
\put(0,-5){$1$} \put(40,-5){$2$} \put(80,-5){$3$} \put(90,40){$4$}
%premiers vertex
\thicklines \multiput(0,10)(40,0){3}{\circle{10}}
% lignes entre les vertex
\multiput(5,10)(40,0){2}{\line(1,0){30}}
%1 vertex du dessus
\put(80,45){\circle*{10}}
%ligne vers le haut
\put(76,15){\line(0,1){30}} \put(79,15){\line(0,1){29}}
\put(82,15){\line(0,1){30}} \put(85,15){\line(0,1){30}}
\end{picture}
}
\end{center}
\label{second}
\end{figure}

The Dynkin diagram of $A_1^{+++}$ is depicted in Fig.\ref{second}.
To get a signature change due to the action of the Weyl reflection
generated by $\alpha_4$ we need sign($\Omega K^3{}_4$) $\neq$
sign($\Omega R^{3,4}$) which is not possible since we started from
$(1,3,+)$. The symmetric tensor $R^{a,b}$ is the representation
[2,0,0] of $A_3$ that occurs at level  one. The only possible
signature is the Minkowskian one, \bea (1,3,+) .\eea

\subsection{$D_{D-2}^{+++}$}

\subsubsection{$D>6$}
\noindent They are two simple roots not belonging to the gravity line,
namely $\a_D$ and $\a_{D+1}$ (see Figure \ref{veryalg}). Given a
signature $(T,S,\varepsilon)$, the first sign in the set
$\varepsilon$ is associated to the generator\footnote{The tensor
$R^{a_1a_2}$ is in the representation $[0,1,0,\dots,0,0]$ of
$A_{D-1}$ that occurs at level $(1,0)$, the tensor $R^{a_5...a_D}$
is in the representation $[0,\dots,0,1,0,0,0]$ of $A_{D-1}$ that
occurs at level $(0,1)$ \cite{Kleinschmidt:2003mf}. The level $(l_1,l_2)$ of a
root $\a$ of $D_{D-2}^{+++}$ can be read in its
 decomposition  in terms of the simple roots $\a = m_1 \a_1+... + m_{D-1} \a_{D-1} + l_1 \a_D + l_2
\a_{D+1}$.  } $R^{a_1a_2}$ and the second one to the generator
$R^{a_5...a_D}$ (see Eqs. (\ref{signa}) and (\ref{signb})). We will analyse the possible signature changes
due to $W_{\a_D}$ and $W_{\a_{D+1}}$.

1) The Weyl reflection generated by the root $\a_{D}$ will never
\textit{non trivially} change the signature. Indeed, its
 action on the simple root $\a_{D-2}$ and the corresponding
action of $U_{W_{\a_D}}$ on the simple generator  $R^{D-2\, D}$ are,

\begin{center}
\begin{tabular}{lrcllrcl}
$W_D$:& $\a_{D-2}$ & $ \leftrightarrow$ & $ \a_{D-2} + \a_{D}$ &
$U_{W_D}$: & $K^{D-2}{}_{D-1}$ & $\leftrightarrow$ & $\sigma
R^{D-2D}.$
\end{tabular} \end{center}

\noindent Therefore to obtain a change of signature one needs (see Eq.(\ref{newinvolve})):
$\mathrm{sign}( \Omega K^{D-2}{}_{D-1}) \neq \mathrm{sign}( \Omega
R^{D-2D})$. We have
$$\mathrm{sign} ( \Omega R^{D-1D} )=
-\mathrm{sign}(\Omega R^{D-2D}) . \mathrm{sign}(\Omega
K^{D-2}{}_{D-1}),$$  thus $\mathrm{sign}(\Omega R^{D-1D}) = +$
implies that the coordinates $D-1$ and $D$ are of different nature because $R^{D-1\, D}$
satisfies Eq.(\ref{signa}). So even if
$\Omega'K^{D-2}{}_{D-1} \neq \Omega K^{D-2}{}_{D-1}$ we will not
have a \textit{non trivial} signature change but just an exchange
between the nature of the coordinates $D-1$ and $D$.

2) We now consider the action of the Weyl reflection generated
by the root $\a_{D+1}$. Its action on $\a_4$ is
\begin{center}
\begin{tabular}{lrcllrcl}
$W_{D+1}$:& $\a_{4}$ & $ \leftrightarrow$ & $ \a_4 + \a_{D+ 1}$ &
$U_{W_{D+1}}$: & $K^4{}_5$ & $\leftrightarrow$ & $ \rho
R^{46...D}.$
\end{tabular} \end{center}

\noindent  In order to have a signature change, we need sign($ \Omega K^4{}_5)
\neq$ sign($ \Omega
 R^{46...D})$. This leads to two possibilities explicitly given in
 Table \ref{conddnD}. The condition \textbf{A} means that the nature of the coordinates
 4 and 5 are different \textit{and} that if we start from $( T
, S, \pm , +)$ (resp. $(T  ,S ,\pm ,-)$) there is an even (resp. odd)
number of times in $\{4,6,...,D\}$ . Whereas condition \textbf{B}
means that that the nature of the coordinates
 4 and 5 are the same \textit{and} that if we start from $( T , S , \pm  ,
+)$ (resp. $(T ,S ,\pm ,-)$) there is an odd (resp. even) number of
times in $\{4,6,...,D\}$.

\begin{table}
\caption{\small  Conditions on $\Omega$'s leading to non-trivial
signature change under the Weyl reflection $W_{\alpha_D}$ for
$S_{D_{D-2}^{++}}$ }
\begin{center}
\begin{tabular}{|c|cc|}
\hline & sign($\Omega K^4{}_5$) & sign($\Omega R^{46...D}$) \\
\hline \textbf{A} &+ & - \\
 \textbf{B}  &- & + \\
\hline
\end{tabular}
\end{center}
\label{conddnD} \end{table}

\begin{itemize} \item By analogy with the $A_{D-3}^{+++}$ case, we start from the signature $(T,S,+,+)$ where $T$ is an odd number.

\textbf{A.} There is an odd number of times in $\{1,2,3,5\}$, i.e
1 or 3. These time coordinates can be distributed as in Table
\ref{newsignAdn}. The new
signatures $\Omega^\prime$, also given in Table \ref{newsignAdn},
are deduced from the fact that the action of the Weyl reflection
generated by $\alpha_{D+1}$ will change the nature of the
coordinates greater than 4.

\begin{table}
\caption{ \small $\Omega$'s leading to  signature changes under
the Weyl reflection $W_{\alpha_{D+1}}$ and the related new signatures
$\Omega^\prime$'s (in the case \textbf{A}).}
\begin{center}
\begin{tabular}{|c|ccccc|c|c|c|}
\hline   & 1 & 2 & 3 & 4 & 5  &$\Omega^\prime$ & $T\geq $ & $S \geq$\\
 \hline i. & s & s& t & t & s  &$(S,T ,+,(-)^D)$& 3 & 3\\
 ii. & s & s & s & s & t  & $(S-4, T + 4,+,(-)^D)$ & 1 & 4\\
 iii. & s & t & t & s & t  &$(S,T ,+,(-)^D)$& 3 & 2\\
\hline \end{tabular}
\end{center}
\label{newsignAdn}
\end{table}

\textbf{B.} There is an even number of times in $\{1,2,3,5\}$ i.e
0, 2 or 4. 4 is excluded here because we want that 1 is
space--like. These time coordinates can be distributed as in Table
\ref{newsignBdn}. The new signatures are also given in Table
\ref{newsignBdn}.

\begin{table}
\caption{$\Omega$'s leading to  signature changes under the Weyl
reflection $W_{\alpha_{D+1}}$ and the related new signatures
$\Omega^\prime$'s (in the case \textbf{B}).}
\begin{center}
\begin{tabular}{|c|ccccc|c|c|c|}
\hline   & 1 & 2 & 3 & 4 & 5  & $\Omega^\prime$ & $T\geq$& $S \geq$\\
\hline i. & s & s & s & s & s  &$(S-4 , T + 4,+,(-)^D)$& 1  & 5\\
 ii. & s & s & t & t & t & $(S, T ,+,(-)^D)$& 3 & 2\\
 iii. & s & t & t & s & s  & $(S, T ,+,(-)^D)$& 3 & 3\\
\hline \end{tabular}
\end{center}
\label{newsignBdn}
\end{table}

\noindent $\Rightarrow$ if $D$ is even, from a signature
$(T,S,+,+)$ we can reach the signatures $(S,T,+,+)$ (if $T \geq 3$
and $S \geq 3$) and $(S-4,T+4,+,+)$ (if $T \geq 1$ and $S \geq 5$
).

\noindent $\Rightarrow$ if $D$ is odd, from a signature
$(T,S,+,+)$ we can reach the signatures $(S,T,+,-)$  (if $T\geq 3
$ and $S \geq 2$) and $(S-4,T+4,+,-)$ (if $T \geq 1$ and $S \geq
4$).

\item From the signature $(T,S,+,-)$ where $T$
is an even number and $S$ odd ($D$ is odd). By the same procedure,
we can conclude that the signatures $(S,T,+,+)$ can be reached (if
$T\geq 2 $ and $S \geq 3$) and the signatures $(S-4,T+4,+,+)$ (if
$S \geq 5$).

\end{itemize}

\noindent \textbf{Summary}: We get the same signatures as the ones
of pure gravity as expected since $W_{\a_D}$ does not change the
signature and only $W_{\a_{D+1}}$ has a non-trivial action,
\bea (1+4n , D-1-4n,+,+) \  && 0 \leq n \leq [ \frac {D-3}{4} ] \nn \\
( D-1-4n, 1 + 4n , +, (-)^D) \ &&  1< n \leq   [ \frac {D-1}{4} ]
, \eea where $[x]$ is the integer part of $x$.
We could have acted with the Weyl reflection generated by the graviphoton lying at level $(1,1)$ instead of acting with $W_{\a_{D+1}}$ and we would have obtained the same results. The sign of the kinetic term of the graviphoton
agrees with Eq.(\ref{confin}).

\subsubsection{$D=6$}
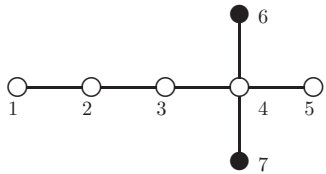
\begin{figure}[h]
\caption{ \small Dynkin diagram of $D_4^{+++}$}
\begin{center}
\scalebox{.7}{
\begin{picture}(180,60)
%non
\put(5,-5){1} \put(45,-5){2} \put(85,-5){3} \put(140,-5){4}
\put(165,-5){$5$} \put(140,45){$6$}\put(140,-35){$7$} \thicklines
%les vertex
\multiput(10,10)(40,0){5}{\circle{10}}
%les lignes
\multiput(15,10)(40,0){4}{\line(1,0){30}}
%un vertex du dessus
\put(130,50){\circle*{10}} \put(130,15){\line(0,1){30}}
%un vertex en dessous
\put(130,-30){\circle*{10}} \put(130,5){\line(0,-1){30}}
\end{picture}
}
\end{center}
\label{third}
\end{figure}

\noindent They are two simple roots out of the gravity line,
namely $\a_6$ and $\a_{7}$ (see Figure \ref{third}).
 Given a
signature $(T,S,\varepsilon)$, the first sign in the set
$\varepsilon$ is associated to the generator $R^{a_1a_2}$ and the second one to the generator
$\tilde{R}^{a_1a_2}$.

We can immediately conclude that no signature changes are
possible. Indeed the Weyl reflections generated by  the two simple
roots not belonging to the gravity line cannot change the signature in the
same way as the Weyl reflection $W _{\a_D}$ cannot do it in the
previous section.

\subsection{$E_6^{+++}$}

They are two simple roots not belonging to the gravity line,
$\a_8$ and $\a_9$ (see Figure \ref{veryalg}). The non trivial actions of the corresponding Weyl
reflections $W_8$, $W_9$ are

\begin{center}
\begin{tabular}{crclcrcl}
 $W_8$:  & $\a_{5}$ & $ \leftrightarrow$ & $ \a_5 + \a_{8}$& $U_{W_8}$: & $K^5{}_6$ &
$\leftrightarrow$ & $ \rho R^{578}$  \\
         & $\a_{9}$ & $ \leftrightarrow$ & $ \a_9 + \a_{8}$&  & $R$ &
$\leftrightarrow$ & $ \sigma \tilde{R}^{678}$ \\
 $W_9$:  & $\a_{8}$ & $ \leftrightarrow$ & $ \a_8 + \a_{9}$& $U_{W_9}$: & $ R^{678}$&
$\leftrightarrow$ & $ \delta \tilde{R}^{678}$.   \end{tabular}
\end{center}
The tensor  $R^{abc}$ is the representation
$[0,0,1,0,0,0,0]$ of $A_{7}$ that occurs at level\footnote{The
level $(l_1,l_2)$ of a root $\a$ of $E_6^{+++}$ can be read in its
 decomposition  in terms of the simple roots $\a = m_1 \a_1+... + m_7 \a_7 + l_1 \a_8 + l_2
\a_9$. } $(1,0)$, the tensor $R$ is the representation
[0,0,0,0,0,0,0] of $A_7$ that occurs at level $(0,1)$ and $
\tilde{R}^{abc}$ is the representation $[0,0,1,0,0,0,0]$ of $A_7$
that occurs at level $(1,1)$  \cite{Kleinschmidt:2003mf}. The sign of the kinetic term of
$\tilde{R}^{abc}$ can be deduced from the ones of $R^{abc}$ and
$R$, it is the product of these two signs.

The Weyl reflection generated by $\a_9$ exchanges the signs of
$R^{abc}$ and $\tilde{R}^{abc}$.

To obtain a signature change from $(1,7,+,+)$\footnote{The first
sign characterises  $R^{abc}$ and the second one characterises $R$ (see Eqs.(\ref{signa}) and (\ref{signb})).}, we must act with
the Weyl reflection generated by $\alpha_8$ and there  must be an
odd number of time coordinates in the following subset: 6,7,8. In
all of this cases, the new signature is $(2,6,-,-)$. Now we can
start from this new signature and act with the Weyl reflection
generated by $\alpha_9$ to get the signature $(2,6,+,-)$ or with
the one generated by $\alpha_8$ to get $(5,3,+,+)$. Acting again with the Weyl reflections generated
by  $\alpha_8$ (vertical arrows) and $\alpha_9$ (horizontal arrows) on
these signatures,  we can conclude that all the signatures are

\begin{center}
\begin{tabular}{ccc}
$(1,7,+,+)$ & & \\
$\downarrow$ & & \\
$(2,6,-,-)$ & $\rightarrow$ & $(2,6,+,-)$ \\
$\downarrow$ & & $\downarrow$\\
$(5,3,+,+)$ & & $(3,5,-,+)$ \\
$\downarrow$ & & $\downarrow$\\
$(6,2,-,-)$ & $\rightarrow$ & $(6,2,+,-).$ \\
\end{tabular}
\end{center}
Theses signatures are the expected ones from the gravity, i.e.
$(1,7,+,+)$, $(3,5,-,+)$ and $(5,3,+,+)$, plus new ones.

\subsection{$E_7^{+++}$}

They are two simple roots not belonging to the gravity line, $\a_9$ and
$\a_{10}$. The non trivial actions of the Weyl reflections $W_9$,
$W_{10}$ are

\begin{center}
\begin{tabular}{rrclrrcl}
 $W_9$:  & $\a_{8}$ & $ \leftrightarrow$ & $ \a_8 + \a_{9}$& $U_{W_9}$: & $K^8{}_9$ &
$\leftrightarrow$ & $ \rho R^{8}$  \\
 $W_{10}$:  & $\a_{6}$ & $ \leftrightarrow$ & $ \a_6 + \a_{10}$& $U_{W_{10}}$: & $ K^{6}{}_7$&
$\leftrightarrow$ & $ \delta R^{689}.$   \end{tabular}
\end{center} The tensor  $R^{a}$ is the representation
$[1,0,0,0,0,0,0,0]$ of $A_{8}$ that occurs at level\footnote{The
level $(l_1,l_2)$ of a root $\a$ of $E_7^{+++}$ can be read in its
 decomposition  in terms of the simple roots $\a = m_1 \a_1+... + m_7 \a_7 + m_8 \a_8 +
 l_1
\a_9 + l_2 \a_{10}$. }  $(1,0)$, the tensor $R^{abc}$ is the
representation [0,0,1,0,0,0,0,0] of $A_8$ that occurs at level
$(0,1)$.

 The Weyl reflection generated by the
root $\alpha_9$ will change the signature if  $\mathrm{sign}
(\Omega K^8{}_9) \neq \mathrm{sign}(\Omega R^8)$. The Weyl
reflection generated by the root $\alpha_{10}$ will change the
signature if $\mathrm{sign} (\Omega K^6{}_7) \neq
\mathrm{sign}(\Omega R^{689})$. With these rules we get the
following signatures (a horizontal arrow represents the action of
the Weyl reflection generated by $\alpha_{10}$ and a vertical one
the action of the Weyl reflection generated by $\a_9$),

\begin{center}
\begin{tabular}{cccccccc}
$(1,8,+,+)$ & $\rightarrow$ & $(2,7,-,-)$ & $\rightarrow$ &
$(5,4,+,+)$ & $\rightarrow$& $(6,3,-,-)$\\
$\downarrow$ & & $\downarrow$ & & $\downarrow$ & & $\downarrow$ \\
$(0,9,-,-)$ & $\rightarrow$ & $(3,6,+,+)$ & $\rightarrow$ &
$(4,5,-,-)$ & $\rightarrow$& $(7,2,+,+).$ &
\end{tabular}
\end{center} The signs
in the above signatures refer to the kinetic terms of $R^a$ and
$R^{abc}$ (see Eqs.(\ref{signa}) and (\ref{signb})).

\subsection{$E_8^{+++}$}

The signatures are given in \cite{Englert:2004ph}. To be complete, we recall
them here, \bea (1,10,+) \  \rightarrow \ (2,9,-) \ \rightarrow \
(5,6,+) \ \rightarrow (6,5,-) \ \rightarrow (9,2,+). \eea The sign
refers to the sign of the kinetic term of $R^{abc}$ which is the
representation [0,0,1,0,0,0,0,0,0,0] of $A_{10}$ that occurs at
level 1.

\section{Non simply laced algebras}
\subsection{$B_{D-2}^{+++}$}

The only non trivial action of the Weyl reflection generated by the short root
$\a_D$ is

\begin{center}
\begin{tabular}{rrclrrcl}
 $W_D$:  $\a_{D-1}$ & $\leftrightarrow$ & $\a_{D-1} + 2
\a_D$ & $U_{W_D}$: $K^{D-1}{}_D$ & $\leftrightarrow$ & $\sigma
R^{D-1D}.$   \end{tabular}
\end{center} The tensor  $R^{ab}$ is the representation
$[0,1,0,,\dots,0]$ of $A_{D-1}$ that occurs at
level\footnote{The level $(l_1,l_2)$ of a root $\a$ of
$B_{D-2}^{+++}$ can be read in its
 decomposition  in terms of the simple roots $\a = m_1 \a_1+...  + m_{D-1} \a_{D-1} +
 l_1
\a_D + l_2 \a_{D+1}$. } $(2,0)$ \cite{Kleinschmidt:2003mf}. To get a signature
change with this Weyl reflection we need $\mathrm{sign}( \Omega
K^{D-1}{}_D ) \neq \mathrm{sign} ( \Omega R^{D-1D}) $ which is
impossible since the sign of the kinetic term of $R^{ab}$ is given by Eq.(\ref{signa}).

We are therefore left with one simple long root, namely $\a_{D+1}$. The Weyl reflection $W_{\a_{D+1}}$ will
clearly produce the same signature changes as $W_{\a_{D+1}}$ does
for $D_{D-2}^{+++}$. Therefore all possible signatures are the
ones found for $D_{D-2}^{+++}$, i.e the signatures of pure
gravity.
\bea (1+4n , D-1-4n,+,+) \  && 0 \leq n \leq [ \frac {D-3}{4} ] \nn \\
( D-1-4n, 1 + 4n , +, (-)^D) \ &&  1< n \leq   [ \frac {D-1}{4} ]
, \eea where $[x]$ is the integer part of $x$. The first sign
refers to the kinetic term of $R^{a}$ and the second to the one of
$R^{a_5...a_D}$. $R^a$ is the representation [1,0,...,0] of
$A_{D-1}$ that occurs at level $(1,0)$ and $R^{a_5...a_{D}}$ the
representation [0,...,0,1,0,0,0] that occurs at level (0,1)
\cite{Kleinschmidt:2003mf}.

\subsection{$C_{q+1}^{+++}$}

Clearly, if the Weyl reflections associated to $\a_4$ and $\a_5$
do not change the signature, there will be no signature changes.
Let us first have a look at $\a_5$,
\begin{center}
\begin{tabular}{crclcrcl}
 $W_5$:  & $\a_{4}$ & $\leftrightarrow$ & $\a_{4} +
\a_5$ & $U_{W_5}$: & $R^4$ & $\leftrightarrow$ & $\rho \tilde{R}^4$  \\
         & $\a_{6}$ &  $\leftrightarrow$ & $\a_6+\a_5$ & &  $\tilde{R}$ &
$\leftrightarrow$ & $ \sigma \tilde{\tilde{R}}.$   \end{tabular}
\end{center}  The tensor  $R^{a}$ is the representation
$[1,0,0]$ of $A_{3}$ that occurs at level\footnote{The level
$(l_{q+1},l_q,...,l_1)$ of a root $\a$ of $C_{q-1}^{+++}$ can be
read in its
 decomposition  in terms of the simple roots $\a = m_1 \a_1+ m_2 \a_2+ m_3 \a_3 + l_{q+1} \a_4 +
 ...+l_1
\a_{q+4}$. } $(1,0,...,0)$, the tensor $\tilde{R}^a$ is the
representation [1,0,0] of $A_3$ that occurs at level
$(1,1,0,...,0)$, the tensor $\tilde{R}$ is the representation
$[0,0,0]$ of $A_3$ that occurs at level (0,0,1,0,...,0) and $
\tilde{\tilde{R}}$ is the representation $[0,0,0]$ of $A_3$ that
occurs at level $(0,1,1,0,...,0)$. Because we started with a
signature where all the generators have the sign of their kinetic
term positive, we will not reach new signatures (\ie here meaning
new signs for the kinetic term of $R^a$ or $\tilde{R}$) with this
reflection.

Let us now look at $\a_4$,

\begin{center}
\begin{tabular}{crclcrcl}
 $W_4$:  &  $\a_{5}$ & $\leftrightarrow$ & $\a_{4} +
\a_5$  & $U_{W_4}$: &  $R$ & $\leftrightarrow$ & $\tilde{R}^4$  \\
         &$\a_{3}$ &  $\leftrightarrow$ & $\a_3+2\a_4$ & &   $K^3{}_4$ &
$\leftrightarrow$ & $R^{34}.$  \end{tabular}
\end{center}  The tensor $R$ is the representation [0,0,0] of $A_3$ that
occurs at level (0,1,0,...,0), $R^{ab}$ is the representation
[0,1,0] of $A_3$ that occurs at level (2,0,...,0). To get a
signature change we need $ \Omega (K^3{}_4) \neq \Omega (R^{34})$
which is impossible because we have Eq.(\ref{signa}) for $R^{34}$.

\subsection{$F_4^{+++}$}

There are two short simple roots not belonging to  the gravity line, namely $\a_6$
and $\a_7$. The generators associated to these simple roots are
given in Table \ref{f4} respectively at level\footnote{The level
$(l_1,l_2)$ of a root $\a$ of $F_4^{+++}$ can be read in its
 decomposition  in terms of the simple roots $\a = m_1 \a_1+... + m_5 \a_5 +
 l_1
\a_6 + l_2 \a_{7}$. } (1,0) and (0,1) \cite{Kleinschmidt:2003mf}.
\begin{table}
\caption{\small Level decomposition of $F_4^{+++}$  }
\begin{center}
\begin{tabular}{|ccc|}
\hline
$(l_1,l_2)$&$A_5$ weight&Tensor\\
\hline (0,1)&[0,0,0,0,0]&$R$\\
(1,0)&[1,0,0,0,0]&$R^{a}$\\
(1,1)&[1,0,0,0,0]&$\tilde{R}^{a}$\\
(2,0)&[0,1,0,0,0]&$R^{ab}$\\
\hline
\end{tabular}
\end{center}
\label{f4} \end{table} The non trivial action of the Weyl
reflection $W_{\a_6}$ (and of the corresponding conjugaison by a
group element $U_{W_{\a_6}}$) on the simple roots (and on the
corresponding simple generators) are,

\begin{center}
\begin{tabular}{crclcrcl}
 $W_6$:  &  $\a_{5}$ & $\leftrightarrow$ & $\a_{5} +2
\a_6$  & $U_{W_6}$: &  $K^5{}_6$ & $\leftrightarrow$ & $R^{56}$  \\
         &$\a_{7}$ &  $\leftrightarrow$ & $\a_7+\a_6$ & &   $R$ &
$\leftrightarrow$ & $\tilde{R}^{6}.$  \end{tabular}
\end{center}
$W_{\a_6}$ can not change the signature because it needs sign($
\Omega K^5{}_6) \neq$ sign$(\Omega R^{56}$) which is
impossible since the sign of the kinetic term for $R^{ab}$
is always positive. Moreover $W_{\a_6}$ can neither change the
sign of the kinetic term of $R$ because we start from a signature
such that all kinetic terms are characterised by Eq.(\ref{signa}). The action of   $W_{\a_7}$
on the simple roots is,
\begin{center}
\begin{tabular}{crclcrcl}
 $W_7$:  &  $\a_{6}$ & $\leftrightarrow$ & $\a_{6} +
\a_7$  & $U_{W_7}$: &  $R^6$ & $\leftrightarrow$ & $\tilde{R}^{6}.$
\end{tabular}
\end{center}
This reflection can only \textit{a priori} change the sign of the
kinetic term for $R^a$. This sign can change if sign($\Omega R^6$)
$\neq $ sign($\Omega \tilde{R}^6$) which is impossible since we
started from a signature with all the signs given by Eq.(\ref{signa}).
Therefore the only possible signature is \bea (1,5,+,+), \eea where
the first sign refers to  $R^a$ and the second
to $R$.

\subsection{$G_2^{+++}$}

There is only one simple root not belonging to the gravity line, namely
$\a_5$.
\begin{table}
\caption{\small Level decomposition of $G_2^{+++}$}
\begin{center}
\begin{tabular}{|ccc|}
\hline $l$&$A_4$ weight&Tensor\\
\hline 1 & [1,0,0,0] & $R^a$ \\
3&[1,1,0,0]&$\tilde{R}^{abc}$\\
\hline \end{tabular}
\end{center}
\label{g2} \end{table} The only non trivial action of the Weyl
reflection generated by $\a_5$ on the simple roots is (see Table 13)

\begin{center}
\begin{tabular}{lrcllrcl}  $W_5$:$\a_{4}$ & $\leftrightarrow$ & $\a_{4} +3
\a_5$ & $U_{W_5}$: & $K^4{}_5$ & $\leftrightarrow$ &
$\tilde{R}^{455}.$
\end{tabular}
\end{center}
To get a signature change we need sign($ \Omega K^4{}_5$) $\neq$
sign($\Omega (\tilde{R}^{455}$), i.e. 5 is a time coordinate
(because we start with the sign of
$\tilde{R}^{abc}$ given by Eq.(\ref{signa})). The new signature is the euclidian one
$(0,5,-)$. These are the only  signatures  we can reach in agreement with  pure gravity in $D=5$,

\bea (1,4,+) \ \ (0,5,-), \eea where the sign refers to the kinetic
term of $R^a$.
%%%%%%%%%%%%%%%%%
%%%%%%%%%%%%%%%
\newpage
\cleardoublepage
\pagestyle{empty}
\addcontentsline{toc}{part}{Conclusion}
\part*{Conclusion}
\pagestyle{myheadings}
\markboth{CONCLUSION}{}
\cleardoublepage
%%%%%%%%%%%%%%%%%%
%%%%%%%%%%%%%%%%%
%%\include{conclusion}

\vspace*{7cm}

This thesis deals with recent aspects in the investigation of the symmetry structure of gravity. First, we addressed  questions within the framework of the \emph{Cosmological Billiards}. The billiard picture is derived within General Relativity, in studying the dynamics of the gravitational field (and other fields) in the vicinity of a space--like singularity. Of course, 
a satisfactory study of the gravitational field in this regime would require the knowledge of a quantum theory of gravity. In the quest for such a theory, it is however natural to examine General Relativity close to the regime where it breaks down. Moreover, the symmetries exhibited in a classical analysis might be also symmetries of the full quantum theory, even if the BKL analysis breaks down. The study of Cosmological Billiards suggests that Kac--Moody algebras might play an important role in a more fundamental theory of gravity. This observation deserves a more in--depth analysis. In this perspective, we obtained the following results: 

\noindent -- Our study of homogeneous cosmologies confirms the restoration of chaos when non--diagonal elements of the 
metric are kept in $5 \leq D \leq 10$ \cite{deBuyl:2003za}.  When considering particular initial conditions, we found some Lorentzian Kac--Moody subalgebras of the algebras 
relevant in the generic, \ie non homogeneous, context. 
We stressed the importance of having a better understanding of the momentum constraints in the generic case since in the very particular case of some empty $D=4$ 
homogeneous cosmologies, the billiard shape might depend on a gauge choice [but the finiteness, or infiniteness, of the billiard volume is not affected by a gauge choice]. 
We hope that these simplified models could provide a laboratory for the proof of the BKL conjecture in the chaotic cases. 

\noindent -- The shape of the billiards contains rather ``coarse--grained'' information about gravitational theories: it is invariant under dimensional reduction, it is not sensitive 
to Chern--Simons terms and it does not depend on the complete field content of the theories. From the billiard shape, we were able to derive information about the oxidation endpoint
of the $D=3$ cosets of non--split groups coupled to gravity \cite{deBuyl:2003ub}. More precisely, we gave the highest dimensions for all cases and showed how to find the field content
of these oxidation endpoints. 

\noindent -- We next turned to hyperbolic Kac--Moody algebras. These algebras are of particular interest since they characterise chaotic Kac--Moody billiards. For any Kac--Moody algebra, we 
addressed the questions of the existence of a Lagrangian reproducing this algebra and the maximal dimension in which such a Lagrangian exists. We completely answered these
questions and gave explicitly the minimal field content of the Lagrangians in the highest dimensions
\cite{deBuyl:2004md}. The maximal rank for hyperbolic algebras is 10 (and  only four hyperbolic algebras do exist); we confirmed the oxidation endpoint for $CE_{10}$ (spacetime dimension, field content and Lagrangian) that is the only rank 10 hyperbolic algebra \emph{a priori }not related to a string/M--theories.    
Our analysis restricts to the Weyl groups of hyperbolic algebras, but the search for Lagrangians invariant under these algebras [which are not over--extended algebras] has not been considered. Non--linear realisations based on these algebras  might give a hint on how to handle this problem.

As stated above, one of  most attractive features of Cosmological Billiards lies in the beautiful symmetric structure of gravitational theories they reveal and which is believed to be preserved in a quantum regime, although the BKL behaviour might be destroyed. Attempts to make this symmetry manifest and to go beyond this billiard picture have been suggested by the construction of actions invariant under Kac--Moody algebras. Reference \cite{Damour:2002cu} proposed an action invariant under $E_{8(8)}^{++}$ ---which is the case relevant to $D=11$ supergravity--- that describes a geodesic motion on the coset space $E_{8(8)}^{++}/K(E_{8(8)}^{++})$. 
The equations of motion resulting from this action have been compared, level by level, to the equations of motion of the (bosonic sector of the)
$D=11$ supergravity. The matching is perfect up to level $3^-$ or equivalently height 30 (which is much better than the height 1 roots accessed in the billiard picture). The inclusion of fermions was however lacking in this construction but we filled this gap in \cite{deBuyl:2005mt}. In the meanwhile, we showed that: 
 
\noindent -- The Dirac field is compatible
with the hidden symmetries that emerge upon toroidal dimensional
reduction to three dimensions, provided one appropriately fixes its
Pauli couplings to the $p$--forms \cite{deBuyl:2005zy}.  We have only considered the
split real form for the symmetry (duality) group in three
dimensions, but similar conclusions would apply to the
non--split forms.
We have also indicated
that the symmetry considerations reproduce some well known
features of supersymmetry when supersymmetry is available.

\noindent --  The Dirac field is compatible
with the conjectured infinite--dimensional symmetry $\cG^{++}$ and a perfect matching was found with the non--linear $\s$--model equations
minimally coupled to a $(1+0)$ Dirac field, up to the levels where
the bosonic matching works \cite{deBuyl:2005zy}.

\noindent --  The Dirac fermions get frozen \emph{\`a la limite BKL}. Moreover when the fermionic currents are treated as ordinary classical fields,  chaos will be destroyed ---whenever it is present in the bosonic theory--- 
in agreement with the
findings of \cite{BK}.  A direct group theoretical
interpretation of this result is that the motion in the Cartan subalgebra becomes  time--like \cite{deBuyl:2005zy}. 

\noindent -- The gravitino field of
11--dimensional supergravity is compatible with the conjectured
hidden $E_{10(10)}$ symmetry  up to the  level reached in the bosonic
sector. 
In $\s$--model terms, the
supergravity action is given by the (first terms of the) action for
a spinning particle on the symmetric space $E_{10(10)}/K(E_{10(10)})$, with
the internal degrees of freedom in the `spin 3/2' representation of
$K(E_{10(10)})$ (modulo the 4--fermion terms) \cite{deBuyl:2005mt}.
This action takes the same form as the action for a Dirac spinor
but with the appropriate `spin 3/2' representation. We can thus analyse
its dynamics in terms of the conserved $K(E_{10(10)})$ currents along
the same lines as for the spin 1/2 case and conclude that the BKL limit
holds.
We have treated explicitly the case of $D=11$ supergravity, but a similar analysis applies to the other supergravities,
described also by infinite--dimensional Kac--Moody algebras (sometimes
in non--split forms \cite{Henneaux:2003kk,Fre':2005sr}).
The inclusion of fermions in the $\cG^{+++}$--invariant actions could
be considered along the same lines. 

Actions invariant under $\cG^{+++}$ have been proposed in \cite{Englert:2003py} that encompass two non equivalent actions invariant under $\cG^{++}$,  $S_{\cG^{++}_B}$ and $S_{\cG^{++}_C}$ \cite{Englert:2004ph}. The action $S_{E_{8(8)\, B}^{++}}$ has been shown to possess ``extremal brane'' solutions that form orbits under the Weyl group identifiable with the U--duality orbits existing in M--theory \cite{Englert:2003py,Englert:2004it}. This
strongly suggests a general group--theoretical origin of
`dualities' for all $S_{\cG^{++}_B}$ --theories transcending string theories and
supersymmetry \cite{Englert:2003py,Englert:2004it,Englert:2003zs}. One aspect of the dualities is that they can change 
the spacetime signature  \cite{Hull:1998vg,Hull:1998fh,Hull:1998ym}. 
In this context, we extended to all
$S_{\cG^{++}_B}$--theories the analysis of signature changing Weyl reflections \cite{deBuyl:2005it}.
 We found for all the
$S_{\cG^{++}_B}$--theories all the possible signatures $(T,S)$ ---where $T$
(resp. $S$) denotes the number of time--like (resp. space--like)
directions--- related by Weyl reflections of $\cG^{++}$ to the Lorentzian 
$(1,D-1)$ signature. Along with the different
signatures the signs of the kinetic terms of the relevant fields
have been  discussed. 
In the context of string theory, the special cases of $D_{24}^{++}$ and
$B_{8}^{++}$ are interesting.

The attempts to reformulate ``M--theories'' as non--linear realisations are very instructive. They do not
include 
space(time) as  a \emph{basic }ingredient, rather it is hoped to be encoded in 
the dynamics. To understand how,  a proof of the conjecture that  certain
 infinite ``tower'' of representations encode the spatial derivatives  the (supergravity) fields \cite{Damour:2002cu}
 would be illuminating. 
But it seems that a better  knowledge of Kac--Moody algebras
is needed 
to tackle this statement.
A first step in this direction could be to seek solutions to the Kac--Moody invariant actions that are identifiable as  supergravity solutions that depend on two spacetime coordinates. A better understanding of infinite--dimensional \emph{subalgebras }might be of great important in that
context. 
More generally,  difficulties arise in the interpretation of fields of increasing level (for instance, for $E_{10(10)}$, no dictionary has been established for fields of level $\ell > 3^-$).
Various interesting propositions have been put forward for how to understand the structure of these higher level fields. 
First, the conjecture mentioned above about the spatial derivatives. Furthermore, these higher level fields may also be interpreted as candidates for higher spin fields because of their index structure (see for instance \cite{deBuyl:2004ps} and references therein). Recently, some very promising results have also been obtained 
regarding the incorporation of quantum corrections into the structure of the Kac--Moody algebras related to string and M--theory \cite{Damour:2005zb,Lambert:2006he,Damour:2006ez}. 
 More precisely, for M--theory the leading contribution of the 8th derivative correction is associated with the fundamental weight 
 $\Lambda_{10}$ conjugate to the exceptional simple root of $\mf{e}_{10(10)}$. Subdominant terms at the same order in 
 derivatives then contributes by the addition of positive roots to this weight. This opens up the tantalising possibility that M--theoretical 
 corrections are encoded in (non--integrable) representations of $e_{10(10)}$. Since the dominant weights, appearing as lowest weights in the representations, correspond to \emph{negative levels }these results indicate that the original $E_{10(10)}$--invariant action [based on a Borel--type decomposition] is probably insufficient to incorporate the full structure of M--theory.
Reference \cite{Brown:2004jb} also deals with imaginary roots.

Whatever M--theory will be, we believe that it will reflect the deep connections between dimensional reduction, asymptotic dynamics of the gravitational field in the vicinity of a space--like singularity, hidden symmetries, dualities, non--local transformations and infinite--dimensional 
Lie algebras.

%%%%%%%%%%%%%%%%
%%%%%%%%%%%%%%%%
\newpage
\appendix
\cleardoublepage
%%%%%%%%%%%%%%
%%%%%%%%%%%%%%%%
%%\include{app_lie}

\chapter{Simple Lie Algebras}
\markboth{SIMPLE LIE ALGEBRAS }{}
\label{lie}

This appendix is based on \cite{Fuchs:1997jv} and \cite{Humphreys:1980dw}. 
For more details and examples, see \cite{BdB}.

\section{Structure of Simple Lie Algebras}

\subsubsection{Basic Definitions}

A \emph{complex Lie algebra} ${\mf{g}}$ is a vector space
equipped with a binary operation $[ \ , \ ]$ called a commutator, mapping
${\mf{g}}\times {\mf{g}}$ into ${\mf{g}}$ with
the following properties; $ \forall \ x,y,z \ \in {\mf{g}}$
and $ \forall \ \a, \b \in \mathbb{C}$,
\beq
&1.& \mathrm{Antisymmetry}:  [x,y] = - [y,x]\,; \nn \\
&2.&  \mathrm{Bilinearity}: [ \a x + \b  y , z] = \a [x,z] + \b [y,z]\,; \nn \\
&3.& \mathrm{Jacobi \ identity} : [[x,y],z] + [z,[x,y]] +
[y,[z,x]] =0 \,. \nn
\eeq
A Lie algebra $\mf{g}$ can be specified by a set of generators
$\{ J^b \}$ and their commutations relations
$$ [J^b,J^c] = f^{bc}_{\hspace{.3cm}d} J^d\,.$$
The \emph{dimension of the algebra} is the dimension of the
underlying vector space. The constants $f^{bc}_{\hspace{.3cm}d}$
are the \emph{structure constants}.

\noindent \emph{Simple Lie algebras} are non-abelian Lie algebras that
contain no proper\footnote{A proper subset of a set is a set which
is not the set itself neither the empty set.} ideal. An ideal $\mf{i}$ 
of ${\mf{g}}$ is a subspace of ${\mf{g}}$ such that $\ [\mf{i}, \mf{g}] 
\subset \mf{i}\,$.

\noindent \emph{Semi-simple Lie algebras} are Lie algebras that contain no
abelian ideal and that are not empty. Every semi-simple Lie algebra
can be decomposed into the direct sum of its simple ideals (see
theorem in Section \textbf{5.2} of \cite{Humphreys:1980dw}).

\noindent \emph{Levi theorem}: Every Lie algebra can be decomposed into the
direct sum of simple Lie algebras and solvable algebras (a
solvable algebra is a Lie algebra such that the series $\
L^{(0)}=L , \ L^{(1)} = [L^{(0)},L^{(0)}], \
L^{(k)}=[L^{(k-1)},L^{(k-1)}]$ stops for some $k$).

\noindent A \emph{linear representation} $\G$ of a Lie algebra ${\mf{g}}$ 
in the vector space $V$ is an homomorphism\footnote{$ \forall x, y  \in \ 
{\mf{g}}$ and $\forall \ v \in V$, $ [\G(x),\G(y)]\cdot v = \G([x,y])\cdot 
v$ and $\G(\a x + \b y) \cdot v =  \a \G(x)\cdot v + \b \G(x)\cdot v $}
\begin{eqnarray}
    \G : \ {\mf{g}} \rightarrow \mf{gl}(V) : x \rightarrow
\G(x)\,,\quad \G(x): V \rightarrow V: v \rightarrow \G(x)\cdot v\,.
\end{eqnarray}
The linear operator $\G(x)$ is sometimes simply denoted by $x\,$.

\subsubsection{The Cartan-Weyl Basis and Roots}

\noindent Only \emph{semi-simple} \emph{complex} Lie algebras are
treated. In the \emph{Cartan-Weyl basis}, the generators are constructed 
as follows. We first find the maximal set of commuting generators
$H^i\,$, $i = 1,...,r\,$, such that $ad_{H^i}$ is semi-simple for all $i$ 
(an endomorphism is called semi-simple if in a suitable basis it can be 
expressed by means of a diagonal matrix). This set of operators forms the 
\emph{Cartan subalgebra}\footnote{An equivalent definition can be found in
Section \textbf{15.3} of \cite{Humphreys:1980dw}. It is also shown that all
Cartan subalgebras of a given finite-dimensional semi-simple Lie algebras
are isomorphic (see the Corollary in Section \textbf{15.3.} (Section 
\textbf{8.1.} for definition) and Section \textbf{16.4}).} $\mf{g}_{\circ}$. 
The {\emph{rank}} of $\mf{g}$ is defined to be the dimension $r$ of its Cartan
subalgebra.

Because the matrices $ad_{H^i}$ can be simultaneously diagonalized, the 
remaining generators are chosen to be those particular combinations $E^{\a}$ 
of the $J^a$'s that satisfy the following eigenvalue equations
$$ ad_{H^i} E^{\a} = [H^i,E^{\a} ] = \a(H^i) E^{\a} $$
where the $\a(H^i)$, also noted $\a^i$, are the (not all vanishing) components
of elements of $\mf{g}_{\circ}^{\star}$, the dual of the Cartan subalgebra
$\mf{g}_{\circ}\,$.
The non-zero elements $\a$ of $\mf{g}_{\circ}^{\star}$ such that the set
$$ {\mf{g}}_{\a} = \{~ X \in {\mf{g}} ~~\arrowvert ~~ [H^i,X ]=
\a^i \, X ~~~ \forall ~ H^i \in \mf{g}_{\circ} \ \}
$$ is not empty are called {\emph{roots}}.
The set of all roots of $\mf{g}$ is denoted $\Delta$ and
is called the \emph{root system} of $\mf{g}$. $E^{\a}$ is the
corresponding \emph{ ladder operator}. To get the commutation relations 
between the ladder operators, one observes that the Jacobi identity implies
\beq
\ [H^i, [ E^{\a}, E^{\b} ] ] = (\a^i + \b^i) [E^{\a},E^{\b}]
\ . 
\nn 
\eeq
If $\a + \b \ \in \ \Delta$, the commutator $ \ [E^{\a},E^{\b}] $
must be proportional to $E^{\a+\b}$. It certainly must vanish if $\a +\b$
is not a root. When $\a = - \b$, the commutator $ [
E^{\a},E^{-\a}]$ commutes with all $H^i$, which is possible only
if it is a linear combination of the generators of the Cartan
subalgebra (see Eq.(\ref{Comalpha}) for precise relation). Therefore, the full 
set of commutation relations in the Cartan-Weyl basis is
\beq
\begin{tabular}{|rll|}
\hline
 \ & \ & \ \\
 $\ [H^i,H^j]$ & = \ 0  & \\
 $\ [H^i,E^{\a} $]  & = $\ \a^i E^{\a}$  & \\
 $\ [E^{\a},E^{\b}]$ &= $\ N_{\a,\b} E^{\a+\b}
 \hspace{.7cm} $for   some $  N_{\a,\b} \in \mathbb{C}\quad$
 & if$ \  \a + \b \in \Delta$ \,,\\ 
  &  $= \ \sum_{i=1}^r a_i H^i \hspace{1cm}$ for \ some \ 
$a_i \in \mathbb{C}$ & 
   if $\a=-\b\ $,, \\
   & = \ 0 & otherwise \,.\\
 & &\\
 \hline
 \end{tabular}
\label{cwcomm}
 \eeq

\subsubsection{The Killing Form and Properties of Roots}

In order to know the Cartan-Weyl basis more explicitly, more must
be said about the root system $\Delta$. The Killing form plays an 
important role in this perspective. 

\noindent A bilinear and symmetric form on $\mf{g}$, called the 
\emph{Killing form}, can be defined by 
\begin{center}
\begin{tabular}{|c|} \hline \\
$ K (x,y) :=  \mathrm{tr}(\mathrm{ad}_x \ 
\mathrm{ad}_y) \hspace{1.5cm} \forall x,
\ y \in {\mf{g}}$\,. \\
\\ \hline \end{tabular} 
\end{center}
The Killing form possesses important properties: (i) its invariance 
(under the adjoint action) $ K(x,[y,z] ) = K([x,y],z) ~~ \forall ~ x,y,z \in
{\mf{g}}$; (ii) it is non-degenerate (Cartan's 
criterion); (iii)  the restriction of the Killing form to 
$\mf{g}_{\circ}$ is non-degenerate. 

\noindent The fundamental role of the Killing form is to establish an 
isomorphism between $\mf{g}_{\circ}$ and 
$\mf{g}_{\circ}^{\star}\,$. To every $\g \in \mf{g}_{\circ}^{\star}$ 
corresponds, up to a normalization factor $c_{\g}$, an element $H^{\g} \in 
\mf{g}_{\circ}$
through
\beq
\label{Halpha}
\g(h) = c_{\g}
K(h,H^{\g}) \hspace{1cm} \forall\; h \in \mf{g}_{\circ}\,.
\eeq
To avoid the normalization factor, we also define
$h_{\g} := c_{\g} H^{\g}\,$. With this isomorphism, the Killing form can 
be transferred into a (non-degenerate) inner product on 
$\mf{g}_{\circ}^{\star}\,$:
\beq
\begin{tabular}{|c|} \hline \\
$ (\g,\b) :=  K(h_{\b},h_{\g}) = c_{\b} c_{\g} K(H^{\b},H^{\g})\,.$ \\ \ \\
\hline \end{tabular}
\label{13.15}
\eeq
Using the Killing form, the commutation relation between $E^{\a}$
and $E^{-\a}$ can be established. One can show that,
\beq 
\label{Comalpha}
\begin{tabular}{|c|}   \hline \\
$[E^{\a},E^{-\a}] = c_{\a} K(E^{\a},E^{-\a}) H^{\a}
= K(E^{\a},E^{-\a}) h_{\a}\,.$ \\  \\
\hline 
\end{tabular}
\eeq

\noindent Here are listed some important results
concerning the root system $\Delta$ of (finite-dimensional) semi-simple
complex Lie algebras.
\begin{itemize}
\item[$\star$] The roots span all  $\mf{g}_{\circ}^{\star}\,$: 
span${}_{\mathbb{C}}(\Delta)=\mf{g}_{\circ}^{\star}\,$
\item[$\star$] The root spaces $\mf{g}_{\a}$ are one-dimensional 
($\mf{g}_{\a}=$ span$_{\mathbb{C}}\{E^{\a}\}$), \emph{i.e.} the roots are 
non-degenerate;
\item[$\star$] The only multiples of $\a\in\Delta$ which also belong to 
$\Delta$ are $\pm\a\;$;
\item[$\star$] On $\mf{g}_{\circ}(\mathbb{R})$, the \emph{real} vector space
with basis $\{H^i\}_{i=1}^r$ (or $\{h_{\b_i}\}_{i=1}^r$), we have 
$\a(H^i)\in\mathbb{R}$ $\;\forall\;i$ and $\forall\;\a\in\Delta$.
\item[$\star$] $K(h,h')$ is \emph{real} for all $h$,
$h'\in\mf{g}_{\circ(\RR)}$. Moreover, for all $h\in
  \mf{g}_{\circ}(\mathbb{R})$, $K(h,h)\geqslant 0$ and $K(h,h)=0$ implies $h=0$.
\end{itemize}
{\emph{The Killing form thus provides an {\textbf{Euclidean}} metric on the 
root space $\mf{g}^{\star}_{\circ}(\mathbb{R})$, so that we have
$\mf{g}^{\star}_{\circ}(\mathbb{R})\cong \mathbb{R}^{\mbox{r}}$}}. This 
highly non-trivial fact is one of the key ingredients in the classification of 
all finite-dimensional simple Lie algebras.

\subsubsection{Simple Roots, Cartan Matrix and Dynkin Diagram}

Let $h_{\b_1}$, $h_{\b_2}$, $\ldots h_{\b_r}$ be a set of $r$ linearly 
independent elements of $\mf{g}_{\circ}$, where $\b_i\in\Delta$, 
$1\leqslant i\leqslant r\,$. It can be shown that every non-zero root
$\a$ of $\Delta$ can be written in the form $\a = \sum_{j=1}^{r} k_i\b_j\,,
\quad  k_i \in \mathbb{Q} \,,\quad i=1,\ldots, r$. A non-zero root of $\Delta$ 
is said to be {\emph{positive}} (with respect to this basis) 
if the first non-vanishing coefficient of the set $\{k_i\}_{i=1}^r$  is 
positive. One then writes $\a>0$. The set of 
all positive (resp. negative) roots is denoted $\Delta_+$ (resp. $\Delta_-$). 
A \emph{simple root} is a positive root that cannot be written as the sum of 
two positive roots. If ${\mf{g}}$ has rank $r$ then ${\mf{g}}$
possesses precisely $r$ simple roots $\{\a^{(i)}\}_{i=1}^{r}$. They form a 
basis of the dual space $\mf{g}_{\circ}^{\star}\,$ such that every positive 
root $\a$ can be written as $\a=\sum_{i=1}^r k_i\a^{(i)}\in\Delta_+$ where the
 $k_i$'s are \emph{non negative integers}. The \emph{level} or \emph{height} 
of $\a$ is the positive integer $k=\sum_{i=1}^r k_i\,$.

A \emph{lexicographic ordering of roots} can be introduced: let $\a$ and $\b$ 
be any two roots of $\Delta\,$, then if $\a-\b$ is positive, $(\a-\b)>0\,$, 
one writes $\a>\b\,$.

The Cartan matrix $\mathbf{A}$ of ${\mf{g}}$ is an $r\times r$ matrix 
whose elements $A^{ij}$ are defined in terms of the simple roots 
$\{\a^{(i)}\}_{i=1}^{r}$ of ${\mf{g}}$ by
\begin{eqnarray}
\begin{tabular}{|c|}   \hline \\
$A^{ij}={\displaystyle{\frac{2(\a^{(i)},\a^{(j)})}{(\a^{(j)},\a^{(j)})}}}\,,\quad 
i,j=1,\ldots,r\,$. \\  \\
\hline \end{tabular} \label{13.61}
\end{eqnarray}
The Cartan matrix $\textbf{A}$ possesses the following properties:  $A^{ii}=2$
$\forall$ $i\in\{1,\ldots,r\}$; for $i\neq j$ the only possible values 
of $A^{ij}$ are $0$, $-1$, $-2$ or $-3$;  $A^{ij}=0$ if and only if 
$A^{ji}=0\,$. 

The \emph{coroot} of the simple root $\a^{(i)}$ is
$$ \a^{(i)\vee} := {2 \a^{(i)} \over (\a^{(i)},\a^{(i)})}\,.$$
There is a \emph{highest root} $\Theta$, the unique root for which, in the 
expansion $\sum m_i \a^{(i)}$, the height (or level) $\sum
m_i$ is maximized.

One depicts a so-called \emph{Dynkin diagram} for each simple
complex Lie-algebra ${\mf{g}}$ where each simple root
$\a^{(i)}$ of ${\mf{g}}$ is assigned a point (a vertex) and 
max$\{|A^{ij}|, |A^{ji}|\}$ lines are drawn between
the vertices corresponding to $\a^{(i)}$ and $\a^{(j)}$ with an 
arrow towards the shortest root, \ie if $\sqrt{\a_{(i)}, \a_{(i)}} < \sqrt{\a_{(j)}, \a_{(j)}}$ the 
arrow is directed toward $\a_{(i)}$. 

\subsubsection{Remaining Commutation Relations}

If $\a, \  \b, \ \g, \  \d, \ \a+\b \in \D$, and $p, \ q$ are such that the 
$\a$-string\footnote{If $\a$ and $\b$ are roots, the $\a$--string of roots containing $\b$ is the set of all roots of the form $\b + k \a$, where $k$ is an integer.} containing $\b$ is $\b-p\,\a$, $\ldots$, $\b$, $\ldots$, 
$\b+q\,\a$ it can be shown that the $N_{\a,\b}$'s of 
Eq.(\ref{cwcomm}) are such that 
\begin{itemize}
\item[(a)]  $N_{\a,\b}\neq 0\,$ ;
\item[(b)]  $N_{\b,\a}=-N_{\a,\b}$ ;
\item [(c)]if  $\a+\b+\g=0$, $ N_{\a,\b}K_{\g} =
  N_{\b,\g}K_{\a} = N_{\g,\a}K_{\b} $  where $ K_\a = K(E^\a, E^{-\a})$;
  \item[(d)] if $\a , \, \b , \,Ê\g, \, \d$ are such that the sum of  two of 
them is never zero, and if $\a+\b+\g+\d=0$, then
$N_{\a,\b}N_{\g,\d}K_{\a+\b}+N_{\b,\g}N_{\a,\d}K_{\b+\g}+N_{\g,\a}N_{\b,\d}K_{\a+\g}=0$
    \item[(e)] 
$N_{\a,\b}N_{-\a,-\b} = -\frac{1}{2}(\a,\a)\frac{K_{\a}K_{\b}}{K_{\a+\b}}q(p+1)$;
\item[(f)]
with $K(E^{\a},E^{-\a})$ taking {\emph{any}} assigned value $K_{\a}$ for 
each
pair of roots $\a$, $-\a\in\Delta$, the basis elements of ${\mathfrak{g}}$ 
may
be chosen so that {\emph{either}} $N_{\a,\b}=N_{-\a,-\b}$ $\forall$ $\a$, $\b$
$\in\Delta$ {\emph{or}} $N_{\a,\b}=-N_{-\a,-\b}$ $\forall$ $\a$, $\b$
$\in\Delta$.
\end{itemize}
To fix the commutations relations (\ref{cwcomm}, \ref{Comalpha}) up to signs, 
one can for instance take

\noindent 1. 
$K_\a = 1\ \forall \ \a \in \D$ (one denotes the ladder operators normalised 
accordingly by  $e_\a$) and \newline
\noindent 2. $N_{-\a,-\b} = - N_{\a,\b} \ \forall 
\a, \b \ \in \D$. With these conventions, all $N_{\a,\b}$ are real and obey
\beq 
N_{\a,\b}{}^2  = {1\over 2} (\a,\a)q(p+1) \ .
\nn
\eeq
3. Moreover, one chooses the Cartan 
generators chosen to be the $h_\a$ such that $\a(h) = K(h,h_\a) \ \forall \
h \in \mf{g}_\circ$. 
 
\noindent With these normalisations, the commutation relations write  
\beq
\ [h_\a,h_\b] &=& 0 \ , \nn \\
\ [h_\a, e_\b] &=& \b(h_\a) e_\b \ , \nn \\
\ [e_\a, e_\b ] &=& \pm \sqrt{{1\over 2} (\a,\a)q(p+1)} e_{\a+\b} \ \ \ 
\mathrm{if} \ \a+\b \in \D \ , \nn \\
                &=& h_\a  \ \ \hspace{3.6cm} \mathrm{if} \ \b =-\a \ , \nn\\
&=& 0 \ \ \hspace{3.8cm} \mathrm{ortherwise} \ . 
\label{commreal}
\eeq
Two other choices are often used in the literature, the Chevalley basis 
and the orthonormal basis.

\subsubsection{The Chevalley Basis and Serre Relations}

The Chevalley basis is a particular case of Cartan-Weyl basis which is the 
one most useful for studying representations. In this section, a construction of 
finite dimensional simple Lie algebras based on Cartan matrices and called the 
\emph{Chevalley--Serre construction}  is presented. A {Cartan matrix} $A\,  $ is an 
$r\times r$ matrix such that (i) $A_{ii} = 2$, (ii) $A_{ij} = 0, -1,-2,-3$ for $i \neq j$, 
(iii) $A_{ij} = 0 \iff A_{ji}Ê= 0$ for $i \neq j$, (iv) det($A) > 0$,  (v) $A$ is indecomposable, \ie
there is no renumbering of the indices which would bring $A$ to a block diagonal form.
Any finite dimensional simple Lie algebra $\mf{g}(A)$ can be built out of a  Cartan matrix $A$, 
see \emph{e.g.} \cite{Humphreys:1980dw}.
The algebra $\mf{g}(A)$ is generated by 3 $r$, $\{e_i, \, f_i, \, h_i\}_{i=1,...,r} $ obeying the 
following commutation relations, 
\begin{center}
\begin{tabular}{|rl|}
\hline
 & \\
 $[h^i,h^j]$&$= \ 0\,, $  \\
 $[h^i,e^{j} ] $ &$ = \ A^{ji} e^{j}\,, $  \\
 $[h^i,f^{j} ] $ &$ = -\ A^{ji} f^{j}\,, $  \\
 $[e^{i},f^{j}] $&$= \ \d^{ij} h^i \,. $ \\
 & \\
 \hline
 \end{tabular}
\label{chevalley}
 \end{center}
Moreover, their multi-commutators ({\it the remaining generators}) are subject
 to the so-called 
\emph{Serre relations},
\begin{eqnarray}
 {(\mathrm{ad}_{e^i})}^{1-A^{ji}} e^j &=& 0 \,,\nn \\
 {(\mathrm{ad}_{f^i})}^{1-A^{ji}} f^j &=& 0 \,.\nn
\end{eqnarray}
It can be shown that the algebra $\mf{g}(A)$ admits a \emph{triangular decomposition}, 
\beq 
\mf{g}(A) = n_- \oplus \mf{g}_\circ \oplus n_+ \, , \nn
\eeq
where $\mf{g}_\circ$ is the Cartan subalgebra generated by the $h_i$, $n_-$ is nilpotent and 
generated by the multicommutators of the $f_i$ (elements of the form $\ [f_i,[f_j,[...,[f_k,f_l]]]]$) and 
$n_+$ is generated by multicommutators of the $e_i$.  As mentioned above, the roots are not degenerate therefore to each multicommutator $e_\a$ of the $e_i$ correspond one positive root $\a$ and 
to each multicommutator $f_\a$ of the $f_i$ correspond one negative root $-\a$. 

\noindent This choice of basis corresponds to the following normalisation choices, 

\noindent 1. The $h^i$ are equal to the $H^{(\a^i)}$ of Eq.(\ref{Halpha}) with $c_{\a^{i}}=(\a^{(i)},\a^{(i)})/2$, \ie they are normalised such that the equation $\a^{(i)}(h)
=  1/2 \ (\a^{(i)},\a^{(i)}) K(h, h^i)$ is satisfied. \newline
2. The $ e^i $ and  $f^i$  are the previously introduced $E^{\a^{(i)}} $ and $E^{-\a^{(i)}}$ generators  normalised such that 
$K_{\a^{(i)}} = K(E^{\a^{(i)}}, E^{-\a^{(i)}}) = 2 / (\a^{(i)}, \a^{(i)})$. They 
obey  Eq.(\ref{Comalpha}) which, with these normalisations,  rewrites $\ [e_i, f_i ] = h_i$.\newline
3. The normalisation of the multicommutators $e_\a$ is chosen such that $K(e_\a , f_\a ) = 2 / (\a,\a)$.  

\noindent \textbf{Remark }: The Cartan-Chevalley
involution $\tau$ reads 
\beq \tau(h_i) = -h_i, \hspace{1cm}Ê\tau(e_i) = - f_i \hspace{.5cm}Êand \hspace{.5cm}Ê
\tau(e_\alpha) = (-1)^{ht(\alpha)} f_\alpha \, . \label{tau} \eeq 

\subsubsection{Chevalley Basis up to signs \label{conventions}} 

A slightly different convention is the same as here above 
 except that we take the ``negative'' generators $f_i$
with the opposite sign. The only
relation that is modified is $[e_i, f_j] = - \d_{ij} \, h_i $.
The sign convention for $f_i$ simplifies somewhat the form of the
generators of the maximal compact subalgebra . The Cartan-Chevalley
involution reads $\tau(h_i) = -h_i$, $\tau(e_i) = f_i$, $\tau(f_i)
= e_i$ and extends to the higher height root vectors as
$\tau(e_\alpha) = f_{\alpha}$, $\tau(f_{\alpha}) = e_{\alpha}$.  A basis of the
maximal compact subalgebra is given by $k_\alpha = e_{\alpha} +
f_{\alpha}$. It is convenient to define $g^T = - \tau(g)$ for
any Lie algebra element $g$.
The invariant bilinear form on the Lie algebra is given for the
Chevalley-Serre generators by 
\be K(h_i,h_j) = \frac{2 \,
A_{ji}}{(\alpha_i \vert \alpha_i)} = \frac{2 \, A_{ij}}{(\alpha_j
\vert \alpha_j)}, \;\; K(h_i,e_j)= K(h_i, f_j)=0, \; \; K(e_i,f_j)
= - \frac{2 \d_{ij}}{(\alpha_i \vert \alpha_i)} 
\ee and is
extended to the full algebra by using the invariance relation
$K(x,[y,z]) = K([x,y],z)$. Here, the $\alpha_i$'s are the simple
roots. The induced bilinear form in root space is denoted by $( \cdot
\vert \cdot )$ and given by $(\alpha_i \vert \alpha_j) = \frac{2
A_{ij}}{(\alpha_i \vert \alpha_i)}$. The numbers $(\alpha_i \vert
\alpha_i)$ are such that the product $A_{ij} \, (\alpha_i \vert
\alpha_i)$ is symmetric and they are normalized so that the
longest roots have squared length equal to $2$.   One gets \be
K(h_i,e_\alpha)= K(h_i, f_\alpha) = 0 = K(e_\alpha, e_\beta) =
K(f_\alpha, f_\beta), \; \; K(e_\alpha, f_\beta) = - N_\a
\d_{\alpha \beta}, \label{norma} \ee where the coefficient
$N_\a$ in front of $\d_{\alpha \beta}$ in the last relation
depends on the Cartan matrix (and on the precise normalization of
the root vectors corresponding to higher roots -- e.g., $ N_\a =
\frac{2}{(\a \vert \a)}$ in the Cartan-Weyl-Chevalley basis).

\subsubsection{Orthonormal Basis }

\noindent The Killing form of ${\mf{g}}$ provides an inner product for the 
real vector space $\mf{g}_{\circ}(\mathbb{R})$. Consequently a 
basis may be set up in $\mf{g}_{\circ}(\mathbb{R})$ that is 
orthonormal with respect to the Killing form, and this basis is also 
a basis $\mf{g}_{\circ}(\mathbb{R})$. Thus there exist $r$ elements 
$H_i$, $i=1,\ldots, r$ of
$\mf{g}_{\circ}(\mathbb{R})$ such that
\begin{eqnarray}
K(H_i,H_j)=2 \d_{ij}\,,\quad i,j=1,2,\ldots,r\,.
\label{13.56}
\end{eqnarray}
It follows that $K(\vec{a}, \vec{b}) = 2 \vec{a} .
\vec{b}$ for $\vec{a}$, $\vec{b}$ in $\mf{g}_\circ$ ($\vec{a} = \sum_i a^i
H_i$, $\vec{b} = \sum_i b^i H_i$) and $(\vec{\a} \vert \vec{\b}) =
\frac{1}{2} \vec{\a} . \vec{\b}$ for $\vec{\a}$, $\vec{\b}$ in the
dual space. Here, $\vec{a} . \vec{b} = \sum_i a^i b^i$ and
$\vec{\a} . \vec{\b} = \sum_i \a_i \b_i$ (and $\a_i, \b_i$
components of $\vec{\a}$, $\vec{\b}$ in the dual basis).

\section{Highest Weight Representations}

The key idea in the analysis of finite-dimensional representations of simple 
Lie algebras is to reduce the problem to the representation theory of the
 algebra ${\mf{sl}}(2)$. This is possible because each $H^{\a}$ together with 
$E^{\a}$ and $E^{-\a}$ span a ${\mf{sl}}(2)$ subalgebra in the Chevalley
basis. Note that we will only be interested in \textit{finite dimensional}
representations of \textit{simple} Lie algebras. In this section,  we use
the Chevalley basis.

\subsubsection{Representations of ${\mf{sl}}(2)$}

The generators  $h, \ e$ and $f$ of ${\mf{sl}}(2)$ satisfy the relations 
\beq \label{Comsu2} [h,e] = 2 e \hspace{1cm} [h,f]=-2f
\hspace{1cm} [e,f]=h \eeq 
For all integer $p \geqslant 0$ there exists a
unique, up to isomorphism, irreducible representation of dimension
$p+1$ of maximal weight $p$. One can always choose a basis $\{
v_0,...,v_p \}$ of the representation space such that $e$, $f$ and
$h$ acts as in Eq.(\ref{Repsu2}) with  $j=0,...,p$.
\beq \label{Repsu2}
h \ v_j = (p - 2 j) \ v_j \hspace{1cm} e \
v_j = j(p-j+1 ) \ v_{j-1} \hspace{1cm} f \ v_j = v_{j+1} \,.
\eeq

\subsubsection{Weights of a Representation}

For any simple Lie algebra ${\mf{g}}$, with the convention corresponding to 
the Chevalley basis, each $H^{\a}$ together with $E^{\a}$ and $E^{-{\a}}
\equiv F^{\a}$ span an ${\mf{sl}}_2$ subalgebra. 
These generators obey the ${\mf{sl}}(2)$ commutation relation given by Eqs. 
(\ref{Comsu2}) with the following identifications,
\beq H^\a \rightarrow h
\hspace{1cm} E^\a \rightarrow e \hspace{1cm} E^{-\a} \rightarrow f\,
\label{ident}
\eeq
One learns therefore that for an arbitrary representation $V$ of ${\mf{g}}$, a 
basis $\{ \arrowvert \l \rangle \}$ can always be found such that the whole 
Cartan subalgebra acts diagonally on it. Indeed, ${\mf{sl}}(2)$ acts on $V$
and the representation theory of this algebra tell us that there
exists a basis of $V$ such that $H^\a$ acts diagonally on it. The
 $h^i$'s (= $H^{\a^{(i)}}$'s) spanning the Cartan subalgebra all commute and 
therefore the whole Cartan subalgebra acts diagonally on a well chosen basis 
of $V$
\begin{eqnarray}
h^i \arrowvert \l \rangle = \l^i \arrowvert \l \rangle\,.
\label{wcbasis}
\end{eqnarray}

The eigenvalues $\l^i $ can be collected into an $r$-dimensional
vector which is called a \emph{weight}. This weight can be seen
as an element of $\mf{g}_{\circ}^{\star}$, $\l (h^i) =\l^i$.

Moreover, the weight $\l^i$ are integers. Indeed, the representation
$V$ provides a representation of the $\mf{sl}_{\a^{(i)}}(2)$
subalgebra spanned
by the Chevalley generators $\{H^{\a^{(i)}},E^{\pm\a^{(i)}}\}\,$.
The latter representation of $\mf{sl}_{\a^{(i)}}(2)$ can be
decomposed into a direct sum of irreducible representations of 
$\mf{sl}_{\a^{(i)}}(2)$  discussed in the previous section. 
Thus the diagonal elements of
$H^{\a^{(i)}}$ must all be integers, which means that
$\l(H^{\a^{(i)}})$ is an integer. We have
$\l^i=\l(h^i)=\l(H^{\a^{(i)}})\,$. From this, one concludes that
$(\l,\a^{\vee})$ is an integer for any $\a\in\Delta$. One just
repeats the previous argument for any subalgebra
$\mf{sl}_2(\a)$ of $\mf{g}$, $\a\in\Delta\,$, and by
noting that $\l(H^{\a})=\frac{2}{(\a,\a)}
\l(h_{\a})=\frac{2(\l,\a)}{(\a,\a)}=(\l,\a^{\vee})\,$.

Suppose that $\a$ is a root of ${\mf{g}}$ and $\l$ is a weight of some
representation of ${\mf{g}}$, then the ``$\a$-string of weights containing 
$\l$'' is the set of all 
weights of that representation of the form $\l+k\a$, where 
$k$ is an integer.

Let $\a$ be a non-zero root of ${\mf{g}}$ and $\l$ a weight of some
representation $V$ of ${\mf{g}}$. Then there exist two non-negative 
integers $p$ and $q$  (which depend on $\a$ and $\l$) such that $\l+k\a$ is in 
the $\a$-string containing $\l$ for \emph{every} integer $k$ that satisfies 
the relations $-p\leqslant k \leqslant q$. Moreover, $p$ and $q$ are such that
\begin{eqnarray}
    p-q = \frac{2(\l,\a)}{(\a,\a)}=(\l,\a^{\vee})\,.
    \label{pq}
\end{eqnarray}

\subsubsection{Fundamental Weights}

The dual basis of the coroot basis is denoted $ \{ \Lambda_{(i)}\}_{i=1}^r$
and its elements obey
\beq \label{funweig}
(\Lambda_{(i)},\a^{(j)\vee} ) = \d_{i}^j\,.
\eeq
The $\Lambda_{(i)}$ are called \emph{fundamental weights}.

One defines the \emph{quadratic form matrix} $G=(G_{ij})$ with lower indices 
by $G_{ij}:=(\Lambda_{(i)},\Lambda_{(j)})$. In terms of the Cartan matrix 
$A^{ij}$, ${(G)}_{ij}={(A^{-1})}_{ij}\frac{(\a^{(j)},\a^{(j)})}{2}$. It
inverse $G^{-1}$ is written $G^{ij}$ in components and is called the
\emph{symmetrized Cartan matrix},  $G^{ij}=(\a^{(i)\vee},\a^{(j)\vee})$. The 
symmetrized Cartan matrix coincides with the restriction of the Killing form 
to the Cartan subalgebra $\mf{g}_{\circ}$. 

In order to describe the inner products on the  root and weight spaces 
explicitly, one expresses roots and weights through their components
with respect to the basis of simple coroots and the Dynkin basis
$\cb^{\star}:=\{\Lambda_{(i)}\;|\;i=1,2,\ldots,r\}$, respectively. 
Thus one writes
\begin{eqnarray}
    \l = \sum_{i=1}^r \l_i\a^{(i)\vee}=\sum_{j=1}^r \l^j\Lambda_{(j)}\quad
    \mbox{with} \quad
    \l_i=(\l,\Lambda_{(i)})\,,\quad \l^i = (\l, \a^{(i)\vee})\,.
    \label{6.45}
\end{eqnarray}
The coefficients $\l^i$ are called \emph{Dynkin labels}. As a consequence of $(\Lambda_{(i)},\a^{(j)\vee})=\delta_i^j$ one has
\begin{eqnarray}
    {(\a^{(j)\vee})}_i = \d_i^j = {(\Lambda_{(i)})}^j\,,\quad
    {(\a^{(j)})}_i = \frac{(\a^{(j)},\a^{(j)})}{2}\d_i^j\,.\label{6.47}
\end{eqnarray}
The quadratic form matrix and the symmetrized Cartan matrix serve to lower and raise
indices, respectively:
\begin{eqnarray}
    \l_i=\sum_{j=1}^r G_{ij}\l^{j}\,,\quad \l^i=\sum_{j=1}^r G^{ij}\l_{j}\,.
    \label{6.48}
\end{eqnarray}
The matrix $G^{ij}$ provides an inner product on the weight space; for simple Lie algebras, this product is in fact euclidean, \emph{i.e.} its signature is $(r,0)$.
Combining (\ref{6.47}) and the expression of $G^{ij}$ in terms of the Cartan matrix,
it follows that
\begin{eqnarray}
    {(\a^{(i)})}^j \equiv \sum_{k=1}^r {(\a^{(i)})}_k G^{kj} = {(A)}^{ij}\,.
    \label{6.51}
\end{eqnarray}
In words: {\emph{the components of the simple roots in the Dynkin
basis coincide with the rows of the Cartan matrix}.

The Dynkin labels are the eigenvalues of the Chevalley generators
of the Cartan subalgebra in the basis were Eq. (\ref{wcbasis}) holds.
Indeed 
\beq h^i \arrowvert \l \rangle &=& \l(h^i)|\l \rangle \nn \\
         &=& \sum_j \l^j \Lambda_{(j)}(h^i) \arrowvert \l \rangle \nn \\    
          &=& \sum_j \l^j \frac{2}{|\a^{(i)}|^2} \Lambda_{(j)}(h_{\a^{(i)}})
          \arrowvert \l \rangle \nn \\
   &=& \l^i \arrowvert \l \rangle\, . \nn
\eeq Recall that  the eigenvalues
$\l^i$ of $h$ are integers. They can be computed by using Eq. (\ref{pq}).

\subsubsection{Weights and their Multiplicities}

A weight $\Lambda$ of a representation of ${\mf{g}}$ is said to be a 
\emph{Highest
  weight $\Lambda$} if $\Lambda$ is such that $\Lambda >\l$ for every other 
weight $\l$.

Any finite-dimensional irreducible representation has a unique
highest weight state $|\Lambda \rangle $ which is completely
determined by its Dynkin labels. The importance of the highest weight of a 
representation stems from the fact that \emph{each irreducible representation 
is uniquely and completely specified by its highest weight, all of its 
properties, such as its dimension and the other weights being easily deducible 
from it}.
Moreover, there is a straightforward
procedure for constructing every possible highest weight.
The highest weight state is such that
$ E^{\a}|\Lambda \rangle = 0,\quad \forall \a > 0 $. The existence
of such a highest weight state is obvious (for finiteness of the
representation) but the uniqueness is not trivial.

More precisely, for every irreducible representation $\mathbf{\G}$ of a 
semi-simple complex Lie algebra ${\mf{g}}$ the highest weight $\Lambda$ can 
be written as
\begin{eqnarray}
    \Lambda = \sum_{i=1}^r n^i \Lambda_{(i)}\,,\quad
    n^i\in\mathbb{Z}_+\,,\quad 1\leqslant i\leqslant r
    \label{5.12}
\end{eqnarray}
where $\{\Lambda_{(i)}\}_{i=1}^r$ is the set of fundamental weights of
${\mf{g}}$. Moreover, to {\emph{every}} set of non-negative integers
$\{n^i\}_{i=1}^r$ there exists an irreducible representation of ${\mf{g}}$
with highest weight $\Lambda$ given by Equation (\ref{5.12}), and this 
representation is unique up to equivalence. This highest weight can be 
encoded into a \emph{Young Tableau}. A Young tableau is set of boxes that encodes the Dynkin labels.  For example, the representation of $A_{10} = \mf{sl}(11,\CC)$ the weight  $\Lambda = \sum_{i=1}^{10}\l^i \Lambda_{(i)}$ such that $\l^i = [0, \ 0, \ 1, \ 0, \ 0 , \ 0 , \ 0 , \ 0 , \ 0, \ 1]$  is encoded in the following Young tableau (the first column contains 8 boxes),
\begin{center}
\scalebox{.9}{\begin{picture}(30,30)(0,-10)
\multiframe(0,10)(10.5,0){1}(10,10){}
\multiframe(10.5,10)(10.5,0){1}(10,10){}
\multiframe(0,-0.5)(10.5,0){1}(10,10){}
\multiframe(0,-18)(10.5,0){1}(10,17){$ $} \put(4,-13.5){$\vdots$}
\multiframe(0,-28.5)(10.5,0){1}(10,10){}
\end{picture}}
\end{center}
The dimension $d$ of such a representation is given by the following ``Weyl's 
dimensionality formula''
\begin{eqnarray}
d = \prod_{\a\in\Delta_+}{(\Lambda+\rho,\a)
\over (\rho,\a)}\,,\quad\quad \r=\sum_{i}^r \Lambda_{(i)}  \ ,
\label{weyldim}
\end{eqnarray}
where $$ \rho = \sum_i \Lambda_{(i)} = {1 \over 2} \sum_{\a \in \Delta_+} \a\,
.$$
For any simple complex Lie algebra ${\mf{g}}$, Weyl's formula (\ref{weyldim})
can be rewritten as
\begin{eqnarray}
    d = \prod_{\a\in\Delta_+}\Big[\Big( \sum_{i=1}^r n^ik_i^{\a}w_i/
     \sum_{j=1}^r k_j^{\a}w_j\Big) + 1  \Big]\,,
     \label{weyldim2}
\end{eqnarray}
where $\a= \sum_{i=1}^r k_i^{\a}\a^{(i)}\,$, $\Lambda=\sum_{i=1}^r n^i 
\Lambda_{(i)}\,$, and $w_i$ is the ``weight''\footnote{$w_i = w (\a^{(i)},
\a^{(i)})$ where the constant $w$ is equal to min$_i\{
\arrowvert(\a^{(i)},\a^{(i)})\arrowvert^{-1} \}$}. 
Although less concise, Eq.(\ref{weyldim2}) is much easier to apply than 
Eq.(\ref{weyldim}).

The set of eigenvalues of all the states in $V_{\Lambda}$ is called
\emph{the weight system} and denoted $\Omega_{\Lambda}\,$.
All the states in the representation space
$V_{\Lambda}$ are obtained by the action of the lowering operators of
$\mf{g}$ as
$$ E^{-\b} E^{-\g}\cdots E^{-\eta} |\Lambda \rangle\,, ~~ \mathrm{where} ~~
\b,\g, \ldots, \eta \ \mathrm{ ~~~are~~ simple ~~roots}\,. $$
As one can learn from the representation theory of $\mf{sl}(2)$, 
each weight $\l \in
\Omega_\Lambda$ is necessarily of the form $\Lambda - \sum n_i
\a^{(i)}$ with $n_i \in \mathbb{Z}_+$. The sum $\sum n_i $ is
called the \emph{level} of $\l$ and an analysis level by level can
be done. Let us focus on the action of a specific simple root
$\a^{(i)}\,$. As $E^{\a^{(i)}} | \Lambda \rangle = 0$ and $h^i
|\Lambda \rangle = \Lambda^i |\Lambda \rangle$, the representation
theory of $\mf{sl}(2)$
 tells us that $E^{-\a^{(i)}} |\Lambda \rangle $ is non vanishing iff 
$\Lambda^i \geqslant 1$, more precisely one can act $\Lambda^i$ times with 
$E^{-\a^{(i)}}$ on $|\Lambda \rangle$. With this criterion, the systematic
construction of all weights in the representation can be done by means of the
following algorithm.
We start with the highest weight
$\Lambda =(\Lambda^1,...,\Lambda^r)$ in the Dynkin basis.
For each positive Dynkin label $\Lambda^i>0$ we
construct the sequence of weights $\Lambda-\a^{(i)}$, ..., $\Lambda-\Lambda^i 
\a^{(i)}$
which belong to $\Omega_{\Lambda}$. The process is repeated with $\Lambda$
replaced by each of the weights just obtained. 

The \emph{Freudenthal recursion formula} gives the multiplicity
of the weight $\l$ in the representation $\Lambda$
\begin{eqnarray}
\fbox{$\displaystyle
[(\Lambda + \rho,\Lambda + \rho) - (\l + \rho, \l + \rho) ]
 \mathrm{mult}_{\Lambda}(\l)
= 2 \sum_{\a >0} \sum_{k=1}^{\infty} (\l + k \a , \a )
\mathrm{mult}_{\Lambda}(\l + k\a) $}
\ .
\label{15.17}
\end{eqnarray}
%%%%%%%%%%%%
%%%%%%%%%%%%%%
\cleardoublepage
%%%%%%%%%%%%%%%
%%%%%%%%%%%%%%%
%%\include{app_real_alg}
\chapter{Real forms of complex semi--simple Lie algebras}
\markboth{REAL FORMS OF COMPLEX SEMI--SIMPLE LIE ALGEBRAS}{}

Let us recall some definitions,  

\noindent $V^\CC$: Let $V$ be a vector space over $\RR$.  $ V^{\CC} := V \otimes_\RR \CC $
is called the complexification of $V$. One has dim$_\RR V =$ dim$_\CC V^{\CC}$. \newline
$W^\RR$: Let $W$ be a vector space over $\CC$. Restricting the definition of 
scalars to $\RR$ then leads to a vector space $W^\RR$ over $\RR$ and 
dim$_\CC W = 1/ 2$ dim$_\RR W^\RR$. \newline

\noindent \emph{Real form of $\mf{g}$ }: Let $\mf{g}$ be a Lie algebra over $\CC$. A \emph{real form }of $\mf{g}$ is a subalgebra $\mf{g}_\circ$ of the real Lie algebra $\mf{g}^\RR$ such that 
$$ \mf{g}^\RR = \mf{g}_\circ \oplus i \mf{g}_\circ \hspace{2cm} \mathrm{ (direct \, sum \, of \, vector \, 
spaces)} \, . $$
In the rest of the appendix, only real forms of  \emph{semi--simple } complex Lie algebras are considered. There are two special real forms, the \emph{split real form} and the
\emph{compact real form}: 
\begin{itemize}
\item Let $\D$ be the root system of $\mf{g}$. Then for each $\a \in \D$ it is possible to choose generators $e_\a \in 
\mf{g}_\a$ such that $\forall \a, \ \b \in \D$: 
\begin{itemize}
\item $\ [e_\a, e_{-\a} ] = h_\a $ 
\item $\ [e_\a, e_\b ] = N_{\a,\b} e_{\a+\b}$ if $\a+ \b \in \D$ with 
$N_{\a,\b}$ real
\item $\ [e_\a, e_\b ] = 0$ if $\a + \b \neq 0$ and $\a +\b \notin  \D$ .
\end{itemize} Defining $\mf{h}_\circ = \{ H \in \mf{h} | \ \a(H)
\in \RR \ \forall \a \in \D \}$, the \emph{split real form} is 
$$ \mf{g}_\circ^{split} = \mf{h}_\circ \oplus_{\a \in \D} \RR e_\a \ .$$ 
\item The \emph{compact real form} $\mf{u}_\circ$ is constructed from the 
split real one,
\beq \mf{u}_\circ = \sum_{\a \in \D} \RR \ ih_\a + \sum_{\a \in \D} \RR 
(e_\a - e_{-\a}) + \sum_{\a \in \D} \RR \ i(e_\a +e_{-\a}) \ . 
\label{crf}
\eeq
\end{itemize}

\subsubsection{Cartan Decomposition}

\noindent An automorphism $\s : \mf{g} \rightarrow \mf{g}$ such that $\s^2 
=1 $ is called an involution. A \emph{Cartan involution} $\th$ of a real semi-simple
Lie algebra is an involution such that $K_\th$ is positive define. $K_\th$ is 
defined by $K_\th(Z,Z') :=-K(Z,\th Z')$ for all $Z,\, Z' \in \mf{g}$ and $K$ is the Killing form of $\mf{g}$. It can be shown that every real 
semi-simple Lie algebra has a Cartan involution. 

\noindent A vector space decomposition
\beq
\mf{g}_\circ = \mf{k}_\circ \oplus \mf{p}_\circ
\label{cartanreal}
\eeq  
of $\mf{g}_\circ$ is called Cartan decomposition if \ (i)
the following bracket laws are satisfied $\ [\mf{k}_\circ ,\mf{k}_\circ ] 
\subseteq \mf{k}_\circ $, $\ [\mf{k}_\circ,\mf{p}_\circ ] \subseteq 
\mf{p}_\circ $, $\ [\mf{p}_\circ,\mf{p}_\circ] \subseteq \mf{k}_\circ$ and 
(ii) the Killing form $K$ is negative definite on $\mf{k}_\circ$ 
and positive definite on $\mf{p}_\circ$. 

\noindent Let $\th$ be a Cartan involution of $\mf{g}_\circ$. Then the 
eigenspace $\mf{k}_\circ$ of eigenvalue +1 and the eigenspace $\mf{p}_\circ$ 
of eigenvalue -1 correspond to a Cartan decomposition. Conversely, one can 
associate a Cartan involution in a natural way to any Cartan decomposition. 

\subsubsection{Iwasawa Decomposition}

\noindent Let $\mf{a}_\circ$ be a maximal abelian subspace of $\mf{p}_\circ$. 
For $\l \in \mf{a}_\circ^\star$, let 
$$ \mf{g}_{\circ,\l} := \{X\in \mf{g}_\circ | \mathrm{ad}_H(X) \ = \ \l(H) X \ \ 
\forall H \in \mf{a}_\circ \} \ .$$
If $\mf{g}_{\circ,\l} \neq 0$ and $\l \neq 0$, $\l $ is called a restricted 
root of $\mf{g}_\circ$. $\S$ denotes the set of restricted roots. 
The restricted roots and restricted root spaces have the following 
properties: (i) $\mf{g}_\circ = \mf{g}_{\circ,0} \oplus_{\l \in \S}
\mf{g}_{\circ,\l}$ is an orthogonal sum,  (ii) $\ [\mf{g}_{\circ,\l},
\mf{g}_{\circ,\m} ] \subseteq \mf{g}_{\circ,\l+\m}$, (iii) $\th
\mf{g}_{\circ,\l} = \mf{g}_{\circ,-\l}$, (iv) $\l \in \S \ \Rightarrow \ 
-\l \in \S$, (v) $\mf{g}_{\circ,0} = \mf{a}_\circ \oplus \mf{m}_\circ$ orthogonally, 
where\footnote{$Z_{\mf{b}}(\mf{c}) = \{X \in \mf{b} \ | \ [X,Y] = 0 \ \forall \
Y \in \mf{c} \}$.} $\mf{m}_\circ = Z_{\mf{k}_\circ}(\mf{a}_\circ)$.

\noindent A notion of positivity can be introduced on $V$, the vector space 
containing an abstract reduced\footnote{An abstract root system $\D$ is said
  to be reducible if $\D$ admits a non trivial disjoint decomposition $\D=
\D'\cup \D''$ with every elemen of $\D'$ orthogonal to every element of 
$D''$.} root system $\D$. A positivity notion is introduced on 
$\mf{a}_\circ^\star$ and the set of positive restricted roots is denoted 
$\S^+$. The subalgebra $\mf{n}_\circ = \oplus_{\l \in \S^+} \mf{g}_{\circ,\l}$ 
is such that  
$\mf{n}_\circ$ is nilpotent, $\ [\mf{a}_\circ,\mf{n}_\circ ] = \mf{n}_\circ$
and $\mf{a}_\circ \oplus \mf{n}_\circ$ is a solvable algebra. Every 
$\mf{g}_\circ$ possesses an \emph{Iwasawa decomposition}, i.e. $\mf{g}_\circ$ 
can be decomposed into a direct sum of vector space directs as follows,
\beq \mf{g}_\circ = \mf{k}_\circ \oplus 
\mf{a}_\circ \oplus \mf{n}_\circ \ . 
\label{iwasawareal}
\eeq 

\noindent The \emph{rank} of $\mf{g}_\circ$ is defined to be the dimension 
of any \emph{Cartan subalgebra} $\mf{h}_\circ \equiv \mf{g}_{\circ,0}$ of $\mf{g}_\circ$. This is 
well defined since $\mf{h}_\circ$ is a Cartan subalgebra of $\mf{g}_\circ$ iff
$\mf{h}_\circ^\CC$ is a Cartan subalgebra of $\mf{g}$. Let $t_\circ$ be a 
maximal abelian subset of $\mf{m}_\circ$, then it can be shown that 
$\mf{h}_\circ 
= t_\circ \oplus \mf{a}_\circ $.

\noindent Using $\mf{h}_\circ = t_\circ \oplus \mf{a}_\circ  $ as a Cartan 
subalgebra of $\mf{g}_\circ$, we built the set $\D(\mf{g},\mf{h})$ of 
roots of $\mf{g}$ with respect to the Cartan subalgebra $\mf{h}$. It is 
possible to introduce a notion of positivity compatible with the one of 
$\S$.

\noindent An Iwasawa decomposition is unique up to conjugation by 
internal automorphisms $Int(\mf{g}_\circ)$ of $\mf{g}_\circ$. But a Cartan subalgebra $\mf{h}_\circ$ is conjugate
via $Int(\mf{g}_\circ)$ to a $\th$ stable Cartan subalgebra. Therefore, it
is sufficient to study stable Cartan subalgebras. Let $\mf{h}_\circ$ be a 
$\th$ stable Cartan subalgebra of $\mf{g}_\circ$. Then $\mf{h}_\circ = t_\circ
\oplus \mf{a}_\circ$ with $t_\circ \subseteq \mf{k}_\circ$ and $\mf{a}_\circ
\subseteq \mf{p}_\circ$. One can show that all roots of $(\mf{g},\mf{h})$ are 
real valued on $i t_\circ \oplus \mf{a}_\circ  $. The roots taking real (resp.
purely imaginary) values on $\mf{h}_\circ$ are called real (resp. imaginary). 
dim $t_\circ$ is called compact dimension and dim $\mf{a}_\circ$ non compact 
dimension. 

\noindent A $\th$ stable Cartan subalgebra is called \emph{maximally 
noncompact} if its noncompact dimension is maximal. It can be shown that 
$\mf{h}_\circ$ is maximally compact iff $t_\circ$ is maximally abelian in 
$\mf{t}_\circ$. The action of $\th$ on a root $\a$ is given by
$$ (\th(\a))(h) = \a(\th(h)) \, . $$

\subsubsection{Satake Diagram}

\noindent 
A \emph{Satake diagram} of the real simple Lie algebra $\mf{g}_\circ$ with \emph{maximally non compact} Cartan subalgebra 
$\mf{h}_{\circ} = t_\circ \oplus \mf{a}_\circ$ consists in the Dynkin diagram of $\mf{g}$ (the root lattice of $\mf{g}$ is denoted $\Delta$) with additional decoration:  
\begin{itemize}
\item Black nodes stand for roots invariant  under the Cartan involution $\th$, \ie if the root $\a_i$ is 
attached to a black node, one has $\th(\a_i) =  \a_i$. This condition is equivalent to ask that $\a_i$ vanishes on $\mf{a}_\circ$, indeed $\th(\a)(x) = \a(x)$ for $x \in\mf{a}_\circ $ gives $\a(\th(x)) = 
-\a(x) = \a(x)$.
\item White nodes stand for roots \emph{not } invariant under the Cartan involution $\th$.
\item White nodes can be connected by arrows. If the node $\a_i$ is connected to the node $\a_j$ it means that 
$ \a_i - \th(\a_i ) = \a_j - \th(\a_j)$. This condition is equivalent to ask that $\a_i$ and $\a_j$ take the 
same value on $\mf{a}_\circ$.
\end{itemize}
Note that (i) the restricted root system $\Sigma$ is the set of roots $\Delta_1 = \Delta - \Delta_0$ where
$\Delta_0 = \{ \a \in \Delta \arrowvert \th(\a) = \a \}$ (ii) the set of roots $\Delta_s$ anti--invariant under $\th$ is orthogonal to $\Delta_0 $. 

Let $\s$ be the involution determining the real form $\mf{g}_\circ$. It can be shown that 
$\s = \th \cdot \tau$ where $\tau$ is the involution fixing the maximally compact real form $\mf{u}_\circ$ 
given by (\ref{tau}). In order to determine the real form $\mf{g}_\circ$, the action of $\th$ on the generators of $\mf{g}$ is needed: 
if $\a \in \Delta_0$, the involution $\th$ acts on 
$e_\a, \, f_\a, \, h_\a$ as $\th(h_\a) = h_\a, \, \th(e_\a) = e_\a$ and $\th(f_\a) = f_\a$, \ie $\th = +1$;  if 
$\a \in \Delta_s$, $\th$ acts on $e_\a, \, f_\a, \, h_\a$ as $\th(h_\a) = -h_\a$, $\th(e_\a) = -f_\a$; finally 
if $\a \in \Delta_1 - \Delta_s$, then $\th (E_\a) =  \rho E_{\th(\a)}$ where $E_{\a} = e_\a$ if $\a$ is a positive root and $E_\a = f_{-\a}$ is $\a$ is a negative root. The sign $\rho$ is fixed by the fact that $\s(e_\a) = + e_{\b}$ for some positive root $\b \in \Delta_1 - \Delta_s$.  
\begin{quote} \emph{Example 1.}: The most simple examples are the real forms of $\mf{a}_1 = \mf{sl}(2,\CC)$, where the two different Satake diagrams 
are shown in Figure \ref{satakea1}.
\begin{center}
\begin{figure}[h]
\scalebox{.5}{
\begin{picture}(180,60)
\put(354,-10){$A I $} 
  \put(494,-10){$ A II$} 
\thicklines 
\put(360,10){\circle{10}}
\put(500,10){\circle*{10}}
\end{picture} }
\caption{ \label{satakea1} {\small Satake diagrams for the real forms of $\mf{sl}(2,\CC)$. }}
\end{figure}
 \end{center}
For $A I$, since $\a_1$ is not black and therefore not invariant under $\th$, one concludes that $\th(\a_1) = - \a_1$. It follows that $\s(f_1) = \th(-e_1) = f_1$, $\s(e_1) = \th(-f_1) = e_1$ and $\s(h_1) =h_1$.   Therefore $A I$ is generated by 
$e_{\a_1}, \, f_{\a_1}$ and $h_{\a_1}$ and $AI = \mf{sl}(2,\RR)$. For $AII$, the only node is black therefore the involution fixing the real Lie algebra is $\tau$ and $AII = \mf{su}(2)$. 
\end{quote}
\begin{quote} \emph{Example 2.}: The Satake diagram depicted in the  Figure \ref{satake} encodes one 
real form of the complex simple Lie algebra $\mf{a}_2$. 
\begin{figure}[h]
  \centering
\begin{picture}(0,0)%
\epsfig{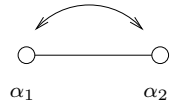}%
\end{picture}%
\setlength{\unitlength}{2763sp}%
\begingroup\makeatletter\ifx\SetFigFont\undefined%
\gdef\SetFigFont#1#2#3#4#5{%
  \reset@font\fontsize{#1}{#2pt}%
  \fontfamily{#3}\fontseries{#4}\fontshape{#5}%
  \selectfont}%
\fi\endgroup%
\begin{picture}(1433,894)(3451,-4244)
\put(3451,-4186){\makebox(0,0)[lb]{\smash{{\SetFigFont{8}{9.6}{\familydefault}{\mddefault}{\updefault}{\color[rgb]{0,0,0}$\a_1$}%
}}}}
\put(4651,-4186){\makebox(0,0)[lb]{\smash{{\SetFigFont{8}{9.6}{\familydefault}{\mddefault}{\updefault}{\color[rgb]{0,0,0}$\a_2$}%
}}}}
\end{picture}%
  \caption{\small Satake diagram of type $AIII$ for $\mf{a}_2$}
  \label{satake}
\end{figure}
 The involution $\th$ acting on the roots can be read from this diagram: $\a_1- \th(\a_1) = \a_2 - \th(\a_2)$. 
 As $\th(\a_1)$ is not $\a_1$ (not a black node) nor in $\{-\a_1, \, \a_2, \, \a_1+\a_2, \, -\a_1-\a_2\}$ since this would imply that $\th(\a_2) $ is not a root, one concludes that $\th(\a_1) = -\a_2$. At the level of the 
 generators, one has therefore $\th(e_1) = \rho f_2$, $\th(e_2) = \rho' f_1$, ... and finally 
 $\s(e_1) = e_2$, $\s(f_1) = f_2$ and $\s(h_1) = h_2$. The generators of $\mf{g}_\circ$ are those invariant under $\s$, \ie 
 \beq h_1 + h_2 \, , \hspace{1cm}Êi(h_1-h_2) \, \hspace{1cm} e_1+e_2 \, \hspace{1cm} i(e_1-e_2) \, , \nn 
 \\
 f_1 + f_2 \, , \hspace{1cm}Êi(f_1-f_2) \, , \hspace{1cm}Êie_3 \, , if_3 \, . \nn \eeq 
 The subalgebra $\mf{a}_\circ$ is generated by $h_1+h_2$ and the restricted root system is : 
 $\a$ with multiplicity 2 and $2\a$ with multiplicity 1 (where $\a$ is such that $\a(h_1+h_2)   = 1$) and the corresponding negative roots.

 \end{quote}
%%%%%%%%%%%%
%%%%%%%%%%%%%
\cleardoublepage
%%%%%%%%%%%%
%%%%%%%%%%%%
%%\include{app_kac_moody}

\chapter{Kac--Moody Algebras}
\markboth{KAC--MOODY ALGEBRAS}{}
\label{km}

\indent This appendix is mainly inspired by \cite{deBuyl:2004ps}. We refer to \cite{Kac:1990gs} for more details. \newline
 Let us define a Kac--Moody algebra $\mf{g}$  via its
Dynkin diagram, with $n$ nodes and links between these nodes, that encodes a  generalised Cartan matrix $A$. The
algebra is a Lie algebra with Chevalley generators $h_i, \ e_i, \
f_i$ ($i=1,\ldots,n$) obeying the following relations 

\begin{eqnarray}
\ [ e_i ,f_j ] &=& \d_{ij} h_i \nonumber \\
\ [h_i ,e_j] &=& A_{ij} e_j \nonumber \\
\ [h_i ,f_j] &=& -A_{ij} f_j \nonumber \\
\ [h_i ,h_j] &=& 0 \label{serrerelationskm}
\end{eqnarray}

\noindent where $A_{ii}=2$ and $-A_{ij}$ ($i \neq j$) is a
non-negative integer related to the number of links between the
$i^{\mathrm{th}}$ and $j^{\mathrm{th}}$ nodes. The so-called Cartan
matrix $A$ in addition satisfies $A_{ij}=0 \Leftrightarrow A_{ji}=0$.
The generators must
also obey the Serre relations,

\begin{eqnarray}
(\mathrm{ad}\,e_i)^{1-A_{ij}} e_j &=& 0 \nonumber \\
(\mathrm{ad}\,f_i)^{1-A_{ij}} f_j &=& 0 \label{serrerel}
\end{eqnarray}
A root $\a$ of the algebra is a non-zero linear form on the Cartan
subalgebra $\mf{h}$ (= the subalgebra generated by the $\{ h_i \ \arrowvert
\  i = 1,... ,n \}$)\footnote{Here we assume the Cartan matrix $A$ to
  be non-degenerate.} such that
$$\mf{g}_{\a} = \{ x \in \mf{g} \ \arrowvert [ h, x ] =
\a (h) x \ \forall h \in \mf{h} \}$$ is not empty. $\mf{g}$ can be
decomposed in the following triangular form,

\beq \mf{g} = \mf{n}_- \oplus \mf{h} \oplus \mf{n}_+ \label{triangkm} \eeq  or according to the root spaces,
$$ \mf{g} = \oplus_{\a \in \Delta} \mf{g}_{\a} \oplus \mf{h}$$ where
$\mf{n}_-$ is the direct sum of the negative roots spaces, $\mf{n}_+$ of the
positive ones and $\mf{h}$ is the Cartan subalgebra. The dimension of
$\mf{g}_{\a}$ is called the multiplicity of $\a$. These
multiplicities obey the Weyl--Kac character formula

$$ \Pi_{\a \in \Delta_+} (1 - e^{\a})^{\mathrm{mult}\a} = \sum_{w
\in W}  \epsilon(w) e^{w(\rho)-\rho} $$

\noindent The sum is over the Weyl group which in the case of
interest here is infinite. $\epsilon(w)$ is the parity of $w$ and
$\rho$ is the Weyl vector. This formula cannot be solved in closed
form in general. 

Our interest is focussed here on a class of Kac--Moody algebras
called \textit{very-extensions} of simple Lie algebras. They are
the natural extension of the over-extended algebras which are
themselves extensions of the affine algebras. The procedure for
constructing an affine Lie algebra $\mf{g}^{+}$ from a simple one $\mf{g}$
consist in the addition of a node to the Dynkin diagram in a
certain way which is related to the properties of the highest root
of $\mf{g}$. One may also further increase by one the rank of the
algebra $\mf{g}^{+}$ by adding to the Dynkin diagram a further node
that is attached to the affine node by a single line. The
resulting algebra $\mf{g}^{++}$ is called the over--extension of $\mf{g}$. The
very-extension, denoted $\mf{g}^{+++}$, is found by adding yet another
node to the Dynkin diagram that is attached to the over-extended
node by one line \cite{Gaberdiel:2002db}.

The algebras $\mathfrak{g}$ 
build on a symmetrisable Cartan matrix $A$ have been classified according to properties of their eigenvalues
\begin{itemize}
\item if $A$ is positive definite,  $\mathfrak{g}$  is a finite
dimensional Lie algebra;
\item if $A$ admits one null eigenvalue and the others are all strictly positive,  $\mathfrak{g}$  is an Affine Kac--Moody algebra;
\item if $A$ admits one negative eigenvalue and all the others
are strictly positive,  $\mathfrak{g}$  is a Lorentzian Kac--Moody algebra.
\end{itemize}
A Kac--Moody algebra such that the deletion of one node from its
Dynkin diagram gives a sum of finite or affine algebras is called an
\textit{hyperbolic} Kac--Moody algebra. These algebras are all known; in particular,  there exists  no hyperbolic algebra with rang higher than 10. 
%%%%%%%%%%%
%%%%%%%%%%%%
\cleardoublepage
%%%%%%%%%%%
%%%%%%%%%%%%
%%\include{app_hamiltonian_gr}

\chapter{Hamiltonian Formulation of General 
Relativity}
\markboth{HAMILTONIAN FORMULATION OF GENERAL RELATIVITY}{}
\label{hamilton}

The Hamiltonian formulation of General Relativity requires a breakup 
of spacetime into space and time. This can be achieved by foliating
spacetime by Cauchy hypersurfaces\footnote{Let $M$ be the spacetime. 
A Cauchy hypersurface is an achronal set $\S$ such that $D(\S) = M$. 
A subset $S \subset M$ is said to be achronal if there exist no 
$p,q \in S$ such that $q \in I^+(S)$. Recall that $I^+(S)$ is the 
chronological future of $S$, $I^+(S) = \cup_{p \in S}I^+(p)$ and $I^+(p) \ = 
\ \{ q \in M |$ there exists a future directed timelike curve $\l(t)$ with 
$\l(0) = p$ and $\l(1) = q$ \} and that $D(\S) = D^+(\S) \cup D^-(\S)$ 
where $D^\pm(\S) = \{ p \in M |$ every past/futur inextendible causal curve 
through $p$ intersects $\S \}$. } 
$\S_t$ parametrised by a global time function $t$. 
\begin{figure}[h] 
  \centering 
\begin{picture}(0,0)%
\includegraphics{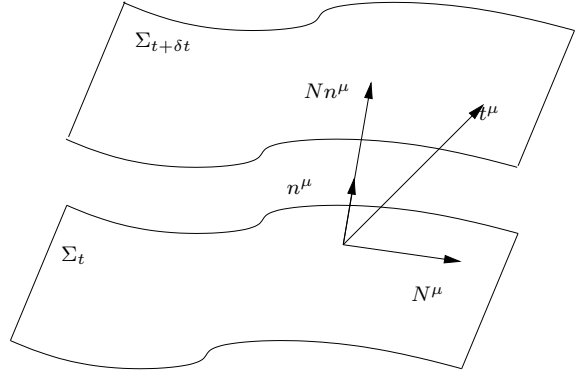}%
\end{picture}%
\setlength{\unitlength}{3108sp}%
\begingroup\makeatletter\ifx\SetFigFont\undefined%
\gdef\SetFigFont#1#2#3#4#5{%
  \reset@font\fontsize{#1}{#2pt}%
  \fontfamily{#3}\fontseries{#4}\fontshape{#5}%
  \selectfont}%
\fi\endgroup%
\begin{picture}(4524,2955)(2329,-4123)
\put(2746,-3256){\makebox(0,0)[lb]{\smash{{\SetFigFont{8}{9.6}{\familydefault}{\mddefault}{\updefault}{\color[rgb]{0,0,0}$\Sigma_t$}%
}}}}
\put(5536,-3571){\makebox(0,0)[lb]{\smash{{\SetFigFont{8}{9.6}{\familydefault}{\mddefault}{\updefault}{\color[rgb]{0,0,0}$N^{\mu}$}%
}}}}
\put(6076,-2131){\makebox(0,0)[lb]{\smash{{\SetFigFont{8}{9.6}{\familydefault}{\mddefault}{\updefault}{\color[rgb]{0,0,0}$t^{\mu}$}%
}}}}
\put(4681,-1951){\makebox(0,0)[lb]{\smash{{\SetFigFont{8}{9.6}{\familydefault}{\mddefault}{\updefault}{\color[rgb]{0,0,0}$N n^{\mu}$}%
}}}}
\put(4546,-2716){\makebox(0,0)[lb]{\smash{{\SetFigFont{8}{9.6}{\familydefault}{\mddefault}{\updefault}{\color[rgb]{0,0,0}$n^{\mu}$}%
}}}}
\put(3331,-1546){\makebox(0,0)[lb]{\smash{{\SetFigFont{8}{9.6}{\familydefault}{\mddefault}{\updefault}{\color[rgb]{0,0,0}$\Sigma_{t+\delta t}$}%
}}}}
\end{picture}%
  \caption{\small{Slicing of spacetime by hypersurfaces $\S_t$.}}
  \label{slice} 
\end{figure} 
Let $n_\m$ be the unit normal vector field to $\S_t$ (this vector
is therefore proportional to $\nabla_\m t$, $n_\m = \a \nabla_\m t$). 
The spacetime metric $g_{\m\n}$ induces, on each 
$\S_t$, a spatial metric $h_{\m\n}$:
\beq h_{\m\n} = g_{\m\n} + n_\m n_\n \ . \label{smetric}\eeq
Let $t^\m$ be a vector field such that the covariant derivative of the 
time function $t$ along its direction equals one, $t^\m\nabla_\m t = 1$. 
This vector field may be interpreted as the ``flow of time''. It is  
useful to decompose $t^\m$ into two vectors, one perpendicular and one 
tangent to $\S_t$, $t^\m = N n^\m + N^\m$. This defines respectively the lapse 
function\footnote{
The second equality of Eq.(\ref{lapse}) is obtained by using $n_\m = \a 
\nabla_\m t$ which implies by the first equality of Eq.(\ref{lapse}) that
$-\a t^\m \nabla_\m t = N $ (and $ \a = -N$). Then $n_\m = -N \nabla_\m t$ 
which contracted with $n^\m$ gives the desired equality.} 
$N$ and the shift vector $N_\m$, through
\beq N & =& -t^\m n_\m = (n^\m \nabla_\m t)^{-1}\ , \label{lapse}\\
     N_\m &=& h_{\m \n}t^\n \ .  \nn \eeq

\subsubsection{$d$--dimensional tensors}

It is natural to choose coordinates appropriate to the $d$+1 dimensional breakup and to introduce $d$--tensors. Let the spacetime indices $\m$ decompose as $ \m = \{ 0,a\}$ where $\{a\}$ label coordinates on the surfaces $\S_t$ and the $0$th index refers to $t$. Instead of considering the induced metric $h_{\m\n}$ as a spacetime tensor, the 
$d$--dimensional metric $^{\sst{(d)}}h_{ab} \equiv g_{ab}$ ($a=1,\dots,d$)  can be used. Since $n_a = 0 $ where $a \in $ $\{ 1,...,d-1\}$, the relationship between these two tensors is given by  
\beq 
h_{\m \n} = \left( \begin{array}{cc} 
0 & 0 \\
0 & ^{\sst{(d)}}h_{ab} \end{array} 
\right) \, .
\eeq
For notational convenience, we will simply denote $^{\sst{(d)}}h_{ab}$ by $h_{ab}$. 
More generally,  a spacetime tensor  
$T^{\m_1...\m_m}{}_{\n_1...\n_n}$ which satisfies
\beq 
h^{\r_1}{}_{\m_1}...h^{\r_m}{}_{\m_m}
h^{\n_1}{}_{\s_1}...h^{\n_n}{}_{\s_n}T^{\m_1...\m_m}{}_{\n_1...\n_n}
= T^{\r_1...\r_m}{}_{\s_1...\s_n} \ , 
\nn 
\eeq 
has its components with one temporal index that vanish. Therefore, one can  
express it as a spatial tensor and in this case one will denote it $^{\sst{(d)}}T^{m_1...m_m}{}_{n_1...n_n}$, or more simply $T^{m_1...m_m}{}_{n_1...n_n}$. 
As we have seen, $h_{\m\n}$ is such a tensor. Note that the given of 
$g_{\m\n}$ is equivalent to that of $\{ g_{ab}, \ N,\ 
N_a \}$, 
\beq 
g_{\m\n} = \left( \begin{array}{cc} N_a N^a -N^2 & N_b \\
N_a & g_{ab}  \end{array} \right) \ 
\label{decmet} 
\eeq 
and that $N_\m$ is also 
a spatial tensor in this sense explained above. 

\subsubsection{Extrinsic curvarture}

An important notion, illustrated by Fig.(\ref{extcurvfi}), is the extrinsic 
curvature because it is related to the ``time derivative'' of $h_{\m\n}$. 
\begin{figure}[h]
  \centering
\begin{picture}(0,0)%
\includegraphics{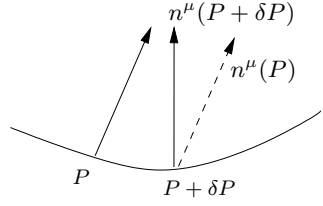}%
\end{picture}%
\setlength{\unitlength}{3108sp}%
\begingroup\makeatletter\ifx\SetFigFont\undefined%
\gdef\SetFigFont#1#2#3#4#5{%
  \reset@font\fontsize{#1}{#2pt}%
  \fontfamily{#3}\fontseries{#4}\fontshape{#5}%
  \selectfont}%
\fi\endgroup%
\begin{picture}(2499,1555)(3409,-4241)
\put(3916,-4111){\makebox(0,0)[lb]{\smash{{\SetFigFont{8}{9.6}{\familydefault}{\mddefault}{\updefault}{\color[rgb]{0,0,0}$P$}%
}}}}
\put(4636,-4201){\makebox(0,0)[lb]{\smash{{\SetFigFont{8}{9.6}{\familydefault}{\mddefault}{\updefault}{\color[rgb]{0,0,0}$P+ \delta P$}%
}}}}
\put(4681,-2806){\makebox(0,0)[lb]{\smash{{\SetFigFont{9}{10.8}{\familydefault}{\mddefault}{\updefault}{\color[rgb]{0,0,0}$n^{\mu}(P  + \delta P)$}%
}}}}
\put(5176,-3256){\makebox(0,0)[lb]{\smash{{\SetFigFont{9}{10.8}{\familydefault}{\mddefault}{\updefault}{\color[rgb]{0,0,0}$n^{\mu}(P)$}%
}}}}
\end{picture}%
  \caption{\small{The extrinsic curvature measures the variation of the unit
normal vector to $\S_t$ projected on $\S_t$.}}
  \label{extcurvfi}
\end{figure} If the hypersurfaces $\S_t$ are flat, the extrinsic curvature 
 clearly vanishes. More formally, the extrinsic curvature $K_{\m\n}$ is, 
\beq K_{\m\n} = h_\n{}^\r \nabla_\r n_\m \nn \eeq
where $h_\n{}^\r = g^{\r\m}h_{\n\m}$ ensures that $K_{\m\n}$ is purely
spacelike\footnote{$h_\n{}^\r$ is a projector on the hypersurface $\S_t$. If 
$v^{\m}$ is orthogonal to $n^{\n}$ then $v^{\m} =  v^\n h_\n{}^\m$.},
i.e. $ K_{\m0} = 0 $ and $K_{0\n} = 0$, so that one can naturally define 
$^{\sst (d)}K_{ab}Ê\equiv K_{ab}$.
To derive other expressions of the extrinsic curvature, it is easier to keep spacetime tensors. In particular, one is interested to know its expression in terms of the  ``time derivative'' of 
$h_{\m\n}$,
\beq \dot{h}_{\m\n} \equiv h_\m{}^\a h_\n{}^\b \cl_t h_{\a\b} \ ,\nn 
\eeq 
where $\cl_t$ stands for 
the Lie 
derivative\footnote{The Lie derivative of a tensor 
$T^{\a_1...\a_i}{}_{\b_1...\b_j}$ with respect to $v^\m$ is 
$\cl_v T^{\a_1...\a_i}{}_{\b_1...\b_j}
= v^\m \nabla_\m T^{\a_1...\a_i}{}_{\b_1...\b_j} 
-\sum_{k=1}^i T^{\a_1...\g...\a_i}{}_{\b_1...\b_j} \nabla_\g v^{\a_k} 
+\sum_{l=1}^j T^{\a_1...\a_i}{}_{\b_1...\g...\b_j} \nabla_{\b_l}v^\g$.} along $t^\m$.
In this vain, an important property of
the extrinsic curvature is its symmetry, $K_{\m\n} = K_{\n\m}$ which can be 
shown using the Frobenius theorem (see Appendix B of \cite{Wald:1984rg}).  
Using the symmetry of $K_{\m\n}$, it is easy to show that 
\beq K_{\m \n} = \frac{1}{2}\cl_n h_{\m\n} \, .\nn \eeq Starting from this expression, one get\beq 
 K_{\m \n} &=& \frac{1}{2}\cl_n h_{\m\n} \nn \\
 &=&  {1\over 2} (n^\rho \nabla_\rho h_{\m\n} + h_{\m\rho} \nabla_\nu n^\rho
 + h_{\rho \n} \nabla_\m n^\rho) \nn \\
 &=&  {1\over 2\, N} (Nn^\rho \nabla_\rho h_{\m\n} + h_{\m\rho} \nabla_\nu (Nn^\rho)
 + h_{\rho \n} \nabla_\m (Nn^\rho)) \nn \\
 &=& {1\over 2\, N}h_\m{}^\rho h_\m{}^\s (\cl_t h_{\rho \s} - \cl_N h_{\rho \s} ) \nn \\
  &=& \frac{1}{2}N^{-1}(\dot{h}_{\m\n} -D_\m N_\n-D_\n
N_\m) \ ,
\label{extcurv}
\eeq
where  $D_\m$ stands for the covariant
derivative on $\S$. The covariant derivative on $\S$ of a tensor $T$ is 
given by 
\beq 
D_\m 
T^{\a_1...\a_i}{}_{\b_1...\b_j} = h^{\a_1}{}_{\g_1}...h^{\a_i}{}_{\g_i}
h^{\d_1}{}_{\b_1}... h^{\d_j}{}_{\b_j} \nabla_\m
T^{\g_1...\g_i}{}_{\d_1...\d_j} \ ,  
\nn 
\eeq 
see chapter 10 of \cite{Wald:1984rg}. 
Note that the expression (\ref{extcurv}) is purely spatial, therefore one has
\beq
K_{ab} = \frac{1}{2}N^{-1}(\dot{h}_{ab} -D_a N_b-D_b
N_a) \ .
\label{extcurvs}
\eeq

\subsubsection{Einstein--Hilbert Lagrangian}

General Relativity possesses gauge symmetries therefore not all
variables are dynamical. The set $\{ h_{ab}, \ N,\ N_a\}$ is very
convenient because the equations of motion for $h_{ab}$ contain second order
derivatives with respect to time; those for $N$ and $N_a$ contain 
only first order derivatives with respect to time and are constraints. 
Moreover, the time derivative of $h_{ab}$ 
 appears in the Einstein-Hilbert Lagrangian only through $K_{ab}$. 
The time derivatives of $N$ and $N_a$ do
not appear since they are not dynamical (they just  describe how to pass from 
one hypersurface to the next).

In order to introduce the Hamiltonian formalism, one needs the expression of the 
Einstein-Hilbert Lagrangian density $ L = \sqrt{-g} R$ in terms of $N,\ N_a, 
\ h_{ab}$ and their derivatives. It is
easy to show that $\sqrt{-g} = \sqrt{h}N$. To get the spacetime scalar 
curvature $R$ in appropriate variables it is a less trivial. One uses the 
fact\footnote{
The first equality of (\ref{frsst}) is obtained by using twice the definition 
(\ref{smetric}). For the second one, one uses the 
\emph{Gauss--Codacci relations} see chapter 10 of \cite{Wald:1984rg},
\beq 
^{(d)}R_{\m\n\r}{}^{\s} = h_\m{}^\a h_\n{}^\b h_\r{}^\g h^\s{}_\d
R_{\a\b\g}{}^\d - K_{\m\r}K_\n{}^\s + K_{\n\r} K_\m{}^\s 
\nn 
\eeq
contracted with $h^{\m\r}$ and $\d^\n{}_\s$. } 
that
\beq 
\label{frsst} 
\frac{1}{2} R_{\m\n\a\b}h^{\m\a}h^{\n\b} 
                       &=& R + 2R_{\m\n}n^\m n^\n  \nn  \\ 
                       &=&  ^{(d)}R + K^2 - K_{\m\n}K^{\m\n} 
                       \eeq  
where $R_{\m\n\a\b}$ and $R_{\m\n}$ are the 
spacetime Riemann and Ricci tensors; $K = K^\m{}_\m$; $ ^{(d)}R$ is the
spatial scalar curvature. 
Then one evaluates\footnote{
The first and second equalities of (\ref{secst}) come from the definitions 
of the Ricci and Riemann tensors. The third uses integration by parts and the 
last one the definition of the extrinsic curvature.} 
$R_{\m\n}n^\m n^\n$,
\beq 
R_{\m\n}n^\m n^\n  &=& R_{\m\g\n}{}^\g n^\m n^\n \nn \\
 & = & -n^\m (\nabla_\m \nabla_\g - \nabla_\b \nabla_m) n^\g \nn \\
 &=& (\nabla_\m n^\m)(\nabla_\g n^\g) - (\nabla_\g n^\m ) (\nabla_\m n^\g)
- \nabla_\m(n^\m \nabla_\g n^\g) + \nabla_\g (n^\m \nabla_\m n^\g) \nn \\
&=& K^2 - K_{\m\n}K^{\m\n} - \nabla_\m(n^\m \nabla_\n
n^\n) + \nabla_\m(n^\n \nabla_\n n^\m) \ . 
\label{secst} 
\eeq
With total derivatives neglected, the Einstein-Hilbert Lagrangian density is, 
\beq  
L  = \sqrt{h}N (^{(d)}R + K_{ab}K^{ab} - K^2) \ . \label{ehlag}
\eeq

\subsubsection{Hamiltonian Formalism}

All ingredients are now in order to get the Hamiltonian. One just needs the
momentum canonically conjugate to $h_{\m\n}$, which is
\beq  
\p^{ab} = { \partial L \over \partial \dot{h}_{ab} } = 
\sqrt{h}(K^{ab}
-  K h^{ab}) \ . \label{conmom} \eeq The Hamiltonian density is 
\beq H &=& \p^{ab} \dot{h}_{ab} - L \nn \\
         &=& \sqrt{h}  \{N(-^{(d)}R + h^{-1}\p^{ab}\p_{ab}-\frac{1}{d-1}
         h^{-1}\p^2) \nn \\
         &-& 2N_a(D_b(h^{-1/2}\p^{ba})) \nn \\
         &+&  2D_a(h^{-1/2}N_b \p^{ab}) \}  \nn
\eeq 
where $\p = \p^a{}_a$. Variation of $H$ with respect to $N$ and 
$N_a$ yields respectively the Hamiltonian and momentum constraints, 
\beq 
H&=&^{(d)}R+h^{-1}\p^{ab} \p_{ab}-\frac{1}{d-1}h^{-1}\p^2 \approx 0
\nn \  , \\
H^a &=& D_b (h^{-1/2}\p^{ab}) \approx 0 \ . \nn 
\eeq

\subsubsection{Hamiltonian Formalism and the Iwasawa Decomposition} 

\noindent This section is a complement to the section \ref{moredetails} (to which we refer for the notations) and is devoted to derive the Hamiltonian formalism  
after the \emph{Iwasawa decomposition of the 
spatial metric} and the choice a \emph{spacetime slicing such that $N^\m = 0$}. 
One needs the Einstein-Hilbert Lagrangian density expressed in terms 
of $\{ \cN^a{}_b, \ \b^a, \  N, \ \dot{\cN}^a{}_b, \ \dot{\b}^a \}$. 
The extrinsic curvature $K$ (\ref{extcurvs}) in these variables, and in matricial notation, is 
\beq 
K = \frac{1}{2}N^{-1}({}^T\dot{\cN} \cA^2 \cN 
                          +2 {}^T\cN \dot{\cA} A \cN
                          +  {}^T\cN \cA^2 \dot{\cN})\ . 
\nn
\eeq
The Einstein-Hilbert Lagrangian density (\ref{ehlag}) is 
\beq
L  &=& \sqrt{g}N (^{(d)}R + \mathrm{tr}(\mathrm{g}^{-1} K)^2  
- (\mathrm{tr}(\mathrm{g}^{-1}K))^2)
\nn \\
   &=& \sqrt{g}N (^{(d)}R + \frac{1}{4}N^{-2}
       \mathrm{tr}(\cN^{-1}\cA^{-2}{}^T\cN^{-1}(
                             {}^T\dot{\cN} \cA^2 \cN 
                          +2 {}^T\cN \dot{\cA} \cA \cN
                          +  {}^T\cN \cA^2 \dot{\cN}))^2 \nn \\
 & -&  \frac{1}{4}N^{-2}
(\mathrm{tr}(\cN^{-1}\cA^{-2}{}^T\cN^{-1}(
                             {}^T\dot{\cN} \cA^2 \cN 
                          +2 {}^T\cN \dot{\cA} A \cN
                          +  {}^T\cN \cA^2 \dot{\cN})))^2 ) \ ,
\nn 
\eeq
where $g$ is the determinant of the spatial metric and g is the spatial 
metric. After expending the above expressing and noticing that 
the tr$(\dot{\cN}\cN)$  vanishes\footnote{
as well as tr$(\dot{\cN}\cN \Lambda)=0$ where $\Lambda$ is any diagonal matrice
and in particular  tr$(\dot{\cN}\cN)^2=0$.} 
due to the form (\ref{ns}) of the $N$'s, one finds 
\beq 
L &=& {\sqrt{g} \over N}  ( N^2{}^{(d)}R %\nn \\
 + ( \mathrm{tr}(\cA^{-1} \dot{\cA})^2  - (\mathrm{tr}\dot{\cA}\cA^{-1})^2 
%\nn \\
 + \frac{1}{2} \mathrm{tr}(\cN^{-1}\cA^{-2}{}^T\cN^{-1}{}^T\dot{\cN} \cA^2 \dot{\cN}) \nn \\
&=& {\sqrt{g} \over N}  ( N^2{}^{(d)}R 
 + (\sum_a (\dot{\b}^a)^2 - (\sum_a
\dot{\b}^a)^2
+ \frac{1}{2} \sum_{b<d}e^{2(\b^b-\b^d)}\cN^{-1a}{}_b
\cN^{-1c}{}_b \dot{\cN}^d{}_c \dot{\cN}^d{}_a) \ .
\nn
\eeq
One can compute the momenta conjugated to the $\b$'s and $\cN$'s, they
are
\beq 
\label{NNmomenta} 
{P^i}_\aaa &=& 
\frac{\partial\mathcal L}{\partial 
\dot{\cN}^\aaa{}_i} = \sum_{b<a} e^{2(\beta^b - \beta^a)} 
{\dot\cN}^\aaa{}_j {\cN^{-1j}}_\bb {\cN^{-1i}}_\bb   \\
\label{Bmomenta} 
\pi_\aaa &=&  2 \ \tilde{N}^{-1}  G_{\aaa \bb}\dot{\b}^\bb 
\eeq 
where $G^{\aaa \bb}$ is the inverse of the auxiliary 
space metric $G_{\aaa \bb}$, 
\beq 
G^{\aaa \bb}\pi_\aaa \pi_\bb = \sum_a \pi_\aaa^2 - {1\over d-1} (\sum_a \pi_\aaa)^2 \ .
\nn
\eeq
The Hamiltonian density is
\beq 
H [\b, \cN, \pi, P] &=&  \pi_\aaa \dot{\b}^\aaa + P^i{}_\aaa \dot{\cN}^\aaa{}_i  -L\nn \\
H [\b, \cN, \pi, P]   &=& {\tilde{N} \over 4} G^{\aaa \bb}\pi_\aaa \pi_\bb \nn \\
  & &  +{1\over 4 \tilde{N}} \mathrm{Tr}(PA^{-2}{}^TP{}^TN A^2 N) \nn \\ 
  & & - g \tilde{N} ^{(d)}R \ .
\label{hamiwa} 
\eeq

%%%%%%%%%%%
%%%%%%%%%%
\cleardoublepage
%%%%%%%%%%%%
%%%%%%%%%%%%%
%%\include{app_cartan_formalism}
\chapter{The Cartan Formalism}
\markboth{THE CARTAN FORMALISM}{}
\label{cartan}

This appendix is inspired essentially from \cite{Ryan:1975jw} and 
\cite{Carroll:1997ar}. Usually, one uses a coordinate basis on spacetime $M$. 
However, it is often useful to consider a non-coordinate basis. Instead of 
choosing in the tangent 
space $T_p$ at each point $p$, a basis which is $e_{\mu}=\partial_\m$, one 
can take a basis $e_{\sst{(\m)}}$. This basis $e_{\sst{(\m)}}$ is numbered by a index 
in brackets 
to emphasize the fact that this is a general basis, \ie not a coordinate basis. Of course one can express the new basis in terms of the old one, 
\beq 
e_{\sst{(\m)}} = e_{{\sst{(\m)}}}{}^\n e_{\n}
\nn 
\eeq 
and vice-versa, 
\beq 
e_{\n} = e_\n{}^{\sst{(\m)}} e_{\sst{(\m)}} \ .
\nn
\eeq 
The inner product of the vector basis give the components of the metric with
respect to this basis, 
\beq 
g_{\sst{(\m)(\n)}} = g(e_{\sst{(\m)}},e_{\sst{(\n)}}) \ . 
\nn
\eeq 
It is in particular possible to 
choose the basis such that $g_{\sst{(\m)(\n)}}= \eta_{\m\n}$. In these cases, the $e_{\sst{(\m)}}$'s  
are called {\it vielbeins}. Here we will consider the general case. Similarly, 
one can introduce a basis $\th^{\sst{(\m)}}$ of one forms in $T^\star_p$ dual to
$e_{\sst{(\m)}}$, 
\beq 
\th^{\sst{(\m)}} e_{\sst{(\n)}} = \d^\m{}_\n \ .
\nn
\eeq 
It is obvious that this basis is 
related to the basis $\th^{\m} = dx^\m$ (which is dual to $e_{\m}$)
through $\th^{\sst{(\m)}} = e^\sst{(\m)}{}_\m \th^{\m}$. Let $T$ be any tensor in 
$\underbrace{T_p \times ... \times T_p}_{ i } \times
\underbrace{T_p^\star ... \times T_p^\star}_j$. In the coordinate basis, one 
has $ T = T^{\m_1...\m_i}{}_{\n_1...\n_j} e_{\m_1} ... e_{\m_i}
\th^{\n_1}...\th^{\n_j}$. One can of course express this tensor in the new
basis $ T = T^{\sst{(\m_1)}...\sst{(\m_i)}}{}_{\sst{(\n_1)}...\sst{(n_j)}} e_{\sst{(\m_1)}} ... e_{\sst{(\m_i)}}
\th^{\sst{(\n_1)}}...\th^{\sst{(\n_j)}}$. The components of $T$ in the two basis are related by 
\beq 
T^{\sst{(\m_1)}...\sst{(\m_i)}}{}_{\sst{(\n_1)}...\sst{(n_i)}} = e^{\sst{(\m_1)}}{}_{\m_1}...  e^{\sst{(\m_i)}}{}_{\m_i}
e_{\sst{(\n_1)}}{}^{\n_1}...  e_{\sst{(n_i)}}{}^{\n_i} T^{\m_1...\m_i}{}_{\n_1...\n_j} \ .
\nn
\eeq

\noindent The commutator $\  [ \ , \ ]$ of two vector fields $U$ and $V$ is 
defined to be
\beq 
\ [ U,V ] = UV-VU \ . 
\nn
\eeq
\noindent The Lie derivative $\pounds_{U}$ of a vector field $V$ along $U$ is
\beq 
\pounds_U V = [U,V] .
\nn
\eeq
\noindent The structure constants $C^\mm{}_{\nnn \rrr}$ in a given basis are defined to 
be the components of the Lie derivative of $e_{\sst{(\n)}}$ along $e_{\sst{(\m)}}$, 
\beq 
\pounds_{e_{\sst{(\m)}}} e_{\sst{(\n)}} = C^\rrr{}_{\mm \nnn} e_{\sst{(\rho)}} \ .
\nn
\eeq 
It can be shown that they are also the components of the two form 
$d\theta^{\sst{(\m)}}$, 
\beq 
d \theta^{\sst{(\m)}} = - \frac{1}{2} C^\mm{}_{\nnn \rrr} \theta^{\sst{(\n)}} 
\wedge \theta^{\sst{(\rho)}} \ .
\label{strucconst}
\eeq
In the special case of a coordinate basis, the structure constants therefore
vanish. \newline

\subsubsection{Covariant Derivative}

\noindent The covariant derivative $\nabla_U$ maps a tensor field $T$ into 
another tensor field $\nabla_U T$ of the same rank. The covariant derivative 
of any tensor along any vector field is given once one knows 
$\nabla_{e_{\sst{(\m)}}} e_{\sst{(\n)}}$. The \emph{connection coefficients} 
$\Gamma^{\rrr}{}_{\mm \nnn}$ are the components of $\nabla_{e_{\sst{(\m)}}} e_{\sst{(\n)}}$, 
\beq 
\nabla_{e_{\sst{(\m)}}} e_{\sst{(\n)}}  = \Gamma^\rrr{}_{\mm \nnn} e_{\rrr} \ . 
\label{concoef}
\eeq
In order to make this connection unique, one requires $\nabla_U g = 0$. With 
the definition of the connection forms 
\beq 
\omega^\sst{(\m)}{}_\nnn = \Gamma^\mm{}_{\nnn \rrr}\theta^{\rrr} \, , 
\label{connforms}
\eeq  
this condition 
reads
\beq 
\nabla_U g = 0 \ \ \Leftrightarrow \ \ dg_{\sst{(\m)} \nnn} = \omega_{\sst{(\m)} \nnn}+\omega_{\nnn \sst{(\m)}} \ .
\label{covderg}
\eeq
One also requires \emph{no torsion}. It can be shown that the no torsion 
condition is equivalent to $C^\mm{}_{\nnn \rrr} =
\Gamma^\mm{}_{\rrr \nnn}-\Gamma^\mm{}_{\nnn \rrr }$. This last relation implies the \emph{first 
Cartan equation}, 
\beq 
d\theta^{\sst{(\m)}} = \omega^{\sst{(\m)}}{}_{\sst{(\n)}} \wedge \theta^{\sst{(\n)}} \ .
\label{firstcartaneq}
\eeq
Using three times the first Cartan equation
(\ref{firstcartaneq}) 
\beq 
C_{\mm \nnn \rrr} &=& \Gamma_{\mm \rrr \nnn }-\Gamma_{\mm \nnn \rrr} \nn \\
C_{\nnn \rrr \mm} &=& \Gamma_{\nnn \mm \rrr }-\Gamma_{\nnn \rrr \mm} \nn \\
C_{\rrr \mm \nnn } &=& \Gamma_{\rrr \nnn \mm}-\Gamma_{\rrr \mm \nnn} \nn  \, ,
\eeq
where three of the $\Gamma$'s are replaced by $\Gamma_{\mm \nnn \rrr} = - \Gamma_{\nnn \mm \rrr} + 
e_{\rrr} g_{\sst{(\m)} \nnn}$ (equality
implied by the condition (\ref{covderg})) gives
\beq 
C_{\mm \nnn \rrr }&=& -\Gamma_{\rrr \mm \nnn} + e_{\sst{(\n)}}g_{\sst{(\m)} \rrr}-\Gamma_{\mm \nnn 
\rrr} \nn \\
C_{\nnn \rrr \mm}&=& -\Gamma_{\mm \nnn \rrr } + e_{\rrr}g_{\nnn \sst{(\m)}}-\Gamma_{\nnn \rrr \mm } \nn \\
C_{\rrr \mm \nnn}&=& -\Gamma_{\nnn \rrr \mm} + e_{\sst{(\m)}}g_{\rrr \nnn}-\Gamma_{\rrr \mm \nnn} \nn  \ ,
\eeq
Substracting the frist equation to the two last ones, one get the connection 
coefficients in a general basis,  
\beq 
\Gamma^\mm{}_{\nnn \rrr} &=& \frac{1}{2} g^{\sst{(\m)}\ddd}(g_{\ddd \nnn,\rrr} + g_{\ddd\rrr,\nnn}-g_{\nnn \rrr,\ddd}) \nn \\
&+& \frac{1}{2}(-C^\mm{}_{\nnn \rrr}+g_{\nnn \ddd}g^{\sst{(\m)}\sss}C^{\ddd}{}_{\sss\rrr}+g_{d\rrr}g^{\sst{(\m)}\sss} C^\ddd{}_{\sss\nnn}) \
,
\label{connection}
\eeq
where $g_{\sst{(\m)} \nnn ,\rrr} = e_{\rrr} g_{\sst{(\m)}\nnn}$.
  
\subsubsection{Curvature} 

\noindent The \emph{curvature operation} $R$ on two vector fields $U$ and $V$ 
is defined by
\beq 
R(U,V) = \nabla_U \nabla_V - \nabla_V \nabla_U - \nabla_{ [U,V]} \ .
\eeq
$R(U,V)$ is a tensor of covariant rank one and contravariant rank one. One 
defines a tensor of covariant rank three and contravariant rank one which 
operates on three vector fields $U, \ V, \ W$ and one differential form 
$\omega$ to produce a function,
\beq R(U,V,W,\omega) = \omega(R(U,V)W) \ .
\nn
\eeq
Consider $R(e_{\sst{(\m)}},e_{\sst{(\n)}})$ acting on $e_{\rrr}$, the components of the 
Riemann curvature tensor $R^\ddd{}_{\rrr \mm \nnn }$, 
\beq R(e_{\sst{(\m)}},e_{\sst{(\n)}}) e_{\rrr} = R^\ddd{}_{\rrr \mm \nnn} e_{\ddd} \ . 
\nn
\eeq
It is a direct computation\footnote{
The components $R^\ddd{}_{\rrr \mm \nnn}$ of the Riemann tensor are given by 
\beq 
R(e_{\sst{(\m)}},e_{\sst{(\n)}})(e_{\rrr}) = R^\ddd{}_{\rrr \mm \nnn} e_{\ddd} \ .
\nn
\eeq 
Inserting the definition of the Riemann tensor $R$, one gets
\beq 
\nabla_{e_{\sst{(\m)}}} [\nabla_{e_{\sst{(\n)}}} e_{\rrr}] - 
\nabla_{e_{\sst{(\n)}}} [\nabla_{e_{\sst{(\m)}}} e_{\rrr}] - \nabla_{[e_{\sst{(\m)}},e_{\sst{(\n)}}]}
e_{\rrr} = R^{\ddd}{}_{\rrr \mm \nnn} e_{\ddd}  
\nn
\eeq
using further the defintion of the connection components (\ref{concoef}) and
the fact that $\ [e_{\sst{(\m)}},e_{\sst{(\n)}}] = -1/2 C^\rrr{}_{\mm \nnn} e_{\rrr}$ together with 
the linearity of the covariant derivative $\nabla_{[e_{\sst{(\m)}},e_{\sst{(\n)}}]} e_{\rrr}
= -1/2 C^\ddd{}_{\mm \nnn } \Gamma^\sss{}_{\rrr \ddd} e_{\sss}$, one deduces that
\beq
R^{\ddd}{}_{\rrr \mm \nnn} e_{\ddd}  &=& (\nabla_{e_{\sst{(\m)}}} \Gamma^\ddd{}_{\rrr \nnn})  e_{\ddd} -
(\nabla_{e_{\sst{(\n)}}} \Gamma^\ddd{}_{\rrr \mm})  e_{\ddd}  \nn \\
&+&
\Gamma^\ddd{}_{\rrr \nnn } \Gamma^\sss{}_{\ddd \mm} e_{\sss} -
\Gamma^\ddd{}_{\rrr \mm } \Gamma^\sss{}_{\ddd  \nnn } e_{\sss}  \nn \\
&+&
\frac{1}{2} C^\ddd{}_{\mm \nnn} e_{\ddd} \wedge e_{\rrr} 
\nn
\eeq
and finally noting that the $\Gamma$'s are functions which implies that 
$(\nabla_{e_{\sst{(\n)}}} \Gamma^\ddd{}_{\rrr \mm}) = e_{\sst{(\n)}} \Gamma^\ddd_{\rrr \mm}$, one
obtain Eq.(\ref{Riemann}).
} 
to show that,
\beq 
R^\ddd{}_{\rrr \mm \nnn} = \Gamma^\ddd{}_{\rrr \nnn , \mm}-\Gamma^\ddd{}_{\rrr \mm , \nnn} + \Gamma^\sss{}_{\rrr \nnn}
\Gamma^\sss{}_{\sss \mm}- \Gamma^\sss{}_{\rrr \mm}\Gamma^\ddd{}_{\sss \nnn} - C^\sss{}_{\mm \nnn}\Gamma^\ddd{}_{\rrr \sss} \
\label{Riemann}
\eeq 
where $\Gamma^\ddd{}_{\rrr \nnn , \mm} = e_{\sst{(\m)}} \Gamma^\ddd{}_{\rrr \nnn}$. \newline

\subsubsection{The Second Cartan Equation}

\noindent The \emph{curvature forms} are defined by
\beq 
\theta^\sst{(\m)}{}_\nnn = d\omega^\mm{}_\nnn + \omega^\sst{(\m)}{}_\rrr \wedge \omega^\rrr{}_\nnn \ ,
\nn
\eeq
and possess the nice property to obey the \emph{second Cartan equation}, 
\beq 
\theta^\sst{(\m)}{}_\nnn = {1 \over 2} R^\mm{}_{ \nnn \rrr \ddd} \theta^\rrr \wedge \theta^\ddd \ . 
\label{seccartaneq}
\eeq
%%%%%%%%%
%%%%%%%%%%
\cleardoublepage
%%%%%%%%%%%%%%
%%%%%%%%%%%%
%%\include{app_dim_reduction}
\chapter{Dimensional Reduction}
\markboth{DIMENSIONAL REDUCTION}{}
\label{app_dim_reduction}

This appendix, based on \cite{reddim}, reviews the dimensional reduction procedure. Firstly by reducing
gravity from $D+1$ dimensions to $D$ dimensions in great details. Next, the one--dimensional reduction of a $p$--form 
is considered. Repeating the procedure, general formulas to reduce gravity coupled to a $p$--form from $D+n$ to $D$ dimensions
can be deduced. These formulas are summarised and can 
be easily used to reduce in dimension actions of the general form (\ref{keyaction}). Here, we consider only 
\emph{toroidal} dimensional reduction. 

\subsubsection{Dimensional Reduction of Gravity: from $D+1$ to $D$}

\noindent The basic idea of the dimensional reduction on a circle is that one of the coordinates no longer 
lives on $\RR$ but rather takes its values on a circle $S^1$.  
The main assumption of dimensional reduction is that the reduced fields are supposed to be independent of the compactified coordinate.  This 
assumption is equivalent to throwing away the massive modes of the fields appearing in the Fourier decomposition of the $(D+1)$--dimensional fields 
along the compactified coordinate \cite{reddim}.  If the $D$--dimensional coordinates are 
$\{ x^{\hm} \} _{\m=0,...,d}$ and 
one compactifies on $x^d = z $, this singles out the latter coordinate. Naively, one is tempted to reinterpret 
the $D$--dimensional metric $\hat{g}_{\hm\hn}$ as the following $D-1$ fields,  a metric $g_{\m\n}=\hat{g}_{\m\n}$,
a one--form $\cA_\m = \hat{g}_{\m z}$ and 
a scalar field $\phi=\hat{g}_{zz}$
where $\m,\n=0,...,d-1$. 
However, these fields do not have the right transformation properties to be interpreted as suggest here above: the "metric" do not 
the right transformation properties under $D$--dimensional 
general coordinate transformations, the "one---form" do not possess $U(1)$ gauge invariance. 
As shown in \cite{reddim}\footnote{
We follow the same notation except for $\a$ which is replaced by $\a /2$.
},
a clever ansatz for the metric is 
\beq 
d\hs^2 &=& e^{\a \phi} ds^2 + e^{\b \phi}(dz + \cA)^2 \, , \nn \\
\hat{g}_{\m\n} &=& e^{\a \phi} g_{\m\n} + e^{\b\phi}\cA_\m \cA_\n \nn \\
\hat{g}_{\m z} &=& e^{\b \phi} \cA_\m \nn \\
\hat{g}_{zz} &=& e^{\b \phi} \, , \nn
\eeq
or at the vielbein level, 
\beq 
\hat{e}^a &=& e^{{\a \over 2} \phi} e^a \nn \\
\hat{e}^z &=& e^{{\b \over 2 }Ê\phi }Ê(dz + \cA) \, . 
\nn 
\eeq 
With this choice, $g_{\m\n}$, $\cA_\m$ and $\phi$ possess the right transformation properties under $D$--dimensional 
general coordinate transformations and local $U(1)$ gauge  transformations to be interpreted as notations suggest. 
The rest of this section is devoted to obtain the Einstein--Hilbert Lagrangian,
\beq 
\cL = \sqrt{-\hat{g}} \hat R \, , 
\nn
\eeq
in terms of the reduced fields  $g_{\m\n}, \ \cA_\m$ and $\phi$.  In order to compute the Ricci scalar $\hat R$ one needs 
the structure constants (\ref{strucconst}) which can be directly read here under, 
\beq 
d\hat{e}^a &=& {\a \over 2} d\phi \wedge \hat{e}^a + e^{{\a \over 2} \phi} (-{1\over 2} C^a{}_{bc}e^b \wedge e^c) \nn \\
 &=& - {1 \over 2}Êe^{-{\a \over 2}\phi}(-\aÊ\partial_b \phi \delta^a{}_c  + C^a{}_{bc} ) \hat{e}^b\wedge \hat{e}^c
\nn \\
d\hat{e}^z &=& {\b \over 2} d\phi \wedge e^z+ e^{{\b \over 2}\phi} d\cA \nn \\
                    &=&  {\b \over 2} e^{-{\a \over 2}\phi} \partial_b \phi \hat{e}^b \wedge \hat{e}^z + {1 \over 2}e^{({\b \over 2}-\a) \phi} \cF_{bc} \hat{e}^b\wedge \hat{e}^c \, ,
\nn
\eeq
where $\partial_b \phi \equiv \th_b \phi = e^{-{\a \over 2}\phi} \hat{\th}_b \phi$, 
 $\th_b$ and $\hat \th_{\hb}$ are the inverse $D$-- and $(D+1)$--dimensional vielbeins. $\cF_{ab} = (\partial_a \cA_b -\partial_b\cA_a)$ are the components of $\cF = d\cA = 
 {1 \over 2} \cF_{ab}e^a\wedge e^b$. 
Therefore the structure constants are 
\beq 
\hat{C}^a{}_{bc} &=&e^{-{\a \over 2}\phi}(-{\a \over 2}Ê(\partial_b \phi \delta^a{}_c-\partial_c \phi \delta^a{}_b)  + C^a{}_{bc} )  \nn \\
\hat{C}^z{}_{bz} &=&  - {\b \over 2} e^{-{\a \over 2}\phi} \partial_b \phi  \nn \\
\hat{C}^z{}_{bc} &=& - e^{({\b \over 2}-\a) \phi} \cF_{bc} \,Ê. \nn
\eeq
The  connection coefficients (\ref{concoef}) are straightforwardly obtained by using (\ref{connection})
where $g_{ab} = \eta_{ab}$, 
\beq 
\hat \G^a{}_{bc} &=& e^{-{\a \over 2}\phi}(\G^a_{bc} + {\a \over 2}Ê(\partial_b \phi \d^a{}_c - \partial^a \phi \eta_{bc})) \nn \\
\hat \G^a{}_{bz} &=& -{1 \over 2}e^{({\b \over 2} - \a)\phi} \cF^a{}_b \nn   \\
\hat \G^a{}_{zb }&=&  {1\over 2} e^{({\b \over 2}-\a) \phi} \cF^a{}_b \nn \\ 
\hat \G^z{}_{bz} &=&  {1\over 2} \b e^{-{\a \over 2}\phi} \partial_b \phi \, . \nn 
\eeq
The connection forms (or \emph{spin connections}) are then deduced, 
\beq 
\hat \o^a{}_b &=& \o^a{}_b +{\a \over 2}Êe^{-{\a \over 2}\phi} (\partial_b \phi \, \hat e^a - \partial^a \phi \eta_{bc} \, \hat  e^c) -{1 \over 2}e^{({\b \over 2} - \a)\phi}\cF^a{}_b \, \hat e^z\nn \\
\hat \o^a{}_z &=&  {1\over 2} \b e^{-{\a \over 2}\phi} \partial^a \phi \,  \hat e^z + {1\over 2} e^{({\b \over 2}-\a) \phi} \cF^a{}_b \, \hat e^b \, . \nn
 \eeq
Using the second Cartan equation (\ref{seccartaneq}), one gets the components of the Riemann tensor (\ref{Riemann}). The Ricci scalar 
is then easily obtained and the Einstein--Hilbert Lagrangian rewrites, 
\beq 
\cL = \sqrt{-\hat g} \hat R = 
\sqrt{-g}(R - {1\over 2} \partial_\m \phi \partial^\m \phi - {1 \over 4} e^{-(D-1)\a \phi}Ê \cF_{\m\n}\cF^{\m\n}) \, , \nn
\eeq
where $\b$ is chosen to be $\b = -2(D-2)\a$ in order to get the Einstein--Hilbert Lagrangian in $D$ dimensions and $\a$
is chosen such that 
$\a^2 = 2 /((D-1)(D-2))$ to normalise the dilaton as usual \cite{reddim}.

\subsubsection{Gravity coupled to a $p$--form: from $D$ to $D-1$}

\noindent One considers now the $p$--form Lagrangian, 
\beq 
\cL_p  = -{\sqrt{-\hat g}  \over 2 p!} \hat F^{\sst{(p)} \, 2}  \, , 
\nn 
\eeq
where $\hat F^{\sst{(p)}}=d \hat A^{\sst{(p-1)}}$. The ansatz for the reduction of the potential $A^{\sst{(p-1)}}$  is
\beq \hat A^{\sst{(p-1)}}(x^{\hm}) = A^{\sst{(p-1)}}(x^{\m})  + A^{\sst{(p-2)}}(x^{\m}) \w dz \, , \nn 
\eeq
and the "clever" one for the field strength $F^{\sst{(p)}}$ \cite{reddim} is, 
\beq 
\hat F^{\sst{(p)}} = F^{\sst{(p)}}+ F^{\sst{(p-1)}}   \w (dz + \cA) \, , \nn 
\eeq
where
\beq
F^{\sst{(p)}} &=& dA^{\sst{(p-1)}} - dA^{\sst{(p-2)}} \w \cA \, \nn \\
F^{\sst{(p-1)}} &=& d A^{\sst{(p-2)}} \, . \nn 
\eeq
In the vielbein basis, the components of $\hat F^{\sst{(p)}}$ are
\beq 
\hat F^{\sst{(p)}}_{ \sst{(a_1)}\dots \sst{(a_{p})}} &=& e^{-p{\a \over 2} \phi} F^{\sst{(p)}}_{\sst{(a_1)}\dots \sst{(a_{p})}} \, , \nn \\
\hat F^{\sst{(p)}}_{\sst{(a_1)}\dots \sst{(a_{p-1})(z)}} &=& e^{(D-p-2){\a \over 2} \phi}ÊF^{\sst{(p-1)}}_{ \sst{(a_1)}\dots \sst{(a_{p-1})}} \, . \nn 
\eeq 
Therefore the Lagrangian re--expresses as 
\beq 
\cL_p  = - {\sqrt{-g}  \over 2 p!} e^{-(p-1)\a \phi} F^{\sst{(p)} \, 2}
- {\sqrt{- g}  \over 2 (p-1)!} e^{(D-p-1)\a \phi} F^{\sst{(p-1)} \, 2} \, . 
\nn 
\eeq

\subsubsection{General Formula for Dimensional Reduction}

\noindent The $(\hat D = D+n)$--dimensional Lagrangian is 
\beq
{\cal L}_{\hat{\sst D}} = \sqrt{-\hat g} \, R-
{1\over 2 \, p!} \hat F^{\sst{(p)}\, 2}\ .
\label{Ddimlag}
\eeq
The metric will be reduced using the standard Kaluza--Klein ansatz
which, in the notation of \cite{reddim,Cremmer:1997ct}, is
\beq
ds_{\hat{\sst D}}^2 = e^{\vec s\cdot\vec\phi}\, ds_{\sst D}^2 +
     \sum_{i=1}^n e^{2\vec\g_i\cdot\vec\phi}\, (h^i)^2\ ,
\label{metricreduction}
\eeq
where
\beq
h^i &=& dz^i + \cA^i_{\0 j}\, dz^j + \cA_\1^i= \td\g^i{}_j\, 
 (dz^j + \bar \cA_\1^j)
\, ,
\nn \\
\td\g^i{}_j &=& \delta^i_j + \cA^i_{\0 j} \, , \nn \\
\bar \cA_\1^i &=& \g^i{}_j \cA_\1^j \, . \label{redmetric}
\eeq
We define also
$\g^i{}_j = (\td \g^{-1})^i{}_j$, as in \cite{Cremmer:1997ct}.
The constant vectors $\vec s$ and $\vec \g_i$ are given by
\beq
\vec s = (s_1,s_2,\ldots, s_n)\ ,\qquad
\vec\g_i = \fft12 \vec s -\fft12 \vec f_i
\, ,
\nn 
\eeq
where
\beq
s_i = \sqrt{\fft{2}{(\hat D -1 -i)(\hat D-2-i)}}\ ,\qquad
\vec f_i = \Big(\underbrace{0,0,\ldots, 0}_{i-1}, (\hat{D}-1-i) s_i, s_{i+1},
s_{i+2}, \ldots, s_n\Big)\ .
\eeq
The potential $A^{\sst{(p-1)}}$ is reduced according to the standard
procedure
\beq
A^{\sst{(p-1)}}\longrightarrow A^{\sst{(p-1)}} + A^{\sst{(p-2)}}{}_{ i}\, dz^i 
+ \fft12 A^{\sst{(p-3)}}{}_{ij}\, dz^i\wedge dz^j \cdots \ .
\label{reda}\eeq
After reduction on the $n$--torus, the Lagrangian in $D$
dimensions is given by
\beq
{\cal L} &=& R\, {*\oneone} -\fft12 {*d\vec\phi}\wedge d\vec\phi 
-\fft12 e^{\vec a\cdot\vec\phi}\, {*F^{\sst{(p)}}}\wedge F^{\sst{(p)}} -\fft12 
\sum_i e^{\vec a_i\cdot\vec\phi}\, {*F^{\sst{(p-1)}}{}_{i} }\wedge
F^{\sst{(p-1)}}{}_{i}\nn\\
&& -\fft12 \sum_{i<j} e^{\vec a_{ij}\cdot\vec\phi}\, 
{*F^{\sst{(p-2)}}{}_{ij} }\wedge F^{\sst{(p-2)}{}_{ij}} -\cdots\nn\\
&& 
-\fft12 \sum_{i_1<i_2<\cdots <i_{p-1}} 
e^{\vec a_{i_1i_2\cdots i_{p-1}}\cdot\vec\phi}\, 
{*F^{\sst{(1)}}{}_{i_1i_2\cdots i_{p-1}} }\wedge
 F^{\sst{(1)}}{}_{i_1i_2\cdots i_{p-1}}\nn\\ 
&&-\fft12 \sum_i e^{\vec b_i\cdot\vec\phi}\, {*\cF_\2^i}\wedge \cF_\2^i
-\fft12 \sum_{i<j} e^{\vec b_{ij}\cdot\vec\phi}\, {*\cF^i_{\1 j}}\wedge
\cF^i_{\1 j} .\label{dmnlag}
\eeq
The dilaton vectors are given by
\beq
&&\vec a = -(p-1)\, \vec s\ ,\qquad 
\vec a_i = \vec f_i-(p-1)\, \vec s\ ,\qquad
\vec a_{ij} = \vec f_i + \vec f_j -(p-1)\, \vec s\ ,\cdots\nn\\
&&\vec a_{i_1\cdots i_{p-1}} = \vec f_{i_1} + \vec f_{i_2} + \cdots 
 + \vec f_{i_{p-1}} -(p-1)\, \vec s \ ,\nn\\
&& \vec b_i = -\vec f_i\ ,\qquad \vec b_{ij} = -\vec f_i + \vec f_j 
\ .\label{dvec}
\eeq
The field strengths are given by
\beq
F^{\sst{(q)}}{}_{i_1 i_2\cdots i_{p-q}} &=& \g^{j_1}{}_{i_1}\,
\g^{j_2}{}_{i_2} \cdots \g^{j_{p-q}}{}_{i_{p-q}}\, 
\bar F^{\sst{(q)}}{}_{j_1 j_2\cdots j_{p-q}} \nn \\
\cF^i &=& \td\g^i{}_j \bar{\cF}^j_\2\  \nn \\
 \cF^i{}_j &=& \g^k{}_j d \cA^i{}_k \, , 
\nn
\eeq
with
\beq
\bar F^{\sst{(p)}} &=& dA^{\sst{(p-1)}} - dA^{\sst{(p-2)}}{}_{i}\, \bar \cA_\1^i +
  \fft12 dA^{\sst{(p-3)}}{}_{ij}\, \bar \cA_\1^i \, \bar \cA_\1^j\nn\\
&& -\fft16 dA^{\sst{(p-4)}}{}_{ijk}\, \bar \cA_\1^i \, \bar \cA_\1^j\, \bar \cA_\1^k +
\cdots\ ,\nn\\
\bar F^{\sst{(p-1)}}{}_{i} &=&  dA^{\sst{(p-2)}} + dA^{\sst{(p-3)}}{}_{ij}\, \bar \cA_\1^j +
\fft12 dA^{\sst{(p-4)}}{}_{ijk}\, \bar \cA_\1^j \, \bar \cA_\1^k\nn\\
&& +
\fft16 dA^{\sst{(p-5)}}{}_{ijk\ell}\, \bar \cA_\1^j \, \bar \cA_\1^k\, \bar \cA_\1^\ell 
+\cdots\ ,\nn\\
&&\cdots\cdots\nn\\
\bar F^{\1}{}_{ i_1\ldots i_{p-1}} &=& dA^\0{}_{ i_1\ldots i_{p-1}}
\ .\label{fs1} 
\eeq 
and 
\beq 
\bar{\cF}_\2^i &=& d\bar{\cA}_\1^i\ , \nn \\
\bar \cA_\1^i &=& \g^i{}_j \cA_\1^i \ . 
\nn
\eeq

\subsubsection{Dualisation}

\noindent If we take the case where $D=n+3$, so that the reduction goes down
to three dimensions, the Lagrangian will be simply
\beq
{\cal L}_3 &=& R\, {*\oneone} -  \fft12{*d\vec\phi}\wedge d\vec\phi -
\fft12
\sum_i e^{\vec b_i\cdot \vec \phi}\, {*\cF}_\2^i\wedge \cF_\2^i \nn\\
&& -\fft12
\sum_{i<j} e^{\vec b_{ij} \cdot \vec\phi}\, {*\cF^i_{\1 j}}\wedge
\cF^i_{\1 j}
-\fft12 \sum_{i_1<\cdots<i_{p-2}} e^{\vec
a_{i_1\cdots i_{p-2}}\cdot \vec \phi}\, {*F}^\2_{ i_1\cdots i_{p-2}}
\wedge F^\2_{ i_1\cdots i_{p-2}}\nn\\
&& -\fft12 \sum_{i_1<\cdots<i_{p-1}}
e^{\vec a_{i_1\cdots i_{p-1}}\cdot \vec \phi}\, {*F}^\1_{ i_1\cdots
i_{p-1}} \wedge F^\1_{ i_1\cdots i_{p-1}}\ .  
\eeq 
This is obtained from (\ref{dmnlag}) by dropping all field strengths
associated with forms of degree higher than 2.  We may then follow the
standard procedure for dualising the 1--form potentials  $\cA_\1^i$ and
$A^\1_{ i_1\cdots i_{p-2}}$ to axionic scalars 
$\chi_i$ and $\psi^{j_1\cdots j_{p-2}}$, by introducing the axions as 
Lagrange multipliers
for the Bianchi identities for the 2--form field strengths (see, for
example, \cite{Cremmer:1997ct}).   Upon doing so, we arrive at the purely
scalar three-dimensional Lagrangian
\beq
{\cal L}_3 &=& R\, {*\oneone} -\fft12{*d\vec\phi}\wedge d\vec\phi
- \fft12
\sum_i e^{-\vec b_i\cdot \vec\phi}\, {*\cG}_{\1 \,  i}\wedge \cG_{\1 \, i} \nn\\
&&-\fft12
\sum_{i<j} e^{\vec b_{ij} \cdot \vec\phi}\, {*\cF^i_{\1 j}}\wedge
\cF^i_{\1 j}
-\fft12 \sum_{i_1<\cdots<i_{p-2}} e^{-\vec
a_{i_1\cdots i_{p-2}}\cdot \vec \phi}\, {*G}^{\1 \, i_1\cdots i_{p-2}}
\wedge G^{\1 \, i_1\cdots i_{p-2}}\nn\\
&& -\fft12 \sum_{i_1<\cdots<i_{p-1}}
e^{\vec a_{i_1\cdots i_{p-1}}\cdot \vec \phi}\, {*F}^\1_{ i_1\cdots
i_{p-1}} \wedge F^\1_{ i_1\cdots i_{p-1}}\ ,
\label{d3scallag}
\eeq
where
\beq
F^\1_{ i_1\cdots i_{p-1}} &=& \gamma^{j_1}{}_{i_1} 
\cdots \gamma^{j_{p-1}}{}_{i_{p-1}}\, dA^\0_{ j_1 \cdots j_{p-1}}
\ ,\nn\\
G^{\1 \, i_1\cdots i_{p-2}} &=& \td\gamma^{i_1}{}_{j_1} 
\cdots \td\gamma^{i_{p-2}}{}_{j_{p-2}}\, d\psi^{j_1 \cdots j_{p-2}}
\ ,\nn\\
\cG_{\1 i} &=& \gamma^j{}_i(d\chi_j - \sum_{k_1<\cdots< k_{p-2}}
A^\0_{ k_1\cdots k_{p-2}j}\, d\psi^{k_1\cdots k_{p-2}})\ ,\nn\\
\cF^i_{\1 j} &=& \gamma^k{}_j\, d\cA^i_{\0 k}\ .
\label{d3genkkmod}
\eeq

%%%%%%%%%%
%%%%%%%%%%
\cleardoublepage
%%%%%%%%%%%
%%\include{app_conventions}
\chapter{Notations and Conventions}
\markboth{NOTATIONS AND CONVENTIONS}{}

\indent {\color{white}Êfdjkmq}

$\star 1 = \sqrt{-g} d^Dx$

$F^{\sst(p)} = dA^{\sst(p)}$

$^{\sst (p)}F = {1 \over (p+1)!} F_{\m_1\dots \m_{p+1}Ê} dx^{\m_1} \wedge \dots \wedge dx^{\m_{p+1}} \, \, $ 

$\star ^{\sst (p)}F = {1 \over (p+1)! (D-p-1)!} 
\epsilon_{\n_1\dots \n_{D-p-1}}{}^{\m_1\dots \m_{p+1}} 
F_{\m_1\dots \m_{p+1}} dx^{\n_1}\dots dx^{\n_{D-p-1}}$

$r$ rank of an algebra

$n$ number of dilatons

$d$ spatial dimension

$D$ spacetime dimensions

$g$ determinant of the spatial metric

g matrix of the spatial metric

$N$ lapse

$N_a$  shift

$ a, b, c$ curved spatial indices in a coordinate basis OR indices for reduced dimensions 

$\m \n ...$ curved spacetime indices in a coordinate basis

$ (a), \, (b), \ ... $ flat spatial indices in a general basis

$(\m), \, (\n), \, ... $ flat spacetime indices in a general basis

BKL  Belinsky, Khalatnikov and Lifshitz

$A_n^{++} = A_n^{\wedge \wedge} = EA_n$ 
as well as for the B, C, D. 

$E_n^{++} = E_n^{\wedge \wedge}$ for $n=6,7,8$;  $E_8^{++}Ê = E_{10}$, $E_8^{+++}Ê= E_{11}$ 

$G_2^{++} = G_2^{\wedge \wedge}Ê$, $F_4^{++}Ê= F^{\wedge \wedge}_4$  

$\cG$ finite dimensional simple Lie Group

$\cG^{+}$ Affine Lie group 

$\cG^{++}$ Overextended simple Lie Group

$\cG^{+++}$ Very extended simple Lie Group

$\mf{g}$ Complex Lie algebra

$\mf{g}_\circ$ Real form of the complex Lie algebra $\mf{g}$

$\mf{h} $ Cartan subalgebra of $\mf{g}$ 

$\mf{h}_\circ = \mf{g}_{\circ,0}$ Cartan subalgebra of $\mf{g}_\circ$

%%%%%%%%%
%%%%%%%%
\pagestyle{empty}
\cleardoublepage

\cleardoublepage
\pagestyle{empty}

\addcontentsline{toc}{chapter}{Bibliography}
\markboth{BIBLIOGRAPHY}{}
\begin{thebibliography}{100}

\bibitem{Andersson:2000cv}
L.~Andersson and A.~D. Rendall.
\newblock Quiescent cosmological singularities.
\newblock {\em Commun. Math. Phys.}, 218:479--511, 2001.

\bibitem{Argurio:1998cp}
R.~Argurio.
\newblock Brane physics in {M}-theory.
\newblock 1998.
\newblock Ph{D} {T}hesis.

\bibitem{Argurio:1997gt}
R.~Argurio, F.~Englert, and L.~Houart.
\newblock Intersection rules for p-branes.
\newblock {\em Phys. Lett.}, B398:61--68, 1997.

\bibitem{Arnowitt:1962hi}
R.~Arnowitt, S.~Deser, and C.~W. Misner.
\newblock The dynamics of general relativity.
\newblock 1962.

\bibitem{Banks:1998vs}
T.~Banks, W.~Fischler, and L.~Motl.
\newblock Dualities versus singularities.
\newblock {\em JHEP}, 01:019, 1999.

\bibitem{Barrow:1998rm}
J.~D. Barrow and M.~P. Dabrowski.
\newblock String cosmology and chaos.
\newblock 1998.

\bibitem{Barrow:1985hy}
J.~D. Barrow and J.~Stein-Schabes.
\newblock {K}aluza-{K}lein mixmaster universes.
\newblock {\em Phys. Rev.}, D32:1595--1596, 1985.

\bibitem{BK}
V.~A. Belinsky and I.~M. Khalatnikov.
\newblock Effect of scalar and vector fields on the nature of the cosmological
  singularity.
\newblock {\em Sov. Phys. JETP}, 36:591--597, 1973.

\bibitem{Belinsky:1988mc}
V.~A. Belinsky and I.~M. Khalatnikov.
\newblock On the influence of matter and physical fields upon the nature of
  cosmological singularities.
\newblock {\em Sov. Sci. Rev.}, A3:555, 1981.

\bibitem{Belinsky:1970ew}
V.~A. Belinsky, I.~M. Khalatnikov, and E.~M. Lifshitz.
\newblock Oscillatory approach to a singular point in the relativistic
  cosmology.
\newblock {\em Adv. Phys.}, 19:525--573, 1970.

\bibitem{BKL2}
V.~A. Belinsky, I.~M. Khalatnikov, and E.~M. Lifshitz.
\newblock Construction of a general cosmological solution of the {E}instein
  equation with a time singularity.
\newblock {\em Sov. Phys. JETP}, 35:838, 1972.

\bibitem{Belinsky:1982pk}
V.~A. Belinsky, I.~M. Khalatnikov, and E.~M. Lifshitz.
\newblock A general solution of the {E}instein equations with a time
  singularity.
\newblock {\em Adv. Phys.}, 31:639--667, 1982.

\bibitem{Belinsky:1971nt}
V.~A. Belinsky and V.~E. Zakharov.
\newblock Integration of the {E}instein equations by the inverse scattering
  problem technique and the calculation of the exact soliton solutions.
\newblock {\em Sov. Phys. JETP}, 48:985--994, 1978.

\bibitem{Bergshoeff:1981um}
E.~Bergshoeff, M.~de~Roo, B.~de~Wit, and P.~van Nieuwenhuizen.
\newblock Ten-dimensional {M}axwell-{E}instein supergravity, its currents, and
  the issue of its auxiliary fields.
\newblock {\em Nucl. Phys.}, B195:97--136, 1982.

\bibitem{Bergshoeff:1996ui}
E.~Bergshoeff, M.~de~Roo, M.~B. Green, G.~Papadopoulos, and P.~K. Townsend.
\newblock Duality of type ii 7-branes and 8-branes.
\newblock {\em Nucl. Phys.}, B470:113--135, 1996.

\bibitem{BdB}
N.~Boulanger and S.~de~Buyl.
\newblock Semi-simple {L}ie algebras and representation.
\newblock Lecture given at the first Modave Summer School in Mathematical
  Physics.

\bibitem{Breitenlohner:1986um}
P.~Breitenlohner and D.~Maison.
\newblock On the {G}eroch group.
\newblock {\em Ann. Poincare}, 46:215, 1987.

\bibitem{Breitenlohner:1987dg}
P.~Breitenlohner, D.~Maison, and G.~W. Gibbons.
\newblock Four-dimensional black holes from {K}aluza-{K}lein theories.
\newblock {\em Commun. Math. Phys.}, 120:295, 1988.

\bibitem{Brown:2004jb}
J.~Brown, O.~J. Ganor, and C.~Helfgott.
\newblock {M}-theory and {E}(10): {B}illiards, branes, and imaginary roots.
\newblock {\em JHEP}, 08:063, 2004.

\bibitem{Carroll:1997ar}
S.~M. Carroll.
\newblock Lecture notes on general relativity.
\newblock 1997.

\bibitem{Chamseddine:1980sp}
A.~H. Chamseddine and H.~Nicolai.
\newblock Coupling the {SO}(2) supergravity through dimensional reduction.
\newblock {\em Phys. Lett.}, B96:89, 1980.

\bibitem{Chapline:1982ww}
G.~F. Chapline and N.~S. Manton.
\newblock Unification of {Y}ang-{M}ills theory and supergravity in ten-
  dimensions.
\newblock {\em Phys. Lett.}, B120:105--109, 1983.

\bibitem{Bar}
D.~F. Chernoff and J.~D. Barrow.
\newblock Phys. Rev. Lett. {\bf 50}, 134 (1983).

\bibitem{Chitre}
D.~M. Chitre.
\newblock Ph. D. Thesis, University of Maryland, 1972.

\bibitem{Coussaert:1993ti}
O.~Coussaert and M.~Henneaux.
\newblock Bianchi cosmological models and gauge symmetries.
\newblock {\em Class. Quant. Grav.}, 10:1607--1618, 1993.

\bibitem{Cremmer:1978ds}
E.~Cremmer and B.~Julia.
\newblock The {N}=8 supergravity theory. 1. the {L}agrangian.
\newblock {\em Phys. Lett.}, B80:48, 1978.

\bibitem{Cremmer:1979up}
E.~Cremmer and B.~Julia.
\newblock The {SO}(8) supergravity.
\newblock {\em Nucl. Phys.}, B159:141, 1979.

\bibitem{Cremmer:1997ct}
E.~Cremmer, B.~Julia, H.~Lu, and C.~N. Pope.
\newblock Dualisation of dualities. {I}.
\newblock {\em Nucl. Phys.}, B523:73--144, 1998.

\bibitem{Cremmer:1998px}
E.~Cremmer, B.~Julia, H.~Lu, and C.~N. Pope.
\newblock Dualisation of dualities. {II}: {T}wisted self-duality of doubled
  fields and superdualities.
\newblock {\em Nucl. Phys.}, B535:242--292, 1998.

\bibitem{Cremmer:1999du}
E.~Cremmer, B.~Julia, H.~Lu, and C.~N. Pope.
\newblock Higher-dimensional origin of {D} = 3 coset symmetries.
\newblock 1999.

\bibitem{Cremmer:1978km}
E.~Cremmer, B.~Julia, and J.~Scherk.
\newblock Supergravity theory in 11 dimensions.
\newblock {\em Phys. Lett.}, B76:409--412, 1978.

\bibitem{Dabrowski:2001zk}
M.~P. Dabrowski.
\newblock Kasner asymptotics of mixmaster {H}orava-{W}itten and pre - big bang
  cosmologies.
\newblock {\em Nucl. Phys. Proc. Suppl.}, 102:194--200, 2001.

\bibitem{Damour:2005ef}
T.~Damour.
\newblock Chaos, symmetry and string cosmology.
\newblock In *Shifman, M. (ed.) et al.: From fields to strings, vol. 2*
  923-966.

\bibitem{Damour:2004gn}
T~Damour.
\newblock Cosmological singularities, billiards and {L}orentzian {K}ac- {M}oody
  algebras.
\newblock 2004.

\bibitem{Damour:2005mr}
T.~Damour.
\newblock Cosmological singularities, {E}instein billiards and {L}orentzian
  {K}ac-{M}oody algebras.
\newblock 2005.

\bibitem{Damour:2005pe}
T.~Damour.
\newblock Poincare, relativity, billiards and symmetry.
\newblock 2005.

\bibitem{Damour:2002fz}
T.~Damour, S.~de~Buyl, M.~Henneaux, and C.~Schomblond.
\newblock {E}instein billiards and overextensions of finite-dimensional simple
  {L}ie algebras.
\newblock {\em JHEP}, 08:030, 2002.

\bibitem{Damour:2006ez}
T.~Damour, A.~Hanany, M.~Henneaux, A.~Kleinschmidt, and H.~Nicolai.
\newblock Curvature corrections and {K}ac-{M}oody compatibility conditions.
\newblock 2006.

\bibitem{Damour:2000wm}
T.~Damour and M.~Henneaux.
\newblock Chaos in superstring cosmology.
\newblock {\em Phys. Rev. Lett.}, 85:920--923, 2000.

\bibitem{Damour:2000th}
T.~Damour and M.~Henneaux.
\newblock Oscillatory behaviour in homogeneous string cosmology models.
\newblock {\em Phys. Lett.}, B488:108--116, 2000.

\bibitem{Damour:2000hv}
T.~Damour and M.~Henneaux.
\newblock E(10), {BE}(10) and arithmetical chaos in superstring cosmology.
\newblock {\em Phys. Rev. Lett.}, 86:4749--4752, 2001.

\bibitem{Damour:2001sa}
T.~Damour, M.~Henneaux, B.~Julia, and H.~Nicolai.
\newblock Hyperbolic {K}ac-{M}oody algebras and chaos in {K}aluza-{K}lein
  models.
\newblock {\em Phys. Lett.}, B509:323--330, 2001.

\bibitem{Damour:2002cu}
T.~Damour, M.~Henneaux, and H.~Nicolai.
\newblock E(10) and a 'small tension expansion' of {M} theory.
\newblock {\em Phys. Rev. Lett.}, 89:221601, 2002.

\bibitem{Damour:2002et}
T.~Damour, M.~Henneaux, and H.~Nicolai.
\newblock Cosmological billiards.
\newblock {\em Class. Quant. Grav.}, 20:R145--R200, 2003.

\bibitem{Damour:2002tc}
T.~Damour, M.~Henneaux, A.~D. Rendall, and M.~Weaver.
\newblock Kasner-like behaviour for subcritical {E}instein-matter systems.
\newblock {\em Annales Henri Poincare}, 3:1049--1111, 2002.

\bibitem{Damour:2005zs}
T.~Damour, A.~Kleinschmidt, and H.~Nicolai.
\newblock Hidden symmetries and the fermionic sector of eleven- dimensional
  supergravity.
\newblock 2005.

\bibitem{Damour:2004zy}
T.~Damour and H.~Nicolai.
\newblock Eleven dimensional supergravity and the {E}(10)/{K}({E}(10))
  sigma-model at low {A}(9) levels.
\newblock 2004.

\bibitem{Damour:2005zb}
T.~Damour and H.~Nicolai.
\newblock Higher order {M} theory corrections and the {K}ac-{M}oody algebra
  {E}(10).
\newblock {\em Class. Quant. Grav.}, 22:2849--2880, 2005.

\bibitem{D'Auria:1997cz}
R.~D'Auria, S.~Ferrara, and C.~Kounnas.
\newblock N = (4,2) chiral supergravity in six dimensions and solvable {L}ie
  algebras.
\newblock {\em Phys. Lett.}, B420:289--299, 1998.

\bibitem{deBuyl:2003ub}
S.~de~Buyl, M.~Henneaux, B.~Julia, and L.~Paulot.
\newblock Cosmological billiards and oxidation.
\newblock {\em Fortsch. Phys.}, 52:548--554, 2004.

\bibitem{deBuyl:2005zy}
S.~de~Buyl, M.~Henneaux, and L.~Paulot.
\newblock Hidden symmetries and {D}irac fermions.
\newblock {\em Class. Quant. Grav.}, 22:3595--3622, 2005.

\bibitem{deBuyl:2005mt}
S.~de~Buyl, M.~Henneaux, and L.~Paulot.
\newblock Extended {E}(8) invariance of 11-dimensional supergravity.
\newblock {\em JHEP}, 02:056, 2006.

\bibitem{deBuyl:2005it}
S.~de~Buyl, L.~Houart, and N.~Tabti.
\newblock Dualities and signatures of {G}++ invariant theories.
\newblock {\em JHEP}, 06:084, 2005.

\bibitem{deBuyl:2004ps}
S.~de~Buyl and A.~Kleinschmidt.
\newblock Higher spin fields from indefinite {K}ac-{M}oody algebras.
\newblock 2004.

\bibitem{deBuyl:2003za}
S.~de~Buyl, G.~Pinardi, and C.~Schomblond.
\newblock {E}instein billiards and spatially homogeneous cosmological models.
\newblock {\em Class. Quant. Grav.}, 20:5141--5160, 2003.

\bibitem{deBuyl:2004md}
S.~de~Buyl and C.~Schomblond.
\newblock Hyperbolic {K}ac {M}oody algebras and {E}instein billiards.
\newblock {\em J. Math. Phys.}, 45:4464--4492, 2004.

\bibitem{deWit:1985iy}
B.~de~Wit and H.~Nicolai.
\newblock Hidden symmetry in d = 11 supergravity.
\newblock {\em Phys. Lett.}, B155:47, 1985.

\bibitem{deWit:1986mz}
B.~de~Wit and H.~Nicolai.
\newblock d = 11 supergravity with local {SU}(8) invariance.
\newblock {\em Nucl. Phys.}, B274:363, 1986.

\bibitem{deWit:2000wu}
B.~de~Wit and H.~Nicolai.
\newblock Hidden symmetries, central charges and all that.
\newblock {\em Class. Quant. Grav.}, 18:3095--3112, 2001.

\bibitem{deWit:1992up}
B.~de~Wit, A.~K. Tollsten, and H.~Nicolai.
\newblock Locally supersymmetric {D} = 3 nonlinear sigma models.
\newblock {\em Nucl. Phys.}, B392:3--38, 1993.

\bibitem{deWit:1991nm}
B.~de~Wit and A.~Van~Proeyen.
\newblock Special geometry, cubic polynomials and homogeneous quaternionic
  spaces.
\newblock {\em Commun. Math. Phys.}, 149:307--334, 1992.

\bibitem{deWit:1992wf}
B.~de~Wit, F.~Vanderseypen, and A.~Van~Proeyen.
\newblock Symmetry structure of special geometries.
\newblock {\em Nucl. Phys.}, B400:463--524, 1993.

\bibitem{Demaret:1988sg}
J.~Demaret, Y.~De~Rop, and M.~Henneaux.
\newblock Chaos in nondiagonal spatially homogeneous cosmological models in
  space-time dimensions $<=$ 10.
\newblock {\em Phys. Lett.}, B211:37--41, 1988.

\bibitem{Demaret:1985js}
J.~Demaret, J.~L. Hanquin, M.~Henneaux, and P.~Spindel.
\newblock Cosmological models in eleven-dimensional supergravity.
\newblock {\em Nucl. Phys.}, B252:538--560, 1985.

\bibitem{Demaret:1986ys}
J.~Demaret, J.~L. Hanquin, M.~Henneaux, P.~Spindel, and A.~Taormina.
\newblock The fate of the mixmaster behavior in vacuum inhomogeneous
  {K}aluza-{K}lein cosmological models.
\newblock {\em Phys. Lett.}, B175:129--132, 1986.

\bibitem{Dietz}
W.~Dietz and C.~Hoenselars~(eds).
\newblock Solutions of Einstein's equations: techniques and results, Springer,
  1984.

\bibitem{Ehlers}
J.~Ehlers.
\newblock Les th\'eories relativistes de la gravitation.
\newblock CNRS, Paris, 1959.

\bibitem{Elitzur:1997zn}
S.~Elitzur, A.~Giveon, D.~Kutasov, and E.~Rabinovici.
\newblock Algebraic aspects of matrix theory on {T}**d.
\newblock {\em Nucl. Phys.}, B509:122--144, 1998.

\bibitem{Englert:2004ph}
F.~Englert, M.~Henneaux, and L.~Houart.
\newblock From very-extended to overextended gravity and {M}-theories.
\newblock {\em JHEP}, 02:070, 2005.

\bibitem{Englert:2004ug}
F.~Englert and L.~Houart.
\newblock From brane dynamics to a {K}ac-{M}oody invariant formulation of
  {M}-theories.
\newblock 2004.

\bibitem{Englert:2003py}
F.~Englert and L.~Houart.
\newblock {G}+++ invariant formulation of gravity and {M}-theories: {E}xact
  {BPS} solutions.
\newblock {\em JHEP}, 01:002, 2004.

\bibitem{Englert:2004it}
F.~Englert and L.~Houart.
\newblock {G}+++ invariant formulation of gravity and {M}-theories: {E}xact
  intersecting brane solutions.
\newblock {\em JHEP}, 05:059, 2004.

\bibitem{Englert:2003zs}
F.~Englert, L.~Houart, A.~Taormina, and P.~West.
\newblock The symmetry of {M}-theories.
\newblock {\em JHEP}, 09:020, 2003.

\bibitem{Englert:2003pd}
F.~Englert, L.~Houart, and P.~West.
\newblock Intersection rules, dynamics and symmetries.
\newblock {\em JHEP}, 08:025, 2003.

\bibitem{Feingold:2003es}
A.~J. Feingold and H.~Nicolai.
\newblock Subalgebras of hyperbolic {K}ac-{M}oody algebras.
\newblock 2003.

\bibitem{Fischbacher:2005fy}
T.~Fischbacher.
\newblock The structure of {E}(10) at higher {A}(9) levels: A first algorithmic
  approach.
\newblock {\em JHEP}, 08:012, 2005.

\bibitem{Fre:2003ep}
P.~Fre et~al.
\newblock Cosmological backgrounds of superstring theory and solvable algebras:
  {O}xidation and branes.
\newblock {\em Nucl. Phys.}, B685:3--64, 2004.

\bibitem{Fre':2005sr}
P.~Fre, F.~Gargiulo, and K.~Rulik.
\newblock Cosmic billiards with painted walls in non-maximal supergravities:
  {A} worked out example.
\newblock {\em Nucl. Phys.}, B737:1--48, 2006.

\bibitem{Fuchs:1997jv}
J.~Fuchs and C.~Schweigert.
\newblock Symmetries, {L}ie algebras and representations: {A} graduate course
  for physicists.
\newblock Cambridge, UK: Univ. Pr. (1997) 438 p.

\bibitem{Gaberdiel:2002db}
M.R. Gaberdiel, D.~I. Olive, and P.~C. West.
\newblock A class of {L}orentzian {K}ac-{M}oody algebras.
\newblock {\em Nucl. Phys.}, B645:403--437, 2002.

\bibitem{Gasperini:2002bn}
M.~Gasperini and G.~Veneziano.
\newblock The pre-big bang scenario in string cosmology.
\newblock {\em Phys. Rept.}, 373:1--212, 2003.

\bibitem{Geroch:1970nt}
R.~Geroch.
\newblock A method for generating solutions of {E}instein's equations.
\newblock {\em J. Math. Phys.}, 12:918--924, 1971.

\bibitem{Goroff:1985th}
M.~H. Goroff and A.~Sagnotti.
\newblock The ultraviolet behavior of {E}instein gravity.
\newblock {\em Nucl. Phys.}, B266:709, 1986.

\bibitem{Hawking:1969sw}
S.~W. Hawking and R.~Penrose.
\newblock The singularities of gravitational collapse and cosmology.
\newblock {\em Proc. Roy. Soc. Lond.}, A314:529--548, 1970.

\bibitem{Helgason}
S.~Helgason.
\newblock Differential Geometry, {L}ie Groups and Symmetric Spaces, Graduate
  Studies in Mathematics vol.34, AMS 2001 printing.

\bibitem{Henneaux:1980ft}
M.~Henneaux.
\newblock Bianchi type {I} cosmologies and spinor fields.
\newblock {\em Phys. Rev.}, D21:857--863, 1980.

\bibitem{Henneaux:1981vr}
M.~Henneaux.
\newblock Bianchi universes and spinor fields.
\newblock {\em Annales Poincare Phys. Theor.}, 34:329--349, 1981.

\bibitem{MHBJ}
M.~Henneaux and B.~Julia.
\newblock unpublished.

\bibitem{Henneaux:2003kk}
M.~Henneaux and B.~Julia.
\newblock Hyperbolic billiards of pure {D} = 4 supergravities.
\newblock {\em JHEP}, 05:047, 2003.

\bibitem{Henneaux:1988gg}
M.~Henneaux and C.~Teitelboim.
\newblock Dynamics of chiral (selfdual) p forms.
\newblock {\em Phys. Lett.}, B206:650, 1988.

\bibitem{Henry-Labordere:2002dk}
P.~Henry-Labordere, B.~Julia, and L.~Paulot.
\newblock {B}orcherds symmetries in {M}-theory.
\newblock {\em JHEP}, 04:049, 2002.

\bibitem{Henry-Labordere:2002xh}
P.~Henry-Labordere, B.~Julia, and L.~Paulot.
\newblock Real {B}orcherds superalgebras and {M}-theory.
\newblock {\em JHEP}, 04:060, 2003.

\bibitem{Henry-Labordere:2003rd}
P.~Henry-Labordere, B.~Julia, and L.~Paulot.
\newblock Symmetries in {M}-theory: {M}onsters, inc.
\newblock 2003.

\bibitem{Howe:1983sr}
P.S. Howe and P.~C. West.
\newblock The complete {N}=2, d = 10 supergravity.
\newblock {\em Nucl. Phys.}, B238:181, 1984.

\bibitem{Hull:1998ym}
C.~M. Hull.
\newblock Duality and the signature of space-time.
\newblock {\em JHEP}, 11:017, 1998.

\bibitem{Hull:1998vg}
C.~M. Hull.
\newblock Timelike {T}-duality, de {S}itter space, large {N} gauge theories and
  topological field theory.
\newblock {\em JHEP}, 07:021, 1998.

\bibitem{Hull:1998fh}
C.~M. Hull and R.~R. Khuri.
\newblock Branes, times and dualities.
\newblock {\em Nucl. Phys.}, B536:219--244, 1998.

\bibitem{Hull:1994ys}
C.~M. Hull and P.~K. Townsend.
\newblock Unity of superstring dualities.
\newblock {\em Nucl. Phys.}, B438:109--137, 1995.

\bibitem{Humphreys:1980dw}
J.~E. Humphreys.
\newblock Introduction to {L}ie algebras and representation theory. (3rd
  print., rev.).
\newblock New York, Usa: Springer ( 1980) 171p.

\bibitem{Isenberg:2002jg}
J.~Isenberg and V.~Moncrief.
\newblock Asymptotic behavior of polarized and half-polarized {U}(1) symmetric
  vacuum spacetimes.
\newblock {\em Class. Quant. Grav.}, 19:5361--5386, 2002.

\bibitem{Ivashchuk:1994tu}
V.~D. Ivashchuk, A.~A. Kirillov, and V.~N. Melnikov.
\newblock Stochastic behavior of multidimensional cosmological models near a
  singularity.
\newblock {\em Russ. Phys. J.}, 37:1102--1106, 1994.

\bibitem{Ivashchuk:1994fg}
V.~D. Ivashchuk and V.~N. Melnikov.
\newblock Billiard representation for multidimensional cosmology with
  multicomponent perfect fluid near the singularity.
\newblock {\em Class. Quant. Grav.}, 12:809--826, 1995.

\bibitem{Julia:1980gr}
B.~Julia.
\newblock Group disintegrations.
\newblock Invited paper presented at Nuffield Gravity Workshop, Cambridge,
  Eng., Jun 22 - Jul 12, 1980.

\bibitem{Julia:1986qq}
B.~Julia.
\newblock Induced representations and supersymmetric covariant derivatives.
\newblock In *Annecy 1986, Proceedings, Higher Dimensional Theories and
  Expected Physics*, 2p.

\bibitem{Julia:1981wc}
B.~Julia.
\newblock Infinite {L}ie algebras in {P}hysics.
\newblock Invited talk given at Johns Hopkins Workshop on Current Problems in
  Particle Theory, Baltimore, Md., May 25-27, 1981.

\bibitem{Julia:1982gx}
B.~Julia.
\newblock {K}ac-{M}oody symmetry of gravitation and supergravity theories.
\newblock Invited talk given at AMS-SIAM Summer Seminar on Applications of
  Group Theory in Physics and Mathematical Physics, Chicago, Ill., Jul 6-16,
  1982.

\bibitem{J}
B.~L. Julia.
\newblock Lectures in Applied Mathematics, AMS-SIAM, vol 21 (1985), p.355.

\bibitem{J''}
B.~L. Julia.
\newblock Strings, branes and dualities, Lectures at M Carg\ ese School 1997,
  Kluwer (1999). They contain a review of the `Silver M rules of group
  disintegrations''. They also contain a general conjecture M on the triple
  extensions like $E^{\wedge \wedge \wedge}$ M (paper in preparation).

\bibitem{Kac:1990gs}
V.~G. Kac.
\newblock Infinite dimensional {L}ie algebras.
\newblock Cambridge, UK: Univ. Pr. (1990) 400 p.

\bibitem{Keurentjes:2002xc}
A.~Keurentjes.
\newblock The group theory of oxidation.
\newblock {\em Nucl. Phys.}, B658:303--347, 2003.

\bibitem{Keurentjes:2002rc}
A.~Keurentjes.
\newblock The group theory of oxidation. {II}: {C}osets of non-split groups.
\newblock {\em Nucl. Phys.}, B658:348--372, 2003.

\bibitem{Keurentjes:2002vx}
A.~Keurentjes.
\newblock Oxidation = group theory.
\newblock {\em Class. Quant. Grav.}, 20:S525--S532, 2003.

\bibitem{Keurentjes:2004bv}
A.~Keurentjes.
\newblock E(11): {S}ign of the times.
\newblock {\em Nucl. Phys.}, B697:302--318, 2004.

\bibitem{Keurentjes:2004xx}
A.~Keurentjes.
\newblock Time-like {T}-duality algebra.
\newblock {\em JHEP}, 11:034, 2004.

\bibitem{Keurentjes:2003yu}
A.~Keurentjes.
\newblock The topology of {U}-duality (sub-)groups.
\newblock {\em Class. Quant. Grav.}, 21:1695--1708, 2004.

\bibitem{Keurentjes:2003hc}
A.~Keurentjes.
\newblock U-duality (sub-)groups and their topology.
\newblock {\em Class. Quant. Grav.}, 21:S1367--1374, 2004.

\bibitem{Keurentjes:2005jw}
A.~Keurentjes.
\newblock Poincare duality and {G}+++ algebras.
\newblock 2005.

\bibitem{KLL}
I.~M. Khalatnikov, E.~M. Lifshitz, and I.~M. Lifshitz.
\newblock {\em Sov. Phys. JETP}, 32:173, 1971.

\bibitem{Kirillov1993}
A.~A. Kirillov.
\newblock Sov. Phys. JETP {\bf 76}, 355 (1993).

\bibitem{Kirillov:1994fc}
A.~A. Kirillov and V.~N. Melnikov.
\newblock Dynamics of inhomogeneities of metric in the vicinity of a
  singularity in multidimensional cosmology.
\newblock {\em Phys. Rev.}, D52:723--729, 1995.

\bibitem{Kleinschmidt:2004dy}
A.~Kleinschmidt and H.~Nicolai.
\newblock E(10) and {SO}(9,9) invariant supergravity.
\newblock {\em JHEP}, 07:041, 2004.

\bibitem{Kleinschmidt:2003mf}
A.~Kleinschmidt, I.~Schnakenburg, and P.~West.
\newblock Very-extended {K}ac-{M}oody algebras and their interpretation at low
  levels.
\newblock {\em Class. Quant. Grav.}, 21:2493--2525, 2004.

\bibitem{Kleinschmidt:2003jf}
A.~Kleinschmidt and P.~West.
\newblock Representations of {G}+++ and the role of space-time.
\newblock {\em JHEP}, 02:033, 2004.

\bibitem{Kramer}
D.~Kramer~et al.
\newblock Exact solutions of {E}instein's field equations.
\newblock Cambridge University Press.

\bibitem{Lambert:2006he}
N.~Lambert and P.~West.
\newblock Enhanced coset symmetries and higher derivative corrections.
\newblock 2006.

\bibitem{Lambert:2001gk}
N.~D. Lambert and P.~C. West.
\newblock Coset symmetries in dimensionally reduced bosonic string theory.
\newblock {\em Nucl. Phys.}, B615:117--132, 2001.

\bibitem{LL}
L.~D. Landau and E.~M. Lifshitz.
\newblock Th\' eorie des champs.
\newblock 4\`eme \' edition, Editions Mir, Moscou (1989).

\bibitem{MacC1}
J.~MacCallum, M. A. H.and~Patera and P.~Winternitz.
\newblock Subalgebras of real three and four dimensional {L}ie algebras.
\newblock {\em J. Math. Phys.}, 18:1449, 1977.

\bibitem{Mc}
M.~A.~H. MacCallum.
\newblock General Relativity: An Einstein Centenary Survey, ed S. W. Hawking
  and W. Israel (Cambridge University Press 1979.

\bibitem{MacC2}
M.~A.~H. MacCallum.
\newblock On the enumeration of the real four-dimensional {L}ie algebras.
\newblock In A.L. Harvey, editor,.

\bibitem{Maison:1978es}
D.~Maison.
\newblock Are the stationary, axially symmetric {E}instein equations completely
  integrable?
\newblock {\em Phys. Rev. Lett.}, 41:521, 1978.

\bibitem{Marcus:1983hb}
N.~Marcus and J.~H. Schwarz.
\newblock Three-dimensional supergravity theories.
\newblock {\em Nucl. Phys.}, B228:145, 1983.

\bibitem{Matschull:1994vi}
H.~J. Matschull and H.~Nicolai.
\newblock Canonical treatment of coset space sigma models.
\newblock {\em Int. J. Mod. Phys.}, D3:81--91, 1994.

\bibitem{Miemiec:2004iv}
A.~Miemiec and I.~Schnakenburg.
\newblock Killing spinor equations from nonlinear realisations.
\newblock {\em Nucl. Phys.}, B698:517--530, 2004.

\bibitem{Misnerb}
C.~W. Misner.
\newblock D. Hobill et al. (Eds),{\em Deterministic chaos in general
  relativity}, (Plenum, 1994),p.~317 [gr-qc/9405068].

\bibitem{Mizoguchi:1997si}
S.~Mizoguchi.
\newblock E(10) symmetry in one-dimensional supergravity.
\newblock {\em Nucl. Phys.}, B528:238--264, 1998.

\bibitem{Mizoguchi:1998wv}
S.~Mizoguchi and N.~Ohta.
\newblock More on the similarity between {D} = 5 simple supergravity and {M}
  theory.
\newblock {\em Phys. Lett.}, B441:123--132, 1998.

\bibitem{Mizoguchi:1999fu}
S.~Mizoguchi and G.~Schroder.
\newblock On discrete {U}-duality in {M}-theory.
\newblock {\em Class. Quant. Grav.}, 17:835--870, 2000.

\bibitem{Nahm:1977tg}
W.~Nahm.
\newblock Supersymmetries and their representations.
\newblock {\em Nucl. Phys.}, B135:149, 1978.

\bibitem{Nicolai:1986jk}
H.~Nicolai.
\newblock D = 11 supergravity with local {SO}(16) invariance.
\newblock {\em Phys. Lett.}, B187:316, 1987.

\bibitem{Nicolai:1987kz}
H.~Nicolai.
\newblock The integrability of {N}=16 supergravity.
\newblock {\em Phys. Lett.}, B194:402, 1987.

\bibitem{Nicolai:1991kx}
H.~Nicolai.
\newblock A hyperbolic {L}ie algebra from supergravity.
\newblock {\em Phys. Lett.}, B276:333--340, 1992.

\bibitem{Nicolai:2005su}
H.~Nicolai.
\newblock Gravitational billiards, dualities and hidden symmetries.
\newblock 2005.

\bibitem{Nicolai:2003fw}
H.~Nicolai and T.~Fischbacher.
\newblock Low level representations for {E}(10) and {E}(11).
\newblock 2003.

\bibitem{Nicolai:2004nv}
H.~Nicolai and H.~Samtleben.
\newblock On {K(E}(9)).
\newblock {\em Q. J. Pure Appl. Math.}, 1:180--204, 2005.

\bibitem{Obers:1998fb}
N.~A. Obers and B.~Pioline.
\newblock U-duality and {M}-theory.
\newblock {\em Phys. Rept.}, 318:113--225, 1999.

\bibitem{Pasti:1996vs}
P.~Pasti, D.~P. Sorokin, and M.~Tonin.
\newblock On {L}orentz invariant actions for chiral p-forms.
\newblock {\em Phys. Rev.}, D55:6292--6298, 1997.

\bibitem{Patera:1977hg}
J.~Patera and P.~Winternitz.
\newblock Subalgebras of real three-dimensional and four-dimensional {L}ie
  algebras.
\newblock {\em J. Math. Phys.}, 18:1449--1455, 1977.

\bibitem{Paulot:2006zp}
L.~Paulot.
\newblock Infinite-dimensional gauge structure of d = 2 {N} = 16 supergravity.
\newblock 2006.

\bibitem{reddim}
C.~N. Pope.
\newblock {K}aluza-{K}lein Theory,
  $http://faculty.physics.tamu.edu/pope/ihplec.pdf$.

\bibitem{Ratcliffe}
J.~G. Ratcliffe.
\newblock Foundations of hyperbolic manifolds.
\newblock Springer-Verlag (1994).

\bibitem{Rendall:2001nx}
A.~D. Rendall and M.~Weaver.
\newblock Manufacture of {G}owdy spacetimes with spikes.
\newblock {\em Class. Quant. Grav.}, 18:2959--2976, 2001.

\bibitem{Romans:1985tz}
L.~J. Romans.
\newblock Massive {N}=2a supergravity in ten-dimensions.
\newblock {\em Phys. Lett.}, B169:374, 1986.

\bibitem{Ryan:1975jw}
M.~P. Ryan and L.~C. Shepley.
\newblock Homogeneous relativistic cosmologies.
\newblock Princeton, Usa: Univ. Pr. ( 1975) 320 P. ( Princeton Series In
  Physics).

\bibitem{S}
C.~Sa\c{c}lio\u{g}lu.
\newblock Dynkin diagrams for hyperbolic {K}ac-{M}oody algebras.
\newblock J.Phys. A\textbf{22}, 3753 (1989).

\bibitem{Sagnotti:1992qw}
A.~Sagnotti.
\newblock A note on the {G}reen-{S}chwarz mechanism in open string theories.
\newblock {\em Phys. Lett.}, B294:196--203, 1992.

\bibitem{Saha:2003xv}
B.~Saha and T.~Boyadjiev.
\newblock Bianchi type {I} cosmology with scalar and spinor fields.
\newblock {\em Phys. Rev.}, D69:124010, 2004.

\bibitem{Schnakenburg:2004vd}
I.~Schnakenburg and P.~West.
\newblock {K}ac-{M}oody symmetries of ten-dimensional non-maximal supergravity
  theories.
\newblock {\em JHEP}, 05:019, 2004.

\bibitem{Schnakenburg:2001ya}
I.~Schnakenburg and P.~C. West.
\newblock {K}ac-{M}oody symmetries of {IIB} supergravity.
\newblock {\em Phys. Lett.}, B517:421--428, 2001.

\bibitem{Schwarz:1983wa}
J.~H. Schwarz and P.~C. West.
\newblock Symmetries and transformations of chiral {N}=2 {D} = 10 supergravity.
\newblock {\em Phys. Lett.}, B126:301, 1983.

\bibitem{Sen:1994fa}
A.~Sen.
\newblock Strong - weak coupling duality in four-dimensional string theory.
\newblock {\em Int. J. Mod. Phys.}, A9:3707--3750, 1994.

\bibitem{Sen:1994wr}
A.~Sen.
\newblock Strong - weak coupling duality in three-dimensional string theory.
\newblock {\em Nucl. Phys.}, B434:179--209, 1995.

\bibitem{Sneddon}
G.~E. Sneddon.
\newblock J. Phys. A. : Math. Gen. 9 229 (1975).

\bibitem{Steinhardt:2001st}
P.J. Steinhardt and N.~Turok.
\newblock Cosmic evolution in a cyclic universe.
\newblock {\em Phys. Rev.}, D65:126003, 2002.

\bibitem{Teitelboim:1977fs}
C.~Teitelboim.
\newblock Supergravity and square roots of constraints.
\newblock {\em Phys. Rev. Lett.}, 38:1106--1110, 1977.

\bibitem{Veneziano:2003sz}
G.~Veneziano.
\newblock A model for the big bounce.
\newblock {\em JCAP}, 0403:004, 2004.

\bibitem{Vermaseren:2000nd}
J.~A.~M. Vermaseren.
\newblock New features of {FORM}.
\newblock 2000.

\bibitem{Wald:1984rg}
R.~M. Wald.
\newblock General relativity.
\newblock Chicago, Usa: Univ. Pr. ( 1984) 491p.

\bibitem{WZX}
Zhe-xian Wan.
\newblock Introduction to Kac-Moody algebra, World Scientific (1991).

\bibitem{Wesley:2005bd}
D.~H. Wesley, P.J. Steinhardt, and N.~Turok.
\newblock Controlling chaos through compactification in cosmological models
  with a collapsing phase.
\newblock {\em Phys. Rev.}, D72:063513, 2005.

\bibitem{West:2003fc}
P.~West.
\newblock E(11), {SL}(32) and central charges.
\newblock {\em Phys. Lett.}, B575:333--342, 2003.

\bibitem{West:2002jj}
P.~West.
\newblock Very extended {E}(8) and {A}(8) at low levels, gravity and
  supergravity.
\newblock {\em Class. Quant. Grav.}, 20:2393--2406, 2003.

\bibitem{West:2004iz}
P.~West.
\newblock Brane dynamics, central charges and {E}(11).
\newblock 2004.

\bibitem{West:2004kb}
P.~West.
\newblock E(11) origin of brane charges and {U}-duality multiplets.
\newblock {\em JHEP}, 08:052, 2004.

\bibitem{West:2004wk}
P.~West.
\newblock Some simple predictions from {E}(11) symmetry.
\newblock 2004.

\bibitem{West:2000ga}
P.~C. West.
\newblock Hidden superconformal symmetry in {M} theory.
\newblock {\em JHEP}, 08:007, 2000.

\bibitem{West:2001as}
P.~C. West.
\newblock E(11) and {M} theory.
\newblock {\em Class. Quant. Grav.}, 18:4443--4460, 2001.

\bibitem{Witten:1995ex}
E.~Witten.
\newblock String theory dynamics in various dimensions.
\newblock {\em Nucl. Phys.}, B443:85--126, 1995.

\end{thebibliography}
\end{document}